\documentclass[a4paper,11pt]{article}
\usepackage[T1]{fontenc}
\usepackage[utf8x]{inputenc} 
\usepackage{lmodern}
\usepackage{amsmath}
\usepackage{amsthm}
\usepackage{amssymb}
\usepackage{authblk}
\usepackage{graphicx}
\usepackage{enumitem}
\usepackage{xcolor}
\usepackage{dsfont}
\usepackage{mathrsfs}
\usepackage{hyperref} \hypersetup{colorlinks=true, citecolor=blue, linkcolor=black}
 \usepackage[capitalise,nameinlink]{cleveref}
\usepackage{crossreftools}

\usepackage{amsthm}
\usepackage{amssymb}
\usepackage{xparse}
\usepackage{thmtools}
\usepackage{stackrel}
\usepackage{bbold}
\usepackage{relsize}
\usepackage{tikz}

\usetikzlibrary{hobby}

\newcommand{\R}{\mathbb{R}}
\newcommand{\C}{\mathbb{C}}
\newcommand{\Z}{\mathbb{Z}}
\newcommand{\N}{\mathbb{N}}

\newcommand{\dd}{\mathrm{d}}

\renewcommand{\P}{\mathbb{P}}

\renewcommand{\O}{\mathcal{O}}

\newcommand{\JJ}{\mathrm{J}}

\DeclareMathOperator{\supp}{supp}

\DeclareMathOperator{\tr}{tr}

\DeclareMathOperator{\dist}{dist}

\makeatletter
\newtheoremstyle{indented}
{7pt} 
{7pt} 
{} 
{1.5em} 
{\bfseries} 
{.} 
{.5em} 
{} 

\theoremstyle{definition}

\newtheorem{defn}{Definition}[section]
\newtheorem{notation}{Notations}[section]

\theoremstyle{plain}
\newtheorem*{theorem*}{Theorem}

\newtheorem{theorem}{Theorem}

\newenvironment{preuve}{
	\noindent \textbf{Proof. }}{\hfill $\square$\medskip\par}

\newtheorem{prop}[defn]{Proposition}
\crefname{prop}{Proposition}{Propositions}
\newtheorem{corr}[defn]{Corollary}
\crefname{corr}{Corollary}{Corollaries}

\newtheorem{lem}[defn]{Lemma}
\crefname{lem}{Lemma}{Lemmas}
\newtheorem{conj}{Conjecture}
\crefname{conj}{Conjecture}{Conjectures}
\crefname{ax}{Axiom}{Axioms}

\theoremstyle{definition}
\newtheorem{rem}[defn]{Remark} 
\crefname{rem}{Remark}{Remarks}

\crefname{rems}{Remarks}{Remarks}

\makeatletter
\renewcommand*\env@matrix[1][*\c@MaxMatrixCols c]{%
  \hskip -\arraycolsep
  \let\@ifnextchar\new@ifnextchar
  \array{#1}}
\makeatother

\numberwithin{equation}{section}
\usepackage[normalem]{ulem}
\usepackage{todonotes}

\newcommand{\vertiii}[1]{{\left\vert\kern-0.2ex\left\vert\kern-0.2ex\left\vert #1 
        \right\vert\kern-0.2ex\right\vert\kern-0.2ex\right\vert}}

\usepackage{a4wide}
\date{\today}
\title{Universality for free fermions  and
  the local Weyl law for semiclassical Schrödinger operators}
\author{Alix Deleporte\thanks{alix.deleporte@universite-paris-saclay.fr}, Gaultier
  Lambert\thanks{glambert@kth.se}}
\affil{
Universit\'e Paris-Saclay, CNRS, Laboratoire de math\'ematiques
d'Orsay, 91405, Orsay, France\\Institut f\"ur Mathematik, Universit\"at Z\"urich,
  Winterthurerstrasse 190, CH-8057 Z\"urich}

\newcommand{\vide}[1]{}
\newcommand{\new}[1]{#1}
\newcommand{\old}[1]{}

\usepackage{scrtime}

\newcommand{\1}{\mathds{1}}

\newcommand{\Ai}{\mathrm{Ai}}
\renewcommand{\d}{\mathrm{d}}   

\newcommand{\E}{\mathbb{E}}

\renewcommand{\i}{\mathbf{i}}    
\newcommand{\I}{\mathrm{I}}
\newcommand{\J}{\mathrm{J}}
\newcommand{\K}{\widetilde{K}}             
\newcommand{\Ks}{\mathcal{K}}        
\renewcommand{\O}{\mathcal{O}}
\renewcommand{\P}{\mathbb{P}}
\newcommand{\rank}{\operatorname{Rank}}

\newcommand{\x}{\mathbf{x}}     
\newcommand{\X}{\mathrm{X}}

\newcommand{\var}{\operatorname{var}}

\newcommand\blfootnote[1]{%
  \begingroup
  \renewcommand\thefootnote{}\footnote{#1}%
  \addtocounter{footnote}{-1}%
  \endgroup
}

\begin{document}

\maketitle

\blfootnote{MSC2010: 35P20, 
  35Q40, 
  35S30, 
  60F05, 
  60G55, 
  81Q20, 
  82B10 
}

\begin{abstract}
We study local asymptotics for the spectral projector associated to a 
Schrödinger operator $-\hbar^2\Delta+V$ on $\R^n$ in the semiclassical
limit as $\hbar\to0$. We prove local uniform convergence of the rescaled integral kernel of this projector towards a universal model, inside the classically allowed region as well as on its boundary.  
 This implies universality of microscopic fluctuations for the corresponding free fermions (determinantal) point processes, both in the bulk and around regular boundary points. 
Our results apply for a general class of smooth potentials in arbitrary dimension $n\ge 1$. 
\new{These results are complemented by studying both macroscopic and
  mesoscopic fluctuations of the point process.
  We obtain tail bounds for macroscopic linear statistics and, provided
  $n\geq 2$,} a central limit theorem \old{which characterizes the}
\new{for both macroscopic and} mesoscopic \old{fluctuations of these
  Fermi gases} \new{linear statistics} in the bulk.
\end{abstract}

 \tableofcontents

\bigskip

\section{Introduction}
\label{sec:introduction}

Consider a semiclassical Schrödinger operator on $L^2(\R^n)$:
\[
  H_{\hbar}=-\hbar^2\Delta+V , 
\]
where $\Delta$ is the standard (negative) Laplacian, $\hbar>0$ plays the role of
the Planck constant, and
the potential $V:\R^n\to \R$ is locally integrable, bounded from below, and confining (that is,
$V(x)\to +\infty$ as $x\to\infty$). The operator $H_{\hbar}$ is then essentially
self-adjoint with domain ${\mathrm H}^2(\R^n)\cap \{u\in L^2,Vu\in
L^2\}$ and has compact resolvent: it admits a non-decreasing sequence $(\lambda_k)_{k\in
  \N}$ of eigenvalues tending to $+\infty$,
and an associated basis $(v_k)_{k\in \N}$ of
$L^2(\R^n)$ consisting of eigenvectors:
\[
H_{\hbar} v_k = \lambda_k v_k , \qquad \langle v_k,v_\ell\rangle_{L^2} =\delta_{k=\ell} ,  \qquad k,\ell\in\N .
\]
 Note that these eigenvalues and eigenfunctions depend on $\hbar$. We refer to
 the book \cite{helffer_spectral_2013} for
background on the spectral theory of Schrödinger operators. 
Using the eigenfunctions $(v_k)_{k\in \N}$, one can build a \emph{Slater
  determinant} with $N$ particles associated to $H_{\hbar}$, which is the following normalized element of
$L^2(\R^{n\times N})$:
\[
\Psi_N(\x) = \frac{1}{\sqrt{N!}}\det_{N\times N}\big[ v_k(x_j)
\big] , \qquad \x \in \R^{n\times N}.
\]
Physically, $\Psi_N$ represents \old{the ground state}\new{a zero
  temperature state}\footnote{\old{by appropriately choosing the
    particle's number $N$ to avoid \emph{degeneracies},
    cf.~\eqref{det}}\\\new{uniqueness of the zero temperature state is
    ensured when $\lambda_N<\lambda_{N+1}$, a condition which we will enforce throughout this article}.} of a system of $N$ non-interacting  spinless fermions subject to the Hamiltonian $H_{\hbar}$. Using the usual probabilistic
interpretation of quantum mechanics, to $\Psi_N$ we associate a
$N$-point process (a probability measure $\mathbb{P}_N$ on $\R^{n\times N}$) whose
density with respect to the Lebesgue measure is
\begin{equation} \label{slater}
  \mathbb{P}_N[\dd \x]=|\Psi_N(x)|^2=\frac{1}{N!}\Big|\det_{N\times
      N}\left[v_k(x_j)\right]\Big|^2.
      \end{equation}

The purpose of this article is to show that in a thermodynamical
limit where both $N\to +\infty$ and $\hbar\to 0$, the
properties of $\mathbb{P}_N$ are identical
for a large class of potentials $V$. We use methods from
semiclassical analysis, and the fact that $\mathbb{P}_N$ is
a \emph{determinantal point process}.

Realisations $\{x_j\}_{j=1}^N\in \R^{n\times N}$ of $\mathbb{P}_N$ are usually referred
  to as free fermions or non-interacting Fermi gases \cite{Macchi75}
  since the $N$ particles are only subject to the external potential $V$
  and the Pauli exclusion principle.  In the large $N$ limit, this
  induces non-trivial spatial correlations, named \emph{quantum fluctuations}.   
Using cold atoms in optical traps, it is now possible to
experimentally simulate such non-interacting Fermi gases in a general
potential and study these quantum fluctuations \cite{bloch_many-body_2008,giorgini_theory_2008}. 
This has led to a significant interest in the statistical physics
literature regarding the theoretical description and universality of these local
fluctuations. These predictions rely on standard methods from
many-body physics such as \emph{local density approximations}, and on
\emph{random matrix theory}. Indeed, for $V:x\mapsto x^2$ in dimension 1, $\mathbb{P}_N$  corresponds to the law of the eigenvalues of the Gaussian
Unitary Ensemble (GUE), the most studied model of random matrices.
We refer to the reviews \cite{Castin07, dean_noninteracting_2019} and
Section \ref{sec:related-work} for some background on these results.

Except in few specific cases, such as the harmonic oscillator, the eigensystem $(\lambda_k,v_k)_{k\in\N}$ of the Schrödinger operator $H_{\hbar}$ are neither explicit, nor given by induction formulas. 
\new{The main novelty of this paper is to apply semiclassical methods to study (non-interacting) Fermi gases for a general class of potential $V$, a problem which is open in the mathematical literature since \cite{Macchi75}.
This requires to extend, beyond the standard framework, both semiclassical estimates for the spectral projectors of Schrödinger operators and estimates for determinantal processes with general reproducing kernels.}
The greatest source of difficulty in this problem lies in the fact that the ground state of such free fermions are
\emph{gapless}: one has $\lambda_{N+1}-\lambda_N\to 0$ in the
asymptotic regime considered.
Hence, one can only hope to describe the state $\Psi_N$ up to $\O(\hbar)$.
In contrast, for gapped systems where $\lambda_{N+1}-\lambda_N$ is
bounded from below, one can use perturbation theory
to describe the state $\Psi_N$ up to any polynomial precision in $\hbar$. Such cases with recent activity are the Szeg\H{o} or Bergman projector
(see notably \cite{berman_determinantal_2013,berman_determinantal_2014,charles_entanglement_2018} for a fermionic
perspective). 

\subsection{Main results}
\label{sec:main-results}

\subsubsection{Setting and assumptions.}
The function $\Psi_N$ is non-ambiguously defined when
$\lambda_N<\lambda_{N+1}$. To enforce this condition, and to introduce
the asymptotic setting considered, we let $\mu\in \R$, \new{$\hbar>0$, $V\in
L^1_{\rm loc}$ with} \new{$V\stackrel[x\to \pm \infty]{}{\longrightarrow} +\infty$} and
\[
  \Pi_{\hbar,\mu}=\1_{(-\infty,\mu]}(H_{\hbar})
\]
be the spectral projector of $H_{\hbar}$ on the interval $(-\infty,\mu]$. 
\old{The energy $\mu$ is called the \emph{Fermi energy} of the system.}
We also denote by $  \Pi_{\hbar,\mu}$ the integral kernel of the previous
operator:
\[
  \Pi_{\hbar,\mu}:(x,y)\mapsto \sum_{\lambda_k\leq \mu} v_k(x) v_k(y).
\]
Then, letting $N=\mathrm{Rank}(\Pi_{\hbar,\mu})$, the probability density associated with the state $\Psi_N$ in \eqref{slater} is:
\begin{equation} \label{det}
\P_N[\d\x]  = \frac{1}{N!}\det_{N\times N}\big[  \Pi_{\hbar,\mu}(x_i,x_j) \big].
\end{equation}
\new{A precise physical description of  $\P_N[\d\x]$ is the joint probability of positions
  of a system of non-interacting fermions, subject to the one-body
  Hamiltonian $H_{\hbar}$, connected to a reservoir with chemical
  potential $\mu$, and at zero temperature equilibrium. The parameters
  we tune are $\mu$ and $\hbar$, then $N$ is determined by $N=\mathrm{Rank}(\Pi_{\hbar,\mu})=\max(j\in \N, \lambda_j\leq
    \mu)$ so that automatically $\lambda_N<\lambda_{N+1}$.}

Since $\Pi_{\hbar,\mu}$ is an orthogonal projection, the random measure $\X : =\sum_{j=1}^N \boldsymbol\delta_{x_j}$, where the configuration $\{x_j\}_{j=1}^N$ follows $\mathbb{P}_N$, is a determinantal point process
with kernel $\Pi_{\hbar,\mu}$.
This means that all the correlation functions of $\X$ are given by determinants of the kernel $\Pi_{\hbar,\mu}$; we refer to Appendix \ref{sec:dpp} for background on determinantal point processes.

Let us now specify the hypothesis on $V$ that we use throughout
this article.

\begin{defn}~\begin{enumerate}[beginpenalty=10000]
\item[{\crtcrossreflabel{(H)}[hyp:weak]}] \new{A couple $(\mu,V)\in \R \times L^1_{\rm loc}(\R^n,\R)$ satisfies \ref{hyp:weak} if there exists $M\in\R$ with $\min V<\mu< M $ such that $\{V\leq M\}$ is compact and $V\in C^{\infty}(\{V< M\},\R)$.}
  \end{enumerate}
\end{defn}


The scope of this article is the following asymptotic regime: \new{$(\mu,V)$
satisfies \ref{hyp:weak} are fixed, and $\hbar\to 0$}. 
\old{This normalization guarantees that} \new{In this regime, one has
  in particular $N<+\infty$ for every fixed $\hbar>0$, and}
$N\to+\infty$ \new{as $\hbar\to 0$}. In fact, \new{if $(\mu,V)$ satisfy \ref{hyp:weak}}, one can prove Weyl's law:
\begin{equation} \label{Z}
\lim_{\hbar\to0} \hbar^n N = \frac{\omega_nZ}{(2\pi)^n} , \qquad Z := \int_{\R^n} \big(\mu- V(x)\big)_+^{n/2} \d x, 
\end{equation}
where $\omega_n:=\frac{\pi^{n/2}}{\Gamma(1+n/2)}$ denotes the volume
of the unit Euclidean ball in $\R^n$ and $(y)_+ = \max\{0,y\}$ for
$y\in\R$; see Proposition \ref{lem:Weyl_law}.

\subsubsection{Pointwise Weyl law and macroscopic fluctuations}
A stronger form of \eqref{Z} is that the density of
states (proportion of particles by unit volume) is given by 
\begin{equation} \label{dos}
  \varrho(x) : = Z^{-1}\big(\mu-V(x)\big)_+^{n/2} , \qquad x\in\R^n .
\end{equation}
Our first result is a probabilistic interpretation of this \emph{pointwise Weyl law}.
Let us denote by $\d_{W}$ the Kantorovich or Wasserstein$_1$ distance on the space on probability measures on $\R^n$:
\[
\d_W(\nu_1,\nu_2) = \sup \left\{  \int f \d(\nu_1-\nu_2) : f \in \mathrm{Lip}_1(\R^n) \right\} . 
\]
This provides a natural metric on the space of probability
measures which is stronger than weak convergence.
Recall that $\X$ denotes the determinantal point process on $\R^n$ associated with the operator
  $\Pi_{\hbar,\mu}=\1_{-\hbar^2\Delta+V\leq \mu}$ and that $N =
  \tr(\Pi_{\hbar,\mu})$ is the particle number.

  \renewcommand\thetheorem{I.1}
  \begin{theorem} \label{thm:LLN}
Let $(\mu,V)$ satisfy \ref{hyp:weak}. There
exists a constant $c>0$ so that for any $\epsilon\in (0,1]$ \new{and
  for any $\hbar\in (0,1]$,}
 \[
 \P_N\left[ \d_{W}(N^{-1}\X,\varrho) \ge  \epsilon \right] \le C_\epsilon \exp(- c N \epsilon^2) . 
 \]  
\end{theorem}

Theorem~\ref{thm:LLN}  is a law of large numbers for the random probability measure 
$N^{-1}\X$, with an exponential rate of convergence.

The set $\{V\le \mu\}$ which supports $\varrho$ is called the \emph{droplet}. The droplet consists in a
\emph{bulk} $\{V<\mu\}$ and an \emph{edge} $\{V=\mu\}$. The real
random variables $\X(f) = \sum_{j=1}^N f(x_j)$, for
$f:\R^n\to \R$, are called \emph{linear statistics}.

Our next theorems describe the fluctuations of linear statistics in the
bulk. They exhibit Gaussian-like tails and, if $n\geq 2$,
converge in distribution to a Gaussian, as a form of Central Limit
Theorem (CLT).

\renewcommand\thetheorem{I.2}
\begin{theorem} \label{thm:gtb}
Let $(\mu,V)$ satisfy \ref{hyp:weak} and let
$f\in C^\infty_c(\{V<\mu\},\R)$. There exists $c>0, \epsilon >0$ so
that for all \new{$\hbar\in (0,1]$ and all} $t \le \epsilon \sqrt{\hbar N}$, 
\[
\P_N\left[  |\X(f) - \E\X(f)| \ge \sqrt{\hbar N} t \right] \le 2 e^{-c t^2} . 
\]
\end{theorem}

This shows that the particles' fluctuations are much smaller than in
the independent case, a phenomenon related to rigidity and
hyperuniformity \cite{torquato_hyperuniformity_2016} of the particles.
For free fermions, this rigidity is entirely due to the ``repulsion'' coming from the exclusion principle.

We expect (Conjecture \ref{conj:var_macro} \new{below}) that the variance of
$\X(f)$ is always of order $N \hbar$ for a sufficiently smooth function $f$, so that Theorem~\ref{thm:gtb} captures the typical size of the fluctuations.
In particular, in dimension $2$ or more, the variance of $\X(f)$ diverges, and it is
a general feature of determinantal point processes (see Corollary
\ref{clt_dpp}) that this implies a CLT.

\renewcommand\thetheorem{I.3}
\begin{theorem} \label{thm:clt}
Let $n\ge 2$, let $(\mu,V)$ satisfy \ref{hyp:weak} and let $f \in C(\R^n,\R)$ with at most exponential growth such that $f \in H^1(\Omega)$ on an open set $\Omega \subset \{V \le \mu\}$ with $\displaystyle \int_{\Omega}  |\nabla f|^2 >0$. 
It holds in distribution as $\hbar\to0$, or equivalently $N\to\infty$, 
\[
\frac{\X(f) - \E \X(f)}{\sqrt{\operatorname{var} \X(f)}} \, \Rightarrow\,  \mathcal{N}_{0,1} . 
\]
\end{theorem}

Even if $f \in C^\infty_c(\{V<\mu\},\R)$, it is not clear that $\var\X(f)$ does not
oscillate because of Remark \ref{rem:remainder_weyl} below, \new{and
  its behaviour is expected to depend strongly on properties of the
  Newtonian dynamics associated with the potential $V$}. The proofs of Theorems \ref{thm:gtb} and \ref{thm:clt} respectively rely
on an upper bound and lower bound for this variance, but these bounds
differ by a factor $\hbar$ (see \eqref{eq:bound_inf_variance} and
\eqref{est_var_macro}): we are able to show that
\[
  c\hbar^{2-n}\leq \operatorname{var} \X(f)=-\tfrac 12
  \tr([\Pi_{\hbar},f]^2)\leq C\hbar^{1-n}.
  \]
This prompts the following conjecture.
\begin{conj}\label{conj:var_macro}
  Let \new{$n\ge 2$ and} $f\in
  C^{\infty}_c(\R^n,\R)$ be nonzero. There
exists $0<c<C$ such that
\[
  c\hbar^{1-n}\leq \operatorname{var} \X(f)=-\tfrac 12 \tr([\Pi_{\hbar},f]^2)\leq C\hbar^{1-n}.
\]
\end{conj}
The constants $c$ and $C$ presumably correspond to (weighted) $H^{\frac 12}$
Sobolev norms, as in the case of the free Laplacian
as in the mesoscopic CLT (Theorem \ref{thr:mesoscopic_bulk_pot}
below). \new{In the physics literature, certain examples of counting statistics (non-smooth test functions) have been considered in} \cite{Smith_21}. 

\new{In dimension $n=1$, we subsequently showed in} \cite{DL23} \new{that under generic conditions on the potential $V$,  for $f\in C^\infty(\R,\R)$ with at most polynomial growth, without any normalization, the statistic $\X(f)$ obeys a CLT as $\hbar\to 0$ and the variance converges to a weighted $H^{1/2}$ square-norm.}

\subsubsection{Universal limit at microscopic scales}

The method of proof of Theorems \ref{thm:LLN}, \ref{thm:gtb} and
\ref{thm:clt} relies on the asymptotics of the integral kernel of
$\Pi_{\hbar}=\1_{(-\infty,\mu]}(H_{\hbar})$ near the diagonal. At the
\emph{microscopic scale}, that is, when zooming to the typical distance between
particles, this kernel converges to a universal limit, which does not
depend on $V$ but only on the dimension $n$ and on whether we are in
the bulk $\{V<\mu\}$ or at the edge $\{V=\mu\}$ of the droplet. This
universal limit  implies convergence in distribution of the point
process at this microscopic scale.

  Given $x_0\in \R^n$ and $\mathcal{U} \in SO_n$, we define the rescaled kernel obtained by zooming around $x_0$ at scale $\epsilon$ as
  \begin{equation}\label{zoom}
  K_{x_0,\epsilon} : (x,y)\mapsto  \epsilon^{n}  \Pi_{\hbar,\mu}(x_0 + \epsilon\, \mathcal{U}^*  x , x_0+\epsilon\, \mathcal{U}^* y) . 
\end{equation}
This is the  kernel of the determinantal point process
\begin{equation}\label{zoom_pp}
  T_{x_0,\epsilon}^* \X = \sum_{j=1}^N
  \boldsymbol\delta_{\epsilon^{-1}\mathcal{U}(x_j-x_0)}.
\end{equation}

\renewcommand\thetheorem{II.1}
 \begin{theorem}\label{thr:micro_limit_kernel}
Let $(\mu,V)$ satisfy \ref{hyp:weak} \new{and $\hbar\in (0,1]$}. \old{For} \new{Let}
$x_0\in \{V<\mu\}$, \new{ let $\epsilon := 2\pi\hbar
  \frac{\omega_n^{-1/n}}{\sqrt{\mu-V(x_0)}}$, let $\mathcal{U}\in
  SO_n$} \old{define}\new{and set} $ K_{x_0,\epsilon}$ as in \eqref{zoom} \old{with
$\epsilon := 2\pi\hbar \frac{\omega_n^{-1/n}}{\sqrt{\mu-V(x_0)}}$ and
$\mathcal{U}\in SO_n$ arbitrary. }
For any compact sets $\mathcal{A} \Subset \{V<\mu\}$ and  $\mathcal{K} \Subset \R^{2n}$, there exists a constant $C>0$ so that
\[
\sup_{x_0 \in \mathcal{A}}    \sup_{(x,y)\in \mathcal{K}}  \big| 
K_{x_0,\epsilon}(x,y) -     K_{\rm bulk}(x,y) \big| \le C\hbar 
\]
where the bulk kernel \new{is $K_{\rm
  bulk}:=\1_{(-\infty,4\pi^2\omega_n^{-\frac 2n}]}(-\Delta)$;} \old{is given by} \new{see
  the explicit formula \eqref{eq:Kbulk} below}. 
\end{theorem}

\begin{rem}\label{rem:remainder_weyl}
This result is optimal in the sense that one cannot in general obtain asymptotics for $K_{x_0,\epsilon}$ beyond $\O(\hbar)$. There is no asymptotic expansion in powers of
$\hbar$ and the error is known to oscillate.
However, our methods can be used to obtain a stronger mode of
convergence: using the ellipticity of the operator, it holds for any multi-indices $\alpha,\beta \in \N_0^n$, 
  \[
   \sup_{x_0 \in \mathcal{A}} \sup_{(x,y)\in \mathcal{K}} |\partial_x^{\alpha}\partial_{y}^{\beta}K_{x_0,\epsilon}(x,y)-\partial_x^{\alpha}\partial_y^{\beta}K_{\rm bulk}(x,y)|=\O(\hbar).
  \]
 Only local $C^0$ convergence is required for our applications to determinantal point processes. 
\end{rem}

\renewcommand\thetheorem{II.2}
 \begin{theorem}\label{thr:micro_edge_kernel}
Let $(\mu,V)$ satisfy \ref{hyp:weak}\new{ and $\hbar\in (0,1]$}. \old{For} \new{Let} $x_0\in \{V =\mu\}\cap \{\nabla V
\neq 0\}$, \new{let} \old{define $ K_{x_0,\epsilon}$ as in
  \eqref{zoom}, with} $\epsilon=  \hbar^{\frac23}|\nabla V(x_0)
|^{-\frac13}$ and  $\mathcal{U} \in SO_n$ such that
$\mathcal{U}(\nabla V(x_0))=|\nabla V(x_0)| \mathrm{e}_1$. \new{Let
  $K_{x_0,\epsilon}$ be as in \eqref{zoom}.}
For any compact sets $\mathcal{A} \Subset \{V=\mu ; \nabla V\neq  0\}$ and  $\mathcal{K} \Subset \R^{2n}$, there exists a constant $C>0$ so that
\[
\sup_{x_0 \in \mathcal{A}}    \sup_{(x,y)\in \mathcal{K}}  \big| 
K_{x_0,\epsilon}(x,y) - K_{\rm edge}(x,y) \big| \le C\hbar^{1/3} 
\]
where the edge kernel \new{ is $K_{\rm
    edge}:=\1_{(-\infty,0]}(-\Delta+x_1)$;} \old{is given by}
\new{see the explicit formula }\eqref{eq:Kedge} below. 
\end{theorem}

Theorems \ref{thr:micro_limit_kernel} and \ref{thr:micro_edge_kernel}
directly imply universality of the point process obtained by zooming at microscopic scales either in the bulk or at the edge of the droplet (see Proposition \ref{prop:wcvg}).

\renewcommand\thetheorem{II.3}
\begin{theorem}\label{prop:micro_limit_process}
  Let $(\mu,V)$ satisfy \ref{hyp:weak}.
  \begin{enumerate}
  \item If $V(x_0)<\mu$ and $\epsilon = 2\pi\hbar
    \frac{\omega_n^{-1/n}}{\sqrt{\mu-V(x_0)}}$,  then for any
    $\mathcal{U} \in SO_n$,  the point process $T_{x_0,\epsilon}^*\X$
    given by \eqref{zoom_pp} converges in distribution as
    $\hbar \to 0$ to the determinantal point process associated with $K_{\rm
      bulk}$.
  \item If  $V(x_0)=\mu$, $\nabla V(x_0)\neq 0$ and $\epsilon=
    \hbar^{\frac23}|\nabla V(x_0) |^{\frac13}$, then if
    $\mathcal{U}(\nabla V(x_0))=|\nabla V(x_0)| \mathrm{e}_1$, the
    point  process $T_{x_0,\epsilon}^*\X$ given by \eqref{zoom_pp} converges in    distribution, as $\hbar \to 0$, to the determinantal point process associated  with $K_{\rm edge}$.
  \end{enumerate}
\end{theorem}

Let us now explain the heuristics behind these results and define the limit objects $K_{\rm bulk}$ and $K_{\rm edge}$.
Following Theorem \ref{thm:LLN} and \eqref{Z}, the typical inter-particle distance around a point $x_0$
in the bulk is approximately $\epsilon= \hbar
  c_n(Z\rho(x_0))^{-\frac 1n}$, where $c_n=2\pi\omega_n^{-\frac 1n}$.
\old{At this scale} \new{After zooming at scale $\epsilon$}, the Schrödinger operator has close to
constant potential:
\[
  \hbar^2\epsilon^{-2}\Delta + V(x_0+\epsilon x)-\mu \simeq
    c_n^2(Z\rho(x_0))^{\frac 2n}(-\Delta+c_n^{-2}).
\]

Hence, the natural candidate for the scaling limit is   \begin{equation}\label{eq:Kbulk}
    K_{\rm bulk}=\1_{(-\infty,c_n^{-2}]}(-\Delta):(x,y)\mapsto \frac{\J_{\frac
        n2}(c_n^{-1}|x-y|)}{\sqrt{\omega_n} |x-y|^{\frac n2}}
  \end{equation}
  where $(\J_{\nu})_{\nu\in \frac{\N}{2}}$ are Bessel functions of the first kind, cf.~\eqref{Picomputation}. The normalization is such that the determinantal point process with kernel $K_{\rm bulk}$ is translation and rotation invariant on $\R^n$ with density 1.

At the boundary of the droplet, the density of states $\varrho$
vanishes. Assuming that $\vec{\gamma} : = \nabla V(x_0) \neq
0$, if we zoom at scale $\epsilon$ and
apply an orthogonal matrix $\mathcal{U}$,
we obtain the approximation

\[
  -\epsilon^{-2}\hbar^2\Delta+V(x_0+\epsilon \mathcal{U}^*x)-\mu\simeq
  \epsilon^{-2}\hbar^2(-\Delta+x_1)
\]
 provided
 \[
   \epsilon=\hbar^{\frac 23}|\nabla V(x_0)|^{-\frac 13}\qquad \qquad
   \mathcal{U}V(x_0)=\|\nabla V(x_0)\|e_1.
 \]
Hence, at the edge the typical inter-particle distance is much larger
than in the bulk and given by $\hbar^{\frac23}|\nabla
V(x_0) |^{-\frac13}$. Then, up to the rotation $\mathcal{U}$, the
natural candidate for the scaling limit is now 
  \begin{equation}\label{eq:Kedge}
    K_{\rm edge}=\1_{(-\infty,0]}(-\Delta+x_1):(x,y)\mapsto
    \int_0^{+\infty}\Ai(x_1+s) \Ai(y_1+s)
    \frac{\J_{\frac {n-1}2}(\sqrt{s}| x^\perp- y^\perp|)}{\big(2\pi |x^\perp- y^\perp|\big)^{\frac{n-1}{2}}} s^{\frac{n-1}{2}} \d s,
  \end{equation}
  where $\Ai$ denotes the standard \emph{Airy function} (cf.~the Appendix~\ref{sec:airy-type-integrals}) and if $n\ge 2$, we decompose $x=(x_1,x^\perp)\in \R\times \R^{n-1}$ and similarly for $y$.
  
The bulk and edge point processes from Theorem~\ref{prop:micro_limit_process} are the natural generalizations of the Sine and Airy point processes in higher dimension $n \ge 2$. 
 The relevance of the determinantal point processes associated with $K_{\rm bulk}$ has first been reckognized in \cite{torquato_point_2008, Torquato_09} under the name ``Fermi-sphere'' processes.
The edge processes associated with $K_{\rm edge}$ were first obtained
in \cite{dean_universal_2015,dean_noninteracting_2016}, and the kernel
asymptotics were proven in the case of the multidimensional harmonic
oscillator in \cite{hanin_scaling_2017}.

\subsubsection{Gaussian field at mesoscopic scales} One can interpolate
between Theorems \ref{thm:LLN}, \ref{thm:gtb} and \ref{thm:clt}, on one hand,
and Theorem \ref{prop:micro_limit_process}, on the other hand, by
considering mesoscopic scales, that is, the behaviour of the rescaled
point process $T^*_{\epsilon,x_0}\X$ for $\hbar\ll \epsilon \ll 1$. In
this case, we obtain convergence of the linear statistics to a Gaussian field with
$H^{\frac 12}$ covariance.

\renewcommand\thetheorem{III.1}
\begin{theorem}\label{thr:mesoscopic_bulk_pot}
Suppose that $n\ge 2$, let $(\mu,V)$ satisfy \ref{hyp:weak} and let  $x_0\in \{V<\mu\}$.
We consider a mesoscopic scale $\epsilon:[0,1]\to [0,1]$ such that
$\hbar^{1-\beta}\leq \epsilon(\hbar)\leq \hbar^{\beta}$ for some $\beta\in(0,1)$.
 Let \[\delta(\hbar) = \frac{\hbar}{\epsilon(\hbar)}
   \frac{1}{\sqrt{\mu-V(x_0)}}\qquad \qquad \text{and}\qquad \qquad
 \sigma_n^2 =  \frac{\omega_{n-1}} {(2\pi)^n}.\]
We define the random distribution
   \[
\X_{\hbar, \epsilon}  : =  \sigma_n^{-1} \delta(\hbar)^{\frac{n-1}{2}} \big( T_{x_0,\epsilon}^*\X  - \E [T_{x_0,\epsilon}^*\X]  \big) . 
   \]
 Then for any $g\in C^{\infty}_c(\R^n,\R)$, it holds    
    \[
   \lim_{\hbar\to0} \E\big[ e^{\X_{\hbar, \epsilon}(g) } \big]  = e^{-\Sigma^2(g)/2} 
   \qquad\text{where}\quad \Sigma^2(g) =  \int_{\R^n} |\widehat{g}(\xi)|^2
|\xi| \d\xi.
    \]
In particular, $\X_{\hbar, \epsilon}$ converges in the sense of finite-dimensional distributions as $\hbar\to0$ to a (centred) Gaussian field $\mathrm{G}$ on $\R^n$ with correlation kernel
\[
\E \mathrm{G}(f) \mathrm{G}(g) =  \int \widehat{f}(\xi) \overline{\widehat{g}(\xi)} |\xi| \d\xi , \qquad f,g \in H^{\frac 12}(\R^n).
\]
\end{theorem}

The variance $\Sigma^2$ is the square of the $H^{\frac 12}$ seminorm, and it enjoys
the following scaling invariance properties:
$\Sigma^2(T_{\epsilon,x_0}f) = \epsilon^{\frac{1-n}{2}} \Sigma^2(f)$ where $T_{\epsilon,x_0}f = f(x_0+\epsilon\, \mathcal{U} \cdot)$ for $x_0 \in \R^n$, $\epsilon>0$,  $\mathcal{U} \in SO_n$ and $f\in H^{\frac 12}(\R^n)$. 
In particular, the Gaussian field $\mathrm{G}$ has the following  invariance properties:
$\epsilon^{\frac{n-1}{2}}T_{\epsilon,x_0}^*G \overset{\rm law}{=} G$.

\medskip

The proof of Theorem~\ref{thr:mesoscopic_bulk_pot} relies on
Corollary \ref{prop:wcvg} which is valid for general
determinantal processes, and it boils down again to obtaining the
asymptotics of the variance
\[
\operatorname{var} \X_{\hbar, \epsilon}(f) = - \frac12  \sigma_n^{-2} \delta(\hbar)^{n-1}  \tr\big( [K_{x_0,\epsilon}, f]^2 \big)  .
\]

These asymptotics are involved and cannot be directly deduced from
Theorem~\ref{thr:micro_limit_kernel}; nor can
the upper bound on the macroscopic variance leading to Theorem
\ref{thm:gtb}. However, all our results are derived using the same method based on a (semi-classical)
expression for the quantum propagator $e^{\i t \frac{H_{\hbar}}{\hbar}}$.

Again, the situation in dimension 1 is special as $\operatorname{var} T_{\epsilon,x_0}^*\X(f)$ is bounded for a smooth function $f$. 
Nonetheless, by analogy with the known case of the harmonic oscillator
$V:x\mapsto x^2$ (cf.~\cite{Lambert_17}), we
expect that Theorem~\ref{thr:mesoscopic_bulk_pot} holds in full
generality.

\begin{rem}
In the formulation of Theorem~\ref{thr:mesoscopic_bulk_pot}, we have
exactly centred the random variable $\X_{\hbar, \eta}$, but its
expectation $ \E [T_{\epsilon,x_0}^*\X]$ admits a simple
equivalent. Indeed, if $\epsilon_0 =  c_n\hbar/\sqrt{\mu-V(x_0)}$ with $c_n$ as in \eqref{eq:Kbulk} denotes the microscopic scale at $x_0$, using \eqref{zoom} with $\mathcal{U} =\mathrm{I}$, we can rewrite for $f\in C_c(\R^n,\R)$, 
\[
\E [T_{\epsilon,x_0}^*\X]  = c_n^{-n} \delta(\hbar)^{-n} \int f(x) K_{\epsilon_0, x_0+ \epsilon x}(0,0) \d x  .   
\]
As a consequence of the uniformity of Theorem~\ref{thr:micro_limit_kernel} with respect to $x_0$, we obtain
\[
\E [T_{\epsilon,x_0}^*\X] = c_n^{-n} \delta(\hbar)^{-n}  \int f(x)  \d x + \O(\eta^{-n} \hbar).
\]
In general, we can only replace $\E [T_{\epsilon,x_0}^*\X]$ by its
leading contribution in Theorem \ref{thr:mesoscopic_bulk_pot} when the scale is small
enough, that is, if $\epsilon(\hbar) \ll \hbar^{\frac{n-1}{n+1}}$. 
\end{rem}

\subsection{Discussion and related works}\label{sec:related-work}

\subsubsection{Hermitian random matrices}
The free fermions point processes studied in this article are basic examples of quantum gases which are exactly solvable due to their determinantal structure\footnote{In a physical context, this \emph{integrable structure} of the $N$-particle density function $\P_N$ arises from applying Wick's Theorem; see e.g.~\cite{Lytvynov02}.}. 
In fact, they were introduced by Macchi \cite{Macchi75} as the first instances of determinantal point processes. 
So far, there has been only few rigorous progress on the statistical properties of free fermions, with the exception of the one-dimensional harmonic oscillator which corresponds  to the eigenvalues of the Gaussian unitary ensemble \cite[Section~2]{Soshnikov_01}. 
The GUE is a central model in random matrix which has been extensively
studied and Theorems~\ref{thm:LLN},
\ref{thr:micro_limit_kernel}, \ref{thr:micro_edge_kernel},
\ref{prop:micro_limit_process} and \ref{thr:mesoscopic_bulk_pot} are
well-known in this context, e.g.~\cite{Forrester10}. 
In particular, in dimension $n=1$, the microscopic limits
  $K_{\rm bulk}$ and $K_{\rm edge}$ from \eqref{eq:Kbulk} and \eqref{eq:Kedge} are the celebrated Sine and Airy kernels
\[
    K_{\rm bulk}(x,y) = \frac{\sin(\pi|x-y|)}{\pi|x-y|} , \qquad      K_{\rm edge}(x,y) = \int_0^{+\infty}\Ai(x+s) \Ai(y+s) \d s = \frac{\Ai(x)\Ai'(y)- \Ai'(x)\Ai(y)}{x-y} . 
\]
 The corresponding point process arise in a range of other contexts (as originally surmised by Wigner) and  have been extensively studied. 
 In particular, they  exhibit well-known integrable structures, related  for instance to the Tracy-Widom distribution and the Kardar-Parisi-Zhang equation, e.g.~\cite{dean_finite-temperature_2015}. 
 
One-dimensional free fermions and the eigenvalues of Hermitian random matrices fall in the same universality class and there is a substantial body of works on the fluctuations of random matrices. 
The closest context being that of unitary-invariant ensembles which are also determinantal processes whose correlation kernels are expressed in terms of orthogonal polynomials \cite{Deift99}. 
Universal asymptotics for these \emph{Christoffel--Darboux} kernels are known under very general assumptions and can be obtained by several different methods; cf.~the surveys \cite{Lubinsky09,EY12}. 
In fact, there are exact mappings between the ground state of one-dimensional  free fermions trapped by specific potentials with $\hbar= N^{-1}$ and the three classical unitary-invariant ensembles. The following table collects the weight $\prod_{j=1}^Nw_N(\lambda_j)$ of the random matrix ensemble together with the corresponding change of variables and Schr\"odinger eigensystem;
{\small
\vspace{-.3cm}
\begin{center}
\begin{tabular}{|c|c| c| c| }
\hline
& Hermite& Laguerre & Jacobi\\ \hline
$w_N(\lambda)$
&$e^{-\lambda^2}$ 
&  $\lambda^{N\alpha}e^{- \lambda}\1_{\lambda>0} ; \, \alpha\ge 0$ 
& $(1-\lambda)^{N\alpha}(1+\lambda)^{N\beta} \1_{\lambda\in(-1,1)} ; \, \alpha,\beta \ge 0 $  \\  \hline
Mapping
& $ x_j = \lambda_j N^{-1/2}$ 
& $x_j =\sqrt{\lambda_j N^{-1}} ; \qquad x_j\in\R_+$  
& $x_j = \arccos(\lambda_j) ; \qquad x_j\in [0,\pi]$\\ \hline
Eigensystem
& $(-\frac{1}{N^2}\Delta+ x^2) H_{N,n}(x)$ 
& $(-\frac{1}{N^2}\Delta+ x^2 + (\alpha^2-\frac{1/4}{N^2})x^{-2}) $ 
& $(-\frac{1}{N^2}\Delta+ \frac{(\alpha \cos \frac x2)^2+ (\beta \sin \frac x2)^2-\frac{1/4}{N^2}}{(\sin x)^2})$    \\
&$ = 2\frac nN H_{N,n}(x) $ 
& $\quad L_{N,n}(x) = 2(\frac{2n+1}N+\alpha) L_{N,n}(x) $
& $ \quad J_{N,n}(x) =  (\frac{n+1/2}{N} + \alpha+\beta )^2 J_{N,n}(x)$  \\
\hline
\end{tabular}
\end{center}
}
There is also a correspondance between free fermions  on the unit circle and the circular unitary ensembles \cite{CMO18}. One can realize Gaussian/circular $\beta$-ensembles for any $\beta>0$ by considering the ground state of certain interacting Fermi gases known as Calogero-Sutherland systems \cite{Forrester98,Smith_21}
(these are the only cases where such exact mappings exist). 
This suggests that more generally, $\beta$-ensembles fall in the universality class of certain interacting 1-dimensional Fermi gases. 

Let us also mention that there is an extensive body of work on \emph{rigidity} and CLT for eigenvalues of random matrices.
For instance the counterpart of Theorem~\ref{thr:mesoscopic_bulk_pot} in dimension 1 is well-known for unitary-invariant ensembles ($\beta=2$) and general $\beta$-ensembles \cite{Lambert_17,Bekerman-Lodhia-18}. 

\subsubsection{Free fermions on $\R^n$}
Recently, there has been a significant activity in theoretical physics on the statistical properties of Fermi gases in general dimensions. 
Let us emphasize again the results from \cite{dean_universal_2015,dean_noninteracting_2016,dean_wigner_2018} which \old{has}\new{have} been of inspiration for our work. 
By explicit asymptotic calculations based on short-time expansion of the quantum propagator (this approach is similar to ours, albeit non rigorous), 
they establish universality of microscopic fluctuations for free Fermi gases in arbitrary dimension at the bulk and (regular) edge points. 
Then,  \cite{LLMS18, Smith_20} explore other \emph{universal local behavior} around a singularity of the external potential, respectively around an interior point where the density of state vanishes. 
We refer to the review \cite{dean_noninteracting_2019} for applications of these results, some further perspectives on the connection with random matrices  and a discussion of the positive temperature regime. 
The article \cite{dean_impuriies_2020} studies the influence of impurities (modelled by delta function potentials) inside a Fermi gas. 
It is  worth mentioning  \cite{Gouraud_hole_2021} on large hole probabilities for Fermi gases confined by rotation-invariant potentials. 
Finally, there are also explicit results on number variance fluctuations, cf.~\cite{Smith_21} and the \emph{perspectives} below.

\subsubsection{Other universality classes for \emph{fermionic systems}}
 In general dimension, a (classical) statistical mechanics model to compare our results are Coulomb gases, also known as one-component plasma or Jellium. 
 They arise in the description of several different physical phenomena, such as superconductivity, the (fractional) quantum Hall effect or as the eigenvalue of random normal matrices \cite{Serfaty21}.
 The prototypical model in this class is the Ginibre ensemble which is also determinantal \cite{ameur_fluctuations_2011}. 
 In fact, the (infinite) Ginibre point process can also be interpreted as the ground state of a (free) quantum system subject to a strong magnetic Laplacian which forces the particles to lie in the lowest Landau level \cite{For98,LLR}.

  Such fermionic systems can be generalized on $\C^n$ or over an integrable compact Kähler manifold for any $n\ge1$. 
In contrast to the Euclidean case considered here, the kernel of such determinantal point processes associated with a magnetic Laplacian is gapped. Hence, they fall in a different universality class. Notably, these Bargmann kernels (linked with Berezin-Toeplitz quantization associated with the Kähler structure) have semiclassical expansion up to $\O(\hbar^\infty)$ which renders asymptotics possible using methods from perturbation theory.
This has been used to study scaling limits in the bulk   \cite{berman_determinantal_2013,berman_determinantal_2014}
and at the edge in the orthogonal polynomial case \cite{hedenmalm_off-spectral_2020}. 
From a probabilistic perspective, another significant difference is that these Ginibre-like process exhibit exponential decay of (spatial) correlations. 
Concretely, to compare with the Ginibre kernel $|K_{\rm Gin}(x-y)| = e^{-|x-y|^2/2}$, the universal limit 
$|K_{\rm bulk}(x,y)|^2 = \frac{\J_1^2(2\sqrt{\pi}|x-y|)}{\pi|x-y|^2}$  decays like $|x-y|^{-3}$ in 2d as $|x-y| \to\infty$. 
For Coulomb gases, besides the long-range interactions, these (super)-exponential decay of correlations is a consequence of a \emph{screening} phenomenon. 

Besides in the determinantal case, the description of the thermodynamical limit of Coulomb gases on $\R^n$ for $n\ge 2$ is still a major open problem \cite{Serfaty21}, though recent progress have been achieved \cite{leble_large_2017,AS21}. 
For two-dimensional Coulomb gases, mesoscopic and macroscopic fluctuations have been studied in \cite{bauerschmidt_local_2017,leble_fluctuations_2018,BBNY19}.
In particular, a CLT for smooth linear statistics holds under general conditions and the limit can be described in terms of the Gaussian free field ($H^1$-noise, in contrast to the $H^{\frac 12}$ noise which arises from Theorem~\ref{thr:mesoscopic_bulk_pot}). We also refer to \cite{rider_noise_2007,ameur_random_2015} for proofs in the Ginibre ensemble and normal matrix models and further on the relationship to the GFF. 
In dimension $n\ge3$, there has been progress in the Hierarchical case \cite{chatterjee_rigidity_2019,ganguly_ground_2020} and for the true model under a ``no phase transition'' assumption \cite{serfaty_gaussian_2020}. 
Hence, Theorems~\ref{thm:clt} and~\ref{thr:mesoscopic_bulk_pot} are
among the few CLT valid for a class of  correlated statistical
mechanics models in arbitrary dimension.  An analogous CLT was obtained
in \cite{bardenet_monte_2020} for another family of determinantal process on the $n$-dimensional hypercube, called multivariate orthogonal ensembles, and interesting applications to numerical quadrature are discussed.

\subsubsection{Semiclassical projector asymptotics.}

In dimension 1, universality of free fermions associated with
Schrödinger operators in the semiclassical limit is heuristically described in the textbook
\cite[Chap. 3.3]{Tao12}, as well as in
\cite{eisler_universality_2013}, based on Wentzel-Kramers-Brillouin (WKB)
approximations. 
In higher dimensions, there is no exact mapping between free fermions
and random matrix models, and furthermore the WKB method is less
powerful (the individual eigenfunctions of $H_{\hbar}$ do not admit
asymptotic \old{forumlas}\new{formulas}). We
refer to \cite{dean_universal_2015,dean_noninteracting_2016} for a
physical derivation of e.g.~the edge scaling limit, based on a short-time expansion
of the heat kernel associated with the Schrödinger operator: $s\mapsto
\exp(-s H_{\hbar})$. This approach can be made rigorous (using the
Tauberian theorem of Hardy-Littlewood-Karamata), but yields non-sharp
kernel asymptotics (only $O(\hbar^{\frac 12})$ in Theorem
\ref{thr:micro_limit_kernel}). Instead, we rely on the description of
the quantum propagator $t\mapsto e^{\i t H_{\hbar}/\hbar}$ (see section \ref{sec:pseud-oper}), which is known to yield
the sharpest forms of Weyl's law.

Semiclassical techniques devoted to the study of spectral projectors
are mostly developed in the homogeneous case of the Laplace-Beltrami
operator over a compact manifold and the Laplace operator on a domain
of $\R^n$ with Neumann or Dirichlet boundary conditions; we refer to
the recent review \cite{ivrii_100_2016}. The
cornerstone is to use a semiclassical Fourier transform to
write a spectral function $f(H_{\hbar})$ of the operator as
\[
  f(H_{\hbar})=\frac{1}{\hbar}\int
  \exp\left(\i\frac{t}{\hbar}H_{\hbar}\right)\widehat{f}\left(\frac{t}{\hbar}\right)\dd
    t.
  \]
  The operator $\exp\left(\i\frac{t}{\hbar}H_{\hbar}\right)$ has a
  physical interpretation: it
  captures the time evolution of a quantum system subject to $H_{\hbar}$, and admits an
  approximation as an integral operator for $t$ in a (fixed) 
  neighbourhood $[-\tau,\tau]$ of $0$; cf.~Proposition~\ref{prop:propagator}.  Even if $f$ is not smooth, one can hope that
  \[
    f(H_{\hbar})\approx \frac{1}{\hbar}\int
  \exp\left(\i\frac{t}{\hbar}H_{\hbar}\right)\widehat{f}\left(\frac{t}{\hbar}\right)\1_{|t|\leq
    \nu}\dd
    t,
  \]
 then perform computations using the the right-hand side. The
 last step consists in recovering $f(H_{\hbar})$ from the asymptotics
 of this frequency
 cutoff, which involves \emph{Fourier Tauberian theorems}.

In the context of the Laplace-Beltrami operator over a smooth compact
manifold, the equivalent of Theorem \ref{thr:micro_limit_kernel} is
well-known, at least on the diagonal $x=y$. A major part of the
literature is devoted to the improvements on the $\O(\hbar)$
remainder, which depend on dynamical assumptions on the geodesic flow. Recent developments
include the study of kernel asymptotics at mesoscopic scales around the diagonal, \cite{canzani_scaling_2015,canzani_Cinf_2018,keeler_logarithmic_2019}.

Using the Tauberian method, the Weyl law for semiclassical Schrödinger
operator \eqref{Z} was obtained in \cite{robert_comportement_1981}, with a
remainder $\O(\hbar)$, under the assumption that $\nabla V \neq 0$
along the whole edge. Asymptotics on the diagonal in the bulk and at a
non-degenerate edge (that is, Theorems \ref{thr:micro_limit_kernel}
and \ref{thr:micro_edge_kernel} at $x=y=0$) have been computed in
\cite{karadzhov_semi-classical_1986} using the preliminary results of
\cite{chazarain_spectre_1980}, under a growth condition of $V$ at infinity. 
In this article's context, the study of the kernel on the diagonal
translates to asymptotics for the expectation of linear statistics of
the Fermi gas. 

  Quantities associated with the semiclassical spectral projector for
  harmonic oscillators were studied in detail in
  \cite{hanin_interface_2020}, with a
  $\O(\hbar^{\frac 12})$ error term (see notably Proposition 1.8 there). The upper bound on the variance
  which leads to Theorem \ref{thm:gtb} (cf. \eqref{est_var_macro}) is similar to the bound in
  trace norm
  obtained in \cite{fournais_optimal_2020}, which also holds under
  hypothesis \ref{hyp:weak}. Remarkably, many recent advances in the
  semiclassical literature about the off-diagonal asymptotics of
  spectral projectors
  \cite{canzani_scaling_2015,canzani_Cinf_2018,hanin_scaling_2017,hanin_interface_2020}
  are also motivated by probabilistic models (and notably the
  behaviour of the nodal sets of random linear combinations of eigenfunctions).

  \subsubsection{A few perspectives}

  Even within the scope of semiclassical Schrödinger operators, the
 asymptotic picture is far from complete.
  For instance, to complement our results, it is of interest to obtain a large deviation principle for the empirical measure and determine how the rate function depends on the potential.
  An analogous question can be raised about the asymptotics of the variance in Theorem~\ref{thm:clt} in relation to Conjecture~\ref{conj:var_macro}. In particular, it is relevant to investigate in which case the normalized variance converges to an $H^{\frac 12}$ norm or whether there can be oscillations. 
  In dimension $1$ where the variance of smooth linear statistics is bounded, the issue of \new{a mesoscopic CLT is still open.}
  \old{the CLT is still open and it is of interest to determine under which conditions a CLT holds at mesoscopic and/or macroscopic scales. One could try to tackle these questions by exploiting the complete
  integrability of such systems (WKB approsimations). }
   
  For fermions point processes, the equivalent of
  Theorem \ref{thr:mesoscopic_bulk_pot} when $f$ is the indicator of a
  smooth set (``How many particles are in this box''), is a topical question since the number variance is a common measure of the entanglement entropy of subsystems, e.g. 
  \cite{leschke_scaling_2014,calabrese_exact_2012,calabrese_entanglement_2012}.
    The gapless case
  (such as semiclassical Schrödinger operators) is linked to a conjecture by Widom concerning
  pseudodifferential operators with rough symbols
  \cite{widom_class_1990,sobolev_pseudo-differential_2013},
 while  the asymptotics of the variance of linear statistics  performed in Section \ref{sec:mesoscopic-scales} relies on the high regularity of $f$, the case of an indicator function falls  outside the scope of
  this article. 
 This problem has been extensively studied recently in the context of random matrix theory based on so-called Fisher-Hartwig asymptotics and it relates optimal particles' rigidity and Gaussian multiplicative chaos, e.g. \cite{Smith_21,claeys_how_2021,fyodorov_statistics_2020}

   In a gapped system (Integer Quantum Hall
  states linked with Berezin-Toeplitz quantization), the so-called \emph{area law} holds and 
  macroscopic CLTs have been rigorously established in \cite{charles_entanglement_2018,fenzl_precise_2020}. 

 Finally one can also mention the ``hard edge'' model of a semiclassical
 Laplacian $-\hbar^2\Delta_{\Omega}$ with
  Dirichlet or Neumann  boundary conditions on a relatively compact open set $\Omega$.
  The behaviour of the point process near the boundary of $\Omega$ is presumed to be very different from the ``soft edge''  case of Theorem \ref{thr:micro_edge_kernel}. A
  complete description of the off-diagonal kernel at the relevant
  scale is an ongoing problem, see for instance the recent upper
  bounds on the size of the off-diagonal kernel
  \cite{ivrii_pointwise_2021}.

  \subsection{Organisation and Notation.}

In Section \ref{sec:lots-not-so}, we describe the basic semiclassical
techniques and objects that are used  throughout this article. We
begin with \emph{Agmon estimates}, which allow us to prove exponential
decay for eigenfunctions of $H_{\hbar}=-\hbar^2\Delta+V$ in the ``forbidden
region'' where the value of $V$ is greater than the eigenvalue. 
 Then, we introduce (semiclassical) pseudodifferential operators and Fourier
Integral operators, which are used to obtain asymptotics for the integral kernel of spectral functions of $H_{\hbar}$.

Section \ref{sec:spectr-proj-bulk} is devoted to the proof of Theorem
\ref{thr:micro_limit_kernel} and its corollaries (Theorem
\ref{thm:LLN}, \ref{thm:gtb}, and the first part of Theorem
\ref{prop:micro_limit_process}). Theorem \ref{thr:micro_limit_kernel}
is proved using the  stationary phase method and Tauberian techniques. Its probabilistic
consequences make use of fine properties of determinantal point
processes.

In Section \ref{sec:edge-asympt-case}, we modify the previous
arguments to study the edge of the droplet, proving Theorem
\ref{thr:micro_edge_kernel} and the second part of Theorem
\ref{prop:micro_limit_process}. The main difference is the treatment
of rapidly oscillating integrals, whose \emph{phase} is degenerate
compared to Section \ref{sec:spectr-proj-bulk}.

Finally, we study mesoscopic fluctuations in Section \ref{sec:mesoscopic-scales} by proving Theorems
\ref{thm:clt} and \ref{thr:mesoscopic_bulk_pot}. Again, we carefully
study the oscillating integrals cut-off at short time, to
obtain asymptotics for the variance of the (rescaled) point process;
then we use concentration inequalities and properties of determinantal
point processes (notably the fact that the variance diverges in dimensions $n\ge 2$) to conclude the proof of the CLT. 

For completeness, we review in the Appendix several notation and basic
facts  which are instrumental to prove our results. \new{We also
  derive several general estimates on determinantal point processes
  and oscillatory}\\ \new{integrals, which we believe may be of independent
  interest, such as Proposition \ref{prop:Laplace}, Corollary
  \ref{clt_dpp}, and} \\ \new{Propositions \ref{prop:spl_xxi}, \ref{prop:sineint}.}
\begin{itemize}[leftmargin=*]
\item Section~\ref{sect:op}  presents our Hilbert space setup and various operator topologies that we use.
\item The usual Laplacian $\Delta$ on  $L^2(\R^n)$ plays a crucial role in our analysis. 
Section~\ref{sec:unbounded-operators} reviews the basic spectral theory for $\Delta$, based on the Fourier transform,  as well as standard elliptic regularity estimates. 
\item  Section~\ref{sec:dpp} provides an introduction to the theory of determinantal point process. We give a general definition (independent of the concept of \emph{correlation kernel}) and review the notion of weak convergence used in Theorem~\ref{prop:micro_limit_process}. 
We also revisit a classical CLT of Soshnikov  \cite{Soshnikov_02} by obtaining  new bounds for the Laplace functional of a general  determinantal point process which might be of independent interest (cf.~Proposition~\ref{prop:Laplace}).
\item Section~\ref{sec:osc-int} gives several versions of the stationary phase lemma, which are tuned to obtain the asymptotics of oscillatory integrals which arise in our proofs. Depending on the regularity of the integrand, we obtain different estimates for the remainder which might be of independent interest. 

\item Section~\ref{sec:airy-type-integrals} reviews basic facts about the Airy function including  its integral representation, asymptotics and its relationship to the spectral theory of the operator \eqref{eq:Kedge}. 
\end{itemize}

\medskip

In the rest of this article we use the following conventions: 
\begin{itemize}[leftmargin=*]
\item
We denote by $x\cdot \xi$ the Euclidean inner-product for
$x,\xi\in\R^n$ and $|x| = \sqrt{x\cdot x}$ for $x\in\R^n$. 
 \item We write $A\Subset \R^n$ to denote a compact subset $A$ of $\R^n$.
\item We denote by $B^n_{x,r} = \{z\in\R^n : |x-z| \le r \}$ the Euclidean ball of radius $r$ around
  $x$ in $\R^n$ and $\omega_n = |B^n_{0,1}|$. 
\item  We denote by $\d x$ the Lebesgue measure on $\R^n$ and $\displaystyle \langle \phi , \varphi \rangle = \int \phi(x) \varphi(x) \d x$ the usual inner-product on $L^2(\R^n)$. 
We denote $\|\phi\| = \sqrt{ \langle \phi , \phi \rangle}$. 
\item  For $\phi\in\mathcal{S}(\R^n)$, we define its
  Fourier transform by 
\[
 \widehat\phi(\xi) = \int_{\R^n} e^{-\i \xi \cdot x}
\phi(x) \frac{\d x}{(2\pi)^{n/2}} , \qquad \xi\in\R^{n} . 
\]
With this convention, the Fourier transform extends to a
unitary operator $\mathcal{F}$ on $ L^2(\R^n)$ and we write 
$\widehat\phi = \mathcal{F} \phi$ for $\phi \in L^2(\R^n)$, as well as 
$\phi\in\mathcal{S}'$, where $\mathcal{S}'$ denotes the space of Schwartz distributions, dual
to the space of Schwartz functions $\mathcal{S}$. 
  \item Given $V:\R^n \to \R$ satisfying \ref{hyp:weak}, $\hbar\in(0,1]$
    and  $\mu\in \R$, we denote 
 \[
H_{\hbar}=-\hbar^2\Delta+V   
\qquad\text{and}\qquad    
\Pi_{\hbar,\mu}=\mathds{1}_{(-\infty,\mu]}(H_{\hbar}) .
\]
Without loss of generality we can always assume that $V\ge 0$ and $\mu \in (0,M)$.
\item If an operator $A$ acting on $L^2(\R^n)$ admits an
  integral kernel, we also denote by $A$ its integral kernel.
Given $f:\R_+ \to\R_+$, $\mathcal{U} \in SO_n$, $\epsilon\in(0,1]$ and
$x_0\in\{V<M\}$, provided that $f(H_{\hbar})$ admits an integral
kernel, we write
 \begin{equation}\label{eq:Kfxmu}
K^{f}_{x_0,\epsilon} : (x,y) \mapsto \epsilon^n
  f(H_{\hbar})(x_0+\epsilon \mathcal{U}^*   x, x_0+ \epsilon \mathcal{U}^*   y) .
\end{equation}
In particular, according to \eqref{zoom}, we simply have $K_{x_0,\epsilon} = K_{x_0,\epsilon}^{\1_{[0,\mu]}}$.  We also let $ K^{f}_{x_0,0} =  f(-\Delta+ V(x_0))$.
 \item Given a multi-index $\alpha \in \N_0^n$ and $f\in S(\R^n)$, we denote
$\partial_x^\alpha f  =  \partial^{\alpha_n}_{x_n} \cdots \partial^{\alpha_1}_{x_1} f$ 
and define the norms for $k\in\N_0$,
\[
\| f\|_{C^k(\Omega)}  = \sup_{x\in\Omega} \sum_{\alpha : |\alpha| \le k} \big| \partial_x^\alpha f (x) \big | 
\]
where $\Omega \subset \R^n$ is any open set and the sum is over all muti-indices $\alpha \in \N_0^n$  with $|\alpha| = \alpha_1+\cdots +\alpha_n \le k$. 
The definition of this norm extends to general functions.  
We also set $\| \cdot\|_{C^k} = \| \cdot\|_{C^k(\R^n)} $.

\item We use the notation $\partial_x = \nabla = (\partial_{x_1}, \cdots, \partial_{x_n}) $. 

\item $C,c>0$ are  constants  which (vary from line to line) only depend on the dimension $n\in\N$ and the potential $V$. 
\item$C_\alpha, c_\alpha>0$ are constants which depend only on $n\in\N$, $V$, the parameter $\alpha$ and vary from line to line.
\item If an element $A$ of a Banach space depends on a
  parameter $\eta>0$, we write for $k\in\N_0$, $A= \O(\eta^{k})$ if
  there exists a constant $C_k$ such that $\| A\| \le C_k \eta^{k}$ as
  $\eta\to 0$.
We further write  $A= \O(\eta^{\infty})$ when $A= \O(\eta^{k})$  for
all $k\in\N_0$. \new{In cases where there is ambiguity about the used
  norm, we indicate it as a subscript of $\O$.}

\item  We denote the commutator of two operators $A,B$ on $L^2(\R^n)$ by $[A,B] =AB-BA$. 
\end{itemize}

\subsection{Acknowledgements}
The authors thank L. Charles for useful discussion.

G.L. research is supported by the SNSF Ambizione grant S-71114-05-01.

A.D. is supported by a CNRS PEPS JCJC grant.

\bigskip

\section{Semiclassical techniques for Schrödinger operators}
\label{sec:lots-not-so}

In this section, we cover basic techniques from  semiclassical
analysis tailored to the study of
Schrödinger operators of the form $H_{\hbar}=-\hbar^2\Delta+V$, in the
semiclassical limit $\hbar \to 0$. Some of these techniques rely on stronger assumptions than \ref{hyp:weak}. The following ancillary hypothesis will be useful in this section.

\medskip

From now on, \new{let $M$ be as in \ref{hyp:weak}} and we assume (without loss of generality) that $V \geq 0$. 

\begin{defn}~\begin{enumerate}[beginpenalty=10000]
\item[{\crtcrossreflabel{(H')}[hyp:strong]}] A potential $V\in
  C^{\infty}(\R^n,\R_+)$ satisfies \ref{hyp:strong} if for every $k\geq
  0$, \old{a constant $0<C_k$}\new{there exists a constant $C_k>0$} such that,
 for all $x\in \R^n$ outside of a compact set, for every $|\alpha|=k$,
  \[
    |\partial^{\alpha}V(x)|\leq C_k|x|^{2};
  \]
  and if there exists $c>0$ such that, outside of a
  compact set,
  \[
    c|x|^{2}\leq V(x).
    \]
  \end{enumerate}
\end{defn}

This section is devoted to the foundational techniques and results
that allow us to study the spectrum of $H_{\hbar}$ as $\hbar\to 0$. Subsection
\ref{sec:first-trace-estim} is relatively elementary: we use operator
bracketing and \emph{Agmon estimates} to prove that, when $(\mu,V)$
satisfies \ref{hyp:weak}, the spectrum of $H_{\hbar}$ below $\mu$
consists of a finite number of eigenvalues, and
that the eigenfunctions tend very quickly to zero in the forbidden
region $\{V>\mu\}$. This is used in Subsection \ref{sec:replacement} to
prove a \emph{replacement principle}; we can change $V$ on $\{V>M\}$ so
that it satisfies \ref{hyp:strong}. Then, in Subsection
\ref{sec:pseud-oper}, we introduce semiclassical pseudodifferential
operators and Fourier Integral Operators. They are a priori suited to the stronger assumptions~\ref{hyp:strong}, but thanks to the replacement principle, we can relax the conditions on $V$ in the forbidden region
Then, under \ref{hyp:weak},
Proposition \ref{prop:propagator_rem_Cinf}  expresses rapidly
oscillating functions of $H_{\hbar}$ of the form
\[
  e^{\i \frac{tH_{\hbar}}{\hbar}}\vartheta(H_{\hbar})   \qquad t\in [-\tau,\tau], 
\]
for a (fixed) small $\tau>0$ and  $\vartheta \in C^\infty_c(\R)$ supported in $(-\infty, M)$, as an integral operator up to a small remainder in $C^{\infty}$-kernel topology.
This will be our main tool to prove the results of Section~\ref{sec:main-results}.

\subsection{Spectral theory and Agmon estimates}
\label{sec:first-trace-estim}

\begin{lem}\label{prop:rough_upper_Weyl}
 \new{Let $(\mu,V)$ satisfy \ref{hyp:weak}}. There exists \new{$M\geq
   \mu$ and $C_M>0$} such
  that for all $\hbar\in (0,1]$,
   \[ {\rm
    Rank}(\1_{H_\hbar \leq M}) \le \new{C_M} \hbar^{-n} .\] 
    Hence, $H_{\hbar}$ has discrete spectrum on $[-\infty,M]$
    consisting of finite rank eigenvalues.
    
 Moreover, if $V$ satisfies~\ref{hyp:strong}, there exists \old{$\gamma>0$
 and} $C>0$ 
 such that for all $\hbar\in (0,1]$ and $\mu>0$,
  \[{\rm Rank}(\1_{H_\hbar \leq \mu} )\leq C\hbar^{-n}\mu^{\old{\gamma}\new{n}} . \]
  \end{lem}
  \begin{preuve}
    We proceed as in \cite{reed_methods_1978}, Theorem XIII.81. We
    start with an upper bound for the Neumann Laplacian $ \Delta_{\mathcal{N}}$ on a hypercube   of size $1$:
    \[
      {\rm Rank}\left(\1_{-  \Delta_{\mathcal{N}}\leq \Lambda}\right)\leq C_n(\Lambda+1)^{\frac{n}{2}},
    \]
    valid for $\Lambda\geq 0$; if $\Lambda<0$ the left-hand side is
    zero.

    Now, for each $j\in \Z^n$, let $V_j$ denote the infimum of $V$ on
    the hypercube $\Omega_j$ centred at $j$ with size $1$, and let
    $\Delta_{\mathcal{N},j}$ denote the Neumann Laplacian on $\Omega_j$. Then, writing
    $L^2(\R^n)=\overline{\bigoplus_{j\in \Z^d}L^2(\Omega_j)}^{L^2}$, the following
    inequality holds in the sense of quadratic forms:
    \[
    H_\hbar=  -\hbar^2\Delta+V\geq \bigoplus_{j\in \Z^d}\left(-\hbar^2\Delta_{\mathcal{N},j}+V_j\right).
    \]
    By \ref{hyp:weak}, there is only a finite number
    of indices $j$ such that $V_j\leq M$. Hence the previous bound implies that
    \[
      {\rm Rank}\left(\1_{-\hbar^2\Delta+V\leq M}\right)\leq
      \new{C_M}{\rm Rank}\left(\1_{-\Delta_{\mathcal{N}}\le \hbar^{-2}M }\right),\]
    which shows that
    \[
     {\rm Rank}\left(\1_{-\hbar^2\Delta+V\leq M}\right) \leq
    C\hbar^{-n}. 
  \]
  If moreover $V$ satisfies \ref{hyp:strong}, then the number of
  indices $j$ such that $V_j\leq \mu$ grows \old{polynomially in
  $\mu$}\new{like $\mu^{\frac n2}$}, \old{as does} \new{and} ${\rm
  Rank}(\1_{-\Delta_{\mathcal{N}}\le \hbar^{-2}\mu })$\new{$\leq
  C\hbar^{-n}\mu^{\frac n2}$ for a universal constant $C>0$}. This concludes the proof.
 \end{preuve}

According to Lemma~\ref{prop:rough_upper_Weyl}, the spectrum of
$-\hbar^2\Delta+V$ below $M$ consists of a finite number of finite
rank eigenvalues. Even though $V$ is not regular on $\{V>M\}$, there
are robust tools to study the decay properties of any eigenfunction with
eigenvalue less than $ M$ on this region.

\begin{prop}\label{prop:domination}
  Let $(\mu,V)$ satisfy \ref{hyp:weak}, \new{let $\delta>0$ such that
    $(\mu+\delta,V)$ satisfies \ref{hyp:weak}} and define
  \[
    \mathrm{f}_{\delta}:\R^n \ni x\mapsto \delta \dist(x,\{V\leq \mu+\delta\}).
    \]
  Let $v$ be a normalised eigenfunction of $H_{\hbar}=-\hbar^2\Delta+V$, with
  eigenvalue $\leq \mu$.
  Then
  \[
    \|e^{\frac{\mathrm{f}_{\delta}}{\hbar}} v\|_{L^2}\leq 1+2\mu/\delta .
    \]
\end{prop}
\begin{preuve}
  The proof is based on the celebrated Agmon estimates \cite{agmon_lectures_2014}. Let $R>0$ and
  let
  \[
    \mathrm{f}_{\delta}^R:x\mapsto \min(\mathrm{f}_{\delta}(x),R).
    \]
  First notice
  that $\mathrm{f}_{\delta}^R$ is bounded, Lipschitz, and such that $|\nabla
  \mathrm{f}_{\delta}^R|= \delta \mathds{1}_{\Omega_R}$ a.e.,
  where
  \[
    \Omega_R=\{x\in \R^n,\dist(x,\{V\leq \mu+\delta\}) \le R/\delta\}.
    \]

  Integrating by parts, for any $u\in H^1(\R^n)$, there holds
  \[
    \left\langle
    u,e^{\frac{\mathrm{f}_{\delta}^R}{\hbar}}H_{\hbar}e^{-\frac{\mathrm{f}_{\delta}^R}{\hbar}}u\right\rangle
    =      \left\langle u H_{\hbar} u \right\rangle  - \|u \nabla \mathrm{f}_{\delta}^R \|_{L^2}^2
    =\left\langle
    u,\left(H_{\hbar}-\delta^2\mathds{1}_{\Omega_R}\right)u\right\rangle.
\]
Take $u=ve^{\frac{\mathrm{f}_{\delta}^R}{\hbar}}$, where $v$ is a normalised
eigenfunction of $H_{\hbar}$ with eigenvalue $\lambda\leq
\mu$\old{,then}\new{. In this case,} both $ v , u \in H^1(\R^n)$ and the last equation reads
\[
  \lambda \int
  e^{2\frac{\mathrm{f}_{\delta}^R}{\hbar}}|v|^2=\hbar^2\int\left|\nabla u\right|^2+\int
  e^{2\frac{\mathrm{f}_{\delta}^R}{\hbar}}\left(V- \delta^2\mathds{1}_{\Omega_R}\right)|v|^2.
\]
In particular, the eigenfunction $v$ satisfies  if $\lambda\le \mu$, 
\[
  \int
  e^{2\frac{\mathrm{f}_{\delta}^R}{\hbar}}\left(V-\delta^2 -\mu\right)|v|^2\leq
  0.
  \]
  We decompose this integral above into two parts: $\{V\leq
  \mu+\delta\}$ where $f_\delta^R=0$; and its complement set where
  $V- \delta^2- \mu \geq  \delta/2$.  We obtain
  \[
    \int_{\{V>\mu+\delta\}}e^{2\frac{\mathrm{f}_{\delta}^R}{\hbar}}|v|^2\leq \frac2\delta \int_{\{V\leq \mu+\delta\}} (\mu+ \delta^2-V)|v|^2\leq 2\mu/\delta 
  \]
 where we  used that $V\ge 1$ and  $\int_{\{V\le \mu+\delta\}}|v|^2 \le 1$. 
  This bound does not depend on $R$, so that by monotone convergence, we conclude that
  \[
    \int_{\{V>\mu+\delta\}}e^{2\frac{\mathrm{f}_{\delta}}{\hbar}}|v|^2\leq2\mu/\delta.
  \]
  Using again $f_\delta^R=0$ on $\{V\leq  \mu+\delta\}$ and $\mathrm{f}_{\delta} =0$ on $\{V\le \mu+\delta\}$, this completes the proof.
\end{preuve}

This implies,  in particular, uniform decay for the integral kernel for (compactly supported)
spectral functions of $H_{\hbar}$.

\subsection{Replacement principle}
\label{sec:replacement}
\new{Hypothesis \ref{hyp:strong} is much more suited to semiclassical
  analysis. One can \emph{replace}, to some
  extent, a potential satisfying \ref{hyp:weak} with one satisfying
  \ref{hyp:strong}, as follows}.

\begin{defn}\label{def:remplacement}
  \new{Let $(\mu,V_1)$ satisfy \ref{hyp:weak}, let $V_2$ satisfy
    \ref{hyp:strong}. Let $M>\mu$, and suppose that
  $V_2=V_1$ on $\{V_1\leq M\}$ and $\{V_2>M\}=\{V_1>M\}$. Then, we say
  that $V_2$ replaces $V_1$ up to $M$.}
\end{defn}

\begin{prop}\label{prop:remplacement}
  \old{Let $(M,V_1)$ and $(M,V_2)$ satisfy \ref{hyp:weak} and suppose that
  $V_2=V_1$ on $\{V_1\leq M\}$ and $\{V_2>M\}=\{V_1>M\}$. Then, we say
  that $V_2$ replaces $V_1$.} \new{Let $V_2$ replace $V_1$ up to $M$ and recall that the trace-norm $\|\cdot\|_{\J^1}$ is defined in Proposition~\ref{prop:op}.}
   Let us denote $H_{j;\hbar}=-\hbar^2\Delta+V_j$ for $j\in\{1,2\}$. 
   Then for any $f\in C^{\infty}_c(\R)$ with support in $[0,M)$, there holds  as $\hbar\to 0$,
   \[
     f(H_{1;\hbar})=f(H_{2,\hbar})+\O_{\J^1}(\hbar^{\infty}).
     \]
\end{prop}

\begin{preuve}
  It suffices to prove that there are constants $C,c>0$ so that  
  \begin{equation} \label{fHbd}
  \|f(H_{1;\hbar}) - f(H_{2;\hbar})
  \|_{L^2\to L^2} \le Ce^{-\frac{c}{\hbar}}  
\end{equation}
  since by Lemma~\ref{prop:rough_upper_Weyl}, both
  $f(H_{j;\hbar})$ have rank $\O(\hbar^{-n})$. In
  addition, it suffices to prove the claim in the case where $f$ takes
  non-negative values  (by linearity).

  We rely again on an Agmon estimate.
Let $\epsilon >0$ be small and $\phi\in C^{\infty}(\R^n,[0,1])$ be a cutoff with $\phi=0$ on $\{V_j\leq
   M-\frac{\epsilon}{2}\}$ and $\phi=2c$ on $\{V_j\geq
   M-\frac{\epsilon}{4}\}$ where $c=c_\epsilon>0$ is small enough such that $|\nabla
   \phi|^2\leq \frac{\epsilon}{4}$. Then, proceeding exactly as in the proof of Proposition~\ref{prop:domination}, for any normalized eigenfunction $v$ of
$H_{j;\hbar}=-\hbar^2\Delta+V_j$ (for either $j\in\{1,2\}$) with eigenvalue $\lambda$,  
 \[
\int e^{2\frac{\phi}{\hbar}}(V_j-|\nabla \phi|^2-\lambda)|v|^2\leq 0.
\]

Let us split  this integral into two parts: on $\{V_j\leq
M-\frac{\epsilon}{2}\}$, one has $e^{2\frac{\phi}{\hbar}}=1$ and
$|\nabla \phi|=0$; on $\{V_j\geq M-\frac{\epsilon}{2}\}$, one has
$V_j-|\nabla \phi|^2-\lambda\geq \frac{\epsilon}{4}$ provided that $\lambda \le M-\epsilon$ and $\phi =2c$. We obtain
\[
e^{\frac{4c}{\hbar}}\int_{\{V_j\geq
  M-\frac{\epsilon}{4}\}}|v|^2\leq \int_{\{V_j\geq
  M-\frac{\epsilon}{2}\}}e^{2\frac{\phi}{\hbar}}|v|^2\leq \frac4\epsilon
\int_{\{V_j< M-\frac{\epsilon}{2}\}}(\lambda-V_j)|v|^2\leq \frac{4M}{\epsilon}.
\]
This yields the following uniform control: for every $\hbar\in(0,1]$, $j\in\{1,2\}$ and every normalised eigenfunction $v$ of  either $H_{j;\hbar}$
with eigenvalue $\le M-\epsilon$,
\[
\int_{\{V_j\geq M-\frac{\epsilon}{4}\}}|v|^2\leq C e^{-4\frac{c}{\hbar}}.
\]
Moreover, using the eigenvalue equation, one has
\[
  \int_{\{M-\frac{\epsilon}{4}\leq V_j\leq M\}}|\Delta v|^2
  \le  2 \frac{\lambda^2}{\hbar^4}    \int_{\{M-\frac{\epsilon}{4}\leq V_j\leq M\}} |v|^2
  + 2   \int_{\{M-\frac{\epsilon}{4}\leq V_j\leq M\}} V_j^2 |v|^2
  \leq C  e^{-3\frac{c}{\hbar}}. 
\]

Let  $\chi\in C^{\infty}(\R^n,[0,1])$ be equal to $1$ on $\{V\leq
M-\frac{\epsilon}{4}\}$ and to $0$ on $\{V\geq M\}$. 
We claim that if $v$ is a normalised eigenfunction of either $H_{j;\hbar}$, then
 $\chi v$ is an almost eigenfunction of both $H_{j;\hbar}$ for $j\in\{1,2\}$ in the sense that 
\begin{equation} \label{aef}
  \| H_{j;\hbar}(\chi v)-\lambda \chi v\|_{L^2} \leq C  e^{-2\frac{c}{\hbar}}.
\end{equation}
First observe that $H_{j;\hbar}(\chi v) = \lambda \chi v + \hbar^2( v \Delta \chi -2 \nabla \chi \cdot \nabla v )$ for $j\in\{1,2\}$ since $V_1=V_2$ on $\supp(\chi)$ so that
\[
  \| H_{j;\hbar}(\chi v)-\lambda \chi v\|_{L^2} \le C \hbar^2 \left( \int_{\{V\geq M-\frac{\epsilon}{4}\}}|v|^2 + \int | \nabla \chi \cdot \nabla v |^2  \right)  .
\]
Second, by an integration by parts and then Cauchy--Schwarz's inequality, 
\[ \begin{aligned}
\left( \int | \nabla \chi \cdot \nabla v |^2 \right)^2
& \le  2 \left( 2 \int v \Delta \chi  (\nabla \chi \cdot \nabla v)  \right)^2 + 2\left( \int v \Delta v |\nabla \chi|^2  \right)^2 \\
& \le C   \left( \int_{\{V\geq M-\frac{\epsilon}{4}\}}|v|^2  \right) \left( \int | \nabla \chi \cdot \nabla v |^2 +  \int_{\{M-\frac{\epsilon}{4}\leq V\leq M\}}|\Delta v|^2 
  \right)  \\
  & \le C e^{-4\frac{c}{\hbar}} \left( \int | \nabla \chi \cdot \nabla v |^2  + C e^{-3\frac{c}{\hbar}} \right).
\end{aligned}\]
This computation shows that
$  \| H_{j;\hbar}(\chi v)-\lambda \chi v\|_{L^2} \le C \hbar^2 e^{-3\frac{c}{\hbar}}$
which proves \eqref{aef}.
Using the resolvent identity and the fact that $\|(H_{j,\hbar}-z)^{-1} \| \le 1/|\Im z|$, this bound implies that for all $z\in \C\setminus \R$,
\[
  \|((H_{1,\hbar}-z)^{-1}-(H_{2,\hbar}-z)^{-1})\chi
  v\|_{L^2} \le \frac{2C}{|{\rm Im}(z)|^2}e^{-2\frac{c}{\hbar}},
\]
Since in addition
\[
  \|(H_{j,\hbar}-z)^{-1}(1-\chi)
  v\|_{L^2}  \le \frac{1}{|{\rm Im}(z)|} \left( \int_{\{V_j\geq M-\frac{\epsilon}{4}\}}|v|^2 \right)^{1/2} \le \frac{C}{|{\rm Im}(z)|} e^{-2\frac{c}{\hbar}}, 
\]
we conclude that for any normalised eigenfunction $v$  of either $H_{j;\hbar}$ with eigenvalue $\le M-\epsilon$ and $z\in \C\setminus \R$,
\begin{equation} \label{resbd}
  \|((H_{1,\hbar}-z)^{-1}-(H_{2,\hbar}-z)^{-1})
  v\|_{L^2} \le \frac{C}{|{\rm Im}(z)|^2}e^{-2\frac{c}{\hbar}} . 
\end{equation}

Now, if  $f\in C^{\infty}_c(\R,\R_+)$ with support in $[0,M)$, using the
spectral resolution
\[
  f(H_{j;\hbar}) : (x,y) \mapsto \sum_{\lambda \in \sigma(H_{j;\hbar})}f(\lambda) v_\lambda(x) v_\lambda(y)
\]
where $(\lambda,v_\lambda)$ are normalised eigenpairs of $H_{j;\hbar}$, we verify that 
\[
  \|((H_{1,\hbar}-z)^{-1}-(H_{2,\hbar}-z)^{-1})f(H_{j;\hbar})
  \|_{L^2\to L^2} \le \max_{\lambda \le M-\epsilon}   \|((H_{1,\hbar}-z)^{-1}-(H_{2,\hbar}-z)^{-1})
  v_\lambda\|_{L^2} \sum_{\lambda \in \sigma(H_{j;\hbar})}  f(\lambda)  ,
\]
where we used that $\supp(f) \subset [0,M-\epsilon]$ for $\epsilon>0$ small enough. 
By Lemma~\ref{prop:rough_upper_Weyl} and \eqref{resbd}, this implies that 
\[
  \|((H_{1,\hbar}-z)^{-1}-(H_{2,\hbar}-z)^{-1})f(H_{j;\hbar})
  \|_{L^2\to L^2} \le  \frac{C}{|{\rm Im}(z)|^2}e^{-\frac{c}{\hbar}}  . 
\]
Using the Helffer--Sj\"ostrand formula \cite{helffer_equation_1989}, it follows that 
\[
  \|(f(H_{1;\hbar}) - f(H_{2;\hbar}))f(H_{j;\hbar})
  \|_{L^2\to L^2} \le Ce^{-\frac{c}{\hbar}}  . 
\]
This shows that 
$  \|f(H_{1;\hbar})^2 - f(H_{2;\hbar})^2
  \|_{L^2\to L^2} \le 2Ce^{-\frac{c}{\hbar}}  $ and it proves \eqref{fHbd} (upon replacing $f^2$ by $f$). 
\end{preuve}

\new{Let us mention that the conclusion of Proposition
  \ref{prop:remplacement} breaks down without regularity hypotheses on
  $f$. In particular in the case where $f=\1_{(-\infty,\mu]}$, which
  is the spectral function of interest, individual eigenfunctions of
  $H_{1;\hbar}$ are $O(e^{-c/\hbar})$-quasimodes of $H_{2;\hbar}$ but
  their energy might shift from just below $\mu$ to just above
  it. Hence, without further assumptions, the
  trace norm of
  $\1_{(-\infty,\mu]}(H_{1;\hbar})-\1_{(-\infty,\mu]}(H_{2;\hbar})$
  can be as large as $\hbar^{-n}$ and the operator norm can fail to
  tend to zero.}

\subsection{Pseudodifferential operators}
\label{sec:pseud-oper}

If the potential $V$ satisfies \ref{hyp:strong}, then the operator
$H_{\hbar}=-\hbar^2\Delta+V$ is an example of a \emph{Weyl
  pseudodifferential operator}. Informally speaking, $H_{\hbar}$ is
the quantum equivalent of the classical energy \[\R^{2n}\ni
(x,\xi)\mapsto V(x)+|\xi|^2\in \R,\] using the quantization rule $\xi \leftrightarrow-i\hbar
\nabla$. We write
\[
  H_{\hbar}=Op_{\hbar}((x,\xi)\mapsto V(x)+|\xi|^2).
\]
Weyl pseudodifferential operators generalise differential operators
with smooth coefficients. The advantage of this setting is that this
class of operators is, up to a small error, preserved by smooth
functional calculus: if $P_{\hbar}$ is a Weyl pseudodifferential
operator and $f\in C^{\infty}(\R)$ has good properties at infinity
then $f(P_{\hbar})$ is also a Weyl pseudodifferential operator.

Pseudodifferential operators are associated with \emph{symbols}, which
are defined as follows.

\begin{defn}\label{def:symbols}
  Let $k,n\in\N$, $m\in \Z$.
  The Fréchet space $S^{m}(\R^{k},\R^{n})$ is defined as the set of
  smooth functions $a:\R^{k}\times \R^{n}\times (0,1]\to \C$
  such that, for any multi-index $\alpha\in \N_0^{k+n}$, there exists $C_{\alpha}>0$ such
  that, for all $(x,\xi,\hbar)\in \R^{k+n}\times (0,1]$, one has
  \[
    |\partial^{\alpha}_{x,\xi}a(x,\xi;\hbar)|\leq
    C_{\alpha}(1+|\xi|^2)^{\frac{m}{2}}(1+|x|^2)^{\frac{m}{2}}.
  \]
  The optimal constants $C_{\alpha}$ are the seminorms on
  $S^{m}$.

  An element $a\in S^{m}(\R^k\times \R^n)$ is said to be
  \emph{elliptic} when there exists $c>0$ such that, for all $(x,\xi,\hbar)\in \R^{ k+n}\times (0,1]$,
  \[
    |a(x,\xi;\hbar)|\geq c(1+|\xi|^2)^{\frac{m}{2}}(1+|x|^2)^{\frac{m}{2}}
    \]
  \end{defn}

  \new{The symbol classes defined here are particular cases of the
    symbol classes considered in the textbook
    \cite{dimassi_spectral_1999}. Indeed, t}he function
  \[
    (x,\xi)\mapsto (1+|\xi|^2)^{\frac m2}(1+|x|^2)^{\frac{m}{2}}
  \]
  is an order function as in Definition 7.4 in
  \cite{dimassi_spectral_1999}.

  \new{A natural element of these symbol classes is the harmonic
    oscillator $x^2+\xi^2$ which belongs to $S^2$. Note that, contrary
  to \emph{microlocal} symbol classes, the small parameter in our
  semiclassical techniques is $\hbar>0$ and not $|\xi|^{-1}$; in
  particular we do not impose that differentiating the symbol with
  respect to $\xi$ improves its decay.}

For instance, if $V$
satisfies \ref{hyp:strong}, then $(x,\xi)\mapsto V(x)+|\xi|^2\in
S^{2}(\R^{n},\R^n)$ is elliptic near infinity.

\begin{defn}\label{def:class_symb}
  An element $a\in S^{m}(\R^{k},\R^{n})$ is called a \emph{classical symbol} if
  there exists a sequence $(a_k)_{k\geq 0}$ of smooth functions in $\R^{k}\times\R^{n}$
  such that for every $k\geq 0$, $a_k$ is an $\hbar$-independent
  element of $S^{m}$ and, for every $\ell \in \N_0$, \[\hbar^{-\ell-1}\left(a-\sum_{k\le \ell}
      \hbar^ka_k\right)\in S^{m}.\]

  \new{The principal symbol of $a$ is then defined as $a_0$.}
\end{defn}

\begin{defn}\label{def:pseudos}
  Let $a\in S^m(\R^n,\R^n)$. The Weyl quantization $Op_{\hbar}(a)$ of $a$ is
  the family of integral operators with distribution-valued kernel
  \[
    \R^{2n}\ni (x,y)\mapsto \frac{1}{(2\pi \hbar)^n}\int_{\R^n}
    e^{\i\frac{(x-y) \cdot \xi}{\hbar}}a\left(\tfrac{x+y}{2},\xi;\hbar\right)\dd \xi.
  \]
  This is a family (indexed by  $\hbar\in(0,1]$) of operators on $L^2(\R^n)$ with dense
  domain. If $a$ is real-valued, then $Op_{\hbar}(a)$ is symmetric. 
\end{defn}

 Weyl pseudodifferential operators obey an exact (smooth) functional calculus. 

\begin{prop}[\cite{dimassi_spectral_1999}, Theorem 8.7]\label{prop:func-calc}
  Let $\vartheta \in C^\infty_c(\R,\R)$.
  Let $P_{\hbar}=Op_{\hbar}(p)$ be a family of self-adjoint pseudodifferential
  operators with symbol $p \in S^{m}(\R^n)$ such that
    $p+i$ is elliptic and $m>0$.
  Then $\vartheta(P_{\hbar}) =Op_{\hbar}(a)$ where $Op_{\hbar}(a)$  is a
  self-adjoint pseudodifferential operator with symbol $a\in
  \mathcal{S}$. Moreover, if $p$ is classical, then
  $a$ is also classical and its principal part is $a_0=
  \vartheta(p)$.

\end{prop}
If $p\in S^{m}$ is real-valued and elliptic near infinity (if,
for instance, $p=|\xi|^2+V$ with $V$ satisfying \ref{hyp:strong}),
then $p+i$ is elliptic.

  We emphasize that this representation of spectral functions
of pseudodifferential operators is \emph{exact}. 
An explicit induction formula for the symbol $a$ (giving, in particular, the
condition on the support of $a$) can be found in
\cite{dimassi_spectral_1999} after Theorem 8.7.
In particular, Schrödinger operators are elliptic in the sense of the last
proposition with $m=2$.

\medskip

As an illustration of the versatility of this functional
calculus and the ``replacement principle'', let us prove a weak form of
the \emph{Weyl law}, that is, obtain formula \eqref{Z}.

\begin{prop}\label{lem:Weyl_law}
 \new{Let $(\mu,V)$ satisfy} \ref{hyp:weak}, \new{recall that $\omega_n$ denotes the volume
of the unit Euclidean ball in $\R^n$} and that $  \Pi_{\hbar,\mu}=\1_{(-\infty,\mu]}(H_{\hbar})$.
  Then, as
  $\hbar\to 0$,
  \[
   N= \tr( \Pi_{\hbar,\mu})\sim (2\pi\hbar)^{-n}\omega_n\int
    (\mu-V(x))_+^{\frac n2}\dd x.
\]
  \end{prop}
  \begin{preuve}
  \new{Let $V_2$ replace $V_1$ up to $M$}.
  \old{Let $V_2$ be a potential which satisfies \ref{hyp:strong} such that $V_2=V$ on $\{V\leq M\}$ and  $\{V_2>M\}=\{V>M\}$.}
 Then, by combining Propositions \ref{prop:remplacement} and
 \ref{prop:func-calc}, for any  $\vartheta\in
 C^{\infty}_c(\R,[0,1])$ supported inside $(0,\tfrac{M+\mu}{2}]$, one has
\begin{equation} \label{trace_est} 
\begin{aligned}
      \tr(\vartheta(H_{\hbar}))
      &=\tr(\vartheta(H_{2;\hbar}))+O(\hbar^{\infty}) \\
      & =  \frac{1}{(2\pi \hbar)^n}\int_{\R^{2n}} a\left(x,\xi;\hbar\right)\dd \xi \dd x +\O_\vartheta(\hbar^\infty) \\
      &= \frac{1}{(2\pi \hbar)^n}\int \vartheta(V(x)+|\xi|^2)\dd x\dd \xi+\O_{\vartheta}(\hbar^{-n+1})
   \end{aligned} 
   \end{equation}
where we used that $a$ is a classical symbol with compact support on
$\R^{2n}$ and principal part $a_0(x,\xi)= \vartheta( V_2(x) +
|\xi|^2)$.

The spectrum of the operator $H_\hbar$ lies in $[\min V, \infty)$, so that if $\vartheta_+\geq \mathds{1}_{[\min(V),\mu]} \geq \vartheta_-$,  then as operators 
$ \vartheta_+(H_{\hbar}) \ge \Pi_{\hbar,\mu} \ge \vartheta_-(H_{\hbar}) $.

Hence, by \eqref{trace_est}, taking the limit as $\hbar\to0$ and  then the infimum over all $\vartheta_+ \ge \mathds{1}_{[\min(V),\mu]}$, we obtain
    \[
      \limsup_{\hbar\to 0}
      (2\pi\hbar)^n\tr(\Pi_{\hbar,\mu})\leq
      \int_{\R^{2n}}  \1_{|\xi| \le (\mu-V(x))_+^{1/2}} \dd \xi \dd x = \omega_n
  \int
      (\mu-V(x))_+^{\frac n2}\dd x.
    \]
   Similarly, by taking a supremum over $\vartheta_-\leq
    \mathds{1}_{[\min(V),\mu]}$, we obtain the other inequality, hence
    the claim. \new{Here, it is very important that for every $E\in
      \R$, the set $\{V(x)+|\xi|^2=E\}$ has zero measure; indeed,
      outside of the measure zero set $\{\xi=0\}$, the symbol
      $V(x)+|\xi|^2$ has no critical points.}
  \end{preuve}
  If  $\vartheta$ is not smooth, then
    $\vartheta(P_{\hbar})$ cannot be written as a pseudodifferential operator
    in a satisfactory way.
However, by a Fourier transform, writing  
\[
  \vartheta(\lambda) = \frac{1}{2\pi\hbar} \int e^{\i \frac{t \lambda}{\hbar}}
  \hat{\vartheta}(\tfrac{\lambda}{\hbar}) \d t,\]
the crucial step to obtain approximation for $\vartheta(P_{\hbar})$ is the study of the operator  $f_{t;\hbar}(H_\hbar)$
 where $f_{t;\hbar}(\lambda) = e^{\i \frac{t\lambda}\hbar} \chi(\lambda)$
oscillates at frequency $\O(\hbar^{-1})$ and $\chi\in
C^{\infty}_c(\R)$ is introduced for technical reasons. In this case, $f_{t;\hbar}(H_\hbar)$ is an approximation of the propagator associated with the Schr\"odinger operator $H_\hbar$.
It can express as an integral operator with an
oscillating phase, but one must replace $e^{\i\frac{(x-y)\cdot \xi}{\hbar}}$ \new{in
Definition~\ref{def:class_symb}} by the
solution of an order 1 partial differential equation called the Hamilton-Jacobi equation\footnote{This equation \eqref{HJE} corresponds to the classical (Lagrangian) dynamics associated with the symbol $p(x,\xi)$ of $P_{\hbar}=Op_{\hbar}(p)$.}.

\begin{prop}[Chapter 10 in
  \cite{dimassi_spectral_1999}, notably equations (10.2), (10.5) and (10.8)]
  \label{prop:propagator}
  Let $m>0$. Let $p\in S^{m}(\R^n,\R^n)$ be independent of $\hbar$ and such
  that $p+i$ is elliptic. Let
  $P_{\hbar}=Op_{\hbar}(p)$. 
Given $\vartheta\in C^{\infty}_c(\R,\R_+)$, there exist $\tau>0$ and a classical symbol $a\in S^0(\R^{2n+1}, \R^n)$ such that 
\[
 e^{\i t P_{\hbar}/\hbar}\vartheta(\old{H_{\hbar}}\new{P_{\hbar}}) = I_{\hbar,t}^{\phi, a}+\O_{\J^1}(\hbar^{\infty})
 \quad\text{uniformly for  $t\in [-\tau,\tau]$,}
 \]
 where $I_{\hbar,t}^{\phi, a}$ is a (non self-adjoint) integral operator with kernel
 \begin{equation} \label{kop}
I_{\hbar,t}^{\phi, a} : (x,y) \mapsto
      \frac{1}{(2\pi \hbar)^n}\int e^{\i \frac{ \phi(t,x,\xi)-y\cdot \xi  }{\hbar}}a(t,x,y,\xi;\hbar)\dd \xi.
    \end{equation}
 \old{and}\new{Here,} 
 \begin{itemize}
\item  $(x,y,\xi) \mapsto a(t,x,y,\xi;\hbar)$ has compact support
  $\mathcal{K} \Subset \R^{3n}$ for all $t\in[-\tau,\tau]$ and
  $\hbar\in(0,1]$.
\item $\mathcal{K}$ is a small $\tau$-neighbourhood of
  $\{(x,x,\xi),\vartheta(p(x,\xi)) > 0\}$.
\item the principal part of $a$  at $t=0$ satisfies, on the diagonal,
\begin{equation} \label{pp}
 a_0(0,x,x,\xi) = \vartheta(p(x,\xi)).
\end{equation}
\item there exists a compact $\mathcal{K}_\tau\Subset \R^{2n+1}$ containing
  $[-\tau,\tau]\times \mathcal{K}$ such that $\phi :
  \mathcal{K}_\tau \to\R $ is the (unique) solution of the initial value problem: 
\begin{equation} \label{HJE}
\phi(0,x,\xi) = x \cdot \xi  \qquad \qquad
    \partial_t\phi(t,x,\xi)=p(x,\partial_x\phi(t,x,\xi)) . 
\end{equation}

  \end{itemize}
\end{prop}
Notice the similarities with Proposition \ref{prop:func-calc}: at
$t=0$ one has $I^{\phi,a}_{\hbar,0}=Op_{\hbar}(a)$. 

 The compact set $\mathcal{K}_\tau$ and the solution $\phi$ of the
  Hamilton-Jacobi equation can be obtained from
  $\mathcal{K},p,\phi|_{t=0}$ by the method of characteristics (see
  \cite{evans_partial_1998}, Section 3.2.4, Theorem 2 for a general
  statement concerning first-order PDEs and Section
  3.2.5.c, Example 6 for an application to Hamilton-Jacobi equations).

\medskip

 Our goal is to apply Propositions \ref{prop:func-calc} and
\ref{prop:propagator} to obtain pointwise estimates for the spectral
projector $\Pi_{\hbar} = \1_{[0,\mu]}(H_\hbar)$, but there are two
small obstacles. First, $H_\hbar=-\hbar^2\Delta+V$ is a
pseudodifferential operator, with elliptic symbol in $S^2$, only when
$V$ satisfies \ref{hyp:strong}. Second, we need to improve the control of the 
remainder in Proposition~\ref{prop:propagator}, from $\J^1$ norm
control to local $C^{\infty}$ norm. This relies on
elliptic estimates (in our case, Proposition
\ref{prop:elliptic_reg_V}). 

The next Proposition will be our main input to prove the results of Section~\ref{sec:main-results}.

 \begin{prop} \label{prop:propagator_rem_Cinf}
\new{Assume that $(\mu,V)$ satisfies \ref{hyp:weak} and let $\vartheta
  \in C^\infty_c((-\infty,M),\R_+)$ with $M>\mu$ as in \ref{hyp:weak}.}
There exists $\tau>0$  so that for $t\in[-\tau,\tau]$, 
\[
\vartheta(H_{\hbar})e^{\i t H_{\hbar}/\hbar} = I_{\hbar,t}^{\phi, a} + R_{\hbar,t}
\]
with $I_{\hbar,t}^{\phi, a}$ as in \eqref{kop}  (under the same assumptions for the classical symbol $a\in S^0$)  and the error term
satisfies $\|R_{\hbar,t}\|_{\J^1} = \O(\hbar^\infty)$
\new{uniformly}\old{and}. \new{Moreover,} for any
 $ \mathcal{K} \Subset \{V<M\}^2$ and every multi-indices $\alpha,\beta \in\N_0^n$,   
\begin{equation} \label{Cinftycontrol}
 \max_{(x,y) \in \mathcal{K}}\sup_{t\in[-\tau,\tau]} \sup_{\hbar \in (0,1]}\big| \partial_x^\alpha \partial_y^\beta R_{\hbar,t}(x,y) \big| = \O_{\alpha,\beta}(\hbar^\infty). 
 \end{equation}
 \end{prop}

 \begin{preuve}
 \new{Let $V_2$ replace $V_1$ up to $M$ and assume that $V_1,V_2\ge 0$}.
 \old{To this end, let $V_1=V$  and $V_2$ satisfying \ref{hyp:strong}
   be equal to $V$ on $\{V\leq M\}$ and such that
   $\{V_2>M\}=\{V>M\}$.} By Proposition
   \ref{prop:remplacement}, it holds for any $\vartheta\in C^{\infty}_c([0,M),\R_+)$, 
   \[
     \|\vartheta(H_{1;\hbar})-\vartheta(H_{2;\hbar})\|_{\J^1}=\O(\hbar^{\infty}).
   \]
   Next, let $\varkappa \in C^{\infty}_c([0,M),\R)$ be equal to
   $\lambda\mapsto \lambda$
   on the support of $\vartheta$. Then for $j\in\{1,2\},$
   \[
     e^{\i tH_{j;\hbar}/\hbar}\vartheta(H_{j;\hbar})=e^{\i t\varkappa(H_{j;\hbar})/\hbar}\vartheta(H_{j;\hbar}) .
   \]
Hence, since $\lambda\mapsto e^{\i\lambda}$ is 1-Lipschitz, by using again Proposition \ref{prop:remplacement}, 
   \[
     \|\vartheta(H_{1;\hbar}) e^{\i tH_{1;\hbar}/\hbar}-\vartheta(H_{2;\hbar}) e^{\i tH_{2;\hbar}/\hbar}\|_{\J^1}=\O(\hbar^{\infty}) ,
   \]

We can deduce from this trace norm estimate,  a (local) $C^{\infty}$ control for the kernel valid in $\{V<M\}$ in the following way. Letting
   $(\lambda^1_k,\phi^1_k)$ and $(\lambda^2_k,\phi^2_k)$ be respective
   spectral resolutions of $H_{1;\hbar}$ and $H_{2;\hbar}$ below $M$,
   one has for $j\in\{1,2\}$,
   \[
     \vartheta(H_{j;\hbar})e^{\i tH_{j;\hbar}/\hbar}(x,y)=\sum_{\lambda^j_k\leq
       M}\vartheta(\lambda_k^j)e^{\i t \lambda_k^j}\phi^j_k(x)\phi^j_k(y). 
   \]
  Let $\vartheta_m(\lambda)= \lambda^{2m}\vartheta(\lambda)$ for $\lambda\ge0$ and $m\in\N_0$.   
In particular, for $(x,y)\in \{V<M\}^2$,  it holds for every $m\in \N$,
   \[
     \vartheta_m(H_{j;\hbar})e^{\i tH_{j;\hbar}/\hbar}(x,y)=(-\hbar^2\Delta_x+V)^m(-\hbar^2\Delta_y+V)^m\sum_{\lambda^j_k\leq
       M}\vartheta(\lambda_k^j)e^{\i t\lambda_k^j}\phi^j_k(x)\phi^j_k(y).
   \]

    Using Proposition \ref{prop:elliptic_reg_V}  and the following  Remark~\ref{rk:ellpiticity2},  for any $\mathcal{K}\Subset \{V<M\}^2$ and every $k\geq0$, there exists $C_k$ such that with $m={k+\lfloor \frac n4\rfloor + 1}$,
\begin{align*}
 &\left\|\vartheta(H_{1;\hbar}) e^{\i
     tH_{1;\hbar}/\hbar}-\vartheta(H_{2;\hbar}) e^{\i
     tH_{2;\hbar}/\hbar}\right\|_{C^{2k}(\mathcal{K})}\\ 
     &\leq C_k
                                           \hbar^{-2k-n}\left\|(-\hbar^2\Delta+V)_x^{m}(-\hbar^2\Delta+V)_y^{m}\left[\vartheta(H_{1;\hbar}) e^{\i
     tH_{1;\hbar}/\hbar}-\vartheta(H_{2;\hbar}) e^{\i
                                           tH_{2;\hbar}/\hbar}\right]\right\|_{L^2(\R^n\times
                                                        \R^n)}\\
 & \le C_k \hbar^{-2k-n}\left\|\vartheta_{m}(H_{1;\hbar}) e^{\i
     tH_{1;\hbar}/\hbar}-\vartheta_{m}(H_{2;\hbar}) e^{\i
   tH_{2;\hbar}/\hbar}\right\|_{\J^1}
\end{align*}
since the $L^2$ norm of a kernel
   operator is equal to the Hilbert-Schmidt norm of the corresponding operator,
   which is smaller than its trace norm (cf.~Proposition~\ref{prop:op}). 
   
   Hence, provided that $\supp\vartheta \subset [0,M)$,
   the conclusions of Proposition \ref{prop:propagator} applied  to the
   symbol $p(x,\xi)=|\xi|^2+V(x)$ also hold if $V$ satisfies 
   \ref{hyp:weak}.
   It remains to prove the claim under the hypothesis
   \ref{hyp:strong} in which case we can use ellipticity. 
   
   Let $\chi:\R\to [0,1]$ be compactly supported inside
   $(0,M)$ and such that
   $\chi=1$ on the support of $\vartheta$.  Given $m\in \N_0$, let
\[
     \chi_m:\lambda\mapsto \lambda^{-m}\chi(\lambda)
     \qquad\text{and}\qquad
     \vartheta_m:\lambda\mapsto \lambda^{2m}\vartheta(\lambda).
\]
     Then, for every $m\in \N$,
   \[
     \chi_m(H_{\hbar})e^{\i t
       H_{\hbar}/\hbar}\vartheta_m(H_{\hbar})\chi_m(H_{\hbar})=e^{\i
       t H_{\hbar}/\hbar}\vartheta(H_{\hbar}).
   \]
   Let us now apply Proposition \ref{prop:propagator} to $e^{\i t
     H_{\hbar}/\hbar}\vartheta_m(H_{\hbar})$. Under the hypothesis
   \ref{hyp:strong}, there exist classical symbols $a_m$ such that for $m\in\N_0$, 
   \[
     e^{\i t
       H_{\hbar}/\hbar}\vartheta_m(H_{\hbar})=I_{t,\hbar}^{\phi,a_m}+R_{t,\hbar}^m,
   \]
   where, since all $\vartheta_m$ have the same support, 
   \[
     \|R_{t,\hbar}^m\|_{\J^1}=\O_m(\hbar^{\infty}) \qquad\text{uniformly for $t\in[-\tau,\tau]$.}
   \]
   In particular,
   \[
     e^{\i t
       H_{\hbar}/\hbar}\vartheta(H_{\hbar})=\chi_m(H_{\hbar})I_{t,\hbar}^{\phi,a_m}\chi_m(H_{\hbar})+
       \widetilde{R}_{t,\hbar}^m 
        \qquad \widetilde{R}_{t,\hbar}^m:= \chi_m(H_{\hbar})R_{t,\hbar}^m\chi_m(H_{\hbar}).
   \]

   Let $\mathcal{K} \Subset \{V<M\}^2$ and $\varkappa\in C^{\infty}_c(\R^{2n},\R_+)$ be a cutoff equal\old{s} to   $1$ on $\mathcal{K}$.
   Applying Proposition \ref{prop:elliptic_reg_V}, it holds for every $k, m \in \N$,
   \[
     \|\widetilde{R}_{\hbar,t}^m\|_{C^{2k}(\mathcal{K})}\leq
     C_k  \hbar^{-2k-n}\|(-\hbar^2\Delta+V)_x^{k+\lfloor \frac
       n4\rfloor+1}(-\hbar^2\Delta+V)_y^{k+\lfloor \frac
       n4\rfloor+1} [\widetilde{R}_{\hbar,t}^m\varkappa]\|_{L^2}, 
   \]

     Setting $m= k+\lfloor \frac n4 \rfloor +1$, by definitions, one has
     \[
       \|(-\hbar^2\Delta+V)_x^{k+\lfloor \frac
         n4\rfloor+1}(-\hbar^2\Delta+V)_y^{k+\lfloor \frac
         n4\rfloor+1}[\widetilde{R}_{\hbar,t}\varkappa]\|_{L^2}\leq
       C\|R_{\hbar,t}\|_{L^2}.
     \]

 Again, the $L^2$-norm of the kernel of $R_{\hbar,t}$ is controlled by the $\J^1$ norm of the corresponding operator; to conclude,
     \[
       \|\widetilde{R}_{\hbar,t}\|_{C^{2k}(\mathcal{K})}=\O(\hbar^{\infty}).\]
     It remains to show that for $t\in[-\tau,\tau]$, 
     \[
       \chi_m(H_{\hbar})I^{\phi,a_m}_{\hbar,t}\chi_m(H_{\hbar}) 
       = I^{\phi,a}_{\hbar,t} + \O(\hbar^\infty)
     \]
for a compactly supported symbol $a$ which satisfies \eqref{pp} and that the error is controlled as in \eqref{Cinftycontrol}. 

     Observe that as $\chi_m\in C^{\infty}_c$, by Proposition
     \ref{prop:func-calc}, there exists a
     classical symbol $b_m \in S^0$ so that 
          \[
       \chi_m(H_{\hbar})=Op_{\hbar}(b_m)
     \]
and for all  $x,y \in \R^n$, 
     \[\begin{aligned}
     \chi_m(H_{\hbar})I^{\phi,a_m}_{\hbar,t}\chi_m(H_{\hbar})(x,y)  =\frac{1}{(2\pi\hbar)^{3n}}\int &
       e^{\frac{\i}{\hbar}\left((x-x_1)\cdot
           \xi_1+\phi(t,x_1,\xi)-x_2\cdot \xi+(x_2-y)\cdot
           \xi_2\right)} \\ 
         &  b_m(\tfrac{x+x_1}{2};\xi_1;\hbar)a_m(t,x_1,x_2,\xi;\hbar)b_m(\tfrac{x_2+y}{2},\xi_2;\hbar)\dd
       x_1\dd \xi_1\dd x_2\dd \xi_2 \dd \xi .
    \end{aligned} \]

For any  $t\in[-\tau,\tau]$,
$(x_1,x_2,\xi) \mapsto a_m(t,x_1,x_2,\xi;\hbar)$ has compact support.  
Hence, for $(x,y)\in\mathcal{K}$, both variables  $\tfrac{x+x_1}{2}$
and $\tfrac{x_2+y}{2}$ lie in a compact subset of $\R^n$ and  one can
use the fact that $a_m
\in \mathcal{S}$ has rapid decay in $\xi$.
This allows us to localize the integral over $\xi_1,\xi_2$ to a fixed compact, up to an error whose $C^k(\mathcal{K})$-norm is $\O_k(\hbar^\infty)$ for every $k\in\N_0$. 

Hence, for fixed $x,y,\xi$, one can apply the stationary phase
 lemma to the previous integral. 
One easily checks that the only critical point of the oscillating phase is
     \[
       (x_1,\xi_1,x_2,\xi_2)=(x,\partial_x\phi(t,x,\xi),y,\xi)
     \]
     and the Hessian is non-degenerate at this point, with determinant $1$. By Proposition
     \ref{prop:spl}, this implies that for $(x,y) \in \mathcal{K}$ and $t\in[-\tau,\tau]$, 
     \[
         \chi_m(H_{\hbar})I^{\phi,a_m}_{\hbar,t}\chi_m(H_{\hbar})(x,y)  = \frac{1}{(2\pi\hbar)^{n}} \int e^{\frac{\i}{\hbar}(\phi(t,x,\xi)-y\cdot
         \xi)}a(t,x,y,\xi;\hbar) \d \xi +R_{\hbar,t}(t,x,y),
     \]
     where
 \begin{itemize}
 \item $a \in S^0$ is a classical symbol such that   for all $t\in[-\tau,\tau]$,  $(x,y,\xi) \mapsto a(t,x,y,\xi;\hbar)$  has a given compact support.
\item for every $k\in\N_0$
\[
\sup_{t\in[-\tau,\tau]}\| R_{\hbar,t}(t,x,y) \|_{C^k(\mathcal{K})} = \O_k(\hbar^\infty)
\]
\item The principal part of $a$ satisfies at $t=0$ on the diagonal,
\[
 a_0(0,x,x,\xi)  =  b_{m,0}(x, \xi)a_{m,0}(0,x,x,\xi)b_{m,0}(x,\xi) .
\]
Here we used the equations for the critical point (in particular $x_1=x_2=x$ on the diagonal) and that
$\xi_1=\partial_x\phi(0,x,\xi) =\xi$ by the Hamilton-Jacobi equation \eqref{HJE}. Since $b_{m,0}(x, \xi) = \chi_m(p(x,\xi))$, 
$a_{m,0}(t,x,x,\xi) = \vartheta_m(p(x,\xi))$ (cf.~ Proposition \ref{prop:propagator}) and, by construction, $\chi_m^2\vartheta_m = \vartheta $, we conclude that
\[
 a_0(0,x,x,\xi)  =  \vartheta(p(x,\xi)) , \qquad p(x,\xi)= |\xi|^2+ V(x). 
\]
 \end{itemize}    
  This concludes the proof.   
   \end{preuve}

Recall that the free fermion point process, denoted by $\X$,  is the
determinantal point process associated with the operator $\Pi_{\hbar,
  \mu} =\mathds{1}_{(-\infty,\mu]}(H_{\hbar})$ and $N= \tr \Pi_{\hbar,
  \mu} $. For fixed $\hbar>0$,
the probability measure $N^{-1}\E\X$ admits a density with respect to the Lebesgue measure on $\R^n$, since it can be expressed using the first $N$ eigenfunctions $(v_j)_{1\leq
  j\leq N}$ of $H_{\hbar}$ by
\[
  N^{-1}\E\X=\frac{1}{N}\sum_{k=1}^N|v_k(x)|^2\dd x,
\]
where each term of the sum belongs to $L^1(\R^n,\R)$ by definition. The \emph{intensity function} is the density of this
probability measure:
\begin{equation} \label{intensity}
  \rho_N(x) : =N^{-1}\Pi_{\hbar,\mu}(x,x)  ,\qquad x\in\R^n.
\end{equation}
As a simple consequence of Proposition~\ref{prop:propagator_rem_Cinf}, we obtain (locally) uniform bounds for the intensity of this point process.

\begin{prop}\label{prop:bound_rho_N} Assume that \new{$(\mu,V)$}
  satisfies \ref{hyp:weak} \new{and let $M>\mu$ as in \ref{hyp:weak}}.
 For any  compact $\mathcal{K} \Subset  \{V<M\}$, there exists a constant $C$ (depending on $\mathcal{K}$ and $\mu$) so that for all $\hbar\in (0,1]$, 
\[
\max_{\mathcal{K}} \rho_N \le C.  
\]
\end{prop}

\begin{preuve}
Recall that $V\ge 0$ and let  $\vartheta\in C^{\infty}_c(\R,[0,1])$ be
  such that  $\supp(\vartheta) \subset (-\infty,M)$ and $\1_{[0,\mu]}\leq \vartheta$. Then, as the spectrum of $H_\hbar$ lies in $[0,\infty)$, 
\[
 \Pi_{\hbar,\mu}= \1_{\{H_\hbar \le \mu\}} \le \vartheta(H_\hbar) 
\]
as operators, so their kernels can be compared pointwise on the diagonal. 
Moreover, by Proposition \ref{prop:propagator_rem_Cinf} with $t=0$, we have 
$\vartheta(H_\hbar)  = I_{\hbar,0}^{\psi, a} + \O(\hbar^\infty)$ where
the kernel of the error is controlled locally uniformly inside $\{V<M\}$. By \eqref{kop}, this implies that for any   $\mathcal{K} \Subset \{V<M\}$,  
\[
 \Pi_{\hbar,\mu}(x,x)  \le   \frac{1}{(2\pi \hbar)^n} \int a(0,x,x,\xi;\hbar) \d \xi + \O(\hbar^\infty)
\]
uniformly for $x\in \mathcal{K}$, where $(x,\xi) \mapsto a(0,x,x,\xi;\hbar) $ is uniformly bounded with compact support (independently of $\hbar\in (0,1]$). 
In particular, there exists $C>0$ such that, for all $x\in
\mathcal{K}$ and all $\hbar\in(0,1]$,
\[
  \Pi_{\hbar,\mu}(x,x)\leq C\hbar^{-n}.
\]
Recalling from Lemma \ref{lem:Weyl_law} that
$\displaystyle     N\sim (2\pi\hbar)^{-n}\omega_n\int
    (\mu-V(x))_+^{\frac n2}\dd x$, this completes the proof. 
  \end{preuve}

  \begin{rem}
  Since $V$ is not supposed to be regular on $\{V>M\}$, the ellipticity
  techniques used to control the $L^{\infty}$ norm from the $L^2$ norm
  do not work there. If $V$ is
  $C^{\infty}$ everywhere, on the other hand, it is possible to obtain
  pointwise equivalents of Proposition \ref{prop:domination}, see for
  instance \cite[Proposition 5.5]{helffer_multiple_1984}.
\end{rem}

To conclude this section and illustrate the methods used in this article, we apply Proposition~\ref{prop:propagator_rem_Cinf} with $t=0$ to derive the \emph{microscopic asymptotics} of the kernel of the operator $\chi(H_{\hbar})$ for a fixed smooth spectral function~$\chi$.

 \begin{prop}\label{prop:convergence_smooth_functions_bulk}
\new{ Assume that $(\mu,V)$ satisfies \ref{hyp:weak}  and let  $\chi\in
   C^\infty_c((-\infty,M),\R_+)$  with $M>\mu$ as in \ref{hyp:weak}.}
 Using the notation \eqref{eq:Kfxmu},  for any compact sets  $\mathcal{A} \subset \{V<M\}$ and  $\mathcal{K}\subset \R^{2n}$, it holds
 \[
   \max_{x_0 \in \mathcal{A}}   \max_{(x,y) \in \mathcal{K}} \big| 
    K^{\chi}_{x_0,\hbar}(x,y)- K^{\chi}_{x_0,0}(x,y)\big| =\O(\hbar) . 
 \]
 \end{prop}
 \begin{preuve}
   By Proposition~\ref{prop:propagator_rem_Cinf} with $t=0$, one can write
   \[
     K^{\chi}_{x_0,\hbar}(x,y;x_0,\hbar)=\frac{1}{(2\pi)^n}\int e^{\i (x-y)\cdot
       \xi}a(x_0+\hbar x, x_0+\hbar y,\xi;\hbar)\dd \xi+ R(x,y; x_0 , \hbar),
   \]
   where the error $R(x,y;x_0,\hbar)$ and all its derivatives are $\O(\hbar^\infty)$ uniformly for $x_0 \in \mathcal{A}$ and $(x,y)\in \mathcal{K}$. 
   Note that in this integral, the phase is independent of the
   parameter $\hbar$ and the symbol of $a$ satisfies
   \[
     a(x_0,x_0,\xi;\hbar)=\chi(V(x_0)+|\xi|^2)+\O(\hbar) . 
   \]
 Upon identifying the kernel of the operator $K_{x_0,0}^\chi$ as the leading term, this concludes the proof.
 \end{preuve}

\bigskip

  \section{The spectral projector in the bulk at microscopic scale}
  \label{sec:spectr-proj-bulk}
  
  This section is devoted to the proof of Theorem
  \ref{thr:micro_limit_kernel} and its consequences.
  
\medskip

Our starting point consists in writing \new{compactly supported} spectral
  functions of the operator $H_{\hbar}=-\hbar^2\Delta+V$ via a
  semiclassical Fourier transform:
  \[
    f(H_{\hbar})= \frac{\chi(H_\hbar)}{\sqrt{2\pi} \hbar} \int
    \old{\hat{\rho}(t/\hbar)}\new{\hat{g}(t)}e^{\frac{\i tH_{\hbar}}{\hbar}}\dd t,
  \]
  where \old{$f:\lambda\mapsto \chi(\lambda) \rho(\hbar^{-1}\lambda)$
    and both $\chi$ and $\hat\rho$ have compact supports}
  \new{$f:\lambda\mapsto \chi(\lambda)g(\hbar^{-1}\lambda)$}. 
 Using the fact that $\chi(H_\hbar) e^{\frac{\i tH_{\hbar}}{\hbar}}$ can be well
  approximated by integral operators \new{for short time $t\in [-\tau,\tau]$}, see Proposition
  \ref{prop:propagator_rem_Cinf}, up to an $\O(\hbar^\infty)$ error,
  \new{provided the support of $\hat{g}$ lies inside $[-\tau,\tau]$, }the integral kernel of this operator has the form:
  \[
\frac{\sqrt{2\pi}}{(2\pi\hbar)^{n+1}}\int
    e^{\i \frac{\phi(t,x,\xi)-y\cdot\xi}{\hbar}}a(x,y,\xi,t;\hbar)\dd \xi ,
  \]
  where  $a$ is a classical symbol and $\phi$ satisfies the following
  Hamilton-Jacobi differential equation:
   \begin{equation}\label{eq:HJ}
    \begin{cases}
      &\partial_t\phi=V(x)+|\partial_x\phi|^2\\
      &\phi|_{t=0}=x\cdot \xi.
    \end{cases}
  \end{equation}
  
  \medskip

One can then obtain the asymptotics of such integrals by applying the
stationary phase method as $\hbar\to0$ using the properties of the
phase $\phi$ and of the symbol $a$ for small $t$. In particular,  we
study in Subsection~\ref{sec:study-phase-function} general properties
of $\phi$, that will be useful in the rest of this article.

The approach described above cannot be directly applied to
the spectral function $\mathds{1}_{(-\infty,\mu]}$, which is not the product of a smooth compactly supported function and the
  Fourier transform of a smooth compactly supported function. 
The idea is to regularize this function by applying a frequency cutoff $\rho_
\hbar$ on scale $\hbar^{-1}$ and consider instead
$ f_{\hbar, \mu}= \vartheta \cdot(\mathds{1}_{(-\infty,\mu]} *\rho_\hbar)$ where
$\vartheta \in C^\infty_c(\R,\R_+)$ is equal to $1$ on $[0,\mu]$. 
 In Subsection~\ref{sec:conv-regul-kern}, we perform the first step of the proof of Theorem \ref{thr:micro_limit_kernel} which consists in obtaining the asymptotics of the regularized kernel associated with such~$ f_{\hbar, \mu}$. 
Then, in Subsection
    \ref{sec:from-regul-kern}, we recover
    $\mathds{1}_{(-\infty,\mu)}(H_{\hbar})$ from its frequency cutoff
    using the Tauberian theorem of Hörmander, concluding the proof of
    Theorem \ref{thr:micro_limit_kernel}. 
    Finally, the probabilistic consequences of Theorem \ref{thr:micro_limit_kernel} for free fermions processes are discussed in Section~\ref{sec:conv-point-proc-1}.

   \subsection{Study of a phase function}
\label{sec:study-phase-function}

\begin{prop}\label{prop:reduction-phase}\new{Let $(\mu,V)$ satisfy
    \ref{hyp:weak} and let $M>\mu$ as in \ref{hyp:weak}.}
  Let $\mathcal{K}\Subset \{V<M\}\times \R^n$. Let
  $\mathcal{K}'\Subset \R^{2n+1}$ be a neighbourhood of
  $\{0\}\times \mathcal{K}$, let $\phi : \mathcal{K}' \to \R$ solve \eqref{eq:HJ} and let
  \[
    \Psi:(t,x,\xi)\mapsto \phi(t,x,\xi)-x\cdot \xi.
  \]
There exists $\tau>0$,
  $\eta\in C^{\infty}([-\tau,\tau]\times \mathcal{K},\R^n)$ with $\det \frac{\dd \eta}{\dd
    \xi}\neq 0$ and $g\in C^{\infty}([-\tau,\tau]\times \R^n,\R)$ such that, for all $(t,x,\xi)\in [-\tau,\tau] \times   \mathcal{K}$, 
  \[
    \Psi(t,x,\xi)=t\big(|\eta(t,x,\xi)|^2+g(t,x)+V(x)\big) , 
  \]
and $g$ has the following Taylor expansion as $t\to0$, 
  \[
    g(t,x)=t^2\frac{|\nabla V(x)|^2}{12}+\O(t^4) . 
  \]
 \end{prop}

\begin{preuve}
  Since $\phi$ solves the Hamilton-Jacobi equation \eqref{eq:HJ}, it is smooth on $\mathcal{K}'$ and, as $t\to0$, 
\begin{equation} \label{Taylor1}
    \phi=x\cdot \xi+t(V(x)+|\xi|^2)+\O(t^2).
  \end{equation}
  In particular,   $\left.\frac{\Psi(t,x,\xi)}{t}\right|_{t=0} = V(x) + |\xi|^2$ is a Morse function of $\xi$. This property is preserved for $t\in[-\tau,\tau]\times \mathcal{K}$  if $\tau>0$ is small enough, that is   $\frac{\Psi}{t}$ admits exactly one non-degenerate critical point with respect to $\xi$, which is a (global) minimum. 
  By the Morse lemma, there exists a smooth change of variables
  $\xi\mapsto \eta$, which is smooth in $(t,x)$ such that
  \[
    \frac{\Psi(t,x,\xi)}{t}=|\eta(t,x,\xi)|^2+g(t,x)+V(x)  . 
  \]
 Using \eqref{eq:HJ}, we can iterate the Taylor expansion of $\phi$ as
 $t\to0$. We obtain
  \[\begin{aligned}
    \phi(t,x,\xi)
    &= x\cdot \xi+t(V(x)+|\xi|^2)+ t^2 \xi \cdot \nabla V(x) + \tfrac{t^3}{3} | \nabla V(x)|^2 +   \tfrac{2}{3}  t^3 \langle \xi , {\rm Hess}(V)(x) \xi  \rangle+ \O((|\xi|+t)t^4)   \\
 &   =x\cdot \xi+t\left(\langle \xi+ \tfrac{t}{2} \nabla V(x),{\rm
        Id}+\tfrac{2}{3} t^2{\rm Hess}(V)(x),\xi+ \tfrac{t}{2} \nabla
      V(x)\rangle + V(x) +\frac{|\nabla V(x)|^2}{12}t^2\right)+\O((|\xi|+t) t^4) . 
  \end{aligned}  \]
Note that  in this expansion, all terms of even
power in $t$ are odd with respect to $\xi$, which explains the error term.
This yields as $t\to0$, 
\begin{equation} \label{etaTaylor}
\eta(t,x,\xi)=\left({\rm
      Id}+\frac{t^2}{3}{\rm Hess}(V)(x)\right)\left(\xi+\frac t2 \nabla V(x)\right)+\O(|\xi| t^4)
      \quad\text{and}\quad 
      g(t,x)=\frac{|\nabla
    V(x)|^2}{12}t^2+\O(t^4) .
\end{equation}
 \end{preuve}

\subsection{Convergence of regularised kernels}
\label{sec:conv-regul-kern}

In this section, we obtain asymptotics for a family of regularisations of the operator $
\1_{(-\infty,\mu]}(H_{\hbar})$ which are essentially obtained by smoothing the function $\old{\mu \mapsto
}\1_{[\new{(-\infty,\mu]}}$ using a frequency cutoff at scale $\hbar^{-1}$.
To define this approximation, we introduce the following parameters:

\todo[inline]{Clarified notation}

\begin{notation} \label{regul-kern}
\new{Let $(\mu_0,V)$ satisfy \ref{hyp:weak}, without loss of generality $V\ge 0$ and let $M>0$ be as in \ref{hyp:weak}.} Let 
\begin{itemize}[leftmargin=*]
\item  $\varkappa\in C^{\infty}_c(\R,[0,1])$ where $\varkappa =1$ on
  $[0,M]$.
\item $\vartheta\in C^{\infty}_c((\mu_0,M), \R_+)$ where $\mu_0$ can be fixed at will later on.
\item $\tau>0$ is a small parameter such that, for $t\in [-\tau,\tau]$:
  \begin{itemize}[leftmargin=*]
  \item One can apply Proposition \ref{prop:propagator_rem_Cinf}.
  \item One can apply Proposition \ref{prop:reduction-phase} after
    fixing $\mathcal{K} \Subset \R^{2n}$ which contains
    $\{(x,\xi), \vartheta(V(x)+|\xi|^2)>0\}$.
  \end{itemize}
\item $\rho\in \mathcal{S}(\R,\R_+)$ is even with  $\int_\R
  \rho(\lambda)\dd \lambda = 1$. $\hat{\rho}$ is supported on
  $[-\tau,\tau]$.
\item $\rho_\hbar = \hbar^{-1} \rho(\hbar^{-1} \cdot)$ for $\hbar\in(0,1]$. 
  \end{itemize}
In the sequel, we treat $\mu$ as a parameter and we also let  $\varkappa_{\mu} :=\varkappa \mathds{1}_{[0,\mu]}$ and  $\vartheta_{\mu} :=\vartheta\mathds{1}_{[0,\mu]}$ for $\mu_0 \le \mu<M$.  
We will consider  spectral functions of $H_{\hbar}$ of the form
\begin{equation*}
 f_{\hbar, \mu}: \sigma \mapsto  \vartheta(\sigma)  \int  \varkappa_\mu(\lambda) \rho_\hbar(\lambda-\sigma) \dd \lambda.
\end{equation*}
\end{notation}

By Proposition~\ref{prop:propagator_rem_Cinf} applied to the operator $\vartheta(H_{\hbar})e^{\i t H_{\hbar}/\hbar} $, it holds
\begin{equation} \label{fop}
f_{\hbar, \mu}(H_\hbar) =  \frac{1}{\sqrt{2\pi}\hbar} \int  \varkappa_\mu(\lambda)   
 \big( I_{\hbar,t}^{\phi, a}  + R_{\hbar,t}\big)     e^{ -\i\frac{t\lambda}{\hbar}}  \hat{\rho}(t)\dd \lambda \dd t . 
\end{equation}
The role of $\varkappa$ is to limit the integral above to a compact
set in $\lambda\in\R$.
In particular, the operator  $f_{\hbar, \mu}(H_\hbar)$ has an integral kernel of the form:
\begin{equation} \label{fkernel} 
 f_{\hbar, \mu}(H_\hbar) : (x,y)
 \mapsto \frac{\sqrt{2\pi}}{(2\pi\hbar)^{n+1}}\int
  e^{\i \frac{\phi(t,x,\xi) - \xi\cdot y
   - t\lambda}{\hbar}}a(x,y,\xi,t;\hbar)\hat{\rho}(t)  \varkappa_\mu(\lambda)  \dd \xi \dd \lambda \dd
  t+ R_{\hbar}(x,y),
\end{equation}
 where $\xi \mapsto a(x,y,\xi,t;\hbar)$  has compact support for $t\in [-\tau,\tau]$
and the error $R_{\hbar}$, as well as all its derivatives, are $\O(\hbar^\infty)$ and controlled uniformly for $\mu\in\R_+$ and  locally uniformly for $x,y \in  \{V< M\}$. 

The goal of this section is to apply the stationary phase method to
the integral kernel \eqref{fkernel} in order to obtain the following
asymptotics for the rescaled kernel $K_{x_0,\hbar}^{f_{\hbar,\mu}}$ of
$f_{\hbar,\mu}(H_{\hbar})$, as defined in
 \eqref{eq:Kfxmu}.

\begin{prop}\label{prop:regularised_kernel_bulk}
Let $\mathcal{K}\Subset \R^{2n}$, let $f_{\hbar,\mu}$ be as in \new{Notations}~\ref{regul-kern} and let
\[\mathcal{A} \Subset \{(x_0,\mu) \in \R^n \times \R_+ : V(x_0)\le \mu_0 <\mu\leq M\}.\] 
Then, it holds uniformly for $(x_0,\mu)\in \mathcal{A}$ and $(x,y)\in \mathcal{K}$, 
\[
K^{f_{\hbar,\mu}}_{x_0,\hbar}(x,y)  =  K^{\vartheta_\mu}_{x_0,0} (x,y) +\O(\hbar). 
\]
  \end{prop}
  
  \begin{preuve}
   By  formula \eqref{fkernel}, it suffices to obtain the asymptotics of the rescaled kernel
   \begin{equation*}
   \mathcal{K} \ni (x,y) \mapsto 
    \frac{1}{(2\pi)^{n+\frac12}\hbar}\int
  e^{\i \frac{ \phi(t, x_0+\hbar x ,\xi)- (x_0+\hbar y)\cdot \xi -    t\lambda }{\hbar}}a(x_0+\hbar x,x_0 +\hbar y,\xi,t;\hbar)\hat{\rho}(t) \varkappa_\mu(\lambda)   \dd \xi \dd \lambda \d t
   \end{equation*}
  where $a \in S^0$ is a classical symbol whose principal part is
  given by \eqref{pp}  at time $t=0$ and $p(x,\xi) =|\xi|^2+V(x)$.
Since the phase $\phi$ is smooth for $t\in[-\tau,\tau]$ and
Proposition \ref{prop:reduction-phase} applies, by a Taylor expansion
and making a spherical change of variables $\eta(\xi)=r(\xi)\omega(\xi)$ where $(r,\omega)\in
\R_+ \times S^{n-1}$,  there exists a classical symbol $b\in S^0$ such that for $(x,y) \in\mathcal{K}$, 
\begin{equation} \label{linphase}
\sqrt{2\pi} e^{\i \frac{\phi(t,x_0+\hbar x,\xi)-(x_0+\hbar y)\cdot \xi
    -t\lambda}{\hbar}}a(x_0+\hbar x,x_0+\hbar y,\xi,t;\hbar)
\hat{\rho}(t)\dd \xi
=e^{\i\frac{\psi(t,r,\lambda)}{\hbar}}b(x,y,r\omega,t;\hbar)r^{n-1} \dd r \dd \omega
  \end{equation}
and
\begin{equation*} 
  \psi(t,r,\lambda)=t(r^2+g(t,x_0)+V(x_0)-\lambda).
\end{equation*}
Since $\hat{\rho}$ and $a$ have compact supports,
the function 
 $(r,t) \mapsto b(x,y,r,\omega,t;\hbar)$ also has compact support.
  Moreover, since $\hat{\rho}(0)= 1/\sqrt{2\pi}$  and $\partial_x \phi(0,x,\xi) = \xi$  (by \eqref{eq:HJ} and the change of variable \eqref{etaTaylor}), the principal part of $b$ satisfies at $t=0$, 
\begin{equation} \label{b01}
    b_0(x,y,\eta,0)=e^{\i (x-y)\cdot \eta} \vartheta(r^2+V(x_0)).
  \end{equation}
  In particular, $r$ is bounded away from $0$ in the previous integral, since
  $\vartheta$ is supported inside $(\mu_0,M)$ and $V(x_0)<\mu_0$. 

This implies that uniformly for $(x,y) \in\mathcal{K}$, 
\[
K^{f_{\hbar,\mu}}_{x_0,\hbar}(x,y) 
 =      \frac{1}{(2\pi)^{n+1}\hbar} \int
  e^{\i \frac{\psi(t,r,\omega,\lambda)}{\hbar}} b(x,y,r\omega,t;\hbar) \varkappa_\mu(\lambda)  r^{n-1}   \dd  t \dd \lambda \dd \omega \dd r+ \O(\hbar^{\infty}) .
\]

We apply the stationary phase method to the previous integral in the variables $(r,t) \in \R \times [-\tau,\tau]$ for a fixed $(\lambda, \omega) \in \R_+^* \times S^{n-1}$.
By \eqref{Taylor1}, the equations for the critical point(s) are
\begin{equation} \label{eq:phase1}
\left\{ \begin{aligned}
  \frac{\partial \psi}{\partial r} &= 2 r t=0 \\
  \frac{\partial \psi}{\partial t}  &=   V(x_0)+r^2 -\lambda + O(t) =0 .  
\end{aligned}
\right.
\end{equation}
These equations have the following consequences:
\begin{itemize}
\item Since $r$ is bounded away from $0$ (otherwise $b=0$), any critical point satisfies $t=0$. 
\item If $\lambda \le V(x_0)$, there is no critical
  points of the phase near the support of the symbol $b$; by Proposition~\ref{prop:nspl}, the integral is $\O(\hbar^\infty)$ with the required uniformity. 
\item If $\lambda>  V(x_0)$, there is a
  unique critical point given by $r_\star(\lambda) :=
  \sqrt{\lambda-V(x_0)}$, and we assume that $r_\star$ is bounded away from $0$.
\end{itemize}

We also verify that $\psi|_{t=0}=0$ and the Hessian of the phase at
the critical point, 
$\rm{Hess}\, \psi  = \left(\begin{smallmatrix}
* & 2r_\star \\ 2r_\star & 0
\end{smallmatrix}\right)$, is not degenerate.
Hence, by applying Proposition~\ref{prop:spl}, we obtain uniformly for  $(x,y) \in \mathcal{K}$, 
\[
K^{f_{\hbar,\mu}}_{x_0,\hbar}(x,y)  =   \frac{1}{(2\pi)^n}\int \frac{r_\star(\lambda)^{n-2}}{2} \int_{S^{n-1}} s\big(x,y, r_\star(\lambda)\omega;\hbar\big)  \varkappa_\mu(\lambda)  \dd \omega \dd \lambda + \O(\hbar^{\infty}) ,
\]
where according to \eqref{b01}, 
 $r\mapsto s\big(x,y, r\omega;\hbar\big) $ is a classical symbol with compact support in $\R_+^*$ and principal part
\[
s_0(x,y, \xi) = e^{\i(x-y)\cdot \xi} \vartheta(|\xi|^2+V(x_0)).
\] 
To conclude, we go back to the original variable $\xi =r_\star(\lambda) \omega$. 
We have $\lambda= |\xi|^2+V(x_0)$ and the Jacobian is $\dd \xi =  r_\star^{n-1}(\lambda) 
                              \frac{\dd r_\star(\lambda) }{\d
                              \lambda} \dd \lambda \dd \omega=
                              \frac{r_\star(\lambda)^{n-2}}{2}  \d
                              \lambda \d \omega$
so that 
\begin{equation*}
K^{f_{\hbar,\mu}}_{x_0,\hbar}(x,y)  =   \frac{1}{(2\pi)^n}\int   \big( e^{\i(x-y)\cdot
    \xi} \vartheta(|\xi|^2+V(x_0)) +\O(\hbar) \big)   
 \varkappa_\mu(|\xi|^2+V(x_0) ) \dd\xi + \O(\hbar^{\infty}),
\end{equation*}
where both errors are controlled uniformly for  $(x,y) \in\mathcal{K}$ and $(x_0,\mu) \in\mathcal{A}$.
Since $\varkappa =1$ on $[1,M]$, $V(x_0) \ge 1$ and $\mu\le M$, we identify the leading term as the kernel of the operator $K^{\vartheta_{\mu}}_{x_0,0}= \vartheta_\mu(-\Delta + V(x_0)) $, this concludes the proof.
\end{preuve}


To recover the asymptotics of $\vartheta_{\mu}(H_{\hbar})$ from that
of $f_{\hbar,\mu}(H_{\hbar})$, we treat $\mu$ as a
parameter and we will rely on the following estimate on the
derivative $\partial_{\mu}f_{\hbar,\mu}(H_{\hbar})$.

\begin{lem} \label{lem:pre-tauber}
Let $\rho, \vartheta$ be as in \new{Notations}~\ref{regul-kern} and set $\zeta_{\hbar,\mu}(\lambda) = \vartheta(\lambda) \rho_{\hbar}(\lambda-\mu)$.
There exists a constant $C>0$  so that for all $x_0 \in \{V\le \mu_0\}$, $\mu\in\R_+$ and $\hbar \in (0,1]$, 
\[
\hbar^n \zeta_{\hbar,\mu}(H_\hbar)(x_0, x_0)  \le  C.
\]

 \end{lem}

\begin{preuve}
By writing $\rho_\hbar$ in terms of its Fourier transform and applying Proposition  \ref{prop:propagator_rem_Cinf}, we obtain
  \[
\hbar^n \zeta_{\hbar,\mu}(H_\hbar)(x_0, x_0)  =
\frac{1}{(2\pi)^{n+1/2}\hbar}\int e^{\i \frac{\phi(t,x_0,\xi)-x_0\cdot \xi-\mu
      t}{\hbar}}a(x_0,x_0,\xi,t;\hbar)\hat{\rho}(t)\dd t \dd
    \xi+\O(\hbar^{\infty}),
  \]
  where $a\in S^0$, $(x_0,\xi) \mapsto a(x_0,x_0,\xi,t;\hbar)$ as a fixed compact support for $t\in \supp(\hat{\rho})$, $\hbar\in(0,1]$ and 
 the error (as well as all its derivative) are controlled uniformly for $x_0\in\{V\le \mu\}$ and $\mu\in \R_+$. 
  
We proceed like in the proof of Proposition~\ref{prop:regularised_kernel_bulk}, by decomposing $\eta(\xi) =r\omega$ for $(r,\omega) \in\R_+ \times S^{n-1}$ and applying the stationary phase method in the variables $(r,t)$. According to \eqref{linphase}--\eqref{b01}, 
  \[
\hbar^n \zeta_{\hbar,\mu}(H_\hbar)(x_0, x_0)=
\frac{1}{(2\pi)^{n+1}\hbar}  \int_{S^{n-1}} \int_{[-\tau,\tau]\times\R_+}  e^{\i\frac{\psi(t,r,\omega,\mu)}{\hbar}}b(x_0,r\omega,t;\hbar) r^{n-1} \dd t \dd r \dd \omega +\O(\hbar^{\infty}),
  \]
  where $\psi(t,r,\omega,\mu)=t(r^2+g(x,t)-\mu)$ and $b\in S^0$ is another classical symbol. 

Like in the proof of Proposition~\ref{prop:regularised_kernel_bulk}, the equations for the critical point(s) are given by \eqref{eq:phase1} with $\lambda=\mu$ and 
the parameter $r$ is bounded away from 0. 
In particular, uniformly for $(x_0,\mu)$ in a (small) neighbourhood of
$\{\mu\leq V(x_0)\}$, there is no critical point and by applying Proposition~\ref{prop:nspl}, 
\[
\zeta_{\hbar,\mu}(H_\hbar)(x_0, x_0)  = \O(\hbar^{\infty}).
\]
On the other hand, if $\mu$ is bounded away from $V(x_0)$, then  there is a unique critical point $(r_\star,t) = (\sqrt{\mu-V(x_0)}, 0)$ and applying Proposition~\ref{prop:spl}, it holds 
  \[
\hbar^n \zeta_{\hbar,\mu}(H_\hbar)(x_0, x_0) =
\frac{r_\star^{n-2}/2}{(2\pi)^{n}}    \int_{S^{n-1}} \big( b_0(x_0,r_\star\omega,0)  + \O(\hbar) \big) \dd \omega  +\O(\hbar^{\infty}).
  \]

In both cases, this proves that there is a constant $C$ so that for $x_0 \in \{V\le \mu_0\}$, $\mu\in
\R_+$ and $\hbar \in (0,1]$, 
\[
\hbar^n \zeta_{\hbar,\mu}(H_\hbar)(x_0, x_0) \le C\big(1+  r_\star^{n-2}(\mu) \big) .
\]
\end{preuve}

\subsection{From the regularised kernel to the projection kernel}
\label{sec:from-regul-kern}

To conclude the proof of Theorem \ref{thr:micro_limit_kernel}, it remains to replace $K^{f_{\hbar,\mu}}_{x_0,\hbar}$ by the rescaled kernel of the projection $\Pi_{\mu,\hbar}=\1_{[0,\mu]}(H_\hbar)$. To
this end, we treat the energy level $\mu$ as a parameter and we rely on the following Tauberian theorem. 

\begin{prop}[Theorem B.2.1 in \cite{safarov_asymptotic_1997}]\label{thm:Taub}
Let $N:[0,\infty) \to [0,\infty)$ be a function with  $N(0)=0$, $N'\ge
0$ and at most polynomial growth. Let $\rho$ be a mollifier as in \new{Notations~\ref{regul-kern}}. If 
 $ N'* \rho_\hbar (\lambda) \le 1 + \lambda^\alpha$ for some $\alpha\in\R_+$, then  $\big| N(\lambda)  - N* \rho_\hbar(\lambda) \big| \le C_\rho \hbar (1 + \lambda^\alpha) $ for all $\lambda\in\R_+$. 
 \end{prop}

This statement follows directly from \cite[Theorem
B.2.1]{safarov_asymptotic_1997} by rescaling the mollifier $\rho$ at
scale $\hbar$ as in the Notations~\ref{regul-kern}. 
 Moreover, the \emph{counting function} $N$ is allowed to depend on the parameter $\hbar$ as long as the condition  $ N'* \rho_\hbar (\lambda) \le C( 1 + \lambda^\alpha)$ holds. 

Let us sketch how this result comes in play. 
Recall that $\vartheta_\mu = \vartheta \1_{[0,\mu]}$ where $\vartheta \in C^\infty_c((\mu_0,M),\R_+)$.
Treating $\mu$ as a parameter, we are interested in the kernel of the operator $K^{\vartheta_{\mu}}_{x_0,\hbar}$. To ease notation, let 
\[
K: \mu \mapsto K^{\vartheta_{\mu}}_{x_0,\hbar}(x,y) =  \hbar^n
\vartheta_{\mu}(H_\hbar)(x_0+\hbar x,x_0+\hbar y).
\]
Then
\[ 
K *\rho_{\hbar} (\mu)=\hbar^n\int
\vartheta_{\sigma}(H_{\hbar})(x_0+\hbar x,x_0+\hbar y)
\rho_\hbar(\mu-\sigma) \d \sigma=   \hbar^n
g_{\hbar,\mu}(H_\hbar)(x_0+\hbar x,x_0+\hbar y),
\]
where using that $\rho$ is even, we defined
\begin{equation} \label{convg}
 g_{\hbar,\mu}(\lambda)  =   \vartheta(\lambda) \int \1_{\{\sigma\le  \mu\}} \rho_{\hbar}(\lambda-\sigma) \d \sigma .  
\end{equation}

This function is essentially the same as $f_{\hbar,\mu}$  as appearing in the
Notations~\ref{regul-kern}.

\begin{prop}\label{prop:from_f_to_g}
Let $\mathcal{K}_0\Subset \{V<M\}^2$.
It holds uniformly for $(\mu,x,y)\in
  \R\times \mathcal{K}_0\times \mathcal{K}_0$,
  \[
g_{\hbar,\mu}(H_\hbar)(x,y)= f_{\hbar,\mu}(H_\hbar)(x,y)+\O(\hbar^{\infty}).
    \]
\end{prop}
\begin{preuve}
  One has
  \[
    f_{\hbar,\mu}(\lambda)-g_{\hbar,\mu}(\lambda)=\vartheta(\lambda)\int
    (1-\varkappa(\sigma))\1_{\{\sigma\leq
      \mu\}}\rho_{\hbar}(\lambda-\sigma)\dd \sigma.
  \]
 Since the supports of $\vartheta$ and $1-\varkappa$ are disjoint and $\rho\in \mathcal{S}$, we obtain
 \[
   \|f_{\hbar,\mu}-g_{\hbar,\mu}\|_{L^{\infty}}=\O(\hbar^{\infty}),
 \]
 and moreover $(f_{\hbar,\mu}-g_{\hbar,\mu})$ is supported inside
 $(\mu_0,\mu_1)$ with $\mu_1 < M$.
Using the spectral resolution of $H_\hbar$ for energy $<\mu_1$, by Lemma~\ref{prop:rough_upper_Weyl}, this implies that
\[
\big\| x \mapsto (f_{\hbar,\mu}-g_{\hbar,\mu})(H_\hbar)(x,x) \big\|_{C^0(\mathcal{K}_0)}
\le C \hbar^{-n}    \|f_{\hbar,\mu}-g_{\hbar,\mu}\|_{L^{\infty}}  
\max_{\lambda \le \mu_1} \| v_\lambda^2\big\|_{C^0(\mathcal{K}_0)}
\] 
where $v_\lambda$ denotes the eigenfunction(s) of  $H_\hbar$  with energy $\lambda$.
Moreover, for any cutoff $\chi \in C^\infty_c(\{V<M\}, \R_+)$ such that $\chi \ge \1_{\mathcal{K}_0}$,  by Lemma~\ref{prop:elliptic_reg}, 
 \[
 \| v_\lambda^2\big\|_{C^0(\mathcal{K}_0)} 
 \le  \| v_\lambda \chi\big\|_{C^0(\R^n)}^2 
 \le C \|(1-\Delta)^{\left\lfloor
        \frac{n}{4}\right\rfloor+1}(v_\lambda \chi)\|_{L^2(\R^n)}
 \]
Using the eigenvalue equation, $-H_\hbar v_\lambda = \lambda v_\lambda$, one verifies that for any $\ell\in\N$, there exists a constant $C_\ell>0$ so that for any $\lambda<M$ and $x \in \supp(\chi)$, 
\[
| (1-\Delta)^\ell (v_\lambda \chi) (x) | \le C_\ell \hbar^{-2\ell}  |v_\lambda(x)| .
\]
 This implies that for any $\lambda<M$ and $\ell\in\N$,  
 \[
\| (1-\Delta)^\ell (v_\lambda \chi) \|_{L^2(\R^n)} = \O( \hbar^{-2\ell})
\qquad\text{and}\qquad
 \| v_\lambda^2\big\|_{C^0(\mathcal{K}_0)}  = \O( \hbar^{-(\left\lfloor
        \frac{n}{2}\right\rfloor+2)})
 \]
 We conclude that 
 \[
\big\|  x \mapsto  (f_{\hbar,\mu}-g_{\hbar,\mu})(H_\hbar)(x,x) \big\|_{C^0(\mathcal{K}_0)} = \O(\hbar^\infty) .
 \]
 Since $f_{\hbar,\mu} ,g_{\hbar,\mu} \ge 0$, by Cauchy-Schwarz's inequality, this completes the proof.
\end{preuve}

 In addition,
  \[
    \partial_\mu(K *\rho_{\hbar})(\mu)=\hbar^n\vartheta(H_{\hbar})\rho_{\hbar}(\mu-H_{\hbar})(x_0+\hbar
    x,x_0+\hbar y),
  \]
  which we studied in Proposition~\ref{lem:pre-tauber}. 

  If $\mu \mapsto K^{\vartheta_{\mu}}_{x_0,\hbar}(x,y)$ would be non-increasing, one could directly
  combine Propositions \ref{prop:from_f_to_g}, \ref{lem:pre-tauber}
  and \ref{prop:regularised_kernel_bulk} with the Tauberian theorem to deduce the asymptotics of this kernel. 
  Nevertheless, using the positivity of $
  K^{\vartheta_{\mu}}_{x_0,\hbar}$ on the diagonal \new{as} an additional trick, we obtain the desired result.
\begin{prop} \label{prop:trunckernel}
Let $\vartheta_\mu$ be as in Notations~\ref{regul-kern}.
For any compact sets 
 $\mathcal{A} \Subset \{(x_0,\mu) \in \R^n \times \R_+ : V(x_0)\le\mu_0<\mu \le M\}$ and $\mathcal{K} \Subset \R^n$, there exists a constant $C$ so that
\begin{equation*} 
\max_{(x_0,\mu)\in \mathcal{A}} \max_{(x,y)\in \mathcal{K}\times \mathcal{K}} \left|  K^{\vartheta_{\mu}}_{x_0,\hbar}(x,y)-  K^{\vartheta_{\mu}}_{x_0,0} (x,y) \right| \le C\hbar . 
\end{equation*} 
\end{prop}
\begin{preuve}
Recall that we set $K_\lambda=K^{\vartheta_{\lambda}}_{x_0,\hbar}$ and define for $x_0, x,y \in \R^n$ and $\mu \in \R_+$, 
\begin{align*}
  N^1_{\hbar,\mu}(x) &= \hbar^n\sum_{\lambda
    \leq \mu}\vartheta(\lambda)|v_{\lambda}(x_0+\hbar
               x)|^2
  \\
  N^2_{\hbar,\mu}(x,y) &= \hbar^{n}\sum_{\lambda
    \leq \mu}\vartheta(\lambda)|v_{\lambda}(x_0+\hbar
               x)-v_{\lambda}(x_0+\hbar y)|^2 
\end{align*}
where  $(\lambda,\phi_\lambda)$ are normalised eigenpairs of $H_{\hbar}$.
Using the spectral resolution of operator $H_\hbar$, it holds 
\begin{equation} \label{square}
N^1_{\hbar,\lambda}(x_0+ \hbar x) = K_\lambda(x,x) 
\qquad\text{and}\qquad
N^2_{\hbar,\lambda}(x,y)  = K_\lambda(x,x) + K_\lambda(y,y)- 2 K_\lambda(x,y).
\end{equation}

The \emph{counting functions} $\lambda \mapsto
N^j_{\hbar,\lambda}(x,y)$ for $j\in\{1,2\}$ satisfy all the assumptions of Proposition~\ref{thm:Taub}. 
Indeed, they are non-decreasing, and we verify that if $\mu_0<\mu_1<M$ and $\hbar$ is sufficiently small,  it holds uniformly for all $\mu\in\R_+$, $x_0 \in \{V\le \mu_0\}$ and $x,y \in \mathcal{K}$, 
\[
N^2_{\hbar,\mu}(x,y)  
\le 4 \max_{x_0 \in \{V<\mu_1\}} N^1_{\hbar,\mu}(x_0)
\le 4 \hbar^n \max_{x_0 \in \{V<\mu_1\}} \vartheta(H_\hbar)(x_0 , x_0) <\infty .
\]
The last bound follows from Proposition~\ref{prop:bound_rho_N}. 
From Proposition \ref{lem:pre-tauber}, the convolution
$\partial_\lambda(K*\rho_{\hbar})$ is uniformly bounded. Hence, the same holds for the derivatives of the counting functions $N^j_{\hbar,\lambda}(x,y)$ for $j\in\{1,2\}$. 

Thus, one can apply the Tauberian theorem to $N^1$ and $N^2$, and 
by linearity, we obtain
\begin{equation*} 
\max_{(x_0,\mu)\in \mathcal{A}} \max_{(x,y)\in \mathcal{K}} \left|  K_{\mu}(x,y)-
\int K_{\lambda}(x,y)  \rho_\hbar(\mu-\lambda) \d \lambda \right| \le C\hbar . 
\end{equation*}

On the other hand, according to Proposition \ref{prop:from_f_to_g}, it also holds for $\mu\in\R_+$, 
\[
\int K_{\lambda}(x,y)  \rho_\hbar(\mu-\lambda) \d \lambda
=  \hbar^n g_{\hbar,\mu}(H_\hbar)(x_0+ \hbar x, x_0 + \hbar y) 
=  \hbar^n f_{\hbar,\mu}(H_\hbar)(x_0+ \hbar x, x_0 + \hbar y)  + \O\big( \hbar^\infty)
\]
with the required uniformity. By Proposition~\ref{prop:regularised_kernel_bulk}, we conclude that
$
K^{\vartheta_{\mu}}_{x_0,\hbar}(x,y) 
=   K^{\vartheta_{\mu}}_{x_0,0}(x,y)  + \O(\hbar).
$
\end{preuve}

We are almost done with the proof of Theorem
\ref{thr:micro_limit_kernel}: it remains to add the kernel of a
pseudodifferential operator.
\begin{prop}
Let $f \in C^\infty_c(\R ,\R_+)$ and let $f_\mu =
f\1_{[0,\mu]}$ for $\mu\in\R_+$. For any compact sets 
 $\mathcal{A} \Subset \{(x_0,\mu) \in \R^n \times \R_+ : V(x_0)<\mu < M\}$ and $\mathcal{K} \Subset \R^{2n}$. There exists a constant $C$ so that for $\hbar \in (0,1]$, 
 \begin{equation*} 
\max_{(x_0,\mu)\in \mathcal{A}} \max_{(x,y)\in \mathcal{K}} \left|  K^{f_\mu}_{x_0,\hbar}(x,y)-  K^{f_\mu}_{x_0,0}(x,y) \right| \le C\hbar . 
\end{equation*} 
\end{prop}

\begin{preuve}
Let us choose $\mu_j$ for $j\in\{0,1\}$  so that $V(x_0)< \mu_0 <\mu_1 \le \mu$ for all $(x_0,\mu) \in \mathcal{A}$ and decompose
\[
f\1_{[0,\mu]} = \chi + \vartheta\1_{[0,\mu]} 
\]
where $\chi \in C^\infty_c(\R,\R_+)$ is equal to 0 on $[\mu_1,\infty)$ and  $\vartheta \in C_c^\infty((\mu_0,M),\R_+)$.  
By linearity, it holds for $\hbar\in[0,1]$, 
\[
 K^{f_\mu}_{x_0,\hbar}=   K^{\chi}_{x_0,\hbar}+  K^{\vartheta_\mu}_{x_0,\hbar} .
\]
Moreover, according to Propositions \ref{prop:trunckernel} and
\ref{prop:convergence_smooth_functions_bulk}, 
there exists a constant $C$ depending only on $(\mathcal{A},\mathcal{K})$ so that for all $(x,y) \in \mathcal{K}$, 
\[
\big| K^{\vartheta_\mu}_{x_0,\hbar}(x,y)-K^{\vartheta_\mu}_{x_0,0}(x,y) \big| \le C\hbar
 \qquad\text{and}\qquad
\big|K_{x_0,\hbar}^{\chi}(x,y) - K^{\chi}_{x_0,0}(x,y) \big| \le C\hbar . 
\]
By combining these estimates, this completes the proof. 
\end{preuve}
To conclude with the proof of Theorem \ref{thr:micro_limit_kernel},
one can choose any $f\in C^{\infty}_c(\R,\R^+)$ equal to $1$ on
$[0,\mu]$.

\begin{rem}
Let us comment on the convergence of derivatives for the rescaled kernel at local scales. 
By choosing $f$ equal to $\lambda\mapsto \lambda^{2k}$ on $[0,\mu]$
allows to prove that the kernel of $
  H_{\hbar}^k\Pi_{\hbar}H_{\hbar}^k$ also admits a scaling limit in the bulk which is expressed in terms of the free Laplacian; 
 for any $k\in\N_0$, as $\hbar\to0$, 
   \[
    (-\Delta_x+V(x_0+\hbar x))^k(-\Delta_y+V(x_0+\hbar
    y))^kK^{\mathds{1}_{[0,\mu]}}_{x_0,\hbar}(x,y)
     \to  (-\Delta_x+V(x_0))^k(-\Delta_y+V(x_0))^k  K^{\mathds{1}_{[0,\mu]}}_{x_0,0}(x,y) 
  \]
  locally uniformly on $\R^{2n}$

By  Proposition \ref{prop:elliptic_reg_V}, this implies the convergence of the kernel $(x,y) \mapsto K^{\mathds{1}_{[0,\mu]}}_{x_0,\hbar}(x,y)$
in the local $C^{2k}$ topology (as well as a rate of convergence $\O(\hbar)$). 
  \end{rem}

\subsection{Concentration inequalities for linear statistics}
\label{sec:conv-point-proc-1}

To conclude this Section, we prove the law of large numbers of Theorem \ref{thm:LLN} and the central limit Theorem \ref{thm:clt}. 

Recall that the free fermion point process, denoted by $\X$,  is the
determinantal point process associated with the operator $\Pi_{\hbar,
  \mu} =\mathds{1}_{(-\infty,\mu]}(H_{\hbar})$ and $N= \tr \Pi_{\hbar,
  \mu} $. 
Its (normalized) intensity is denoted $\rho_N$, \eqref{intensity}.
In particular, for any test function $f: \R^n \to \R_+$, 
$ \E \X(f) = \tr\big[ f  \Pi_{\hbar,\mu} \big] = N \int f \d \rho_N$
where we view $f$ as a (unbounded) positive multiplication operator. 
Let us denote
\[
\mathbf{F} = \left\{ f : \R^n \to \R : f\in {\rm Lip}_1 , f(0)=0  \right\}, 
\]
and recall that the Kantorovich distance is $\d_{\rm W}(\nu,\rho) = \sup \big\{ \int f \d(\nu-\rho) : f\in \mathbf{F} \big\}$ for any probability measures $\nu, \rho$ on $\R^n$.

\begin{lem} \label{lem:dos}
 Let $(\mu,V)$ satisfy the hypothesis \ref{hyp:weak} and let $\varrho = Z^{-1}\big(\mu-V\big)_+^{n/2}$ be the corresponding density of states. 
Then, as $\hbar\to0$ (or equivalently  $N\to\infty$), 
 \[
 \d_{\rm W}\big(\rho_N , \varrho) \to 0. 
\]
\end{lem}

\begin{preuve}
  We begin with the elementary identity
  \[
    \forall f\in \mathbf{F}, \forall x\in \R^n,\quad  |f(x)|\leq |x|.
    \]
Let $\delta>0$ be a small parameter and let $\epsilon(x_0)=2\pi\hbar \frac{\omega_n^{-1/n}}{\sqrt{\mu-V(x_0)}}$ for $x_0\in\{V<\mu\}$ as in Theorem \ref{thr:micro_limit_kernel}. 
Let $\mathrm{f}_{\delta}$ be as in Proposition \ref{prop:domination}
and let $f\in \mathbf{F}$. Then, using the spectral resolution of $\Pi_{\hbar,\mu}$, 
\[
  \left|\frac{1}{N}\int_{V(x)\geq \mu+2\delta}
  f(x_0)\Pi_{\hbar,\mu}(x,x)\dd x\right|\leq 2(1+\mu/\delta)\sup_{V(x)\geq
    \mu+2\delta}|f(x)e^{-2\frac{\mathrm{f}_{\delta}(x)}{\hbar}}|,\]
    and
\[
\sup_{f\in \mathbf{F}}\sup_{V(x)\geq
  \mu+2\delta}|f(x)e^{-2\frac{\mathrm{f}_{\delta}(x)}{\hbar}}|\leq \sup_{V(x)\geq
  \mu+2\delta}|x|e^{-2\frac{\mathrm{f}_{\delta}(x)}{\hbar}},
\]
 is finite for every $\hbar \in (0,1]$ and tends to zero exponentially quickly as
$\hbar\to 0$.

On the other hand, by Theorem
\ref{thr:micro_limit_kernel} and Proposition \ref{lem:Weyl_law}, as
$\hbar \to 0$,
\[\begin{aligned}
  \left|\frac{1}{N}\int_{\{V(x)\leq \mu-\delta\}}f(x)\Pi_{\hbar}(x,x)\dd x-
  \int_{\{V(x)\leq \mu-\delta\}}f(x)\varrho(x)\dd
  x\right|
  &\leq 
\int_{\{V(x)\leq \mu-\delta\}} |x|   \left| \frac{K_{x,\epsilon(x)}(0,0)}{N\epsilon(x)^n} -
  \frac{(2\pi \hbar)^n}{Z\omega_n\epsilon(x)^n} \right|\dd
  x
  \\
  & \to 0 
\end{aligned}\]
where we used that $K_{x,\epsilon(x)}(0,0) \to 1$ uniformly for all $x\in \{V\leq \mu-\delta\}$ which is a compact.

The function $\varrho$ is uniformly bounded, so that
\[
  \limsup_{\delta\to 0}\;\sup_{f\in \mathbf{F}}\int_{\{\mu-\delta\leq V(x)\leq
    \mu+2\delta\}}f(x)\varrho(x)\dd x=0.\]
Similarly, by Proposition \ref{prop:bound_rho_N}, $\rho_N$ is uniformly bounded on
compact sets of $\{V<M\}$, so that
\[
  \limsup_{\delta\to 0}\;\sup_{f\in \mathbf{F}}\;\sup_{\hbar\in(0,1]}\int_{\{\mu-\delta\leq V(x)\leq
    \mu+2\delta\}}f(x)\rho_N(x)\dd x=0.
\]
By combining these estimates, this concludes the proof.
\end{preuve}

As consequences of Lemma~\ref{lem:dos}, the (normalized) intensity
$\rho_N$ converges weakly to the density of states~$\varrho$. To
complete the proof of  the law of large numbers, Theorem
\ref{thm:LLN}, we rely on Lemma \ref{lem:dos} and basic concentration bounds for determinantal processes.

\begin{prop} \label{prop:concentration1}
\new{Let $(\mu,V)$ satisfy the hypothesis \ref{hyp:weak}.}
There exists a small constant $c>0$ so that for any $\epsilon>0$ with $\epsilon\le  \hbar^{-1}$, there exists a (non-increasing) constant $C_\epsilon\ge 1$ and
\[
\P\left[ \d_{\rm W}(N^{-1}\X, \rho_N)  \ge \epsilon \right] \le C_\epsilon \exp \left(- cN \epsilon^2 \right) ,
\]
Moreover, we can choose $C_\epsilon = e^{C \epsilon^{-n}}$ for some universal constant $C>0$.
\end{prop}

\begin{preuve}
Let us denote $\widetilde{\X} := \X -\E \X$ for the recentred Fermion
point process. By \eqref{psiXi}, for any $f\in C(\R^n,\R)$, 
\[ \begin{aligned}
\E\big[ e^{\widetilde{\X}(f)} \big]  &= \det\big(\I+ (e^{f}-1)  \Pi_{\hbar,\mu} \big)  e^{-\tr[f  \Pi_{\hbar,\mu}]} \\
&\le \exp\big( \tr\big[ (e^{f}-1-f)  \Pi_{\hbar,\mu} \big] \big),
\end{aligned}\]
where we used the elementary bound
\[
\det(\mathrm{I}+A)\leq \exp(\tr(A))
\]
valid for all finite-rank operators $A\geq -\mathrm{I}$ (here
$A=\Pi_{\hbar,\mu}(e^f-1)\Pi_{\hbar,\mu}$, is bounded from below by $-\mathrm{I}$).

Moreover, by  Proposition \ref{prop:domination} and using the spectral resolution of $\Pi_{\hbar,\mu}$, it holds for all
$g\in C(\R^n,\R_+)$,
\[
  \tr(g\Pi_{\hbar,\mu})\leq  N(1+2\mu/\delta)\sup_{x\in \R^n}\left(g(x)e^{-2\frac{\rm f_{\delta}(x)}{\hbar}}\right).
\]
Using that $0\leq e^f-1-f\leq f^2 e^f $, we obtain
\[
  \E[e^{\widetilde{\X}(f)}]\leq
  \exp\left(N(1+2\mu/\delta)\sup_{x\in\R^n}\left(f(x)^2 e^{f(x)-2\frac{{\rm
        f_{\delta}}(x)}{\hbar}}\right)\right).
  \]

By rescaling, this shows that for any $\lambda>0$ with $\lambda \ll N \hbar^{-1}$, there exists a constant $C>0$ so that if $\hbar$ is sufficiently small, then 
\[
\sup_{f\in \mathbf{F}}\E\big[ e^{ \lambda N^{-1}\widetilde{\X}(f)} \big]   
\le  \exp \left(\frac C{2N} \lambda^2 \right) . 
\]
Here, we crucially use the fact that elements of $\mathbf{F}$ are
globally $1$-Lipschitz, so that $\sup\left(e^{\lambda
  N^{-1}f-2\frac{\mathrm{f}_{\delta}}{\hbar}}\right)$ is uniformly bounded for
$f\in \mathbf{F}$ and $\lambda<cN\hbar^{-1}$ for $c>0$ small enough.

By Markov's inequality, this yields a Gaussian tail bound for the
random variable $\widetilde{\X}(f)$; for any $\epsilon>0$ and
$\lambda>0$,
\[
  \sup_{f\in \mathbf{F}}\P\big[\widetilde{\X}(f) \ge \epsilon N \big] \le 
  e^{- \epsilon \lambda+ C\lambda^2/2N },
\]
so that, choosing $\lambda= c N\epsilon $ for $c>0$ small enough,
\[
\sup_{f\in \mathbf{F}}\P\big[\widetilde{\X}(f) \ge \epsilon N \big] \le 
e^{- \epsilon \lambda+ c \lambda^2/2N }  = \exp \left(- \tfrac{c}{2} N\epsilon^2 \right)
\]
Upon replacing $f$ with $-f$, we obtain a symmetric inequality.
Hence, it holds for any $\epsilon\le \hbar^{-1}$, 
\begin{equation} \label{gtb}
\begin{aligned}
  \sup_{f\in \mathbf{F}}\P\big[ |\widetilde{\X}(f)| \ge \epsilon N \big] \le 
 2\exp \left(-
  \tfrac c2 N\epsilon^2 \right).
\end{aligned}
\end{equation}

To conclude the proof, it remains to use a compactness argument, but
for this we need to localise the problem in space. Let $\chi\in C_c(\R^n,[0,1])\cap \mathbf{F}$  with $\chi =1$ on the compact $\{V\le M\}$ so that for all $f\in \mathbf{F}$, 
\[
|\widetilde{\X}(f)|  \le |\widetilde{\X}(\chi f)|  + \widetilde{\X}(g) + 2\E\X(g)    , \qquad  g(x)= (1-\chi(x))|x| . 
\]
Note that we used that by linearity
\[
|\widetilde{\X}((1-\chi)f)|  \le \X((1-\chi)|f|)
+ \E \X((1-\chi)|f|) \le \widetilde{\X}(g) + 2\E\X(g)   
\]
as $|f(x)| \le |x|$. 
Since $g\in\mathbf{F}$ and $\supp(\varrho) \subset \{V> M\}$, as in the proof of Lemma~\ref{lem:dos}, $\E\X(g) \to 0$ exponentially quickly as $\hbar\to0$. 
This implies that if $\hbar$ is sufficiently small, then
\[
\P\bigg[ \sup_{f\in\mathbf{F}} |\widetilde{\X}(f)| \ge 4\epsilon N\bigg] 
\le \P\bigg[ \sup_{f\in\mathbf{F}} |\widetilde{\X}(\chi f)| \ge 2\epsilon N\bigg]  + 2 \exp \left(- \tfrac c2 N\epsilon^2 \right).
\]

We now use that the set $\{\chi f : f\in \mathbf{F}\}$ is compact for the uniform topology (by the Arzel\`a--Ascoli theorem) so that for any $\epsilon>0$, there is a finite set  $\mathbf{S}_\epsilon \subset \{\chi f : f\in \mathbf{F}\}$ such that for any $f \in \mathbf{F}$, 
\[
\text{there exists $g\in \mathbf{S}_\epsilon$ so that}\qquad
\big| \widetilde{\X}(\chi f)- \widetilde{\X}(g) \big| \le  N\epsilon,
\]
where we used that the point process $\X$ has $N$ particles. 
Since the estimate \eqref{gtb} is uniform over all Lipschitz functions in $\mathbf{F}$, by a union bound,  this implies that for any $\epsilon>0$ with $\epsilon \ll \hbar^{-1}$, 
\[ \begin{aligned}
\P\bigg[ \sup_{f\in\mathbf{F}} |\widetilde{\X}(f)| \ge 4\epsilon N\bigg]  
 &\le \P\bigg[ \sup_{g\in\mathbf{S}_\epsilon} |\widetilde{\X}(g)| \ge \epsilon N\bigg]   + 2\exp \left(- cN\epsilon^2  \right) \\
 & \le 2 ( |\mathbf{S}_\epsilon| +1 ) \exp \left(- c N\epsilon^2\right) . 
\end{aligned}\]
Since $ \sup_{f\in\mathbf{F}} |\widetilde{\X}(f)| = N  \d_{\rm W}(N^{-1}\X, \rho_N)$, this completes the proof.
The form of the constant $C_\epsilon$ follows from standard continuity arguments. 
\end{preuve}

By combining Lemma~\ref{lem:dos} and
Proposition~\ref{prop:concentration1}, the proof of Theorem
\ref{thm:LLN} is complete.

\medskip

Let us now turn to the proof of Theorem~\ref{thm:clt}. 
This CLT follows from Corollary \ref{clt_dpp} and showing that $\operatorname{var} \X(f)
\to\infty$ as $\hbar\to0$, which holds true in dimension $n\ge 2$.

\medskip

\begin{preuve} [of Theorem~\ref{thm:clt}]
  Let us first remark that, for fixed $f\in C(\R^n,\R)$ with at most
  exponential growth, for $\hbar>0$ small enough, $\X(f)\in L^2$ (as a
  real random variable). Indeed, denoting by $f_+$ and $f_-$ the positive and
  negative parts of $f$, respectively, then $\X(f)=\X(f_+)-\X(f_-)$,
  and $\X$ preserves positivity. Moreover, using the determinantal structure,
  \[
    \mathbb{E}[\X(f_\pm)^2]\leq \tr[\Pi_{\hbar}f_\pm]^2+\tr(\Pi_{\hbar}f_\pm^2),
  \]
  and $\tr(\Pi_{\hbar}g)=\int \Pi_{\hbar}(x,x)g(x)<\infty$ as soon as
  $g \ge 0$ has at most exponential growth,  by Proposition
  \ref{prop:domination} (indeed, ${\rm rank}(\Pi_{\hbar})$ is finite
  and the range consists of eigenfunctions with spacial decay at a
  rate $\hbar^{-1}|x|$ near infinity).

  By Proposition \ref{prop:domination}, if $f$ is supported on $\R^n\setminus\{V\le\mu\}$, then we even have 
$\E  \X(f) = \O(\hbar^\infty)$. 
In this case, by Lemma~\ref{lem:cov},
\[
\operatorname{var} \X(f) \le  \tr\big( f^2 \Pi_{\hbar, \mu}\big)=  \E  \X(f^2)  = \O(\hbar^\infty) . 
\]

Let $\chi \in C^\infty_c(\R^n,[0,1])$ be a cutoff such that  $\chi  \ge \1_{\{V\le\mu\}}$. The previous estimate implies that if $f\in C(\R^n,\R)$, 
\[
 \operatorname{var} \X(f\chi)
= \operatorname{var} \X(f) +  \O(\hbar^\infty) . 
\]

Note that the function $f\chi$ is uniformly bounded on $\R^n$. 
Hence, if we can show that $ \operatorname{var} \X(f\chi)  \to \infty$ as $\hbar\to\infty$, by  Corollary~\ref{clt_dpp}, we obtain as $\hbar\to0$, 
\[
\frac{\widetilde{\X}(f\chi)}{\sqrt{\operatorname{var} \X(f)}} \, \Rightarrow\,  \mathcal{N}_{0,1} . 
\]
where  $\widetilde{\X} := \X -\E \X$. 
Moreover $ \widetilde{\X}(f(1-\chi)) \to 0$ in $L^2$ so that by Slutsky's Lemma, this implies the claim of Theorem~\ref{thm:clt}. So we can assume that $f\in C_c(\R^n,\R_+)$ and then 
\[
\operatorname{var} \X(f) = -  \tfrac12 \tr\big( [f, \Pi_{\hbar,\mu} ]^2\big)
= \tfrac12  \int \big( f(x) - f(y) \big)^2 \big| \Pi_{\hbar,\mu} (x,y) \big|^2 \d x d y . 
\]
For any open set $\Omega \subset \{V<\mu\}$ and any continuous function $\epsilon :\Omega \to (0,1]$, we have a lower-bound, 
\[
\operatorname{var} \X(f)
\ge  \frac{1}{2}  \int \1_{x\in\Omega}  \big( f(x) - f(x+ \epsilon(x) z) \big)^2 \epsilon(x)^{-n} \big|   K_{\epsilon(x), x}(0,z)\big|^2 \d x d z,
\]
where we used the notation \eqref{zoom}. 
By Lusin's theorem, since $\nabla f \in L^2(\Omega)$, by choosing a smaller open subset $\Omega \subset \{V \le \mu\}$, we can assume that $f\in C^1(\Omega)$. 
Then, choosing $\epsilon(x) = 2\pi\hbar \frac{\omega_n^{-1/n}}{\sqrt{\mu-V(x)}}$ and applying Theorem~\ref{thr:micro_limit_kernel}, it holds as $\hbar\to 0$,
\[
\big( f(x) - f(x+ \epsilon(x) z) \big)^2 \epsilon(x)^{-2} \big|   K_{\epsilon, x}(0,z)\big|^2
\to \big( z\cdot \nabla f(x) \big)^2  \frac{\J_{\frac n2}^2(c_n|z|)}{\omega_n |z|^{n}},
\]
for every $x\in \Omega$, 
where we used the expression  \eqref{eq:Kbulk} for the kernel $K_{\rm bulk}$ and $c_n = 2\pi \omega_n^{-1/n}$. 
By Fatou's lemma, this implies that 
\[
\liminf_{\hbar\to0}  \big( \hbar^{n-2} \operatorname{var} \X(f) \big)
\ge  \frac{c_n^2}{2 (2\pi)^n}  \int \1_{x\in\Omega}  \big(\mu-V(x)\big)^{\frac{n-2}{2}}
\big( z\cdot \nabla f(x) \big)^2 \J_{\frac n2}^2(c_n|z|) \frac{\d z}{ |z|^{n}} \d x.
\]

By going to spherical coordinates, this integral factorizes and we obtain
\[
\liminf_{\hbar\to0}  \big( \hbar^{n-2} \operatorname{var} \X(f) \big)
\ge  \frac{n \omega_n}{2 (2\pi)^n}  \int_{\Omega}  \big(\mu-V(x)\big)^{\frac{n-2}{2}}  \big| \nabla f(x) \big|^2 \d x \int_{\R_+} \J_{\frac n2}^2(r) r \d r.
\]

Hence, if $\displaystyle \int_{\Omega}  |\nabla f|^2 >0$, the first integral is positive and using the asymptotics of the Bessel functions as $r\to\infty$, 
\begin{equation} \label{Bessel_asymp}
\J_{\frac n2}(r)  = \sqrt{\tfrac{2}{\pi r}} \cos\big(r-\tfrac{(n+1)\pi}{4})+ \O(r^{-3/2})
\end{equation}
we conclude that 
\begin{equation}\label{eq:bound_inf_variance}
\liminf_{\hbar\to0}  \big( \hbar^{n-2} \operatorname{var} \X(f) \big) = +\infty. 
\end{equation}
\end{preuve}

\section{The spectral projector at the edge at microscopic scale}
\label{sec:edge-asympt-case}

This section is devoted to the proof of Theorem
\ref{thr:micro_edge_kernel}, that is, the asymptotics of the rescaled kernel $ K_{\epsilon, x_0}$ around a point $x_0 \in\R^n$ at the boundary of the droplet; $V(x_0)=\mu$.
We also assume that the point is non-degenerate; $ \nabla V(x_0)\neq 0$. In this case, upon an appropriate scaling $(\epsilon,\mathcal{U})$, the limiting kernel is 
\begin{equation} \label{Kedge}
    K_{\rm edge}:(x,y)\mapsto \1_{(-\infty,0]}(-\Delta+x_1).
\end{equation}
 where $x_1 = x\cdot \mathrm{e}_1$ and   $\mathrm{e}_1$  denotes the first vector of the canonical basis of $\R^n$.  This kernel is given explicitly in terms of Airy and Bessel functions by \eqref{eq:Kedge}.  
 
 \medskip

The method of proof is the same as in  Section \ref{sec:spectr-proj-bulk}; we first prove convergence of a regularised projection kernel and then we apply the Tauberian theorem
(Theorem~\ref{thm:Taub}) to recover the asymptotics of the rescaled
kernel $ K_{\epsilon, x_0}$. The main difference with the proof of
Proposition~\ref{prop:regularised_kernel_bulk}  is that, when dealing
with an oscillatory integral of the form $\int
e^{\i\frac{\phi}{\hbar}}a$, the stationary point $\nabla\phi=0$ is
degenerate (the Hessian is not invertible). This explains why $K_{\rm
  edge}$ involves the Airy function, which is the simplest degenerate
oscillatory integral.

\medskip

We again consider a regularised projection of the form $ f_{\hbar,
  \mu}(H_\hbar)$  as in Notations~\ref{regul-kern} and we denote
\[
  K^{ f_{\hbar,\mu}}_{x_0,\epsilon} : (x,y)\mapsto  \epsilon^{n}   f_{\hbar,\mu}(H_{\hbar})(x_0 + \epsilon\, \mathcal{U}^*  x , x_0+\epsilon\, \mathcal{U}^* y), 
\]
where $\epsilon=  \hbar^{\frac23}|\nabla V(x_0) |^{-\frac13}$ and $\mathcal{U} \in SO_n$ is chosen so that $\mathcal{U}(\nabla V(x_0))=|\nabla V(x_0)| \mathrm{e}_1$ as in Theorem
\ref{thr:micro_edge_kernel}. 

\begin{prop}\label{prop:conv_reg_kern_edge}
  Let $\mathcal{K} \Subset \R^{2n}$ and $\mathcal{A} \Subset \{
  (x_0,\mu)\in \R^{n}\times (0,M), \mu=V(x_0), \nabla V(x_0)\neq 0\}$.
In the above setup, it holds
  \[
   \sup_{x_0 \in \mathcal{A}} \sup_{(x,y)\in \mathcal{K}} \big|
      K^{ f_{\hbar,\mu}}_{x_0,\epsilon} (x,y) - \vartheta(\mu) K_{\rm edge}(x,y)  \big| \le C \hbar^{\frac 13} . 
  \]
\end{prop}

\begin{preuve}
Let us denote by $\vec\gamma=\nabla  V(x_0)$, $\gamma=|\vec\gamma|$ and $\delta =\hbar/\epsilon=\hbar^{1/3} \gamma^{1/3} $.  By \eqref{fkernel}, the rescaled kernel  $K^{f_{\hbar,\mu}}_{\epsilon, x_0}$ is given by
\[
(x,y)
 \mapsto\frac{\epsilon^n}{(2\pi\hbar)^{n+1}}\int
  e^{\i\frac{\phi(t,x_0+ \epsilon \mathcal{U}^* x,\xi)-(x_0+ \epsilon \mathcal{U}^* y)\cdot \xi - \lambda t}{\hbar}} a(x_0 + \epsilon\, \mathcal{U}^*  x , x_0+\epsilon\, \mathcal{U}^* y,\xi,t;\hbar)\hat{\rho}(t)  \varkappa_\mu(\lambda)  \dd \xi \dd \lambda \dd
  t+ \O(\hbar^\infty),
\]
where the error term is uniform for $(x_0,\mu) \in \mathcal{A}$ and  $(x,y)\in \mathcal{K}$. 
As in the proof of Proposition~\ref{prop:regularised_kernel_bulk},  we can rewrite
\[
e^{\i\frac{\phi(t,x_0+ \epsilon \mathcal{U}^* x,\xi)-(x_0+ \epsilon \mathcal{U}^* y)\cdot \xi}{\hbar}} a(x_0 + \epsilon\, \mathcal{U}^*  x , x_0+\epsilon\, \mathcal{U}^* y,\xi,t;\hbar)
    = e^{\i \frac{\Psi(t,x_0,\xi)}{\hbar}} e^{\i \frac{\partial_x\phi(t,x_0,\xi) \cdot \mathcal{U}^* x -\xi \cdot  \mathcal{U}^*y }{\delta}} b(x,y,\xi,t; \delta),
\]
where $\Psi(t,x_0,\xi)$ is as in Proposition~\ref{prop:reduction-phase} and $b$ is a classical symbol with principal part at $t=0$, 
\[
b_{0}(x,y,\xi,0) = \vartheta(\mu+|\xi|^2) .
\]

The critical point for the phase $\Psi(t,x_0,\xi)-\lambda t$ is again given by $(t, \xi, \lambda) = (0,0,\mu)$, but it is degenerate. 
However, according to  Proposition \ref{prop:reduction-phase}, one can rewrite the phase as
\[
 \Psi(t,x_0,\xi) = t\big(|\eta(t,x_0,\xi)|^2+g(t,x_0)+\mu\big) ,  \qquad  \text{where}\quad g(t,x_0)=t^2\frac{\gamma^2}{12}+\O(t^3)
\]
and $\eta(t,x_0,\xi) = \xi + t \vec{\gamma}/2 + O(t^2)$. 
We make a change of variable $\xi \leftarrow \eta$ and use that the Jacobian $|\frac{\d \xi}{\d \eta}| =1+ \O(t)$ is a smooth non-vanishing function for $t\in \supp(\hat{\rho})$. 
Then $\sqrt{2\pi}\hat{\rho}(t)b(x,y,\xi,t;\delta) |\frac{\d \xi}{\d
  \eta}| =  c(x,y,\eta, t;\delta)$ is again a classical symbol, compactly supported in $(t,\eta)$. 
Making a change variable $\lambda \leftarrow (\mu-\lambda)$ and
letting  $\chi_1=\varkappa(\mu-\cdot)$ (according to the Notations~\ref{regul-kern}, the cutoff $\chi_1 \in C^\infty_c(\R,\R_+)$ can be chosen independent of $\mu$ and is equal to $1$ on a neighborhood of $0$), this implies that for $(x,y) \in \mathcal{K}$, 
\[
\K^\mu_{\epsilon, x_0}(x,y)
 = \frac{\epsilon^n}{(2\pi\hbar)^{n+1}}\int \1_{\lambda\ge 0}
  e^{\i  \frac{t \psi(\eta, t, \lambda) }{\hbar}} e^{\i  \frac{\omega(x,y,t,\eta) }{\delta}} c(x,y,\eta,t;\delta) \chi_1(\lambda) \dd \eta \dd \lambda \dd
  t+ \O(\hbar^{\infty})  ,
\] 
where the error term is $C^\infty$ with the required uniformity and the phases are given by
\[
\omega (x,y,t,\eta) =  \partial_x\Psi(t,x_0,\xi(t,\eta)) \cdot \mathcal{U}^* x +\xi(t,\eta)\cdot \mathcal{U}^*(x-y)
\qquad\text{and}\qquad 
\psi(\eta, t, \lambda) = |\eta|^2+g(t,x_0) +\lambda . 
\]
Moreover, the classical symbol $c$ has principal part at $t=0$, for $x,y\in\mathcal{K}$, 
\[
c_{0}(x,y,\eta,0) = \vartheta(\mu+|\eta|^2) .
\]

We claim that  this integral can be localized to the set $\big\{ |t| \le \delta^{1-\alpha}, |\eta| \le  \delta^{1-\alpha} \big\}$ for any small $\alpha \in (0,1)$  up to an arbitrary small error.  
Namely if $\chi_2  \in C^\infty_c(\R^{n+1} ,\R_+) $ is a cutoff which equals to 1 on the unit ball $B_{n+1}$, we will show that the integral
\[
J_{\hbar} = \int   e^{\i  \frac{t \psi(\eta, t, \lambda) }{\hbar}} e^{\i  \frac{\omega(x,y,t,\eta) }{\delta}}  c(x,y,\eta,t;\delta) (1-\chi_2(\delta^{\alpha-1}(t,\eta)))  \dd \eta \dd t
\]
is $\O(\hbar^\infty)$. 
This  relies on the fact that for $t\in \supp(\hat{\rho})$ and $\lambda \in[0,\mu] $,
 \[
 \partial_t[ t \psi(\eta, t, \lambda)] \ge c( |t|^2+|\eta|^2)
 \qquad\text{and}\qquad 
 \partial_t^2[ t \psi(\eta, t, \lambda)] =O(t)  .
\]
Let  $\mathcal{D}_t =  \partial_t\big(\frac{\cdot}{\partial_t[ t \psi(\eta, t, \lambda)]} \big) $. The previous bounds imply that  if $ (t,\eta) \notin \delta^{1-\alpha} B_{n+1}$, then for any $k\in \N$, 
\[
\left| \mathcal{D}_t^k \big(e^{\i  \frac{\omega(x,y,t,\eta) }{\delta}} c(x,y,\eta,t;\delta)(1-\chi(\delta^{\alpha-1}(t,\eta))) \big) \right| \le C_k \delta^{-k(3-2\alpha)} =\O_k\big( \hbar^{-k(1-2\alpha/3)} \big).
\]
Hence, repeated integration by parts shows that for any $k\in\N$, 
\[\begin{aligned}
J_{\hbar} & = (\i \hbar)^k\int   e^{\i  \frac{t \psi(\eta, t, \lambda) }{\hbar}}  \mathcal{D}_t^k \big(e^{\i  \frac{\omega(x,y,t,\eta) }{\delta}} c(x,y,\eta,t;\delta)(1-\chi_2(\delta^{\alpha-1}(t,\eta))) \big)\dd \xi \dd \lambda \dd  t \\
& = \O_k(\hbar^{2k\alpha/3}) .
\end{aligned}\]

This proves that uniformly for all $(x_0,\mu) \in \mathcal{A}$ and $(x,y)\in \mathcal{K}$, 
\[
\K^\mu_{\epsilon, x_0}(x,y)
 = \frac{\epsilon^n}{(2\pi\hbar)^{n+1}}\int \1_{\lambda\ge 0}
  e^{\i  \frac{t \psi(\eta, t, \lambda) }{\hbar}} e^{\i  \frac{\omega(x,y,t,\eta) }{\delta}} c(x,y,\eta,t;\delta)  \chi_1(\lambda) \chi_2(\delta^{\alpha-1}(t,\eta)) \dd \eta \dd \lambda \dd  t+ \O(\hbar^{\infty})  . 
\] 

We can now perform a Taylor expansion of the two phases. By Proposition \ref{prop:reduction-phase}, one has
 \[ \begin{aligned}
 \psi(  \gamma^{1/3} \mathcal{U}^* \eta, t \gamma^{-2/3}, \lambda)  & =  \gamma^{2/3} \big( |\eta|^2+ \frac{t^2}{12} \big) +\lambda + \O(t^3) \\
 t \gamma^{-2/3} \psi(  \gamma^{1/3} \mathcal{U}^*\eta, t \gamma^{-2/3}, \gamma^{2/3} \lambda)    & =
 \tfrac13\big(\tfrac t2 +\eta_1 \big)^3+ \tfrac13\big( \tfrac t2 -\eta_1 \big)^3 +  t \big( |\eta^\perp|^2
+ \lambda   \big) + \O(t^4),
\end{aligned} \]
where we decompose $\eta= (\eta_1, \eta^\perp)$.
In addition, by Proposition~\ref{prop:reduction-phase}, 
$\partial_x\Psi(t,x_0,\xi) = t \vec{\gamma}  + \O(t^2)$ and  $\xi(t,\eta)= \eta- \frac{t}{2} \vec\gamma +\O(t^2) $   so that 
\[ \begin{aligned}
\omega (x,y,t,\eta)  & = \xi(t,\eta)\cdot \mathcal{U}^*(x-y) + \partial_x\Psi(t,x_0,\xi(t,\eta)) \cdot \mathcal{U}^* x \\
& =   \mathcal{U} \eta \cdot (x-y) + \frac t2 \mathcal{U}  \vec{\gamma} \cdot (x+y)+ \O(t^2).
\end{aligned}\]
Since $  \mathcal{U} \vec{\gamma} = \gamma \mathrm{e_1}$, this implies that with the same scaling, 
\[
\omega (x,y, \gamma^{-2/3}t,  \gamma^{1/3}   \mathcal{U}^*\eta)= \gamma^{1/3} \big( \eta^\perp \cdot (x-y) + x_1(\tfrac t2 +\eta_1) + y_1( \tfrac t2 -\eta_1) \big)+ \O(t^2).
\]

Let us also decompose $x=(x_1,x^{\perp})$ and $y=(y_1,y^{\perp})$.
These expansions imply that if we make a change of variables
\[
u =\delta^{-1} (\gamma \tfrac t2 +\eta_1), 
\quad  
v = \delta^{-1} ( \gamma  \tfrac t2 -  \eta_1), 
\quad
z =  \delta^{-1} \eta^\perp 
\quad\text{and}\quad
s = \delta^{-2} ( \lambda  + |\eta^\perp|^2) , \]
then we can rewrite
\begin{equation} \label{phase_expand}
\begin{aligned}
  e^{\i  \frac{t \psi(\eta, t, \lambda) }{\hbar}} e^{\i  \frac{\omega(x,y,t,\eta) }{\delta}} c(x,y,\eta,t;\delta) 
 =  \exp\Big( \i ( \tfrac{u^3}{3} + (x_1 + s) u
+  \tfrac{v^3}{3} + ( y_1 + s) v+  \i  z\cdot (x^\perp-y^\perp) \Big) 
f(x,y,u,v,z,s;\delta)
\end{aligned}
\end{equation}
Note that we used in a crucial way that the errors coming from the expansions of the phases are  given
$\O(\hbar^{-1}t^4)= \O\big( (\frac{u+v}{2})^4\delta\big) $ and $\O(\delta^{-1}t^2)= \O\big( (\frac{u+v}{2})^2\delta\big)$, so that $f$ is again a classical symbol with constant principal part;
\[
f_0(x,y,u,v,z,s) = c_0(x,y,0,0) = \vartheta(\mu) . 
\]

Hence making the  change of variables $(t,\eta, \lambda) \mapsto (u,v,z,s)$ as above (whose Jacobian is given by $\hbar \delta^{n}$) and using \eqref{phase_expand}, we obtain 
\begin{equation}  \label{Airy2}
\begin{aligned}
\K^\mu_{\epsilon, x_0}(x,y)
 =  \frac{1}{(2\pi)^{n+1}}
 \int  & \1_{s\ge|z|^2}\, e^{\i \big( \frac{u^3}{3} + (x_1 + s) u+  \frac{v^3}{3} + (y_1 + s) v+  \i z \cdot (x^\perp -y^\perp) \big)}  \\
&\quad f(x,y,u,v,z,s;\delta)\chi_{3, \delta}(s,u,v,z)  \dd u \dd v \dd z \dd s  + \O(\hbar^{\infty}) 
\end{aligned}
\end{equation}
where the cutoff
\[
\chi_{3, \delta}(s,u,v,z) = \chi_1\big(\delta^{2} (s - |z|^2) \big) \chi_2(\delta^{\alpha} (\tfrac{u+v}{\gamma}, \tfrac{u-v}{2},z )\big) .
\]

Since the cutoffs are arbitrary, we can assume that 
\[
\chi_{3, \delta}(s,u,v,z) = \chi_1\big(\delta^{2} (s - |z|^2) \big) 
\chi_1(\delta^{\alpha} u) \chi_1(\delta^{\alpha} v) \chi_1(\delta^{\alpha} |z|) .
\]

Let us denote $\mathcal{D} = - \partial_{uv}\big(\frac{\cdot}{(u^2+s+1)(v^2+s+1)} \big) $ and observe that performing repeated integrations by parts with respect to $(u,v)$, for any smooth function $g :\R^2 \to \R$ with $\| \partial_u^j \partial_v^\ell g \|_{L^\infty} <\infty$ for all $j,\ell \in \N_0$, it holds for any $k\in\N$,  
\[ \begin{aligned}
&\int e^{\i \big( \frac{u^3}{3} + (x_1 + s) u+  \frac{v^3}{3} + (y_1 + s) v \big)} g(u,v) \chi_1(\delta^{\alpha} u) \chi_1(\delta^{\alpha} v) \d u \d v  \\
&\qquad\qquad = \int  e^{\i \big( \frac{u^3}{3} + (s+1)u+  \frac{v^3}{3} + (s+1)v \big)}  
 \mathcal{D}^k\big( e^{\i (u(x_1-1)+v(y_1-1))} g(u,v)  \chi_1(\delta^{\alpha} u) \chi_1(\delta^{\alpha} v)  \big) \dd u \dd v  \\
 & \qquad\qquad 
 =  \O_k\bigg( \int    \frac{ \d u \d v  }{(u^2+s+1)^k(v^2+s+1)^k} \bigg) 
 =\O_k\bigg( \frac{1}{(1+s)^{2k}}\bigg) . 
\end{aligned}\]

These estimates imply that we can localize the integral \eqref{Airy2} on $\{s\le \delta^{-\alpha} \}$ and expand the symbol $f$ with respect to $\delta$; using that $f_0 = \vartheta(\mu)$ we obtain for any $k\in\N$,  
\[ \begin{aligned}
\K^\mu_{\epsilon, x_0}(x,y)
 & =  \frac{\vartheta(\mu)}{(2\pi)^{n+1}}
 \int  \1_{s\ge|z|^2}\, e^{\i \big( \frac{u^3}{3} + (x_1 + s) u+  \frac{v^3}{3} + (y_1 + s) v+  \i z \cdot (x^\perp -y^\perp) \big)}  \chi_{3, \delta}(s,u,v,z) \chi_1(\delta^{\alpha} s)  \dd u \dd v \dd z \dd s  \\
 &\qquad +  \O_k\bigg(\delta \int \1_{s\ge|z|^2}   \frac{\d z \d s }{(1+s)^{2k}}  \bigg) + \O(\hbar^\infty).
\end{aligned}\]
Moreover, $ \chi_1\big(\delta^{2} (s - |z|^2) \big)  \chi_1(\delta^{\alpha} |z|)  =1$ if  $|z|^2 \le s \le C\delta^{-\alpha}$ and $\delta$ is small enough, so that the leading term on the RHS of \eqref{Airy2} factorizes, 
\[ \begin{aligned}
\K^\mu_{\epsilon, x_0}(x,y) = \vartheta(\mu)
\int_0^\infty \chi_1(\delta^{\alpha} s) & \bigg(  \frac{1}{2\pi} \int e^{\i ( \frac{u^3}{3} + (x_1 + s) u)} \chi_1(\delta^{\alpha} u)  \d u \bigg)
\bigg(  \frac{1}{2\pi} \int  e^{\i \frac{v^3}{3} + (y_1 + s) v}  \chi_1(\delta^{\alpha} v) \d v \bigg) \\
& \bigg(  \frac{1}{(2\pi)^{n-1}} \int_{\sqrt{s}\ge  |z|} e^{  \i z \cdot (x^\perp -y^\perp)} \d z  \bigg) \d s + \O(\delta) . 
\end{aligned}\]

Finally, by \eqref{Picomputation} and Lemma~\ref{lem:Airy},  this implies that for any $k\in\N$, 
\[
\K^\mu_{\epsilon, x_0}(x,y) = \vartheta(\mu)
\int_0^\infty \chi_1(\delta^{\alpha} s)\Big(  \Ai(x_1+s) +\O_k\big( \tfrac{\delta^{2\alpha k}}{(s+1)^k} \big) \Big) 
\Big(  \Ai(y_1+s) + \O_k\big( \tfrac{\delta^{2\alpha k}}{(s+1)^k} \big) \Big) 
\frac{  \J_{\frac{n-1}2}(\sqrt{s}|x^\perp-y^\perp|)}{(2\pi|x^\perp-y^\perp|)^{\frac {n-1}2}}  s^{\frac{n-1}4}   \d s + \O(\delta) 
\]
uniformly for all $(x,y)\in \mathcal{K}$. 
Using the uniform bounds  for $\nu \ge 0$,
$\max_{r\ge 0}\frac{ |\J_{\nu}(\sqrt{s}r)|}{r^\nu}  \le C_\nu s^{\nu/2}$,
$\| \Ai \|_{L^\infty(\R)} <\infty $
and that for any $r\in\R$, 
\begin{equation} \label{Ai_control}
 \int_{0}^\infty |\Ai(r+s)| s^{\nu} \d s  <\infty \quad\text{is of order $\O_\nu(r^{-\infty})$ as $r\to\infty$,}
\end{equation}
(cf.~\eqref{Ai_tail}) we conclude that
\[
\K^\mu_{\epsilon, x_0}(x,y) = \vartheta(\mu) \int_0^\infty  \Ai(x_1+s)  \Ai(y_1+s)\frac{  \J_{\frac{n-1}2}(\sqrt{s}|x^\perp-y^\perp|)}{(2\pi|x^\perp-y^\perp|)^{\frac {n-1}2}}  s^{\frac{n-1}4}   \d s + \O(\delta) ,
\]
where the error term is uniform for all $(x_0,\mu) \in \mathcal{A}$ and $(x,y)\in \mathcal{K}$.
Up to the factor $\vartheta(\mu)$, we identify that this kernel is exactly $K_{\rm edge}(x,y)$ and as $\delta= \O(\hbar^{\frac13})$, this completes the proof. 
\end{preuve}

Like in Section~\ref{sec:from-regul-kern}, we may use the asymptotics
of Proposition~\ref{prop:conv_reg_kern_edge} and the Tauberian theorem~\ref{thm:Taub} in order to obtain the edge asymptotics of the
rescaled projection kernel $ \Pi_{\hbar,\mu}$. This application of the
Tauberian method is more subtle because the counting function changes
regimes precisely at $\mu=V(x_0)$.

Recall that $\vartheta_{\lambda} =\vartheta \mathds{1}_{[0,\lambda]}$  and let us denote 
\[
N_{\epsilon, \lambda}^1(x) :=  \epsilon^{n} \vartheta_{\lambda}(H_\hbar)(x,x) , \qquad \lambda\in \R_+  , x\in\R^n. 
\]
Pay attention that the edge-scaling is different from that of Section~\ref{sec:from-regul-kern} and $\epsilon = \O(\hbar^{\frac23})$. 
The derivative of this function (with respect to $\lambda$) satisfies for $\lambda\in\R_+$, 
\begin{equation} \label{zeta2} 
\int N_{\epsilon, \sigma}^{1\, \prime}(x) \rho_\hbar(\lambda-\sigma) \d \sigma=
 \int N_{\epsilon, \sigma}^1(x) \rho_\hbar'(\lambda-\sigma) \d \sigma=
\epsilon^n   \zeta_{\hbar,\lambda}(H_\hbar)(x,x) , 
 \qquad\text{where}\quad 
 \zeta_{\hbar,\lambda} = \vartheta(\cdot) \rho_{\hbar}(\cdot-\lambda).
\end{equation} 
We expect that $\zeta_{\hbar,\lambda}(H_\hbar)\new{(x,x)}= \O(\hbar^\infty)$ for $\lambda<V(x)$ and  this quantity becomes relevant when $\lambda=V(x)$.  
In this case, by adapting the proof of Proposition~\ref{prop:conv_reg_kern_edge}, the kernel of $\zeta_{\hbar,\lambda}(H_\hbar)$ can be controlled appropriately, locally uniformly (at scale $\epsilon$). We obtain the  following bounds;

\begin{prop} \label{prop:zeta_edge}
Let $\vartheta \in C^\infty_c(\R^n,\R_+)$ be a cutoff and for $c\ge 1$, let 
\begin{equation} \label{cond_zeta}
\mathcal{A}_\hbar\Subset \big\{ x\in \R^n , \lambda \in [0, M) : V(x) \ge \lambda - c \hbar^{\frac 23} , \nabla V(x) \neq 0\big\}
\end{equation}
be any compact set. 
There exists a constant $C>0$ so that for all $(x,\lambda) \in \mathcal{A}_\hbar$,
\[
 \hbar^{\frac{2(n+1)}{3}}  \zeta_{\hbar,\lambda}(H_\hbar)(x,x) \le C. 
\]
\end{prop}

\begin{preuve}
By Proposition~\ref{prop:propagator_rem_Cinf}, it holds for $x\in\R^n$, 
  \[
 \zeta_{\hbar,\lambda}(H_\hbar)(x,x) =
\frac{1}{(2\pi)^{n+1/2}\hbar^{n+1}}\int e^{\i \frac{\phi(t,x,\xi)-x\cdot \xi-\lambda
      t}{\hbar}}a(x,\xi,t;\hbar)\hat{\rho}(t)\dd t \dd
    \xi+\O(\hbar^{\infty}),
  \]
  where $a$ is a classical symbol with principal part 
  $a_0(x,\xi,0) =\vartheta(|\xi|^2+V(x))$ at $t=0$,  the error term is independent of $\lambda\in \R_+$ and locally uniform. 
  According to  Proposition~\ref{prop:reduction-phase}, the phase satisfies 
  \[
\phi(t,x,\xi)-x\cdot \xi-\lambda t
  = t (|\eta|^2 + g(t,x) + V(x) -\lambda),
  \]
  where $ g(t,x)=t^2\frac{|\nabla V(x)|^2}{12}+\O(t^4)$. Hence, by a change of variable $\xi \leftarrow \eta$, we can rewrite 
    \[
 \zeta_{\hbar,\lambda}(H_\hbar)(x,x)=
\frac{1}{(2\pi\hbar)^{n+1}}\int e^{\i \frac{t}{\hbar}(|\eta|^2 + g(t,x) + V(x) -\lambda)} b(x,\eta,t;\hbar)\dd t \dd \eta +\O(\hbar^{\infty}),
  \] 
  where $ b(x,\eta,t;\hbar) = \sqrt{2\pi}\hat{\rho}(t)a(x,\xi,t;\delta) |\frac{\d \xi}{\d \eta}|$  is again a classical symbol. 
By assumptions $V(x) \ge \lambda - c \hbar^{\frac 23}$, so that exactly as in the proof of Proposition~\ref{prop:conv_reg_kern_edge}, we can localize this integral in $(t,\eta)$ at scale $\hbar^{1/3-\alpha}$ for any small $\alpha>0$ up to an error which is $\O(\hbar^\infty)$. This means that for any cutoff $\chi  \in C^\infty_c(\R ,\R_+) $ which equals to~1 on $[-1,1]$, it holds 
    \[
 \zeta_{\hbar,\lambda}(H_\hbar)(x,x)=
\frac{1}{(2\pi\hbar)^{n+1}}\int e^{\i \frac{t}{\hbar}(|\eta|^2 + g(t,x) + V(x) -\lambda)} b(x,\eta,t;\hbar) \chi\big(\hbar^{\alpha-1/3}t\big) \chi\big(\hbar^{\alpha-1/3}|\eta|\big)\dd t \dd \eta +\O(\hbar^{\infty}) 
  \] 
  uniformly for $(x,\lambda) \in \mathcal{A}_\hbar$. 
  Let $\gamma(x) = |\nabla V(x)|/2$. By assumption this function is bounded uniformly from above and below  on $\mathcal{A}_\hbar$ and by making a change of variables $\hbar^{-\frac13}\gamma^{\frac 23} t \leftarrow t $ and $ \hbar^{-\frac 13} \gamma^{-\frac 13} \eta \leftarrow \eta$, we obtain 
   \[
  \hbar^{\frac{2(n+1)}{3}}   \zeta_{\hbar,\lambda}(H_\hbar)(x,x)=
\frac{\gamma(x)^{\frac{n-2}{3}}}{(2\pi)^{n+1}}\int 
e^{\i \big(\frac{t^3}{3} + t |\eta|^2  + \frac{t(V(x) -\lambda)}{(\gamma(x)\hbar)^{2/3}}\big)} 
  f(x,t,\eta; \hbar^{\frac13})
\chi\big(\hbar^{\alpha} \gamma^{-\frac 23}t\big) \chi\big(\hbar^{\alpha} \gamma^{\frac 13}|\eta|\big)\dd t \dd \eta +\O(\hbar^{\infty}) 
  \] 
  where
\[  
  f(x,t,\eta; \hbar^{\frac13})
= e^{\i\O(t^4\hbar^{1/3})}  b(x, \gamma^{\frac 13}\hbar^{\frac 13} \eta, \gamma^{-\frac 23} \hbar^{\frac 13}t;\hbar)  
= \vartheta(V(x)) + \O(\hbar^{\frac 13- 4\alpha}) 
  \]
  uniformly for $x\in \mathcal{A}_\hbar$ and $t,|\eta| \le C \hbar^{- \alpha}$. 

If we further let $\eta = r \omega$ where $(r,\omega)\in \R_+\times S^{n-1}$, this implies that for some $\theta(n)\in \N$, 
     \[ \begin{aligned}
 \hbar^{\frac{2(n+1)}{3}}  \zeta_{\hbar,\lambda}(H_\hbar)(x,x)    =
\frac{\gamma(x)^{\frac{n-2}{3}}  \vartheta(V(x))}{(2\pi)^{n+1}} \int_{S^{n-1}\times\R_+} 
 \chi\big(\hbar^{\alpha} \gamma^{\frac 13} r \big)  r^{n-1}  \int   e^{\i \big(\frac{t^3}{3} + t r^2  + \frac{t(V(x) -\lambda)}{(\gamma(x)\hbar)^{2/3}}\big)} \chi\big(\hbar^{\alpha} \gamma^{-\frac 23}t\big) \dd t \dd \omega\d r \\
 + \O\big(\hbar^{\frac 13- \theta \alpha}\big) ,
 \end{aligned} \]  
where the error term is uniform for  $(x, \lambda) \in \mathcal{A}_\hbar$. 
Since the cutoff  $\chi\in C^\infty_c(\R,\R_+)$  equals to 1 on a neighborhood of 0, by Lemma~\ref{lem:Airy}, it holds uniformly for $u\in\R_+$  and $(x,\lambda) \in \mathcal{A}$, 
\[
\frac{1}{2\pi}\int   e^{\i \big(\frac{t^3}{3} + t u\big)} \chi\big(\hbar^{\alpha} \gamma^{-\frac 23}t\big)\dd t
=  \Ai(u)+ \O(\hbar^\infty) . 
\]

By \eqref{Ai_control}, we conclude that there exists a constant $C>0$ so that 
\[
\sup_{\hbar\in(0,1]} \max_{(x,\lambda) \in \mathcal{A}} \hbar^{\frac{2(n+1)}{3}}  \zeta_{\hbar,\lambda}(H_\hbar)(x,x) \le  C .
 \]
This completes the proof. 
\end{preuve}

Let us now introduce again the counting functions
\begin{align*}
  N_{\epsilon,\lambda}^1(x):=&\epsilon^n\sum_{\sigma\leq
  \lambda}\vartheta(\sigma)|v_{\sigma}(x)|^2\\
  N_{\epsilon,\lambda}^2(x):=&\epsilon^n\sum_{\sigma\leq \lambda}\vartheta(\sigma)|v_{\sigma}(x)-v_{\sigma}(y)|^2,
\end{align*}
where $\epsilon=\hbar^{\frac 23}|\nabla V(x_0)|^{-\frac 13}$.

Proposition~\ref{prop:zeta_edge} has the following consequence for the
diagonal counting function.

\begin{corr} \label{corr:N1} 
  Let $\mathcal{K} \Subset \R^{n}$ and $\mathcal{A} \Subset
  \{(x_0,\mu)\in \R^n\times (0,M), V(x_0)=\mu, \nabla V(x_0) \neq 0\}$.
Then
\[
\max_{(x_0,\mu) \in \mathcal{A}}\max_{x\in\mathcal{K} }\big| N_{\epsilon, \mu}^1(x_0+\epsilon x) - N_{\epsilon, \cdot}^1*\rho_\hbar|_\mu(x_0+\epsilon x) \big|  =\O(\hbar^{\frac13}). 
\]
\end{corr}

\begin{preuve} We work under general assumptions. Let $N$ be as in Theorem~\ref{thm:Taub} and define for $\lambda\in\R_+$, $\breve{N}(\lambda)  := \min(N(\lambda),N(\mu+ \hbar^{1-\alpha}))$ where $\alpha>0$ is a small parameter.
Then $(\breve{N})'*\rho_\hbar \le  N'*\rho_\hbar $ pointwise and for $\lambda\le \mu$ 
\[
(N-\breve{N})*\rho_\hbar (\lambda) \le \hbar^{-1} \int_{\R_+} \rho\big(\tfrac{\sigma-\lambda}{\hbar}) \big( N(\sigma)-N(\mu+ \hbar^{1-\alpha}) \big)_+ \d \sigma  = \O(\hbar^\infty) , 
\]
using that $\rho\in \mathcal{S}$ has superpolynomial decay and $N$ has polynomial growth.
Hence, if we assume that $\displaystyle \max_{[0,\mu+\hbar^{1-\alpha}]}N'*\rho_\hbar  \le C \hbar^{-\beta}$, by  applying Theorem~\ref{thm:Taub} to $C^{-1} \hbar^{\beta}\breve{N}$, 
 we obtain the uniform bound
\begin{equation} \label{Nrho_bound}
\max_{\lambda\le \mu} \big| N   - N* \rho_\hbar \big| \le
\max_{\lambda\le \mu}  \big|   \breve{N} -   \breve{N}* \rho_\hbar \big| + \O(\hbar^\infty)
   \le \O(\hbar^{1-\beta}) . 
\end{equation}

By \eqref{zeta2},  
$\partial_\mu\big(N_{\hbar, \mu}^{1}*\rho_\hbar \big) =  \epsilon^n  \zeta_{\hbar,\mu}(H_\hbar)$, so that by Proposition~\ref{prop:zeta_edge}, there exists a constant $C$ such that
\[
\max_{(x_0,\mu) \in \mathcal{A}} \max_{\lambda\le \mu + \hbar^{2/3}}\max_{x\in\mathcal{K} } \epsilon^n   \zeta_{\hbar,\lambda}(H_\hbar)(x_0 +  \epsilon x) \le C \hbar^{-\frac 23} .
\]
Here we used that since $V$ is smooth on $\{V\le M\}$, 
$\mathcal{A}_\hbar = \big\{ (x_0 +\epsilon x,\lambda) : (x_0,\mu)\in\mathcal{A} , x\in\mathcal{K} , \lambda\le   \mu+\hbar^{2/3} \big\}$ is a compact set of \eqref{cond_zeta}.
Thus, if we apply  the bound \eqref{Nrho_bound} to $\lambda\mapsto N_{\hbar, \lambda}^1(x_0+\epsilon x)$ with $(\alpha,\beta) =(\frac13,\frac23)$, we obtain the claim.
\end{preuve}

Recall that $N_{\epsilon, \mu}^1(x)= \epsilon^n\vartheta_{\mu}(H_\hbar)(x,x)$ with $\epsilon=  \hbar^{\frac23}|\nabla V(x_0) |^{-\frac13}$. 
Using the Notations~\ref{regul-kern}, by \eqref{convg}, we obtain
\[
\int N_{\epsilon, \sigma}^1(x_0 +\epsilon \mathcal{U}^*  x) \rho_\hbar(\mu-\sigma) \d \sigma=    K^{f_{\hbar,\mu}}_{x_0,\epsilon}(x,x)  + \O(\hbar^\infty),
\]
where the error is controlled locally uniformly for $\{(x_0,\mu),V(x_0)=\mu, \nabla V(x_0)\neq 0\}$ and $x \in \R^n$. 
Choosing  $\mathcal{U} \in SO_n$ as in Theorem~\ref{thr:micro_edge_kernel}, we conclude 
by Proposition~\ref{prop:conv_reg_kern_edge} and Corollary~\ref{corr:N1} that 
\[
\big| N_{\epsilon, \mu}^1(x_0 + \epsilon\, \mathcal{U}^*  x) - \vartheta(\mu) K_{\rm edge}(x,x)\big|  =\O(\hbar^{\frac13})
\]
with the same uniformity.

Hence, choosing any cutoff $\vartheta$ equal to 1 on $[1,\mu]$, this implies that the  rescaled projection kernel \eqref{zoom}  obeys the relevant asymptotics (on the diagonal) in the edge case; for any compact sets $\mathcal{A} \subset \{V=\mu ; \nabla V\neq  0\}$ and  $\mathcal{K} \subset \R^{n}$, 
\[
\sup_{x_0 \in \mathcal{A}}    \sup_{x\in \mathcal{K}}  \big| 
K_{x_0,\epsilon}(x,x) - K_{\rm edge}(x,x) \big| \le C\hbar^{1/3}. 
\] 
To complete the proof of Theorem~\ref{thr:micro_edge_kernel}, we can just adapt the proof of Proposition~\ref{prop:trunckernel} using the previous estimates to obtain the relevant off-diagonal asymptotics, 
Note that in this case, the scalings are such that we cannot argue that any derivative  of $K_{x_0,\epsilon}$ converges to that of $K_{\rm edge}$.

\medskip

\begin{preuve}[of Theorem~\ref{thr:micro_edge_kernel}]\
Let us choose a cutoff  $\vartheta\in C^{\infty}_c(\R_+,[0,1])$ such that $\vartheta\ge \1_{[0,\mu]}$ and  let $K_\lambda=K^{\vartheta_{\lambda}}_{x_0,\epsilon} = K^{\1_{[0,\lambda]}}_{x_0,\epsilon}$ where $x_0 \in \mathcal{A}$ and $\epsilon=  \hbar^{\frac23}|\nabla V(x_0) |^{-\frac13}$. This holds for any $\lambda\in[0,\mu]$ since we assume that the potential $V\ge 0$. 
We consider the counting function for $\sigma \ge 0$, 
\[
  N^2_{\epsilon,\sigma}(x,y) = \epsilon^n\sum_{\lambda
    \leq \sigma}\big|\phi_{\lambda}(x_0+\epsilon
               x)-\phi_{\lambda}(x_0+\epsilon y) \big|^2,  
\] 
where  $(\lambda,\phi_\lambda)$ are normalised eigenpairs of $H_{\hbar}$.
Like in the proof of Proposition~\ref{prop:trunckernel}, the linear relationships \eqref{square} hold with $\hbar\leftarrow \epsilon$, so that if we proceed like in the proof of Corollary~\ref{corr:N1} for $  N^2_{\epsilon,\sigma}(x,y)$, we conclude that  for any compact sets $\mathcal{A} \subset \{V=\mu ; \nabla V\neq  0\}$ and  $\mathcal{K} \subset \R^{n}$, 
\[
\sup_{x_0 \in \mathcal{A}}    \sup_{x,y\in \mathcal{K}}  \big| 
K_{x_0,\epsilon}(x,y) - K_{\rm edge}(x,y) \big| \le C\hbar^{1/3} .
\] 
This completes the proof.
\end{preuve}

  \section{Mesoscopic fluctuations}
  \label{sec:mesoscopic-scales}

The goal of this Section is to prove  Theorems~\ref{thr:mesoscopic_bulk_pot} and the Gaussian tail bounds of Theorem~\ref{thm:gtb}. 
The arguments consist in controlling the variance of a mesoscopic, or macrocscopic linear statistics.

\medskip

 A first step is to study the
 model case of the free Laplacian in Section~\ref{sec:case-free-laplacian}, which is helpful
 to understand the general picture.
In this case, we obtain an (optimal) functional CLT as the intensity of the point process $\mu \to\infty$, see Theorem~\ref{thr:mesoscopic_bulk_lapl} below.

  \subsection{CLT for the free Laplacian point process}
  \label{sec:case-free-laplacian}
 
  In this section, we study linear statistics of the determinantal point process $\X_\infty$ associated with the free  Laplacian, that is with the operator
 \[
K_{\rm b , \mu} = \1_{(-\infty,\mu^2]}(-\Delta):(x,y)\mapsto  \mu^{\frac n2} \frac{\J_{\frac
        n2}(\mu |x-y|)}{(2\pi |x-y|)^{\frac n2}},
 \]
 in the regime where $\mu\to\infty$, or equivalently as the intensity $\frac{\mu^{n} \omega_{n}}{(2\pi)^n}$ of the particle tend to infinity,  cf.~\eqref{Picomputation}. 
 We obtain the following central limit theorem. 

 \renewcommand\thetheorem{III.2}
 \begin{theorem} \label{thr:mesoscopic_bulk_lapl}
Suppose that $n\ge 2$ and let $\sigma_n^2 =  \frac{\omega_{n-1}} {(2\pi)^{n}}$. For any $g\in H^{\frac 12}\cap L^{1}(\R^n)$, it holds in distribution as $\mu\to\infty$, 
\begin{equation} \label{def:Sigma}
\X_{\infty,\mu}(g) : = \frac{\X_\infty(g)- \frac{\mu^{n} \omega_{n}}{(2\pi)^n} \int g }{\sigma_n\mu^{\frac{n-1}{2}}} \to  \mathcal{N}_{0,\Sigma^2(f)} \qquad\text{where}\quad \Sigma^2(g) =  \int_{\R^n} |\widehat{g}(\xi)|^2
|\xi| \d\xi . 
\end{equation}
\end{theorem} 

As in Theorem~\ref{thr:mesoscopic_bulk_pot}, the interpretation is that the random process $\X_{\infty,\mu}$ converges in the sense of finite-dimensional distributions as $\mu\to\infty$ to a (centred) Gaussian field $\mathrm{G}$ on $\R^n$ with correlation kernel
\[
\E \mathrm{G}(f) \mathrm{G}(g) =  \int \widehat{f}(\xi) \overline{\widehat{g}(\xi)} |\xi| \d\xi , \qquad f,g \in H^{\frac 12}(\R^n).
\]

Note that the process $\X_{\infty,\mu}$ is exactly centred and the assumptions of Theorem~\ref{thr:mesoscopic_bulk_lapl} are optimal in the sense that  $\X_{\infty,\mu}$ is apriori defined on $L^{1}(\R^n)$ and the asymptotic variance $\Sigma^2(g)$ is finite if and only if $g\in H^{\frac12}$.
In dimension $n=1$, $K_{\rm b , \mu} :(x,y) \mapsto \frac{\sin(\mu |x-y|)}{\pi|x-y|}$ is the sine-kernel from random matrix theory and the counterpart of Theorem~\ref{thr:mesoscopic_bulk_lapl} is a classical result. 
In this case, the CLT holds without (re)normalisation; we refer to \cite{Kac54, Soshnikov_00} for different proofs. 

\medskip

The proof of Theorem~\ref{thr:mesoscopic_bulk_lapl} relies on Corollary~\ref{clt_dpp} and the following Lemma which controls the asymptotics of the variance of $\X_{\infty,\mu}(g)$ for general test functions. 

 \begin{lem}\label{prop:limit-commut-lapl}
For $n\ge 1$ and for every $g\in H^{\frac 12}(\R^n)$, it holds as $\mu\to\infty$,
    \[
\operatorname{var} \X_{\infty}(g) =  - \tfrac12 \tr\big(\big[ g , K_{\rm b , \mu} \big]^2\big) 
 \sim \sigma_n^2\mu^{n-1} \Sigma^2(g).    \]
  \end{lem}
    
    \begin{preuve}
By definition, 
\begin{equation} \label{tr2}
\tr\big(\big[ g , K_{\rm b , \mu} \big]^2\big)  =  - \int  |g(x)-g(y)|^2 |K_{\rm b , \mu}(x,y)|^2 \dd x \dd y . 
\end{equation}
By Plancherel's formula, it holds for any $z\in\R^n$,
 \[
 \int |g(x)-g(x+z)|^2 \d x    = 4 \int |\widehat{g}(\xi)|^2 \sin^2(\tfrac{\xi \cdot z}{2}) \d \xi . 
 \]
 Note that this identity makes sense for any $g\in H^{\frac 12}$ and by a change of variable, this implies that
  \[
\tr\big(\big[ g , K_{\rm b , \mu} \big]^2\big)=-  4 \int |\widehat{g}(\xi)|^2 
   \sin^2(\tfrac{\xi \cdot z}{2})|K_{\rm b, \mu}(z,0)|^2 \dd \xi \dd z.
 \]
 We can use  Plancherel's formula in the same way again using that
 $z\mapsto K_{\rm bulk}(z,0)$ is the Fourier transform of $(2\pi)^{-\frac{n}{2}}\1_{|\cdot|
   \le \mu}$  (see formula \eqref{kernelPi}). This yields 
  \[ \begin{aligned}
\tr\big(\big[ g , K_{\rm b , \mu} \big]^2\big)=  \frac{-1}{(2\pi)^{n}}  \int |\widehat{g}(\xi)|^2 
| \1_{|\zeta| \le \mu} - \1_{|\zeta+ \xi| \le \mu} |^2 \dd \xi \dd \zeta .
\end{aligned} \]
Note that one has for any $\xi\in\R^n$
\[
\frac12 \int | \1_{|\zeta|  \le \mu} - \1_{|\zeta+ \xi|  \le \mu} |^2 \d \zeta = \mu^n \int  ( \1_{|\zeta|  \le 1} - \1_{|\zeta+ \xi/\mu|  \le 1,|\zeta| \le 1} )  \d \zeta  = \mu^n \big| B_{0,1}^n \setminus B_{|\xi|/\mu,1}^n \big| .
\]
This shows that 
\[
\operatorname{var} \X_{\infty}(g) =  \frac{\mu^n}{(2\pi)^{n}} \int |\widehat{g}(\xi)|^2 
\big| B_{0,1}^n \setminus B_{|\xi|/\mu,1}^n \big| d\xi . 
\]
Moreover, the function $r\in[0,\infty) \mapsto  \frac{\big|
  B_{0,1}^{n} \setminus B_{r,1}^n \big|}{r}$ is clearly continuous and
bounded and its value at 0 is $|B_{\R^{n-1}}(0,1)|=\omega_{n-1}$.
Hence, by the dominated convergence theorem, if $g\in H^{1/2}(\R^n)$ that is, 
$\displaystyle \int \widehat{g}(\xi)|^2  |\xi| \d \xi <\infty$,
this concludes the proof.
 \end{preuve}

\begin{rem}
  For the last step of the proof, we use the dominated convergence
      theorem; in fact, one can show that the convergence is monotone,
      using the convexity of Euclidian balls in $\R^n$. It follows that, if $g\in L^2$ but $g\notin
      H^{\frac 12}$, then 
      \[ \operatorname{var} \X_{\infty,\mu}(g) <\infty\quad\forall \mu \qquad\text{but}\qquad
      \ \new{\mu^{1-n}}\operatorname{var} \X_{\infty,\mu}(g) \to+\infty \quad\text{as $\mu\to\infty$.}\]
  \end{rem}
  
\begin{rem}\label{rem:Slobodeckij}
From the Bessel functions asymptotics \eqref{Bessel_asymp} and the
explicit formula for the kernel $K_{\rm b , \mu}$, one can derive an
alternative proof of Lemma \ref{prop:limit-commut-lapl} on the more
restrictive class $g\in C^2_c(\R^n)$.

Observe that, as $\mu|x-y| \to\infty$, by \eqref{Picomputation},
 \[
|K_{\rm b , \mu} (x,y)|^2 = \mu^{n-1} \frac{4 \cos^2\big(\mu |x-y| -\tfrac{(n+1)\pi}{4})}{(2\pi |x-y|)^{n+1}}+ \O\bigg(\frac{\mu^{n-2}}{|x-y|^{n+2}}\bigg) . 
\]
Then, by formula \eqref{tr2}, 
\[
\operatorname{var} \X_{\infty}(g) = \frac12  \int  |g(x)-g(y)|^2 |K_{\rm b , \mu}(x,y)|^2 \dd x \dd y
\]
and these asymptotics (and the Riemann-Lebesgue Lemma) allow us to argue that for any  $g\in C^2_c(\R^n)$, it holds as $\mu\to\infty$, 
\[
\operatorname{var} \X_{\infty}(g) \sim  \mu^{n-1}  \int  \frac{ |g(x)-g(y)|^2}{(2\pi |x-y|)^{n+1}} \d x \d y .
\]

We recover the classical expression of the Sobolev-Slobodeckij $W^{1/2,2}(\R^n)$ seminorm via
a singular integral kernel, which is equivalent to the Fourier space
definition of the $H^{1/2}$ seminorm (see for instance
\cite{stein_singular_2016}, p. 155, Theorem 5). In fact, together
with Lemma \ref{prop:limit-commut-lapl}, this provides a comparison of
the constants in the definitions:
\begin{equation} \label{dual_var}
\sigma_n^2\Sigma^2(g) = \frac{1}{(2\pi)^{n+1}} \int  \frac{ |g(x)-g(y)|^2}{|x-y|^{n+1}} \d x \d y
\end{equation}
where  $\sigma_n^{2} =  \frac{\omega_{n-1}}{(2\pi)^{n}}$ as in Theorem~\ref{thr:mesoscopic_bulk_lapl}.
\end{rem}

\begin{preuve}[of Theorem~\ref{thr:mesoscopic_bulk_lapl}]
Let us first assume that  $g\in H^{\frac 12}\cap L^{1} \cap L^\infty(\R^n)$. We make the extra assumption that $g\in L^\infty$ and $n\ge 2$ in order to apply Corollary~\ref{clt_dpp}. 
Since $\X_\infty$ has constant intensity, $\E\X_\infty(g) = \frac{\mu^{n} \omega_{n}}{(2\pi)^n} \int g $ and Lemma~\ref{prop:limit-commut-lapl} gives the asymptotics of $\operatorname{var} \X_\infty(g)$ which diverge if $g \neq 0$. Hence, by  Corollary~\ref{clt_dpp}, for any $t\in\R,$ as $N\to\infty$
\begin{equation} \label{Laplace_Laplace}
 \E\big[ e^{t \X_{\infty, \mu}(g) } \big]   = \exp\Big(t^2(\tfrac{\Sigma^2(g)}2+ \O(\mu^{-\frac{n-1}{2}}) \Big)
\end{equation}
In this case, the random variable $ \X_{\infty, \mu}(g)$ does not only
converge in distribution, but in the sense of its Laplace transform so
that all its moments converge. Moreover, we can easily remove the
technical condition $g\in L^\infty$ by using the following inequality; for any $t\in\R$,
\[
\big| \E e^{\i t X} - \E e^{\i t Y} \big|  \le |t| \sqrt{\operatorname{var}(X-Y)}
 \]
for any two random variables $X,Y$ defined on the same probability space. 
In particular, if  $g\in H^{\frac 12}\cap L^{1}(\R^n)$ and $\chi\in C^\infty_c(\R^n)$, then  by Lemma~\ref{prop:limit-commut-lapl}, it holds for any $t\in\R$,
\[
\limsup_{\mu\to\infty} \big|  \E\big[ e^{\i t \X_{\infty, \mu}(g) } \big]  -  \E\big[ e^{\i t \X_{\infty, \mu}(\chi) } \big]  \big| = |t| \Sigma^2(g-\chi) .
\]
In addition, $\E\big[ e^{\i t \X_{\infty, \mu}(\chi) } \big]   \to e^{-t^2 \Sigma^2(\chi)/2}$ as $\mu\to\infty$ by \eqref{Laplace_Laplace} so that for any small $\epsilon>0$,  choosing $\chi$ in such as way that 
$\Sigma^2(g-\chi)= \| \chi-g\|_{H^{1/2}} \le \epsilon $ (by density),  this implies that  for any $t\in\R$,
\[
\limsup_{\mu\to\infty} \big|  \E\big[ e^{\i t \X_{\infty, \mu}(g) } \big]  - e^{-t^2 \Sigma^2(g)/2} \big| = (1+t^2)\epsilon .
\]
This establishes that the characteristic function $ \E\big[ e^{\i t \X_{\infty, \mu}(g) } \big]\to \E\big[ e^{\i t  \mathcal{N}_{0,\Sigma^2(g)}} \big]= e^{-t^2 \Sigma^2(g)/2}$ as $\mu\to\infty$ for any fixed $t\in\R$ and this implies Theorem~\ref{thr:mesoscopic_bulk_lapl}. 
\end{preuve}

\subsection{Mesoscopic commutator estimates}
\label{sec:trace-comm-estim}

 The goal of this section is to prove Theorem~\ref{thr:mesoscopic_bulk_pot}. The core of the argument is to obtain the asymptotics for the variance of smooth (mesoscopic) linear statistics of the free fermions point processes.
Recall that for a test function $g\in C^\infty_c(\R^n)$, 
\begin{equation} \label{varsplit}
  \var  T_{x_0,\epsilon}^*\X(g) =  -\tfrac{1}{2}   \tr([ T^*_{x_0,\epsilon}g , \Pi_{\hbar,\mu}]^2),
  \end{equation}
  where $T^*_{x_0,\epsilon}g = g(\tfrac{ \cdot-x_0}{\epsilon})$ and $\Pi_{\hbar,\mu} = \1_{-\hbar^2\Delta+V\leq \mu}$. 
Hence, the proof basically amounts to proving the following expansion for the Hilbert-Schmidt norm of the commutator,
\[
\big\| [ T^*_{x_0,\epsilon}g , \Pi_{\hbar,\mu}] \big\|_{\J^2}^2 =- \tr([ T^*_{x_0,\epsilon}g , \Pi_{\hbar,\mu}]^2) . 
\]

   \begin{prop}\label{prop:limit-commut-V}
  Let $(\mu,V)$ satisfy \ref{hyp:weak} and let $\epsilon :[0,1] \to[0,1]$ be a non-increasing function such that $\hbar^{1-\beta} \le \epsilon(\hbar) \le \hbar^{\beta}$ for some $\beta>0$ and let $\delta(\hbar) = \hbar/\epsilon(\hbar)$. 
There exists $\alpha>0$ so that for any $g\in C^\infty_c(\R^n)$, as $\hbar\to0$, 
  \[
-    \tr([ T^*_{x_0,\epsilon}g , \Pi_{\hbar,\mu}]^2) = 
    2\sigma_{n}(\mu-V(x_0))^{\frac{n-1}{2}}\delta(\hbar)^{1-n}\Sigma^2(g) +    \O(\hbar^{\alpha} \delta(\hbar)^{1-n}), 
    \]
 where $\sigma_n$ and $\Sigma^2$ are as in Theorem~\ref{thr:mesoscopic_bulk_lapl} and the  error term is locally uniform for $(x_0,\mu) \in \big\{ (x,\lambda) \in \R^{n+1} : V(x) < \lambda <M \big\}$.
\end{prop}

The proof of this result is divided in three steps and it is similar to the analysis carried out in the previous sections. The first
step (Proposition \ref{prop:commut-macroscopic-smooth}) is to study the commutator between $T^*_{x_0,\epsilon}g$ and a
smooth compactly supported function of $H_{\hbar}$, using the fact
that such a spectral function is a pseudodifferential operator. The
second step is an estimate of the commutator with a projector over a
spectral window of size $\hbar$ (Proposition
\ref{prop:estimee_J1_locale}). This allows us, in the last step, to
replace the spectral projector $\Pi_{\hbar,\mu}$ by  a
  Fourier Integral Operator using Proposition \ref{prop:propagator} and a frequency cutoff at scale $\hbar$. 
In this case, however, recovering the true
  projector from its mollification does not require to use of a
  Tauberian theorem but is a direct consequence of Proposition~\ref{prop:estimee_J1_locale}.

\begin{prop}\label{prop:commut-macroscopic-smooth}
  Let $(\mu,V)$ satisfy \ref{hyp:weak} and $\epsilon= \delta(\hbar)^{-1}
  \hbar$ where $\delta:(0,1]\to (0,1]$ satisfies $\hbar \leq
  \delta(\hbar)\leq 1.$
  Then, for any $\chi\in C^{\infty}_c((-\infty,M),\R)$, $g\in
  C^{\infty}_c(\R^n)$ and $x_0\in\R^n$, it holds as $\hbar\to0$, 
  \[
    \tr([T^*_{x_0,\epsilon}g ,\chi(H_{\hbar})]^2)=\O(\delta(\hbar)^{2-n}).
    \]
  \end{prop}
  
  \begin{preuve}
We can apply Proposition~\ref{prop:propagator_rem_Cinf} with $t=0$.
In particular, since $\|R_{\hbar,t}\|_{\J^1} = \O(\hbar^\infty)$ and $\|\chi(H_{\hbar})\|_{\J^1}=\O(\hbar^{-n})$, we have 
\[
\big\| [T^*_{x_0,\epsilon}g ,R_{\hbar,0}]^2 \big\|_{\J^1} \le 4  \|g \|_{L^\infty}^2 \|R_{\hbar,0}\|_{\J^1}^2 =\O(\hbar^\infty) 
\]
and 
\[
\big| \tr([T^*_{x_0,\epsilon}g ,\chi(H_{\hbar})] [T^*_{x_0,\epsilon}g , R_{\hbar,0}]) \big| 
\le  \big\| [T^*_{x_0,\epsilon}g ,\chi(H_{\hbar})] \big\|_{\J^1} \big\| [T^*_{x_0,\epsilon}g ,R_{\hbar,0}]\big\|_{\J^1}
 \le 4  \|f \|_{L^\infty}^2 \| \chi(H_{\hbar})\|_{\J^1}\|R_{\hbar,0}\|_{\J^1} =\O(\hbar^\infty) . 
\]
This implies that 
\[
\tr([T^*_{x_0,\epsilon}g ,\chi(H_{\hbar})]^2)
=  \tr( [T^*_{x_0,\epsilon}g , I_{\hbar,0}^{\phi, a}]^2) + \O(\hbar^\infty) \]
where $a\in S^0$ is a classical symbol and $ I_{\hbar,0}^{\phi, a} $ denotes the  pseudodifferential operator
\[
 I_{\hbar,0}^{\phi, a} : (x,y)\mapsto \frac{1}{(2\pi \hbar)^n}\int
    e^{\i\frac{(x-y) \cdot \xi}{\hbar}}a\left(x,y,\xi;\hbar\right)\dd \xi .
\]
Then, we have
\[ \begin{aligned}
 \tr( [T^*_{x_0,\epsilon}g , I_{\hbar,0}^{\phi, a}]^2) 
 & =  \frac{-1}{(2\pi\hbar)^{2n}} \int \big( T^*_{x_0,\epsilon}g (x) - T^*_{x_0,\epsilon}g (y) \big)^2\,  I_{\hbar,0}^{\phi, a}(x,y)  I_{\hbar,0}^{\phi, a}(y,x) \d x \d y \\
 &=   \frac{-1}{(2\pi\hbar)^{2n}} \int   \big( T^*_{x_0,\epsilon}g (x) - T^*_{x_0,\epsilon}g (y) \big)^2     e^{\i\frac{(x-y)\cdot   (\xi-\zeta)}{\hbar}} a\left(x,y,\xi;\hbar\right)a\left(y,x,\zeta;\hbar\right) \d \xi \d\zeta  \d x \d y . 
\end{aligned}\]
We are in
position to apply Proposition \ref{prop:spl_xxi}, the special case of
the stationary phase lemma, in the variables $(y,\zeta)$ keeping $(x,\xi)$ fixed. We obtain, for every $\ell\in \N$,
\begin{multline*}
 \tr( [T^*_{x_0,\epsilon}g , I_{\hbar,0}^{\phi, a}]^2) 
 = \\ \frac{-1}{(2\pi\hbar)^{n}} \int\left(
 \sum_{k=0}^{\ell-1}  \frac{(\i\hbar)^k}{k!} \sum_{|\alpha|=k} 
 \partial^{\alpha}_{y}\partial^{\alpha}_{\zeta}
 \Big(  \big( T^*_{x_0,\epsilon}g (x) - T^*_{x_0,\epsilon}g (y)
 \big)^2  a\left(x,y,\xi;\hbar\right)a\left(y,x,\zeta;\hbar\right)
 \Big) \Big|_{y=x,\zeta=\xi} +R_\ell(x,\xi)\right)\d x \d \xi,
\end{multline*}
where the error $R_\ell(x,\xi) = \O(\hbar^\ell \epsilon^{-\ell}) = \O( \delta^\ell) $ is compactly supported in
$\{|x-x_0|\leq C\epsilon,|\xi|\leq C\}$ for some $C>0$ depending on
the support of $g$.

To conclude the proof, it remains to take $\ell=2$. Indeed, observe that
the terms $k=0$ and $k=1$ in the sum vanish, and it remains
\[
  \frac{-1}{(2\pi\hbar)^n}\int R_2(x,\xi)\dd x \dd
  \xi=\O(\delta^{-n+2}).
  \]
  \end{preuve}

\begin{prop}\label{prop:estimee_J1_locale}
  Let $(\mu,V)$ satisfy \ref{hyp:weak} and $g\in
  C^{\infty}_c(\R^n,\R)$. Let $\epsilon:(0,1] \to (0,\epsilon_0]$ be a small ($\hbar$-dependent) parameter. 
  It holds as $\hbar\to0$, 
    \[
    \big\| T^*_{x_0,\epsilon}g \1_{H_\hbar \in [\mu-\hbar,\mu+\hbar]} T^*_{x_0,\epsilon}g\big\|_{\J^1} =\O\big(\epsilon^n \hbar^{1-n}\big)
  \]
  uniformly for $(x_0, \mu) \in \big\{ (x, \lambda) \in\R^{n+1} : \lambda<M, \,  \supp (T^*_{x,\epsilon_0}g) \subset \{V<\lambda\}  \big\}$.
    \end{prop} 
  
  \begin{preuve}
  Let us fix $(x_0, \mu)$ such that $V(x_0)<\mu$. 
By functional calculus,  we can bound  
\[
\1_{H_\hbar \in [\mu-\hbar,\mu+\hbar]}\le \rho(\hbar^{-1}(H_{\hbar}-\mu))\chi(H_{\hbar}) 
\]
 by choosing appropriate cutoff $\chi \in C^\infty_c(\R,\R_+)$ with $\supp \chi  \subset (V(x_0),M)$ and  a smooth mollifier $\rho$ such that $\supp \hat{\rho} \subset [-\tau,\tau]$  (for a small $\tau>0$ so that one can apply Proposition~\ref{prop:propagator_rem_Cinf}).

For any function $\vartheta :\R^n \to \R_+$ measurable with compact support, 
$ T^*_{x_0,\epsilon}g \vartheta(H_\hbar)  T^*_{x_0,\epsilon}g $ is a positive trace-class operator. 
Hence, by Proposition~\ref{prop:top},
\[\begin{aligned}
    \big\| T^*_{x_0,\epsilon}g \1_{H_\hbar \in [\mu-\hbar,\mu+\hbar]}T^*_{x_0,\epsilon}g\big\|_{\J^1}
    & \le \tr\big[ T^*_{x_0,\epsilon}g  \rho(\hbar^{-1}(H_{\hbar}-\mu))  \chi(H_{\hbar}) T^*_{x_0,\epsilon}g\big] \\
    & = \tr\big[ T^*_{x_0,\epsilon}g^2  \rho(\hbar^{-1}(H_{\hbar}-\mu))  \chi(H_{\hbar}) \big].
\end{aligned}\]

Since $\rho(x) = \frac{1}{\sqrt{2\pi}} \int e^{\i x \cdot t} \hat{\rho}(t) \d t $, by Proposition~\ref{prop:propagator_rem_Cinf}, we have
\[
\rho(\hbar^{-1}(H_{\hbar}-\mu))  \chi(H_{\hbar})
:(x,y)\mapsto\frac{1/\sqrt{2\pi}}{(2\pi\hbar)^n}\int e^{\i\frac{\phi(t,x,\xi)-y\cdot
        \xi-t\mu}\hbar}a(t,x,y,\xi;\hbar)\hat{\rho}(t)\dd \xi
      \dd t+\O_{\JJ^1}(\hbar^{\infty}),
\]
where the error is uniform for all $\mu\in \R$ and $a\in S^0$ is a classical symbol.
This implies that 
\begin{equation} \label{J11bound}
    \big\| T^*_{x_0,\epsilon}g \1_{H_\hbar \in [\mu-\hbar,\mu+\hbar]}T^*_{x_0,\epsilon}g\big\|_{\J^1}
    \le \frac{1}{(2\pi\hbar)^n}\int e^{\i\frac{\Psi(t,x,\xi)-t\mu}\hbar}  \big( T^*_{x_0,\epsilon}g(x)\big)^2 b(t,x,\xi;\hbar)\dd \xi  \dd t \d x + \O(\hbar^\infty)
\end{equation}
where $\Psi$ is as in Proposition~\ref{prop:reduction-phase} and
\[
b(t,x,\xi,\hbar)= a(t,x,x,\xi;\hbar)\hat{\rho}(t)/\sqrt{2\pi} 
\]
has compact support. 
We may assume that the parameter $\epsilon_0$ is sufficiently small so that 
\begin{itemize}\item $ \supp(T^*_{x_0,\epsilon}g) \subset \{V<\mu\}$ for any $\epsilon<\epsilon_0$, 
\item   since $\supp \chi  \subset (V(x_0),M)$, $(r,t) \mapsto b(t,x,r\omega;\hbar)$ has compact support in $ \R_+^* \times [-\tau,\tau]$ for any $x\in \supp(T^*_{x_0,\epsilon}g)$, $\omega \in S^{n-1}$ and $\hbar\in(0,1]$.
\end{itemize}

This is exactly the setting of the proof of Proposition~\ref{prop:regularised_kernel_bulk}.
Hence, we can make a change of variables $\xi=r\omega$ where $(r,\omega)\in
\R_+ \times S^{n-1}$ and apply the stationary phase method to the integral \eqref{J11bound} in the variables $(r,t) \in \R_+^* \times [-\tau,\tau]$ for a fixed $(x, \omega) \in \R^n \times S^{n-1}$.
By \eqref{eq:phase1}, the only critical point is $(r_\star,0)$ where $r_\star(x) := \sqrt{\mu-V(x)}$ and the Hessian of the phase is non-degenerate with $\det{ \rm Hess}\, \psi |_{r_\star} = 4 r_\star^2$. Then, by Proposition~\ref{prop:spl}, there exists a classical symbol $s\in S^0$ so that 
\[
\frac{1}{2\pi\hbar}\int e^{\i\frac{\Psi(t,x,\xi)-t\mu}\hbar}  b(t,x,\xi;\hbar)\dd \xi  \dd t 
= r_\star(x)^{n-2} s(x;\hbar) +\O_x(\hbar^\infty), 
\]
where the error is controlled uniformly for all $x\in \supp(T^*_{x_0,\epsilon}g)$,. 
By \eqref{J11bound}, this yields the bound
\[
    \big\| T^*_{x_0,\epsilon}g \1_{H_\hbar \in [\mu-\hbar,\mu+\hbar]}T^*_{x_0,\epsilon}g\big\|_{\J^1} 
    \le \frac{\mu^{\frac n2-1}}{(2\pi\hbar)^{n-1}}  \int \big( T^*_{x_0,\epsilon}g(x)\big)^2 s(x;\hbar) \d x  +\O(\hbar^\infty) .
\]
Since $\| T^*_{x_0,\epsilon}g\|_{L^2}^2 = \O(\epsilon^n)$, and 
$\displaystyle\max_{x\in\{V\le \mu\}}\big| s(x;\hbar)\big| \le C$ for a constant (independent of $\hbar$), this completes the proof.
  \end{preuve}
 
  Let $ f_{\hbar,\mu}$ be as in Notations~\ref{regul-kern} with $\vartheta=1$ on a neighborhood of $\mu$. 
 If we assume for now the asymptotics of the quantity 
 $\big\| [ T^*_{x_0,\epsilon}g , f_{\hbar,\mu}(H_\hbar)] \big\|_{\J^2}^2$ as $\hbar\to0$, we can prove Proposition \ref{prop:limit-commut-V} and then  Theorem~\ref{thr:mesoscopic_bulk_pot}.
 The remaining parts of the proof are given in the next section.
 
 \medskip
 
\begin{preuve}(of Proposition \ref{prop:limit-commut-V})
Let us denote $\old{M}\new{G} =T^*_{x_0,\epsilon}g$ viewed as a (bounded) self-adjoint multiplication operator. 
By Cauchy--Schwarz's inequality, we can compare
\[\begin{aligned}
\big| \tr([\old{M}\new{G}, \Pi_{\hbar,\mu}]^2) -  \tr([\old{M}\new{G},  f_{\hbar,\mu}(H_\hbar)]^2)  \big|
 = \big| \tr\big([\old{M}\new{G}, \Pi_{\hbar,\mu}- f_{\hbar,\mu}(H_\hbar)] [\old{M}\new{G}, \Pi_{\hbar,\mu}+f_{\hbar,\mu}(H_\hbar)]\big)  \big| \\
 \le  \big\| [\old{M}\new{G}, \Pi_{\hbar,\mu}- f_{\hbar,\mu}(H_\hbar)] \big\|^2_{\J^2}
+ 2  \big\| [\old{M}\new{G}, \Pi_{\hbar,\mu}- f_{\hbar,\mu}(H_\hbar)] \big\|_{\J^2}
 \big\| [\old{M}\new{G},f_{\hbar,\mu}(H_\hbar)] \big\|_{\J^2} . 
\end{aligned}\]

Hence, Proposition \ref{prop:limit-commut-V} follows if we can show there exists a small $\alpha>0$ so that as $\hbar\to 0$, 
\begin{equation} \label{mainJ2}
 \big\| [\old{M}\new{G},f_{\hbar,\mu}(H_\hbar)] \big\|_{\J^2}
 = 2\sigma_n^2(\mu-V(x_0))^{\frac{n-1}{2}}(2\pi)^{-n+1}\delta^{1-n} \Sigma^2(f)  
 + \O(\delta^{1-n}\hbar^{\alpha})
\end{equation}
and 
\begin{equation} \label{restJ2est}
 \big\| [\old{M}\new{G}, \Pi_{\hbar,\mu}- f_{\hbar,\mu}(H_\hbar)] \big\|^2_{\J^2} =  \O(\hbar^{2\alpha}\delta^{1-n}) . 
\end{equation}

 The asymptotics \eqref{mainJ2} follows by combining Propositions~\ref{prop:tr1} and~\ref{prop:tr2} below by using that 
\[
 \big\| [\old{M}\new{G},f_{\hbar,\mu}(H_\hbar)] \big\|_{\J^2} = 2 \Big( \tr\big( f_{\hbar,\mu}^2(H_\hbar)\old{M}\new{G}^2 \big)  - \tr\big( (f_{\hbar,\mu}(H_\hbar)\old{M}\new{G})^2 \big)  \Big) .
\]
Note that the leading term  (of order $\delta^{-n}$) in the expansions of both terms cancel exactly

To prove the estimate \eqref{restJ2est}, observe that since $\supp(\vartheta) \subset [0,M]$ with $\vartheta=1$ on a neighborhood of $\mu$, 
 \[
 \1_{(-\infty,\mu]} - f_{\hbar,\mu}
 =  g_{\hbar,\mu} + \chi
 \qquad\text{where }
 g_{\hbar,\mu} =   \vartheta \big(  \1_{(-\infty,\mu]}   - \varkappa_\mu*\rho_\hbar\big) 
 \text{ and } \chi \in C^\infty_c((-\infty,M),\R_+).
 \]
In particular, by Proposition~\ref{prop:commut-macroscopic-smooth}, this implies that
\begin{align} \notag
  \big\| [\old{M}\new{G}, \Pi_{\hbar,\mu}- f_{\hbar,\mu}(H_\hbar)] \big\|^2_{\J^2} 
&   \le 2  \big\| [\old{M}\new{G}, g_{\hbar,\mu}(H_\hbar)] \big\|^2_{\J^2} 
+ 2  \big\| [\old{M}\new{G}, \chi(H_\hbar)] \big\|^2_{\J^2}   \\
& \label{commbound} \le  2  \big\| \old{M}\new{G}  g_{\hbar,\mu}^2(H_\hbar) \old{M}\new{G} \big\|_{\J^1}
+ \O(\delta^{-n+2}) . 
  \end{align}
 
 Since $\varkappa=1$ on $ [0,M]$ and $\rho \in S$, we can bound for any $\gamma>0$, 
  \[
g_{\hbar,\mu}^2 \le \1_{[\mu-\hbar^{1-\gamma},\mu+\hbar^{1-\gamma}]}+ \O(\hbar^\infty) 
  \]
Thus, as operators, one has 
\[
\old{M}\new{G}  g_{\hbar,\mu}^2(H_\hbar) \old{M}\new{G}   \le  \old{M}\new{G}  \big(\1_{[\mu-\hbar^{1-\gamma},\mu+\hbar^{1-\gamma}]}(H_\hbar)+ \O_{\new{\J^1}}(\hbar^\infty)  \big) \old{M}\new{G} .
\]
Then, \old{if $\|M \|_{\J^2} \le C$}\new{since the operator norm of
  $G$ is bounded}, by the triangle inequality, 
\[
 \big\| \old{M}\new{G}  g_{\hbar,\mu}^2(H_\hbar) \old{M}\new{G} \big\|_{\J^1}  \le \sum_{k} 
  \big\| \old{M}\new{G}  \1_{[\mu_k-\hbar,\mu_k+\hbar]}(H_\hbar) \old{M}\new{G} \big\|_{\J^1} + \O(\hbar^\infty)
\]
where $\mu_k$ is a uniform mesh of the interval  $[\mu-\hbar^{1-\gamma},\mu+\hbar^{1-\gamma}]$ with spacing $2\hbar$. 

Hence, by Proposition~\ref{prop:estimee_J1_locale}, we conclude that 
\[
\big\| \old{M}\new{G}  g_{\hbar,\mu}^2(H_\hbar) \old{M}\new{G} \big\|_{\J^1}   = \O(\epsilon^n\hbar^{1-\gamma-n})
= \O(\hbar^{\beta-\gamma}\delta^{1-n})
\]
This, if $\gamma>0$ is small enough (compared to $\beta$), by \eqref{commbound}, this completes the proof of the estimate  \eqref{restJ2est} for some $\alpha>0$. \end{preuve}

\begin{preuve} [of Theorem~\ref{thr:mesoscopic_bulk_pot}]
According to the convention of Theorem~\ref{thr:mesoscopic_bulk_pot},using \eqref{varsplit} together with Proposition \ref{prop:limit-commut-V}, we obtain 
\[
\var\X_{\hbar,\epsilon}(g) = \sigma_n^{-2} \delta(\hbar)^{n-1} \var T_{x_0,g}^*\X (g)
 = \Sigma^2(g) + \O(\hbar^{\alpha}) . 
\]
In dimension $n\ge 2$, we are therefore in a position to apply Corollary~\ref{clt_dpp}.
Indeed, up to normalization,  $\X_{\hbar, \epsilon}(g)$ corresponds to the linear statistic $\X(T_{x_0,\epsilon}^* g)$ where  $\| T_{x_0,\epsilon}^* g\|_{L^\infty} \le \| g\|_{L^\infty}$. We conclude that as $\hbar\to0$,
\[
\X_{\hbar, \delta}(g) \Rightarrow \Sigma^2(g)^{1/2} \mathcal{N}_{0,1} 
\]
and the convergence holds in the sense of the Laplace transform.
\end{preuve}

\subsection{Mesoscopic commutator estimates with a regularised kernel}

The goal of this section is to obtain the asymptotics \eqref{mainJ2}. 
Note that we can split
\begin{equation}
 \begin{aligned}
 \big\| [T^*_{x_0,\epsilon}g,f_{\hbar,\mu}(H_\hbar)] \big\|_{\J^2}
  =    \tr\big( f_{\hbar,\mu}^2(H_{\hbar}) (T^*_{x_0,\epsilon}g)^2 \big)
-    \tr\big( \big( f_{\hbar,\mu}(H_{\hbar}) T^*_{x_0,\epsilon}g \big)^2 \big) .\end{aligned}
\end{equation}

We obtain separately the asymptotics of both terms in
the expansion  \eqref{varsplit} and show that the leading terms
cancel while computing the correction term.
Let us introduce the useful notation
\[g_\epsilon := \epsilon^{-n} T^*_{x_0,\epsilon}g \qquad \text{for}
  \qquad \epsilon\in (0,1].\] 
 
  \begin{prop} \label{prop:tr1}
\new{Recall Notations~\ref{regul-kern}, and choose $\vartheta=1$ on a neighborhood of $\mu$.}
Let  $\epsilon :[0,1] \to[0,1]$ be a non-increasing function such that $\delta(\hbar) = \hbar/\epsilon(\hbar)$ satisfies  $\delta(\hbar) =\O(\hbar^{\beta})$ for some $\beta \in (0,1]$. 
For any $g\in C^\infty_c(\R^n)$ such that  $\supp(T^*_{x_0,\epsilon} g) \subset  \{V \le \mu_0 \}$ for any $\epsilon\le \epsilon(1)$, 
it holds for any $0<\alpha<\beta$ as $\hbar\to 0$, 
 \[
 \tr\big( f_{\hbar,\mu}^2(H_{\hbar}) T^*_{x_0,\epsilon}g\big) =  
 \frac{|S^{n-1}|}{2(2\pi \delta)^n} \int   g_\epsilon(x)  \vartheta(\lambda)^2 \big(\lambda-V(x)\big)^{\frac{n-2}{2}} \1_{\mu \ge \lambda}  \d x  \d \lambda
  + \O(\hbar^{\alpha} \delta^{1-n}), 
  \]
  where the  error term is locally uniform for $(x_0,\mu) \in \big\{ (x,\lambda) \in \R^{n+1} : V(x) < \lambda <M \big\}$. 
  \end{prop}

\begin{preuve}
As in formula \eqref{fop}, it holds
\begin{equation*} 
f_{\hbar, \mu}^2(H_\hbar) =  \frac{1}{2\pi\hbar^2} \int  \varkappa_\mu(\lambda_1)   \varkappa_\mu(\lambda_2) 
 \big( I_{\hbar,t_1+t_2}^{\phi, a}  + \O_{\J^1}(\hbar^\infty)\big)     e^{ -\i\tfrac{t_1\lambda_1+t_2\lambda_2}{\hbar}}  \hat{\rho}(t_1)  \hat{\rho}(t_2) \dd \lambda_1 \d\lambda_2 \d t_1\d t_2 
\end{equation*}
where this estimate is uniform for $\lambda_1,\lambda_2 \in \supp(\varkappa)$ and $t_1,t_2 \in [-,\tau,\tau]$. 
Making a change of variables
\[
        t= t_1+t_2 \qquad s=\tfrac{t_1-t_2}{2}\qquad
        \lambda =\tfrac{\lambda_1+\lambda_2}{2}\qquad
       \sigma=\lambda_1-\lambda_2,
\]
this implies that       
\begin{equation*} 
   \tr\big( f_{\hbar,\mu}^2(H_{\hbar}) T^*_{x_0,\epsilon}g \big) =  \frac{1}{2\pi\hbar^2(2\pi \delta)^n} \int  
    e^{\i \tfrac{\phi(t,x,\xi) - \xi\cdot x - t\lambda - s\sigma }{\hbar}} g_\epsilon(x) a(x,x,\xi,t;\hbar) 
 \Gamma_{\mu}(t,\Lambda) \d t \d x \d \xi \d \Lambda + \O(\hbar^\infty) 
\end{equation*}
where,  correspondingly, we let
\[
        \Gamma_{\mu}(t,\Lambda) = \hat{\rho}(\tfrac t2+s) \hat{\rho}(\tfrac t2-s)
        \varkappa(\lambda+\tfrac{\sigma}{2})\varkappa(\lambda-\tfrac{\sigma}{2})\1_{|\frac{\sigma}{2}|\leq \mu-\lambda} , \qquad 
          \Lambda = (s,\lambda,\sigma) . 
 \]
 
 Let us make a change of variable $\xi \leftarrow \eta$ as in Proposition \ref{prop:reduction-phase} and decompose $\eta$ in polar coordinates: $\eta=r\omega$ where $(r,\omega)\in
\R_+ \times S^{n-1}$. We obtain
\begin{equation*} 
   \tr\big( f_{\hbar,\mu}^2(H_{\hbar}) T^*_{x_0,\epsilon}g\big) =  \frac{1}{2\pi\hbar^2(2\pi \delta)^n} \int  
    e^{\i  \tfrac{t(r^2 + g(t,x) + V(x) - \lambda) - s\sigma }{\hbar}} g_\epsilon(x) 
b(t,x,r,\omega;\hbar) \Gamma_{\mu}(t,\Lambda) \d t \d x \d r \d \omega \d \Lambda + \O(\hbar^\infty) 
\end{equation*}
where $b\in S^0$ is a classical symbol given by
\[
b(t,x,r,\omega;\hbar) = a(x,x,\xi(t,x,r\omega),t;\hbar)  \left|{\rm Jac}[\xi\leftarrow r \omega ](t,x,r,\omega)\right|.
\]

We now apply the stationary phase method as in the proof of Proposition~\ref{prop:regularised_kernel_bulk} in the variable $(r,t) \in [c,\infty) \times [-2\tau,2\tau]$
for a fixed $(x,\omega,\Lambda) \in \supp(g_\epsilon)\times S^{n-1} \times \R^3$. 
The equation for the critical point(s) are given by  \eqref{eq:phase1} (upon replacing $x_0$ by $x$). 
In particular, we can assume that $\supp(g_\epsilon) \subset  \{V \le \mu_0 \}$ for any $\epsilon\le \epsilon(1)$ and introduce a cutoff $\theta \in C^\infty(\R,[0,1])$ such that   $\theta(\lambda)=0$ for $\lambda\le \mu_0$ and $\theta(\lambda)=1$ for $\lambda \in \supp(\vartheta)$. 
Then, for $\lambda \in \supp(\theta)$, the (unique) critical point is given by 
$(r,t) = (r_\star(x,\lambda),0)$ where 
\begin{equation} \label{def:R} 
r_\star(x,\lambda)  := \sqrt{\lambda-V(x)} . 
\end{equation}

By Proposition~\ref{prop:spl}, we obtain
\begin{equation} \label{sp1}
\frac{1}{2\pi\hbar}  \int e^{\i t \tfrac{ r^2 + g(t,x) + V(x) - \lambda}{\hbar}} b(t,x,r,\omega;\hbar) \Gamma_\mu(t,\Lambda) \d t \d r
=\theta(\lambda) d(x,\lambda,\omega; \hbar) \Gamma_\mu(0,\Lambda) + \O(\hbar^\infty) 
\end{equation}
where $d\in S^0$ is again a classical symbol whose principal part is given by 
\[
d_0(x,\lambda,\omega) =  \frac{1}{2r_\star(x,\lambda)}b_0(0,x,r_\star(x,\lambda),\omega) = \frac{1}{2}a_0(x,x,r_\star(x,\lambda) \omega,0) r_\star(x,\lambda)^{n-2}.
\]
where we used that according to \eqref{etaTaylor}, 
$\xi(0,x,\eta) = \eta$ and ${\rm Jac}[\xi\leftarrow \eta](0,x,\eta)=1$. 
Given the explicit formula for $a_0$, $d_0$ is independent of $\omega$ and by \eqref{def:R}, it is given by 
\begin{equation} \label{symbold0}
d_0(x,\lambda) =  \frac 12\vartheta\big( V(x)+ r_\star(x,\lambda)^2 \big)^2 r_\star(x,\lambda)^{n-2}
=  \frac 12 \vartheta(\lambda) r_\star(x,\lambda)^{n-2} .
\end{equation}

Moreover, the error in \eqref{sp1} is uniform for all $x\in \supp(g_\epsilon)$, $\omega\in S^{n-1}$ and locally uniform for $\Lambda\in\R^3$. Hence, we conclude that 
\begin{equation*} 
   \tr\big( f_{\hbar,\mu}^2(H_{\hbar})T^*_{x_0,\epsilon}g \big) =  \frac{1}{\hbar(2\pi \delta)^n} \int  
    e^{-\i  \tfrac{s\sigma}{\hbar}} g_\epsilon(x) 
d(x,\lambda,\omega;\hbar) \Gamma_{\mu}(0,\Lambda) \theta(\lambda) \d x \d \omega \d \Lambda + \O(\hbar^\infty) .
\end{equation*}

Next, let us observe that because of the cutoff $\theta$ ($\theta(\lambda)=0$ for $\lambda\le \mu_0$ with $\mu_0>0$ and $\varkappa =1$ on  $[0,M]$)
\[
\Gamma_{\mu}(0,\Lambda) \theta(\lambda)
= \hat{\rho}(s)^2 \1_{|\frac{\sigma}{2}|\leq \mu-\lambda}  \theta(\lambda) . 
\]
This allows to compute explicitly the integral wrt~$\sigma$, we obtain
\begin{equation*}  
   \tr\big( f_{\hbar,\mu}^2(H_{\hbar})T^*_{x_0,\epsilon}g\big) =  \frac{1}{\hbar(2\pi \delta)^n} \int  \frac{\sin( 2(\mu-\lambda)s /\hbar)}{s/2\hbar} g_\epsilon(x) 
d(x,\lambda,\omega;\hbar) \hat{\rho}(s)^2\theta(\lambda) \1_{\mu\ge \lambda} \d s \d x \d \omega \d \lambda + \O(\hbar^\infty) .
\end{equation*} 

We can now compute the integral wrt to $(\lambda,s)$ using Proposition~\ref{prop:sineint}.
Since $\lambda \mapsto d(x,\lambda,\omega;\hbar) \theta(\lambda) \1_{\mu\ge \lambda}$ is $L^\infty_c$ (uniformly for $\omega\in S^{n-1}$ and $x \in \supp(g_\epsilon)$)
and $\hat\rho^2 \in C^\infty_c(\R)$ with $\hat\rho(0)=1/\sqrt{2\pi}$, this implies that for any $\gamma\in (0,1]$, 
\begin{equation*}  
   \tr\big( f_{\hbar,\mu}^2(H_{\hbar}) T^*_{x_0,\epsilon}g \big) =  \frac{1}{(2\pi \delta)^n} \int   g_\epsilon(x) 
d(x,\lambda,\omega;\hbar) \theta(\lambda)  \1_{\mu\ge \lambda}   \d x \d \omega \d \lambda + \O_\gamma(\delta^{1-\gamma}) .
\end{equation*} 
Here we also used that $\| g_\epsilon\|_{L^1(\R^n)} <\infty$ is independent of $\epsilon>0$.
Since $\delta\le \hbar^\beta$, the error term is of order $\hbar^{\alpha}$ for any $0<\alpha<\beta$ and we can also replace the symbol $d$ by its principal part  \eqref{symbold0} up to a negligible error.
Since $\theta(\lambda) =1$ for $\lambda \in \supp(\vartheta)$ and $d_0$ is independent of $\omega \in S^{n-1}$, this completes the proof. 
\end{preuve}

  \begin{prop} \label{prop:tr2}
Let $\epsilon :[0,1] \to[0,1]$ be a non-increasing function such that $\hbar^{1-\beta} \le \epsilon(\hbar) \le \hbar^{\beta}$ for some $\beta>0$ and let $\delta(\hbar) = \hbar/\epsilon(\hbar)$. 
There exists $\alpha>0$ so that for any $g\in C^\infty_c(\R^n)$, 
\[ \begin{aligned}
   \tr\big( \big( f_{\hbar,\mu}(H_{\hbar}) T^*_{x_0,\epsilon}g \big)^2 \big) 
&=\frac{|S^{n-1}|}{2(2\pi \delta)^{n} }\int g_\epsilon(x)^2
\vartheta(\lambda)
\big(\lambda-V(x)\big)^{\frac{n-2}{2}}\1_{\lambda\leq \mu}\d x  \d \lambda   \\
 &\quad - \sigma_n^{2}\vartheta(\mu)\delta^{1-n} \big(\mu-V(x_0)\big)^{\frac{n-1}{2}} \Sigma^2(f)   
 + \O(\delta^{1-n}\hbar^{\alpha}),  
   \end{aligned} \]
 where $\sigma_n$ and $\Sigma^2(f)$ are as in Theorem~\ref{thr:mesoscopic_bulk_lapl} and the  error term is locally uniform for $(x_0,\mu) \in \big\{ (x,\lambda) \in \R^{n+1} : V(x) < \lambda <M \big\}$.
  \end{prop}

Recall the notations from \eqref{kop}:
\[
I_{\hbar,t}^{\phi, a}(x,y) = 
      \frac{1}{(2\pi \hbar)^n}\int e^{\i \frac{   \psi(t,x,y,\xi)  }{\hbar}}a(t,x,y,\xi;\hbar)\dd \xi 
       , \qquad       \psi(t,x,y,\xi) := \phi(t,x,\xi)-y\cdot \xi . 
\]
The proof requires the following basic estimates on the decay of this kernel.

\begin{lem} \label{lem:cutoff}
Suppose that $\epsilon \ge \hbar^{1-\beta}$ for $\beta \in (0,1]$. 
Suppose that the symbol $(t,x,y,\xi) \mapsto a(t,x,y,\xi;\hbar)$ is supported on 
$
\big\{t \in [-\tau,\tau] , x, y \in B^n_{0,\epsilon} , \xi \in \mathcal{K}  : |t| \ge C \epsilon \big\}
$
where $\mathcal{K} \Subset \R^n \setminus\{0\}$ and $C$ is a large enough constant. 
Then,
$I_{\hbar,t}^{\phi, a}(x,y) = \O(\hbar^\infty)$ for $(t,x,y) \in \supp(a)$.  
\end{lem}

\begin{preuve}
By \eqref{Taylor1}, we have 
\[
\partial_\xi \psi(t,x,y,\xi)
=(x-y)+ 2 t \xi +\O(t^2) .
\]
Hence, since $|\xi| > c$, $|t| \ge C \epsilon$ and $|x-y| \le 2 \epsilon$   on $\supp(a)$,  on has $\big| \partial_\xi \psi(t,x,y,\xi) \big| \ge  \epsilon$ and for any multi-index $\alpha$ with $|\alpha| \ge 1$,
$\big| \partial_\xi^\alpha \psi(t,x,y,\xi)\big|  \le C_\alpha \epsilon$. 
 In particular, for a multi-index $\alpha$ with $|\alpha|=1$, the differential operator
    \[
      \mathcal{L}_\xi : u \mapsto \partial_{\xi}^\alpha \big(\frac{u}{\partial_\xi^\alpha \psi} \big)
    \]
satisfies for $u \in C^\infty_c(\R^n)$ and every $k\in \N$, 
    \[
      \|\mathcal{L}_\xi^k u \|_{C^0}\leq \frac{C_k}{\epsilon^k}\|u\|_{C^{k}} . 
    \]
    
  By repeated integration by parts, we obtain for every $k\in\N$, 
\[
\int e^{\i \frac{   \psi(t,x,y,\xi)  }{\hbar}} a(t,x,y,\xi;\hbar)\dd \xi 
= (\i\hbar)^k \int e^{\i \frac{   \psi(t,x,y,\xi)  }{\hbar}} \mathcal{L}_\xi^ka(t,x,y,\xi;\hbar) \dd \xi  .
\]  
Since $a \in S^0$ is supported in $\mathcal{K} $, the RHS is of order $\O_k\big((\hbar/\epsilon)^k\big)$.
 Because we assume that  $\epsilon \ge \hbar^{1-\beta}$ for $\beta>0$, this proves that  $I_{\hbar,t}^{\phi, a}(x,y) = \O(\hbar^\infty)$. 
\end{preuve}

\begin{preuve} [of Proposition~\ref{prop:tr2}]
The argument splits in several parts: a reduction step, a first
application of the stationary phase, a rescaling, and then the actual
computation of the leading and subleading terms.

\subsubsection{Reduction steps} 
We can assume that $\epsilon \le \epsilon(1)$ is small enough so that $\supp(g_\epsilon) \subset B^n_{0,c}$ for all $\epsilon\in(0,\epsilon(1)]$ and $V(x) < \mu_0$ for all $x\in B^n_{0,c}$.

Using the representation \eqref{fop} and proceeding as in the proof of Proposition~\ref{prop:commut-macroscopic-smooth} for $t\in[-\tau,\tau]$,  we obtain
  \[\begin{aligned}
   \tr\big( \big( f_{\hbar,\mu}(H_{\hbar}) g_\epsilon \big)^2 \big) 
=  \frac{\epsilon^{2n}}{2\pi \hbar^2} \int   \tr\big( g_\epsilon I_{\hbar,t_1}^{\phi, a} g_\epsilon  I_{\hbar,t_2}^{\phi, a \, \dagger} \big)    e^{ -\i\frac{t_1\lambda_1}{\hbar} -\i\frac{t_2\lambda_2}{\hbar}}  \varkappa_\mu(\lambda_1)   \varkappa_\mu(\lambda_2) \hat{\rho}(t_1) \hat{\rho}(t_2)   \d t_1 \d t_2 \d \lambda_1 \d \lambda_2  + \O(\hbar^\infty)  . 
 \end{aligned}\]
where $ I_{\hbar,t}^{\phi, a \, \dagger}$ denotes the adjoint of  $I_{\hbar,t}^{\phi, a}$ 
and $a\in S^0$ is a classical symbol whose principal part satisfies 
\[
a_0(t,x,y,\xi) = \vartheta(V(\tfrac{x+y}{2}+|\xi|^2)) + \O_{x,y,\xi}(t) , \qquad x ,y \in B^n_{0,c} , \quad    \xi\in\R^n . 
\]
Note that we used that the operator $f_{\hbar,\mu}(H_{\hbar})$ is self-adjoint, though $ I_{\hbar,t}^{\phi, a}$ is not. 
Hence,  the only relevant contribution to the trace  at hand is given by the oscillatory integral
\[
\mathfrak{I}_\hbar=  \frac{1}{2\pi\hbar^2(2\pi\delta)^{2n} } \int e^{\i \tfrac{\psi(\Lambda_1)-\psi(\Lambda_2)}{\hbar}}  g_\epsilon (x) g_\epsilon (y) a(t_1,x,y,\xi_1;\hbar) \overline{a(t_2,y,x,\xi_2;\hbar)} 
\varkappa_\mu(\lambda_1)   \varkappa_\mu(\lambda_2) \hat{\rho}(t_1) \hat{\rho}(t_2)   \d \Lambda_1\d \Lambda_2,    
\]
where we used the shorthand notation
\[ \begin{aligned}
 \psi(\Lambda_j)  = \phi(t_j,x,\xi_j)- \xi_j\cdot y - t_j  \lambda_j  \qquad 
\Lambda_j  = (t_j,x,y,\xi_j,\lambda_j) .  
\end{aligned}\]

Observe that there is a constant $c$ so that for $j\in\{1,2\}$,  the symbols
$\Lambda_j \mapsto g_\epsilon (x) g_\epsilon (y)) a(t_j,x,y,\xi_j;\hbar)$ are supported in 
$\big\{t \in [-\tau,\tau] , x, y \in B^n_{0,c\epsilon} , \xi \in \mathcal{K}   \big\}$
 for a $\mathcal{K} \Subset \R^n \setminus\{0\}$ 
(since $\vartheta(V(x))=0$ for all $x$ in a neighborhood of $\supp(g_\epsilon)$). 
Hence, by Lemma~\ref{lem:cutoff}, we may add a cutoff $\chi(\tfrac{t_1+t_2}{2 \epsilon^{1-\kappa}})$, where $\chi \in C^\infty_c(\R,[0,1])$ satisfies $\chi$ is an even function which equals to $1$ on a neighborhood 0 and $\kappa>0$ is a small parameter,  inside the integral $J_\hbar$ up to an error which is $\O(\hbar^\infty)$. 
Indeed, if $|\tfrac{t_1+t_2}{2 \epsilon }| \ge C $, then either $|t_1| \ge C \epsilon$ or $|t_2| \ge C\epsilon$, and in either case Lemma~\ref{lem:cutoff} applies. 
The choice of cutoff $\chi(\tfrac{t_1+t_2}{2 \epsilon^{ 1-\kappa}})$ is tuned for the sequel of the argument.  
In summary, we have 
\begin{align} \notag
\mathfrak{I}_\hbar=  \frac{1}{2\pi\hbar^2(2\pi\delta)^{2n} } \int & e^{\i \frac{\psi(\Lambda_1)-\psi(\Lambda_2)}{\hbar}}  g_\epsilon (x) g_\epsilon (y)  a(t_1,x,y,\xi_1;\hbar) \overline{a(t_2,y,x,\xi_2;\hbar)}  
 \Gamma_{\mu}(\tfrac{\lambda_1+\lambda_2}{2}, \lambda_1-\lambda_2) \Theta_\epsilon(\tfrac{t_1+t_2}{2},t_1-t_2)   \\
  \label{J1}
 & \d \Lambda_1\d \Lambda_2 + \O(\hbar^\infty)
 \end{align} 
where,  correspondingly, we let
\[
        \Gamma_{\mu}(\lambda,\sigma) =
        \varkappa(\lambda+\tfrac{\sigma}{2})\varkappa(\lambda-\tfrac{\sigma}{2})\1_{|\frac{\sigma}{2}|\leq
          \mu-\lambda}
 \]
and 
\[
\Theta_\epsilon(t,s) = \hat{\rho}(t+\tfrac s2) \hat{\rho}(t-\tfrac s2) \chi(t \epsilon^{ \kappa-1}) . 
\]

In preparation for the stationary phase, we introduce the new variables 
      \[
      r \omega = \tfrac{\xi_1+\xi_2}{2}\qquad
      \zeta = \xi_1-\xi_2\qquad
        t=\tfrac{t_1+t_2}{2}\qquad s=t_1-t_2\qquad
        \lambda=\tfrac{\lambda_1+\lambda_2}{2}\qquad
        \sigma=\lambda_1-\lambda_2,
      \]
where $(r,\omega) \in\R_+\times S^{n-1}$ and we let 
\[
\Psi(t,s,r,\zeta,x, \omega,\lambda) = \phi(t+\tfrac s2, x,
        r\omega+\tfrac \zeta2)-\phi(t-\tfrac s2, x,r\omega-\tfrac
        \zeta 2) -s\lambda .
\]
By making this change of variables in \eqref{J1}, one has 
\begin{equation} \label{J2}
\mathfrak{I}_\hbar=  \frac{1}{\hbar (2\pi\delta)^{2n} }  \int e^{-\i \tfrac{\zeta \cdot y+ t\sigma}{\hbar}}  g_\epsilon (x)g_\epsilon (y)  \mathfrak{L}_\hbar(\Lambda)  \Gamma_{\mu}(\lambda, \sigma) \d \Lambda \d \sigma +\O(\hbar^{\infty}),
\end{equation}
where 
\[
\mathfrak{L}_\hbar(\Lambda)
:=  \frac{1}{2\pi \hbar} \int e^{\i \tfrac{\Psi(s,r,\Lambda)}{\hbar}}  d\left(s,r,\Lambda;\hbar\right) \Theta_\epsilon(t,s) \d s \d r , \qquad  \Lambda = (t,\zeta,x,y,\omega,\lambda)
\]
and we set
\[
d\left(s,r,\Lambda;\hbar\right)   =  r^{n-1} a(t+\tfrac s2, x, y, r\omega+\tfrac \zeta2;\hbar) \overline{a(t-\tfrac s2, y, x, r\omega-\tfrac \zeta2;\hbar)} . 
\]

\begin{rem}  \label{rk:zetadecay}
We can also write the integral \eqref{J2} as 
\[
\mathfrak{I}_\hbar=  \frac{1}{2\pi\hbar^2(2\pi\delta)^{2n} }  \int e^{\i \tfrac{\Psi(t,s,r,\zeta,x, \omega,\lambda) - t\sigma}{\hbar}}  g_\epsilon (x)  \mathfrak{M}_\hbar (t,s,r,\zeta,x,\omega)   \Gamma_{\mu}(\lambda, \sigma)  \Theta_\epsilon(t,s)  \d t\d s \d r   \d \zeta \d x \d \omega \d \lambda \d \sigma + \O(\hbar^\infty),
\]
where
\[
\mathfrak{M}_\hbar (t,s,r,\zeta,x,\omega)  := \int  e^{-\i \tfrac{\zeta \cdot y}{\hbar}} g_\epsilon (y) d\left(s,r,\Lambda;\hbar\right)  \d y .
\]
The point is that since $d\in S^0$ and $g\in C^{\infty}$,  by making
repeated integration by parts, we obtain if $|\zeta| \ge z$
for some $z>0$, then for every $k\in\N_0$, 
\[
\mathfrak{M}_\hbar (t,s,r,\zeta,x,\omega)  = \O_k\big(   z^{-k} \hbar^k \epsilon^{-n-k} \big)
\]
uniformly for $(t,s) \in [-2\tau, 2\tau]$, $x\in  B^n_{0,c}$ and $r\omega \in \mathcal{K}$. 
Recall that $\delta  = \hbar/\epsilon$ and we assume that $\delta \le
\hbar^\beta $ where $\beta >0$. Hence, we conclude that
$\mathfrak{M}_\hbar  = \O(\hbar^\infty)$ if  $|\zeta| \ge
\delta^{1-\kappa}$ for any $\kappa>0$. This argument allows us to
include at will a cutoff $\chi(|\zeta| \delta^{\kappa-1})$ inside the
integral \eqref{J2}  where again  $\chi\in C^{\infty}_c(\R,[0,1])$ is
even and equal to 1 on a neighborhood of 0. Hence
$\tr((f_{\hbar,\mu}(H_{\hbar}),g_{\epsilon})^2)=\mathfrak{I}_{\hbar}'+\O(\hbar^\infty)$,
where
\begin{equation}\label{eq:Jprime}
  \mathfrak{I}_\hbar'=  \frac{1}{\hbar (2\pi\delta)^{2n} }  \int e^{-\i
    \tfrac{\zeta \cdot y+ t\sigma}{\hbar}}  g_\epsilon (x)g_\epsilon
  (y)  \mathfrak{L}_\hbar(\Lambda)  \Gamma_{\mu}(\lambda, \sigma) \chi(|\zeta|\delta^{\kappa-1})\d
  \Lambda \d \sigma.
\end{equation}
\end{rem}

\subsubsection{Stationary phase}

We now apply the stationary phase method to the oscillatory integral
$\mathfrak{L}_\hbar$ with respect to the variables $(r,s)$, keeping $\Lambda = (t,\zeta,x,y,\omega,\lambda)$ fixed. 
The equations for the critical point(s) are
\[\begin{cases}
0 = \partial_r \Psi (s,r,\Lambda) =  \omega \cdot \big( \partial_\xi \phi(t+\tfrac s2, x,
        r\omega+\tfrac \zeta2)- \partial_\xi\phi(t-\tfrac s2, x,r\omega-\tfrac
        \zeta 2)  \big) \\
0 =  \partial_s \Psi (s,r,\Lambda) = \tfrac12 \big( \partial_t \phi(t+\tfrac s2, x,
        r\omega+\tfrac \zeta2)+\partial_t\phi(t-\tfrac s2, x,r\omega-\tfrac
        \zeta 2)  \big) - \lambda
\end{cases}.\]
By \eqref{Taylor1}, using that $(t+\tfrac s2)|r\omega+\tfrac \zeta2|^2- (t-\tfrac s2)|r\omega-\tfrac \zeta2|^2 = (r^2 + \frac{|\zeta|^2}{4}) s + 2 rt\, \zeta\cdot \omega$ and considering even/odd terms in $\zeta$, the phase as the following expansion for small times, 
\begin{equation} \label{Psiexp}
 \Psi (s,r,\Lambda) = x \cdot \zeta +s \big(V(x) + r^2 + \frac{|\zeta|^2}{4}-\lambda\big) + 2rt \zeta \cdot \omega
 + s \O(\varsigma) + \zeta \cdot \O(\varsigma^2)
\end{equation} 
where $\varsigma = \max\big\{ |t|, |s|\big\} \le 2
\tau$.  
Similarly, by expanding the previous equations for small $(t,s)$, we obtain 
\begin{equation} \label{criteq}
\begin{cases}
0 = \partial_r \Psi  (s,r,\Lambda) = 2 r s + 2 \omega \cdot \big( t\zeta + s\, \O(\varsigma) + \zeta \, \O(\varsigma^2) \big)  \\
0 =  \partial_s \Psi  (s,r,\Lambda) =V(x) + r^2 + \frac{|\zeta|^2}{4}-\lambda + \O(\varsigma) 
\end{cases}
\end{equation}

Since $r\ge c_0 \gg \tau$ and we may assume that $|\zeta|\le c$ for any constant $c$ (cf.~Remark~\ref{rk:zetadecay}), these equation have the following consequences:
  \begin{itemize}[leftmargin=.5cm]
\item
if the parameter $\lambda\le \mu_0$, then the second equation has no solution $r \ge c$. In this case, there is no critical point and  by Proposition ~\ref{prop:nspl}, we conclude that $I_\hbar =\O(\hbar^\infty)$. We can therefore include a cutoff $\theta(\lambda)$ in the integrand of \eqref{Jint1} where $\theta \in C^\infty(\R,[0,1])$ satisfies  $\theta(\lambda)=0$ for $\lambda\le \mu_0$ and $\theta(\lambda)=1$ for $\lambda \in \supp(\vartheta)$. 
\item For $\lambda\ge \mu_0$, \eqref{criteq} has a unique solution; $r_{\rm c}>0$ and $s_{\rm c}$ have the following expansion
\begin{equation} \label{crit}
\begin{cases}
r_{\rm c}(\Lambda) = \sqrt{\lambda- V(x)-  \tfrac{|\zeta|^2}{4}}  + \O(\varsigma) \\
s_{\rm c}(\Lambda) =-  r_{\rm c}^{-1}\, t \, \zeta \cdot \omega \big(1+ \O(\varsigma) \big)
\end{cases} .
\end{equation}
\end{itemize}

In particular, the critical point at $t=0$ is explicit, $(r_{\rm c,0}, s_{\rm c,0}) = \big(\sqrt{\lambda- V(x)-  \frac{|\zeta|^2}{4}} ,0 \big)$, as well as the corresponding Hessian matrix. Indeed, using the first equation in \eqref{criteq}, we compute
\[\begin{cases}
\partial_{rr} \Psi (t,s_{\rm c},r_{\rm c},\zeta,x, \omega,\lambda) \big|_{t=0}  
= \big[ s_{\rm c} +  \O(\varsigma) \big]_{t=0} = 0  \\
 \partial_{sr} \Psi (t,s_{\rm c},r_{\rm c},\zeta,x, \omega,\lambda) \big|_{t=0}  
 = \big[2 r_{\rm c}+  \O(\varsigma) \big]_{t=0} 
 =2 r_{\rm c,0} >0
  \end{cases}.\]

This shows that the Hessian is non-degenerate at $t=0$ and its determinant is given by
$\Sigma^2_{\rm c,0} = 4 r_{\rm c,0}^2$. 
This property is preserved in a small neighborhood of $t=0$ (which clearly contains $\supp(\Theta_\epsilon)$), so that we can apply Proposition~\ref{prop:spl}  to the integral $\mathfrak{L}_\hbar$, we obtain
\[
\mathfrak{L}_\hbar(\Lambda)
=e^{\i \tfrac{\Phi(\Lambda)}{\hbar}} e(\Lambda;\hbar)\Theta_\epsilon(t,s_{\rm c})  , \qquad  
\Phi(\Lambda) :=\Psi(t,s_{\rm c}, r_{\rm c},\zeta,x, \omega,\lambda),
\]
where $e\in S^0$ is again a classical symbol with principal part at    time $t=0$ along the diagonal $(x=y)$, 
\begin{equation} \label{symbole0}
\begin{aligned}
e_0(\Lambda_0)
 =  \Sigma_{\rm c,0}^{-2}\,  d_0\left(0,r_{\rm c,0},\Lambda_0\right) 
 = \tfrac12 r_{\rm c,0}^{n-2}(\zeta, x, \lambda)  \vartheta\big(V(x)+| r_{\rm c,0}\omega+\tfrac \zeta2|^2)\big) \vartheta\big(V(x)+| r_{\rm c,0}\omega-\tfrac \zeta2|^2\big)
 \end{aligned}
\end{equation}
 where $\Lambda_0 = (0,\zeta,x,x,\omega,\lambda)$.

Let us observe that $s_{\rm c} = \zeta \cdot \O(t)$ and therefore the control parameter $\varsigma =t$ in the regime that we consider. 
By \eqref{Psiexp} and \eqref{crit}, this implies that the new phase has the following expansion for small $t$, 
\begin{equation} \label{Phiexp}
 \begin{aligned}
\Phi(\Lambda) 
&= x\cdot \zeta +  2 r_{\rm c} \, t \zeta \cdot \omega + s_{\rm c} \O(t) +\zeta \cdot \O(t^2) \\
&= x\cdot \zeta + 2 \sqrt{\lambda - V(x)}  \, t \zeta \cdot \big( \omega  + \O(t)+  \O(|\zeta|^2) \big) .
\end{aligned}
\end{equation}

\medskip

Going back to formula \eqref{eq:Jprime}, we conclude that 
\begin{equation}\label{eq:Jprime_post_statphase}
\mathfrak{I}_\hbar'=  \frac{1}{\hbar(2\pi\delta)^{2n} }  \int e^{-\i \tfrac{\zeta \cdot y+ t\sigma - \Phi(\Lambda)}{\hbar}}  g_\epsilon (x) g_\epsilon (y)e(\Lambda;\hbar)  \Theta_\epsilon(t,s_{\rm c})   \Gamma_{\mu}(\lambda, \sigma) \theta(\lambda)  \chi(|\zeta| \delta^{\kappa-1}) \d \Lambda \d \sigma + \O(\hbar^\infty) . 
\end{equation}

\subsubsection{(Re)scaling}
Let us first observe that because of the cutoff  $\theta \le \1_{[\mu_0,\infty)}$ with $\mu_0>\mu$ and $\varkappa =1$ on  $[0,M]$, we have
\[
\Gamma_{\mu}(\lambda, \sigma) \theta(\lambda) 
=\1_{|\frac{\sigma}{2}|\leq
          \mu-\lambda} \theta(\lambda) . 
\]
This allows to compute the integral with respect to~$\sigma$ in
\eqref{eq:Jprime_post_statphase}, since $\sigma$ does not appear in $\Lambda$.
\[
\int e^{- \i \tfrac{t\sigma}{\hbar}}  \Gamma_{\mu}(\lambda, \sigma)\d \sigma\theta(\lambda)
= \1_{\mu\ge \lambda}\theta(\lambda) \frac{\sin(\frac{2(\mu-\lambda)t}{\hbar})}{t/2\hbar}  .
\]
This implies that 
\begin{equation} \label{J3}
\mathfrak{I}_\hbar'=  \frac{2}{(2\pi\delta)^{2n} }  \int e^{\i \tfrac{\Phi(\Lambda)-\zeta \cdot y}{\hbar}}    g_\epsilon (y)     g_\epsilon (x) e(\Lambda;\hbar) \Theta_\epsilon(t,s_{\rm c})  \frac{\sin(\frac{2(\mu-\lambda)t}{\hbar})}{t} \theta(\lambda)  \chi( |\zeta| \delta^{\kappa-1})   \1_{\mu\ge \lambda} \d \Lambda  + \O(\hbar^\infty)
\end{equation}

We can now make the following change of variables in the integral \eqref{J3}
\[
y\leftarrow x_0 + \epsilon y ,\qquad
x\leftarrow x_0 + \epsilon x ,\qquad
t\leftarrow \epsilon t, \qquad
\zeta \leftarrow \delta \zeta 
\]
where we recall that $\delta =\hbar/\epsilon$. In particular, by the expansion \eqref{Phiexp}, the phase satisfies for $|t| \le \epsilon^{-\kappa}$ and $|\zeta| \le \delta^{-\kappa}$, 
\[
\frac{\Phi (\epsilon t, \delta \zeta, x_0 + \epsilon x , x_0 + \epsilon y,\omega,\lambda) - \delta \zeta \cdot(x_0+\epsilon y)}{\hbar}
=  (x-y)\cdot \zeta +  2 r_\star(x_0,\lambda)  t \zeta \cdot  \big( \omega +  \O(\delta^{2-2\kappa}) \big) + t\,  \O(\epsilon  \delta^{-\kappa})
\]
where $r_\star$ is given by \eqref{def:R}.

Moreover, by \eqref{symbole0} Taylor expansions (in a neighborhood of $t=0$, $\zeta=0$ and the diagonal $x=y$), one obtains 
\[ \begin{aligned}
 & e_0(t,\delta \zeta, x_0 + \epsilon x , x_0 + \epsilon y,\omega,\lambda)  \\
& = \tfrac12 r_{\rm c,0}^{n-2}(\delta \zeta, x_\epsilon, \lambda)  \vartheta(V(x_\epsilon)+| r_{\rm c,0}\omega+\delta\tfrac \zeta2|^2)\big) \vartheta\big(V(x_\epsilon)+| r_{\rm c,0}\omega-\delta\tfrac \zeta2|^2\big)
+\O_\Lambda\big(|t|\big)+\O_\Lambda\big(\epsilon|x-y|\big)  \\
&=  \tfrac12 r_\star(x_\epsilon,\lambda)^{n-2} 
 \vartheta\big(V(x_\epsilon)+r_\star(x_\epsilon,\lambda)^2\big)^2+ \O_\Lambda(\delta^{2-2\kappa})
 +\O_\Lambda\big(|t|\big)+\O_\Lambda\big(\epsilon|x-y|\big)  \end{aligned}\]
where the linear terms in $\zeta$ exactly cancel and we use the shorthand notation
\[
\quad x_\epsilon = x_0 + \epsilon x . 
\]

Finally, since $\hat\rho$ is even with $\hat{\rho}(0) =1/\sqrt{2\pi}$, it holds by \eqref{crit}, 
\[
 \Theta_\epsilon(\epsilon t,s_{\rm c})  =  \tfrac{1}{2\pi} \big(1 +  \O(t^2\epsilon^2)\big) \chi(t \epsilon^{\kappa})  . 
\]

We also emphasize that these expansions are  all uniform for all $x,y \in \supp(g)$, $\omega\in S^{n-1}$, $\lambda \in \supp(\theta)$ and $(x_0,\mu) \in \mathcal{A}$. 
Altogether, this implies that as $\hbar\to0$, 
\[ \begin{aligned}
e^{-\i \tfrac{\Phi (\epsilon t, \delta \zeta, x_0 + \epsilon x , x_0 + \epsilon y,\omega,\lambda) - \delta \zeta \cdot(x_0+\epsilon y)}{\hbar}}   \Theta_\epsilon(\epsilon t,s_{\rm c}) e (\epsilon t, \delta \zeta,x_0 + \epsilon x ,x_0 + \epsilon y,\omega,\lambda ;\hbar) \\ 
=   \tfrac{1}{4\pi} e^{\i (x-y+2 r_\star(x_0,\lambda) t \omega)\cdot \zeta } \big( 
F(x_\epsilon,\lambda)   + \Upsilon_{1;\epsilon,\delta}(\Lambda) +t \Upsilon_{2;\epsilon,\delta}(\Lambda)
+ 
  \left(x-y+2r_\star(x_0,\lambda)t\omega\right)\cdot \Upsilon_{3;\epsilon,\delta}(\Lambda)
\big) \chi(t \epsilon^{\kappa})
\end{aligned}\]
where the main term is given by
\begin{equation*} 
F(x,\lambda) 
=  r_\star(x,\lambda)^{n-2} \vartheta\big(V(\tfrac{x+y}{2})+r_\star(x,\lambda)^2\big)^2 
=  r_\star(x,\lambda)^{n-2} \vartheta(\lambda)^2 , 
\end{equation*}
according to \eqref{def:R},
and the error terms are smooth functions such that
\begin{align*}
\Upsilon_{1;\epsilon,\delta}(t,\zeta,x,y,\omega,\lambda)
&=
\O_{\Lambda}(\delta^{2-2\kappa}) +\O_\Lambda(\hbar) ,\\
\Upsilon_{2;\epsilon,\delta}(t,\zeta,x,y,\omega,\lambda)
&=\O_\Lambda(\epsilon^{1-\kappa}) , \\
 \Upsilon_{3;\epsilon,\delta}(t,\zeta,x,y,\omega,\lambda)
  &    =\O_{\Lambda}(\epsilon) .
\end{align*}
In particular, these errors (as well as their derivatives)   are controlled uniformly   for  $\lambda \in \supp(\theta)$, $\omega\in S^{n-1}$, $x,y\in \supp(g)$,  $(x_0,\mu) \in \mathcal{A}$, $|t| \le \epsilon^{-\kappa}$ and $|\zeta| \le \delta^{-\kappa}$. 

According to \eqref{J3}, this allows to rewrite the main contribution to $\mathfrak{I}_\hbar'$ as 
\begin{equation} \label{eq:Iseconde}
\begin{aligned}
\mathfrak{I}_{\hbar}''=\frac{1/2\pi}{((2\pi)^2\delta)^{n} }  \int &  e^{\i (x-y+ 2r_\star(x_0,\lambda) t \omega)\cdot \zeta }  g(x) g(y) \big[ 
F(x_\epsilon,\lambda)   + \Upsilon_{1;\epsilon,\delta}(\Lambda) +t \Upsilon_{2;\epsilon,\delta}(\Lambda) \\ 
&\quad +
  (x-y+2r_\star(x_0,\lambda)t\omega)\cdot \Upsilon_{3;\epsilon,\delta}(\Lambda)\big]
 \frac{\sin(\frac{2(\mu-\lambda)t}{\delta})}{t}  \chi(t \epsilon^{\kappa}) \chi(|\zeta| \delta^{\kappa})  \theta(\lambda)   \1_{\mu\ge \lambda} \d \Lambda   . 
\end{aligned}
\end{equation}

By Proposition~\ref{prop:sineint},  we argue that the error term
$\Upsilon_{1;\epsilon,\delta}$ does not contribute significantly to
the integral $\mathfrak{I}_\hbar''$. 
Indeed the function 
\[
(\lambda,t) \mapsto \Upsilon_{1;\epsilon,\delta}(\Lambda)\chi(t \epsilon^{\kappa})  \theta(\lambda) 
\]
is smooth with respect to $t$, uniformly bounded with respect to
$\lambda$ with 
\[
  \|(\lambda,t)\mapsto \Upsilon_{1;\epsilon,\delta}(\Lambda)\chi(t\epsilon^{\kappa})\theta(\lambda)\|_{L^{\infty}C^{2}}+
  \|(\lambda,t)\mapsto \theta(\lambda)
  t^{\gamma-1}\partial_t\big(\Upsilon_{1;\epsilon,\delta}(\Lambda)\chi(t \epsilon^{\kappa})\big)\|_{L^{\infty}L^1}\leq
   C\epsilon^{-\gamma\kappa}( \delta^{2-2\kappa}+\hbar),
  \]
uniformly for $\omega\in S^{n-1}$, $x,y\in \supp(g)$,  $(x_0,\mu) \in \mathcal{A}$ and $|\zeta| \le \delta^{-\gamma}$.
Since  $ \hbar^{1-\beta} \le \epsilon, \delta \le \hbar^{\beta}$ and
$\int \chi(|\zeta| \delta^{\kappa})  \d \zeta =\O(\delta^{-n\kappa})$,
by choosing $\kappa$ small enough (depending only on $0<\alpha<\beta$
and the dimension $n$), this shows that the contribution of
$\Upsilon_{1;\epsilon,\delta}$ in \eqref{eq:Iseconde} is $\O(\delta^{1-n}\hbar^{\alpha})$.

Similarly,
$\| (\lambda,t)\mapsto \theta(\lambda) \Upsilon_{2;\epsilon,\delta}(\Lambda) \chi(t \epsilon^{\kappa}) \|_{L^{\infty}H^1} \le C \epsilon^{1-2\kappa}$  so that by Proposition \ref{prop:bound-osc-lowreg-easy}, the
contribution of $\Upsilon_{2;\epsilon,\delta}$ in \eqref{eq:Iseconde} is
$\O(\epsilon^{1-2\kappa}\delta^{-n+1-n\kappa})$, which is
$O(\delta^{-n+1}\hbar^{\alpha})$ for $\alpha>0$ small.

Finally, note that we have tuned the pre-factor of $\Upsilon_{3;\epsilon,\delta}$ so that we can perform an integration  by parts with respect to $\zeta$. 
Hence, the contribution of $\Upsilon_{3;\epsilon,\delta}$ in \eqref{eq:Iseconde} is given by
  \[
\frac{\delta/2\pi}{((2\pi)^2\delta)^n }\int  e^{\i (x-y+
  2R(x_0,\lambda) t \omega)\cdot \zeta }  g(x) g(y) {\rm div}_{\zeta}\Upsilon_{3;\epsilon,\delta}(\Lambda)
 \frac{\sin(\frac{2(\mu-\lambda)t}{\delta})}{t}  \chi(t \epsilon^{\kappa}) \chi(|\zeta| \delta^{\kappa})  \theta(\lambda)   \1_{\mu\ge \lambda} \d \Lambda.
\]
Because of the change of variable $\zeta \leftarrow \delta\zeta$ that we performed, $|{\rm div}_{\zeta}\Upsilon_{3;\epsilon,\delta}(\Lambda)|= \O(\epsilon\delta)$ and 
\[
  \|(\lambda,t)\mapsto  {\rm div}_{\zeta}\Upsilon_{3;\epsilon,\delta}(\Lambda)\chi(t\epsilon^{\kappa})\theta(\lambda)\|_{L^{\infty}C^{2}}+
  \|(\lambda,t)\mapsto \theta(\lambda)
  t^{\gamma-1}\partial_t\big({\rm div}_{\zeta}\Upsilon_{3;\epsilon,\delta}(\Lambda)\chi(t \epsilon^{\kappa})\big)\|_{L^{\infty}L^1}\leq
   C \delta \epsilon^{1-\gamma\kappa} . 
  \]
By Proposition \ref{prop:sineint}, we conclude that the contribution of $\Upsilon_{3;\epsilon,\delta}$ in \eqref{eq:Iseconde} 
 is also of order $O(\delta^{-n+1}\hbar^{\alpha})$ for $\alpha>0$ small with the required uniformity. 

In summary, as $\hbar \to 0$,
 \begin{equation} \label{J4} \begin{aligned}
\mathfrak{I}_\hbar'' =  \frac{1/2\pi}{((2\pi)^2\delta)^{n} }  \int & e^{\i (x-y +2r_\star(x_0,\lambda) t \omega) \cdot \zeta} g(y)g(x) h(x_{\epsilon},y_{\epsilon},\lambda)   \frac{\sin(\frac{2(\mu-\lambda)t}{\delta})}{t}  \chi(t \epsilon^\kappa)  \chi(|\zeta| \delta^{\kappa}) \theta(\lambda)   \1_{\mu\ge \lambda} \d \Lambda  + \O(\delta^{1-n}\hbar^{\alpha}) . 
\end{aligned}
\end{equation}
 
\subsubsection{Leading term}
Let us denote 
\[
F(\zeta ,x, \lambda ; \epsilon) =\frac{ g(x) \theta(\lambda)}{(2\pi)^n} \int e^{-\i y \cdot \zeta}  h(x_{\epsilon},y_{\epsilon},\lambda)   g(y) \d y =
\frac{ g(x) \theta(\lambda)}{(2\pi)^n} r_\star(x_\epsilon,\lambda)^{n-2}  \int \hat{g}(\zeta+ \epsilon \xi) \widehat{v_{x_\epsilon}}(\xi)    \d  \xi  ,
\]
where $v_x : y \mapsto  \vartheta\big(V(\tfrac{x+y}{2}+r_\star(x_0,\lambda)^2\big)^2 $.
Since $g \in C^\infty_c(\R^n)$ and $v_x \in C^\infty_c(\R^n)$ for all $x \in B^n_{0,c}$ and $\lambda \in \supp(\theta)$, we have $F \in S^0(\R^n \times \R^{n+1})$ with compact support wrt $(\lambda,x)$ and 
$F_k(\zeta,x,\lambda) = \O_k(|\zeta|^{-\infty})$ as $\zeta\to\infty$ for every $k\in\N_0$. 
In particular, the leading term on the RHS of \eqref{J4} satisfies
\[ \begin{aligned}
 \mathfrak{I}_\hbar''=  \frac{1/2\pi}{(2\pi \delta)^{n} }  \int e^{\i (x+ 2r_\star(x_0,\lambda) t \omega) \cdot \zeta}  F(\zeta ,x, \lambda ; \epsilon)  \frac{\sin(\frac{2(\mu-\lambda)t}{\delta})}{t} \chi(t \epsilon^\kappa) \chi( |\zeta| \delta^{\kappa})   \1_{\mu\ge \lambda} \d \zeta\d t \d x \d \lambda \d \omega  + \O(\hbar^{\infty}) . 
\end{aligned}\]
Note that we used the decay of $F$ to remove the cutoff $\chi( |\zeta| \delta^{\kappa})$ up to another $\O(\hbar^\infty)$ error. 
We can now compute the integral with respect to $\zeta$ by Fourier's inversion formula, we obtain for any $|t| \le \epsilon^{-\kappa}$ and $x\in \supp(g)$, 
\[ \begin{aligned}
 \int e^{\i (x+ 2r_\star(x_0,\lambda) t \omega) \cdot \zeta}  F(\zeta ,x, \lambda ; \epsilon)  \d \zeta   = g(x) \theta(\lambda)  h\big(x_{\epsilon},(x+2r_\star(x_0,\lambda)t\omega)_\epsilon ,\lambda\big)  g\big(x+2r_\star(x_0,\lambda) t \omega\big) \\
  = g(x) g\big(x+2r_\star(x_0,\lambda) t \omega\big)  r_\star(x_\epsilon,\lambda)^{n-2}   \vartheta\big(V(x_\epsilon +\epsilon t r_\star(x_0,\lambda)\omega)+r_\star(x_\epsilon,\lambda)^2\big)^2 \theta(\lambda)  \\
  =  g(x) g\big(x+r_\star(x_0,\lambda) t \omega\big)  r_\star(x_\epsilon,\lambda)^{n-2}  \vartheta(\lambda) + \O_{x,\lambda,\omega}(\epsilon t)\theta(\lambda) ,
\end{aligned}\]
where we used that $V(x_\epsilon)+ r_\star(x_\epsilon,\lambda)^2 = \lambda$ and that $\theta(\lambda) = 1$ for $\lambda \in \supp(\vartheta)$ to rewrite the leading term.
Using again Proposition \ref{prop:bound-osc-lowreg-easy},  integrating
the error term $\O_{x,\lambda,\omega}(\epsilon t)\theta(\lambda)$ over
$(\lambda,t)$, it contributes as $\O_\gamma(\epsilon \delta^{1-\gamma-n\kappa}) = \O(\delta \hbar^\alpha)$ to the previous integral, upon choosing the parameters $\gamma,\kappa>0$ small enough. 
Hence, from \eqref{J4}, one has
$\mathfrak{I}_{\hbar}''=\mathfrak{N}_{\hbar}+\O(\delta^{1-n}\hbar^{\alpha})$, where 
\begin{equation}  \label{J5}
 \mathfrak{N}_\hbar=  \frac{1/2\pi}{(2\pi \delta)^{n} } \int   g(x) g\big(x+2r_\star(x_0,\lambda) t \omega\big) 
r_\star(x_\epsilon,\lambda)^{n-2} 
 \frac{\sin(\frac{2(\mu-\lambda)t}{\delta})}{t}  \chi(t \epsilon^\kappa)
  \vartheta(\lambda)   \1_{\mu\ge \lambda} \d t \d x \d \lambda \d \omega. 
\end{equation}

To finish the proof, we split the integral \eqref{J5} in two parts:
$\mathfrak{N}_{\hbar}=\mathfrak{R}_{\hbar}+{\rm
  Im}(\mathfrak{S}_{\hbar})$, where 
\[
 \mathfrak{R}_\hbar  =  \frac{1/2\pi}{(2\pi \delta)^{n} }\int   g(x)^2 r_\star(x_\epsilon,\lambda)^{n-2}    \frac{\sin(\frac{2(\mu-\lambda)t}{\delta})}{t}   \chi(t \epsilon^\kappa) \vartheta(\lambda)  \1_{\mu\ge \lambda}  \d t \d x \d \lambda \d \omega  
\]
and 
\begin{equation}\label{eq:S} \mathfrak{S}_\hbar  =  \frac{1/2\pi}{(2\pi \delta)^{n} }  \int   g(x) 
 \frac{g\big(x+2r_\star(x_0,\lambda) t \omega\big)  -g(x)}{t} e^{2 \i \tfrac{(\mu-\lambda)t}{\delta}} r_\star(x_\epsilon,\lambda)^{n-2}        \chi(t \epsilon^\kappa) \vartheta(\lambda)^2  \1_{\mu\ge \lambda} \d t \d x \d \lambda \d \omega . 
\end{equation}
Note that the integrand in  $\mathfrak{R}_\hbar$  is independent of
$\omega$.

According to Proposition \ref{prop:sineint}, it holds for any $0<\gamma \ll \kappa$,  
\[ \begin{aligned}
 \mathfrak{R}_\hbar
 & =   \frac{|S^{n-1}|/2\pi}{(2\pi \delta)^{n} }  \int   g_\epsilon(x)^2 r_\star(x,\lambda)^{n-2}    \frac{\sin(\frac{2(\mu-\lambda)t}{\delta \epsilon^\kappa})}{t}   \chi(t) \vartheta(\lambda)  \1_{\mu\ge \lambda}  \d t \d x \d \lambda  \\
   & =  \frac{|S^{n-1}|}{2(2\pi \delta)^{n} }  \int g_\epsilon(x)^2
   r_\star(x,\lambda)^{n-2} \vartheta(\lambda) \1_{\lambda\leq \mu}\d x  \d \lambda + \O(\epsilon^{\kappa(1-\gamma)}\delta^{1-\gamma}).
\end{aligned}\]

Since $ \hbar^{1-\beta} \le \epsilon, \delta \le \hbar^{\beta}$, by choosing $\gamma$ small enough (depending only on $0<\alpha<\beta$ and $\kappa$), the error term is also $\O(\hbar^\alpha)$. 
By \eqref{J4}  and \eqref{J5}, this shows that 
\begin{equation}  \label{J6}
\mathfrak{N}_\hbar
= \frac{|S^{n-1}|}{2(2\pi \delta)^{n} }\int g_\epsilon(x)^2
r_\star(x,\lambda)^{n-2} \vartheta(\lambda)\1_{\lambda\leq \mu}\d x  \d \lambda 
+ \O(\delta^{1-n}\hbar^{\alpha}). 
\end{equation}
This coincides with the asymptotics in Proposition~\ref{prop:tr1}. 
Hence to finish the proof, it remains to compute the leading
asymptotics of  ${\rm Im}( \mathfrak{S}_\hbar)$.

\subsubsection{Subleading term}
First of all, by Proposition \ref{prop:sineint}, when computing ${\rm
  Im}(\mathfrak{S}_{\hbar})$ from \eqref{eq:S}, we may replace
$r_\star(x_{\epsilon},\lambda)$ with $r_\star(x_0,\lambda)$, up to an error
$\O(\delta^{-n+1-\gamma}
\epsilon^{1-\kappa})=\O(\delta^{-n+1}\hbar^{\alpha})$. It remains to
compute the imaginary part of
\[
   \mathfrak{S}_\hbar'  =  \frac{1/2\pi}{(2\pi \delta)^{n} }  \int   g(x) 
 \frac{g\big(x+2r_\star(x_0,\lambda) t \omega\big)  -g(x)}{t} e^{2 \i
   \tfrac{(\mu-\lambda)t}{\delta}} r_\star(x_0,\lambda)^{n-2}
 \chi(t \epsilon^\kappa) \vartheta(\lambda)^2  \1_{\mu\ge \lambda} \d
 t \d x \d \lambda \d \omega.
 \]

Let us observe that using the invariance by rotation with respect to $\omega \in
S^{n-1}$ of $r_\star(x_{0},\lambda)$, one has
\[
  \int g(x)\omega\cdot
  \partial_xg(x)r_\star(x_{0},\lambda)^{n-1}\frac{1}{t}\chi(t\epsilon^{\kappa})\vartheta(\lambda)^2\1_{\mu\geq
    \lambda}\dd t \dd x \dd \lambda \dd \omega = 0.
\]
In particular, letting
\[
  G:(x,\xi)\mapsto \frac{g(x+2\xi)-g(x)-2\xi\cdot
    \partial_xg(x)}{|\xi|^2},
\]
then $(t,\lambda)\mapsto G(x,r_\star(x_0,\lambda)t\omega)$ is smooth with
uniform controls in $\omega$, and 
\[\begin{aligned}
\delta^{n} \mathfrak{S}_\hbar'   = \frac{1}{(2\pi)^{n+1}}   \int   g(x) 
G\big(x, r_\star(x_0,\lambda) t\omega\big) r_\star(x_0,\lambda)^{n}
 t e^{2 \i \tfrac{(\mu-\lambda)t}{\delta}} \chi(t \epsilon^\kappa)\vartheta(\lambda)^2   \1_{\mu\ge \lambda}   \d t \d x \d \lambda \d \omega.
\end{aligned}\]
We rewrite this as
\[
  \delta^{n-1} \mathfrak{S}_\hbar'   = \frac{\i}{2(2\pi)^{n+1}}   \int   g(x) 
G\big(x, r_\star(x_0,\lambda) t\omega\big) r_\star(x_0,\lambda)^{n}
 \partial_{\lambda}\left(e^{2 \i \tfrac{(\mu-\lambda)t}{\delta}}\right) \chi(t
 \epsilon^\kappa)\vartheta(\lambda)^2   \1_{\mu\ge \lambda}   \d t \d
 x \d \lambda \d \omega.
 \]
We can then perform an integration by parts with respect to $\lambda$  to obtain
\begin{equation}  \label{J7}
\begin{aligned}
\delta^{n-1} \mathfrak{S}_\hbar'  & = \frac{\i \vartheta(\mu)}{2(2\pi)^{n+1}}    \int   g(x) G\big(x,r_\star(x_0,\mu) t\omega\big) r_\star(x_0,\mu)^{n}   \chi(t \epsilon^\kappa) \d t \d x  \d \omega \\
 &\quad -  \frac{\i}{2(2\pi)^{n+1}}  \int   g(x)
 e^{2 \i \tfrac{(\mu-\lambda)t}{\delta}} \partial_\lambda \big[ G\big(x,r_\star(x_0,\lambda) t \omega\big)  \vartheta(\lambda)^2 r_\star(x_0,\lambda)^{n}  \big]  \1_{\mu\ge \lambda}   \chi(t \epsilon^\kappa) \d t \d x \d \lambda \d \omega
\end{aligned}
\end{equation}
The function $\lambda \mapsto \partial_\lambda \big(
G\big(x,r_\star(x_0,\lambda) t, \omega\big)  \vartheta(\lambda)^2
r_\star(x_0,\lambda)^{n}  \big) $ is uniformly bounded with a fixed
compact support and $t\mapsto  \chi(t \epsilon^\kappa)$ is a smooth
function with $\|  \chi(\cdot \epsilon^\kappa) \|_{H^s}  \le C_s \|  \chi(\cdot \epsilon^\kappa) \|_{L^2} \le   
C_{s}\epsilon^{- \frac{\kappa}{2}} $ for any $s\ge 0$ so that, by
Proposition~\ref{prop:bound-osc-lowreg-easy}, the second term on the
RHS of \eqref{J7} is $\O(\hbar^{\alpha})$ for some $\alpha>0$.
Hence, with $r_\star= r_\star(x_0,\mu)$, we obtain 
\[ \begin{aligned}
\delta^{n-1} \mathfrak{S}_\hbar  &  =  \frac{\i r_\star^n \vartheta(\mu)}{2(2\pi)^{n+1}}  \int   g(x) G\big(x, r_\star t\omega\big)  \chi(t \epsilon^\kappa) \d t \d x  \d \omega +\O(\hbar^{\alpha})   \\
 &=  \frac{\i r_\star^{n-1} \vartheta(\mu)}{(2\pi)^{n+1}}  \int   g(x) G(x,\xi) \frac{\chi(|\xi|\epsilon^{\kappa}/r_\star)}{|\xi|^{n-1}}  \d \xi \d x +\O(\hbar^{\alpha})
\end{aligned}\]
where we made a change of variable $r_\star(x_0,\mu) t\omega \leftarrow \xi$. 
 An extra factor 2 comes from the fact that this map is two-to-one $(t\in\R)$.
By symmetry,  we can rewrite 
\[ \begin{aligned}
&\delta^{n-1} \mathfrak{S}_\hbar    =  \frac{\i  r_\star^{n-1} \vartheta(\mu)}{(2\pi)^{n+1}} \int   g(x-\xi) G(x-\xi,\xi) \frac{\chi(|\xi|\epsilon^{\kappa}/r_\star)}{|\xi|^{n-1}}  \d \xi \d x +\O(\hbar^{\alpha}) \\
& =    \frac{-\i r_\star^{n-1} \vartheta(\mu)}{2(2\pi)^{n+1}} \int \bigg(  \bigg( \frac{g(x-\xi)-g(x+\xi)}{|\xi|} \bigg)^2 + \frac{\xi}{4|\xi|^2} \cdot \big( \partial_x g(x-\xi)^2- \partial_x g(x+\xi)^2\big) \bigg) \frac{\chi(|\xi|\epsilon^{\kappa}/r_\star)}{|\xi|^{n-1}}  \d \xi \d x +\O(\hbar^{\alpha})  .
\end{aligned}\]
In particular, the second term vanishes since it is an exact derivative wrt~$x$  of an integrable function of $(x,\xi)$ (here we use that $\chi$ and $g$ have compact support). 
Hence, using the H\"older bound $\big| \chi(t\epsilon^{\kappa})- 1\big| \le c_\kappa t^\kappa \epsilon^{\gamma\kappa}$ for any $\gamma\in(0,1)$ and that $\epsilon^{\gamma \kappa} = \O(\hbar^{\alpha})$ if $\alpha>0$ is small enough,  we obtain
\[
\delta^{n-1} \mathfrak{S}_\hbar  
 =   \frac{-\i  r_\star^{n-1} \vartheta(\mu)}{(2\pi)^{n+1}} \int   \frac{\big(g(x-\tfrac{\xi}2)-g(x+\tfrac{\xi}2) \big)^2}{|\xi|^{n+1}} 
 \d \xi \d x  + \O(\hbar^{\alpha}) . 
 \]
 
 By formulae  \eqref{J6} and \eqref{dual_var}, we conclude that 
 \[
\mathfrak{J}_\hbar
=\frac{n|S^{n-1}|}{2(2\pi \delta)^{n} }\int g_\epsilon(x)^2 r_\star(x,\lambda)^{n-2} \vartheta(\lambda) \d x  \d \lambda  
 - \sigma_n^{2}r_\star^{n-1} \vartheta(\mu)\delta^{1-n}  \Sigma^2(f)   +\O(\delta^{1-n}\hbar^{\alpha}) .
 \]
 Since $  \tr\big( \big( f_{\hbar,\mu}(H_{\hbar}) g_\epsilon \big)^2 \big) = \mathfrak{J}_\hbar + \O(\hbar^\infty)$, this completes the proof.
\end{preuve}

\subsection{Macroscopic commutators}

In this section, we turn to the proof of Theorem~\ref{thm:gtb}. 
Let $g\in C^\infty_c(\{V<\mu\},\R)$ be any smooth function compactly supported in the bulk of the droplet.
Given the bounds for the Laplace functional of a determinantal point process from
Proposition~\ref{prop:Laplace}, the main ingredient of the proof is a (sharp) bound for the variance 
\[
  \var  \X(g) =  -\tfrac{1}{2}   \tr([g , \Pi_{\hbar,\mu}]^2) .
\]
Using estimate \eqref{restJ2est} with $M=g$ and the next proposition, we obtain 
\begin{equation} \label{est_var_macro}
  \var  \X(g) =\O(\hbar^{-n+1}). 
\end{equation}

  \begin{prop}\label{prop:last_step_macro}
   Let $(\mu,V)$ satisfy \ref{hyp:weak} and let $f_{\hbar,\mu}$ be as in Notations~\ref{regul-kern} with $\mu_0<\mu$. 
  Then, for any $g\in C^{\infty}_c(\{V<\mu_0\})$, 
    \[
      \tr([f_{\hbar,\mu}(H_{\hbar}),g]^2)=\O(\hbar^{-n+1}).
      \]
  \end{prop}
  \begin{preuve}
  Using the representation \eqref{fop} and proceeding as in the proof of Proposition~\ref{prop:commut-macroscopic-smooth} for $t\in[-\tau,\tau]$, by bilinearity, we obtain
  \[
   \tr([g , f_{\hbar,\mu}(H_{\hbar})]^2)
=  \frac{1}{2\pi \hbar^2} \int \tr\big( [g , I_{\hbar,t_1}^{\phi, a}]
[g , I_{\hbar,t_2}^{\phi, a}] \big)  e^{ -\i\frac{t_1\lambda_1}{\hbar}
  -\i\frac{t_2\lambda_2}{\hbar}}  \varkappa_\mu(\lambda_1)
\varkappa_\mu(\lambda_2) \hat{\rho}(t_1) \hat{\rho}(t_2) \d t_1\dd t_2
\d \lambda_1\dd \lambda_2    + \O(\hbar^\infty) \]
where we recall that $\varkappa_\mu=\varkappa
\mathds{1}_{(-\infty,\mu]}$, and $a\in S^0$ is a classical symbol.

Hence,  the only relevant contribution to this trace is given by the oscillatory integral
\begin{multline*}
J_\hbar=  \frac{4\pi}{(2\pi\hbar)^{2n+2}} \int e^{\i \frac{\Psi_1(t_1,t_2,x,y,\xi_1,\xi_2,\lambda_1,\lambda_2)}{\hbar}}  g (x)  \big( g(y) -g(x) \big)   b\left(t_1,t_2,x,y,\xi_1,\xi_2 ;\hbar\right) 
\varkappa_\mu(\lambda_1)   \varkappa_\mu(\lambda_2)\\ \d t_1\dd t_2 \d
\lambda_1\dd \lambda_2    \d \xi_1\dd \xi_2   \d x  \d y
\end{multline*}
where $b\in S^0$ is given by
\[
b(t_1,t_2,x,y,\xi_1,\xi_2;\hbar) :=  a(t_1,x,y,\xi_1;\hbar) \overline{a}(t_2,y,x,\xi_2;\hbar)\hat{\rho}(t_1)\hat{\rho}(t_2).
\]
In particular, $b$ has compact support on $\R^{4n+2}$ and, 
  provided $\tau$ is small enough, we can assume that for a fixed
  $\delta>0$, for all  $t\in[-\tau,\tau]$, $x\in\supp(g)$, $y\in\R^n$
  and $\hbar \in (0,1]$,
  the function $\xi \mapsto a(t,x,y,\xi;\hbar)\hat{\rho}(t) $ is
  supported on $\{|\xi|\geq \delta\}$ (cf.~Proposition~\ref{prop:propagator}).

\subsubsection{Stationary phase method}
The phase in $J_{\hbar}$ is given by 
\[
\Psi_1(t_1,t_2,x,y,\xi_1,\xi_2,\lambda_1,\lambda_2) : = \phi(t_1,x,\xi_1)-y\cdot
        (\xi_1-\xi_2)-\phi(t_2,x_1,\xi_2) -t_1\lambda_1+t_2\lambda_2.
\]

To apply the stationary phase method, we make a changes of variables
\[
s = t_1-t_2 ,\qquad
t = \frac{t_1+t_2}{2} ,\qquad
\zeta= \xi_1-\xi_2 , \qquad
r \omega =  \frac{\xi_1+\xi_2}{2} , \qquad 
\varsigma = \lambda_1-\lambda_2 ,\qquad
\sigma = \frac{\lambda_1+\lambda_2}{2} ,
\]
where $r\in\R_+$, $\omega \in S^{n-1}$, $t, s \in[-2\tau,2\tau]$, $\sigma, \varsigma\in \R$ and $\zeta\in\R^n$.
The Jacobian of this map is $r^{n-1}$ and, introducing
\begin{align*}
c(t,s,x,y,\zeta, r, \omega ;\hbar) &=   b\left(t+\tfrac s2,t-\tfrac
  s2,x,y,r\omega+\tfrac{\zeta}{2},r\omega-\tfrac{\zeta}{2} ;\hbar\right)   r^{n-1}\\
\Gamma_\mu(\varsigma , \sigma ) &= \varkappa_\mu(\sigma+\tfrac{\varsigma}{2})   \varkappa_\mu(\sigma-\tfrac{\varsigma}{2})\\
\Psi_2(t,s,x,y,\zeta, r, \omega , \varsigma , \sigma ) &=
\Psi_1(t_1,t_2,x,y,\xi_1,\xi_2,\lambda_1,\lambda_2),
\end{align*}
we obtain
\begin{equation} \label{Jint1}
 \begin{aligned}
J_\hbar=  \frac{4\pi}{(2\pi\hbar)^{2n+2}} \int e^{\i \frac{\Psi_2(t,s,x,y,\zeta, r, \omega , \varsigma , \sigma ) }{\hbar}}  g(x)  \big( g(y) -g(x) \big)   c(t,s,x,y,\zeta, r, \omega ;\hbar)  \Gamma_\mu(\varsigma , \sigma ) 
  \d r \d t\dd s  \d \zeta  \d y  \d x \d \sigma \d \varsigma  \d \omega  .
\end{aligned}
\end{equation}

We now apply the stationary phase method to this intergral in the variables $(y,r,s,\zeta)$ while keeping $(t,x,\sigma,\varsigma,\omega)$ fixed.  The equations for the critical point(s) are given by
\begin{align}
\label{crit1}
&0=\partial_{y} \Psi_2(t,s,x,y,\zeta, r, \omega , \varsigma , \sigma )  =- \zeta\\
\label{crit2}
&0 = \partial_{r}   \Psi_2(t,s,x,y,\zeta, r, \omega , \varsigma , \sigma ) = \omega \cdot \big( \partial_{\xi} \phi(t_1,x, \xi_1)- \partial_{\xi}\phi(t_2,x,\xi_2) \big) \\
\label{crit3}
&0 = \partial_{s} \Psi_2(t,s,x,y,\zeta, r, \omega , \varsigma , \sigma )
= \tfrac12 \big( \partial_t  \phi(t_1,x,\xi_1) + \partial_t  \phi(t_2,x,\xi_2) \big) - \sigma \\
\label{crit4}
&0 =\partial_{\zeta} \Psi_2(t,x,y,\zeta, r, \omega , \varsigma , \sigma ) = \partial_{\xi} \phi(t_1,x, \xi_1)+ \partial_{\xi}\phi(t_2,x,\xi_2)  \big) -  y 
\end{align}
where we recall that
\[
  \xi_1= r\omega + \zeta/2 \qquad \qquad \xi_2= r\omega + \zeta/2
  \qquad \qquad t_1= t +s/2\qquad \qquad t_2= t-s/2.
  \]

\medskip

From these equations, we can now argue that there is at most one
critical point within the support of $c$. 
\begin{itemize}[leftmargin=*]
\item \eqref{crit1} implies that $\zeta=0$ (equivalently $\xi_1=\xi_2= r\omega$).
\item Recall that according to \eqref{Taylor1}, $\partial_{\xi} \phi(t,x, \xi) = x+ t \xi +\O(t^2)$.
Then, \eqref{crit2} and the condition $\zeta=0$ imply that
\[
0=\omega \cdot \big( \partial_{\xi} \phi(t_1,x, r\omega)- \partial_{\xi}\phi(t_2,x,r\omega) \big) = 
 s\big( r + \O(t)  \big). 
\] 
Since $\xi \mapsto a(t,x,y,\xi;\hbar) $ is supported on
$\{|\xi|>\delta\}$, one has $r>\delta$ at least when $\zeta=0$, so that this equation is
  satisfied only if $s=0$ provided $\tau$ is small enough (equivalently $t_1=t_2= t$).
\item Using the conditions $\zeta=t=0$, the equations \eqref{crit3}
  and  \eqref{Taylor1} then yield
\[
\sigma= \partial_t  \phi(t,x,r \omega)  = V(x) + r^2 +\O(t) . 
\]
Like in the proof of Proposition~\ref{prop:regularised_kernel_bulk}, we need to distinguish two cases
\begin{itemize}[leftmargin=.5cm]
\item
 If $\sigma \le V(x) + c_1$, for some constant $c_1
  =c_1(\delta,\tau)$, then this equation has no solutions satisfying
  $r\ge \delta$. In this case, there is no critical point within the
support of $c$ and by Proposition ~\ref{prop:nspl}, we conclude that
$J_\hbar =\O(\hbar^\infty)$. We can therefore include a cutoff
$\theta(\sigma-V(x))$ in the integrand of \eqref{Jint1} where $\theta
\in C^\infty(\R,[0,1])$ satisfies $\1_{[c_2,\infty)} \le \theta \le
\1_{[c_1,\infty)}$ for a small enough $c_2>c_1>\tau$ such that $V(x)+
c_2< \mu_0$ for all $x$ in the support of $g$. 
\item For $\sigma \ge V(x)+ c_1$, this equation has a unique (positive) solution which determines the critical value of $r$ as a function of the other parameters;
\begin{equation} \label{rcrit}
r_\star(t,x,\sigma,\omega) = \sqrt{\sigma- V(x)}  + \O(t)
\end{equation}
\end{itemize}
\item Finally, using the conditions $(s,\zeta,r) = (0,0,r_\star)$, the condition \eqref{crit4} determines the critical value of $y$ as a function  of the other parameters;
\begin{equation} \label{ycrit}
y_\star(t,x,\sigma,\omega)  = \partial_{\xi} \phi(t,x,
r_\star\omega )  = x + { \O(t)} . 
\end{equation}
where we used again the expansions \eqref{Taylor1} and \eqref{rcrit}. 
\end{itemize}

This shows that the (unique) critical point is given by $(y,r,s,\zeta) = (y_\star,r_\star,0,0)$ and using \eqref{crit1}--\eqref{crit4}, we verify that the Hessian of $\Psi_2$ in these variables evaluated at the critical point is given by the matrix
\[
\mathrm{H} =   -
{\small \begin{pmatrix}
0 & 0 & 0 & \mathrm{I} \\
0 & 0 &  \bullet& \bullet \\
0 & \bullet & 0 &0 \\
\mathrm{I}  & \bullet & 0 &0
\end{pmatrix}.}
\]
This matrix is non-degenerate and by expanding over the second and third columns, we obtain
\begin{equation} \label{detH}
 \det \mathrm{H} = \big|  \partial_r \partial_{s} \Psi_2\big|^2 (t,s,x,y_\star,0, r_\star, \omega , \varsigma , \sigma )   =  \big| \omega \cdot \partial_{\xi} |\partial_x\phi|^2 (t,x, r_\star\omega ) \big|^2  = 4 \big(\sigma- V(x) \big)^2+  \O(t^2) 
 \end{equation}
where we used the expansion \eqref{Taylor1} to simplify the quantity for small $t$. 
Let us also observe that at the critical point, one has
\[
\Psi_2(t,x,y_\star,0, r_\star, \omega , \varsigma , \sigma )= -t \varsigma.
\]
Hence, applying Proposition~\ref{prop:spl} to the integral \eqref{Jint1} while  keeping the variables $(t,x,\sigma,\varsigma,\omega)$ fixed, since $g\in C^\infty_c(\R^n)$, we obtain the expansion as $\hbar\to 0$, for any $k\in\N$, 

\begin{equation}
J_\hbar  =  \frac{4\pi}{(2\pi\hbar)^{n+1}} \int  e^{- \i \tfrac{t\varsigma}{\hbar}} 
 e(t,x,\omega,\sigma;\hbar)
\Gamma_\mu(\varsigma , \sigma ) \d t   \d x   \d \omega  \d \sigma \d \varsigma  +\O(\hbar^{\infty}),
\label{Jint2}
\end{equation}
where the (classical) symbol $e$ is smooth satisfies
\[
  e(t,x,\omega,\sigma;0)=g(x)(g(y_\star)-g(x))|\hat{\rho}(t)|^2a(t,x,y_\star,r_\star\omega;0)\overline{a}(t,y_\star,x,r_\star\omega;0)\frac{r_\star^{n-1}}{\sqrt{\det \mathrm{H}}} \theta(\sigma-V(x)).
\]
In particular, the function $ (t,x,\omega,\sigma, \varsigma) \mapsto e(t,x,\omega,\sigma;\hbar)
\Gamma_\mu(\varsigma , \sigma ) $ in \eqref{Jint2} is compactly supported, $L^{\infty}$ with
respect to $\varsigma$ and $C^{\infty}$ with respect to $t$. Hence, by Proposition \ref{prop:bound-osc-lowreg-easy},
subprincipal terms contribute as $\O(\hbar^{-n+1})$, so that
\[
  J_{\hbar}=\frac{4\pi}{(2\pi\hbar)^{n+1}} \int  e^{- \i \tfrac{t\varsigma}{\hbar}} 
 e(t,x,\omega,\sigma;0)
 \varkappa(\sigma+\tfrac{\varsigma}{2})\varkappa(\sigma-\tfrac{\varsigma}{2})\1_{|\frac{\varsigma}{2}|\leq \mu-\sigma}
 \d t   \d x   \d \omega  \d \sigma \d
\varsigma  +\O(\hbar^{-n+1}).
\]
Since $y_\star=x+\O(t)$ and $g$ is smooth, the principal symbol of $e =\O(t)$, that is, 
there exists $F\in C^{\infty}_c$ such that
\[
  e(t,x,\omega,\sigma;0)=tF(t,x,\omega,\sigma).
\]
Integrating by parts with respect to $\varsigma$, since  $\varkappa =1$ on  $[0,M]$, we obtain 
\[ \begin{aligned}
 J_{\hbar}
 & = \frac{4\pi}{(2\pi\hbar)^{n}} \int \sin(\tfrac{2(\mu-\sigma)}{t})
 F(t,x,\omega,\sigma) \1_{\sigma\leq
      \mu} \d   t \dd \sigma\dd x \dd \omega
 \\
 &\quad+ \frac{4\pi}{(2\pi\hbar)^{n}}\int
  e^{2\i\frac{t(\mu-\sigma)}{\hbar}}F(t,x,\omega,\sigma) \Gamma_{\mu}'(\varsigma,\sigma)\dd
    t \dd \sigma\dd x \dd \omega \dd \zeta +\O(\hbar^{-n+1}).
\end{aligned}\]
where 
  \[
    \Gamma'_{\mu}:(\varsigma,\sigma)\mapsto \frac{\dd}{\dd
      \varsigma}\big[\varkappa(\sigma+\tfrac{\varsigma}{2})\varkappa(\sigma-\tfrac{\varsigma}{2})]\1_{|\frac{\varsigma}{2}|\leq \mu-\sigma}. 
    \]
The function $\Gamma'_\mu \in L^\infty$ and $F\in C^\infty_c$, so by applying Proposition \ref{prop:bound-osc-lowreg-easy} to both these integrals, we conclude that $  J_{\hbar} = \O(\hbar^{-n+1})$. This  concludes the proof.
\end{preuve}

\begin{rem}
With extra work, it is possible to refine the proof to obtain the leading asymptotics of   $\tr([f_{\hbar,\mu}(H_{\hbar}),g]^2)$. 
 However, upon comparing these asymptotics with that of 
  $\tr([\Pi_{\hbar,\mu}g]^2)$, there is another a priori
  uncontrolled $\O(\hbar^{-n+1})$ contribution  coming from Lemma \ref{prop:estimee_J1_locale}. In view of
  Conjecture \ref{conj:var_macro}, it is not clear how to bypass this difficulty.
\end{rem}

\begin{rem}
The stationary phase in the proof of Proposition
  \ref{prop:last_step_macro} still holds if the test function $g$ depends on $\hbar$ at sufficiently large scales. In this fashion, one
can give an alternative proof of the mesoscopic commutator estimate from Proposition~\ref{prop:limit-commut-V}  in the case $\epsilon \ge \hbar^{\beta}$ for $ \beta>1/2$. Based on the counterpart of \eqref{Jint2} at mesoscopic scales and Proposition \ref{prop:sineint}, one can recover the
leading asymptotics of $\big\| [T^*_{x_0,\epsilon}g,f_{\hbar,\mu}(H_\hbar)] \big\|_{\J^2}$ when $\epsilon\to 0$ in the previous regime. 
\end{rem}

We are now ready to conclude the proof of Theorem \ref{thm:gtb}.

\medskip

\begin{preuve}(of Theorem \ref{thm:gtb})
  Let $f\in C^{\infty}_c(\{V<\mu\})$ and let
  \[
    \widetilde{\X}(f)=\X(f)-\mathbb{E}[\X(f)].
  \]
  Suppose first that $f\leq 0.69$. Then, by Proposition
  \ref{prop:Laplace} and the estimate \eqref{est_var_macro} applied to the test function $g= (e^f-1)\in C^{\infty}_c(\{V<\mu\})$, 
  \[
 \log   \mathbb{E}[e^{\widetilde{\X}(f)}]= \O\big(  \var  \X(g)\big)
    =\O(N\hbar) 
    \]
  as $\hbar \to 0$.

If now $f\in C^{\infty}_c(\{V<\mu\})$ is arbitrary and $\lambda\leq \|f\|_{L^{\infty}}^{-1}\sqrt{N\hbar}$, by rescaling, one has
  \[
    \mathbb{E}\left[\exp\left(\frac{\lambda\widetilde{\X}(f)}{\sqrt{N\hbar}}\right)\right]\leq \exp(C\lambda^2),
  \]
  where $C$ depends only on $f$. In particular, by Markov's inequality,
  for every $t>0$,
  \[
    \mathbb{P}\left[\widetilde{\X}(f)\geq \sqrt{N\hbar}t\right]=\mathbb{P}\left[\frac{\lambda\widetilde{\X}(f)}{\sqrt{N\hbar}}\geq
      t\lambda\right]\leq \exp(C\lambda^2-t\lambda).
    \]
    The optimal value of $\lambda$ for the right-hand side is
    \[
      \lambda=\frac{t}{2C} ,
    \]
so that provided $t \le 2C  \|f\|_{L^{\infty}}^{-1}\sqrt{N\hbar} $, we obtain
    \[
      \mathbb{P}\left[\widetilde{\X}(f)\geq
        \sqrt{N\hbar}t\right]\leq e^{-\frac{t^2}{2C}} . 
    \]
 Replacing $f$ by $-f$  yields the symmetric inequality. 
\end{preuve}

\newpage
\appendix
\section{Appendix}

\subsection{Compact operators}
\label{sec:compact-operators}
\label{sect:op}

We follow the conventions from \cite[Chapter 1--3]{simon_trace_2005}. Throughout this article, we work on the Hilbert space $L^2(\R^n)$
equipped with the norm $\|\cdot\| = \|\cdot\|_{L^2}$. 
Recall that a (linear) operator $B$ is bounded if 
\[
\| B \| = \sup_{\phi \in L^2, \| \phi\| \le1} \|B \phi\| <+\infty .  
\]

If $B$ is a bounded operator, it is called positive if $\langle
B\phi,\phi\rangle=\langle \phi, B\phi \rangle \ge 0$  for all $\phi
\in L^2(\R^n) $. Then, we write $A \ge B$ if $A$ and $B$ are symmetric
and $(A-B)$ is positive. 
Moreover, we make use \new{of} the following conventions and basic properties. 

\begin{prop} \label{prop:op}~
\begin{enumerate} [label=\roman*$)$]

\item   To any function $g\in L^\infty(\R^n)$, we associate a bounded operator $g: \phi \mapsto g\phi$ with norm $\|g\|_{ L^\infty}$.
In particular, this operator is positive if $g\ge 0$. 

\item If $A$ is a compact operator, we let $\big(\mu_k(A) \big)_{k=1}^{+\infty}$ be its singular values as in \cite[Theorem 1.5]{simon_trace_2005}. 
We define the Schatten norms  for $p\ge 1$,
\[
\|A\|_{\J^p} = \left( {\textstyle \sum_{k=1}^{\infty} \mu_k^p(A) }\right)^{1/p} . 
\]
Note that it holds for any bounded operator $B$ and any  $p\ge 1$, $\|BA\|_{\J^p} , \|AB\|_{\J^p} \le \|B\| \|A\|_{\J^p}$.
\end{enumerate}
\end{prop}

\begin{prop}[Convergence of operators]
For a family of bounded operators $(B_\varepsilon)_{\varepsilon>0}$, we will use the following topology of convergence as $\varepsilon\to0$, 
\begin{enumerate} [label=\roman*$)$]

\item $B_\varepsilon \to 0 $ in the \emph{strong operator topology} if $\lim_{\varepsilon\to0} \| B_\varepsilon\phi\| =0 $ for all $\phi\in L^2$. If $\|B_\varepsilon\| \le C$, it suffices to verify that  $\lim_{\varepsilon\to0} \| B_\varepsilon\phi\| =0 $ for all $\phi\in \mathscr{A}$ where $\mathscr{A}$ is dense in $L^2$. 

\item $B_\varepsilon \to 0 $ in the \emph{operator norm topology}  if $\lim_{\varepsilon\to0} \| B_\varepsilon\| =0 $. 

\item For $p\ge 1$, $B_\varepsilon \to 0 $ in the $\J^p$ \emph{norm topology} if  $\lim_{\varepsilon\to0} \| B_\varepsilon\|_{\J^p} =0 $. 
Note that if $q>p\ge 1$, the  $\J^p$ \emph{norm topology}  is stronger than the $\J^q$ \emph{norm topology}. 
\end{enumerate}

Then, we have $iii) \Rightarrow ii) \Rightarrow i)$. 
\end{prop}

\begin{prop}[Hilbert-Schmidt operators] \label{prop:hsop}~
\begin{enumerate} [label=\roman*$)$]
\item We say that $B$ is  Hilbert-Schmidt if  $\|B\|_{\J^2} <+\infty$, in which case by  \cite[Theorem 2.11]{simon_trace_2005},   $B$ has an integral kernel and 
\[
\|B\|_{\J^2}  = \bigg( \int_{\R^{2n}} |B(x,y)|^2 \d x \d y \bigg)^{1/2} . 
\]

\item By \cite[Theorem 2.15]{simon_trace_2005}, it holds $\|B\|_{\J^2}^2  = {\textstyle \sum_{k=1}^{\infty}}  \| B \phi_k \|^2$ for any  orthonormal basis  $(\phi_k)_{k\in\N}$  of $L^2(\R^n)$. 
\end{enumerate}
\end{prop} 

\begin{prop}[Trace-class operators] \label{prop:top}~
\begin{enumerate} [ label=\roman*$)$]

\item We say that  $B$ is trace-class if  $\|B\|_{\J^1} <+\infty$. 
Moreover by \cite[Theorem 2.12]{simon_trace_2005}, if $B \ge 0$ is trace-class and has a kernel which is continuous on $\R^{2n}$,  and 
\[
\|B\|_{\J^1}  =  \int_{\R^n} B(x,x)  \d x . 
\] 

\item If $B$ is trace-class, we define $\tr B = {\textstyle \sum_{k=1}^{+\infty}   \langle  \phi_k, B\phi_k \rangle}$ for any orthonormal basis of $L^2(\R^n)$, see \cite[Theorem 3.1]{simon_trace_2005}. 
Note that  $\tr(\cdot)$ is a linear operator  with  $|\tr B| \le \|B\|_{\J^1}$ and $\tr B =  \|B\|_{\J^1}$ when $B\ge 0$.  


\item If $B$ is trace-class,  then the Fredholm determinant $\det(1+B)$ is well-defined and it is a continuous function on $\J^1$. By  \cite[Theorem 3.4]{simon_trace_2005}, 
\[
\big| \det(1+B) -\det(1+A)\big| \le \| A-B\|_{\J^1} e^{1+  \| A\|_{\J^1} +\|B\|_{\J^1} } .
\]

\item By  \cite[Corollary 3.8]{simon_trace_2005}, if $A,B$ are bounded operators such that $AB$ and $BA$ are trace-class, then $\tr(AB) = \tr(BA)$ and $\det(1+AB) = \det(1+BA)$. 

\item We say that  $B$ is locally trace-class if for any compact set $\Ks \Subset \R^n$, 
$\|B\1_{\Ks}\|_{\J^1} <+\infty$. 
\end{enumerate}
\end{prop}

Let $A, B, M \ge 0$ be bounded operators. If $B \ge A$ and $BM$ is trace-class, then by \cref{prop:top}, 
\begin{equation} \label{TAB1}
\tr[AM] = \tr\big[\sqrt{M} A \sqrt{M}\big]  \le   \tr\big[\sqrt{M} B \sqrt{M}\big]  =\tr[BM]  , 
\end{equation}
where $\sqrt{M}$ denotes the positive solution of  $\sqrt{M}^2 = M$ -- see \cite[Section 1.1]{simon_trace_2005}. 
Moreover by \cref{prop:op},  if $A$ is a finite-rank operator and $B$ is bounded, then $\mu_k(AB) \le \1_{k\le\rank(A)} \|AB\|$ and 
\begin{equation} \label{TAB2}
\| AB\|_{\J^1} \le \|AB\| \rank(A) . 
\end{equation}

\subsection{Operators bounded from below}
\label{sec:unbounded-operators}In this paper, the usual Laplacian
$\Delta$ on  $L^2(\R^n)$ plays a crucial \new{role} and we review below its basic properties. 

An essentially self-adjoint operator, bounded from below, on a  Hilbert space  $H$ is the
data of a dense subspace $\mathcal{D}(A)\subset H$ and a linear map
$A:\mathcal{D}(A)\to H$ such that $\{(x,Ax),x\in \mathcal{D}(A)\}$ is
closed in $H\times H$ and there exists $C>0$ such that, for
every $x\in \mathcal{D}(A)$, one has $\langle x,Ax\rangle=\langle
Ax,x\rangle\geq -C\|x\|^2_{H}.$

The point is that essentially self-adjoint operators, bounded from below, admit a
spectral decomposition and a functional calculus; we refer to the textbook
\cite{helffer_spectral_2013}.
In particular, since $-\Delta \ge 0$ on $L^2(\R^n)$, under the hypothesis~\ref{hyp:weak}, the Schrödinger operator $H_{\hbar}=-\hbar^2\Delta+V$ is essentially self-adjoint,  bounded from below, on $L^2(\R^n)$. 
Moreover, if $V\ge c$ for a constant $c\ge 0$, then $H_{\hbar}\ge c$. 

\subsubsection{Spectral properties of $-\Delta$ on $L^2(\R^n)$. }
Recall that $\mathcal{F}$ denotes the Fourier transform viewed as a unitary
operator on $L^2(\R^n)$ and that $-\i \nabla =  \mathcal{F}^* A
\mathcal{F}$ where $A$ is the multiplication operator, $A \phi:\xi
\mapsto \xi \phi(\xi)$. 
This property of the Fourier transform renders explicit many
properties of the operator $-\i \nabla$ or its functions.
For instance, we define  $-\Delta$ as a self-adjoint operator on $L^2(\R^n)$ with domain 
\[H^2 = \{ \phi\in L^2 :  \xi \mapsto |\xi|^2 \mathcal{F}\phi(\xi)
  \in L^2(\R^n) \}.\]
Similarly, the classical $L^2$ Sobolev spaces can be defined  using the
Fourier transform. For $s\in\R$, let
\[H^s = \{ \phi\in \mathcal{S}' :  \xi \mapsto (1+|\xi|)^s \mathcal{F}\phi(\xi)
  \in L^2(\R^n) \},\]
where $\mathcal{S}'$ denotes the space of Schwartz distributions. There
are alternative definitions of $H^s$, e.g.~Remark~\ref{rem:Slobodeckij}.

For any $f \in L^\infty(\R)$, we define
$f(-\Delta) =  \mathcal{F}^* f \mathcal{F} $ where $f$ stands for a multiplication operator as in \cref{prop:op} $i)$. 
In particular, $f(-\Delta)$  is a pseudodifferential operator in the sense of \cref{def:pseudos}, with semiclassical parameter $\hbar =1$ and symbol $a= f(|\xi|^2)$. 
Hence, by \cref{prop:hsop} $ii)$ and $v)$, if $f: \R \mapsto [0,\infty)$ satisfies $f(|\xi|^2) \in L^1(\R^n)$, the operator $f(-\Delta)$ is locally trace-class with integral kernel
\begin{equation} \label{kernel_Laplace}
(x,y) \in \R^{2n} \mapsto \frac{1}{(2\pi)^n} \int_{\R^n} e^{\i (x-y)\cdot \xi} f(|\xi|^2) \d \xi . 
\end{equation}
In particular, this shows that for any $\mu>0$ the kernel of the \emph{bulk operator} is given by
\begin{equation} \label{kernelPi}
K_{\rm b , \mu}^{(n)} = \1_{(-\infty,\mu^2]}(-\Delta) : (x,y) \in \R^{2n} \mapsto  \frac{1}{(2\pi)^n} \int_{\R^n} \1_{|\xi|\le\mu}e^{\i (x-y)\cdot \xi} \d \xi . 
\end{equation}
By going to spherical coordinates, one can rewrite
\begin{equation} \label{Picomputation}  \begin{aligned}
K_{\rm b , \mu}^{(n)}(x,y)& =   \frac{1}{(2\pi)^n} \int_0^\mu \int_{S^{n-1}} e^{\i r |x-y| \mathrm{e}_1 \cdot \theta}  r^{n-1}  \d \theta \d r \\
&=  \frac{1}{(2\pi|x-y|)^n}  \int_0^{\mu|x-y|} \int_{S^{n-1}} \cos(r  \theta_1 ) r^{n-1}  d \theta \d r \\
&  =  \frac{\Gamma(\frac{n-1}{2})^{-1}/\sqrt{\pi}}{(\sqrt{2\pi}|x-y|)^n}  \int_0^{\mu|x-y|} r^{n-1}  \bigg( \int_0^\pi \cos(r  \cos\varphi ) (\sin\varphi)^{n-2} \d\varphi  \bigg) \d r   \\
& = \frac{1}{(\sqrt{2\pi}|x-y|)^n} \int_0^{\mu|x-y|}r^{\frac n2}  \J_{\frac n2-1}(r)\d r \\
& =   \mu^{\frac n2}  \frac{  \J_{\frac n2}(\mu|x-y|)}{(2\pi|x-y|)^{\frac n2}}
\end{aligned}
\end{equation}
where we used that the Bessel functions of the first kind are given by for any $\nu\in\R_+$ and $r\in\R_+$, 
\[
\J_{\nu}(r) =\frac{(r/2)^{\nu}}{\sqrt{\pi}\Gamma(\nu+1/2)} \int_0^\pi \cos(r \cos\varphi) (\sin\varphi)^{2\nu} \d\varphi 
\]
and $ \frac{\d}{\d r} \big( r^\nu \J_{\nu}(r) \big) = r^\nu   \J_{\nu-1}( r)$.
In particular, the determinantal point process associated with the operator $K_{\rm b , \mu}^{(n)}$ is both translation and rotation invariant on $\R^n$ with intensity $\frac{\mu^{n} \omega_{n}}{(2\pi)^n}$. 

\medskip

Let us also recall the following
standard bounds.

\begin{lem}\label{prop:elliptic_reg}(Sobolev embeddings) 
  Let $k\in\N_0$. There exists $C_k>0$ such that, for every $f\in
  \mathcal{S}'(\R^n)$, for every $\hbar\in(0,1]$, one has
  \[
    \|f\|_{C^{2k}(\R^n)}\leq C_k\hbar^{-2k-\frac n2}\|(1-\hbar^2\Delta)^{k+\left\lfloor
        \frac{n}{4}\right\rfloor+1}f\|_{L^2(\R^n)}.
    \]
\end{lem}
The most direct argument relies on the
properties of the Fourier transform.

\begin{preuve}
  The Fourier transform of  $g \in L^1$ is continuous with
  \[
    \|\hat{g}\|_{C^0}\leq (2\pi)^{\frac n2}\|g\|_{L^1}.
  \]
Hence, for every multi-index $\alpha\in \N_0^{n}$, one has
  \[
    \|x\mapsto \partial_x^{\alpha}f(x)\|_{C^0}\leq (2\pi)^{\frac n2}\|\xi \mapsto
    \xi^{\alpha}\hat{f}(\xi)\|_{L^1},
  \]
  and, by Cauchy-Schwarz's inequality,
  \[
    \|\xi \mapsto
    \xi^{\alpha}\hat{f}(\xi)\|_{L^1}\leq \|\xi\mapsto
    \xi^{\alpha}(1+\hbar^2|\xi|^2)^{-k-\lfloor \frac n4\rfloor - 1}\|_{L^2}\|\xi\mapsto
  (1+\hbar^2|\xi|^2)^{k+\lfloor \frac n4\rfloor+1}\hat{f}\|_{L^2}.
\]
By the above functional calculus for $-\i \nabla$ and since $\|\xi\mapsto
    \xi^{\alpha}(1+\hbar^2|\xi|^2)^{-k-\lfloor \frac n4\rfloor -
      1}\|_{L^2} \le C \hbar^{-|\alpha|-\frac n2}$ if $|\alpha| \le 2k$, this completes the proof. 
\end{preuve}

By working on a (relatively) compact set, one can, in fact, replace $1$ with any smooth positive potential.

\begin{prop}\label{prop:elliptic_reg_V}
  Let $V:\R^n\to [1,+\infty)$ and $\Omega\subset \R^n$ be a
  relatively compact open set such that $V$ is
  $C^{\infty}$ on a neighbourhood of $\overline{\Omega}$. Then, for every $k\in \N$, there exists $C_k(V)$
  such that, for every $f\in  \mathcal{S}'(\R^n)$ with compact support
  in $\Omega$, for every $0< \hbar\leq 1$,
  \[
    \|f\|_{C^{2k}}\leq C_k(V)\hbar^{-k-\frac n2}\|(V-\hbar^2\Delta)^{k+\lfloor \frac n4\rfloor+1}f\|_{L^2} .
    \]
\end{prop}
\begin{preuve}
According to Lemma~\ref{prop:elliptic_reg}, it suffices to show
  that for every $k\in \N$, there exists $C_k>0$  such that, for any
  $f\in C^{\infty}_c(\Omega)$, uniformly for $0< h\leq 1$,
  \begin{equation}\label{eq:ell_reg_V}
    \|(1-\hbar^2\Delta)^k f\|_{L^2}\leq
    C_k\|(V-\hbar^2\Delta)^kf\|_{L^2(\Omega)} .
      \end{equation}
Then the result follows by density.

  Let us prove \eqref{eq:ell_reg_V} by induction on $k$; the case $k=0$ is trivial with $C_0=1$. 

By assumptions, it holds for
  any $k\in\N_0$ and $f\in C^{\infty}_c(\Omega)$,
    \[
    (V-\hbar^2\Delta)^{k+1}f=(1-\hbar^2\Delta)^{k+1} f+L_{k}f,
  \]
  where the differential operator $L_k$ takes the form
  \[
    L_k=\sum_{|\alpha|\leq 2k}g_{k;\alpha}(x,\hbar)\hbar^{|\alpha|}\partial^{\alpha}
  \]
  and $g_{k;\alpha}$ is bounded, along with its derivatives,
  uniformly as $\hbar\to 0$. Indeed, letting $\widetilde{V}\in
  C^{\infty}_c(\R^n,\R)$ be equal to $V$ on $\overline{\Omega}$, the
  sequence $(L_k)_{k\geq 0}$ satisfy
  \[
    L_0=\widetilde{V}-1 ,  \qquad
    L_{k+1}=(-\hbar^2\Delta+\widetilde{V})L_{k-1}+(\widetilde{V}-1)(-\hbar^2\Delta+1)^k.
  \]
  Hence, there exists $C_k$ such that, for every $f\in
  C^{\infty}_c(\Omega)$,
  \[
    \|(1-\hbar^2\Delta)^{k+1}f\|_{L^2}\leq
    \|(V-\hbar^2\Delta)^{k+1}f\|_{L^2}+C_k\sum_{|\alpha|\leq
      2k}\|\hbar^{|\alpha|}\partial^{\alpha}f\|_{L^2}.
  \]
  Moreover, one directly has
  \[
    \sum_{|\alpha|\leq
      2k}\|\hbar^{|\alpha|}\partial^{\alpha}f\|_{L^2}\leq
    C_k\|(1-\hbar^2\Delta)^kf\|_{L^2},
  \]
  since all operators involved are Fourier multipliers.

  In conclusion, by the induction hypothesis
  \begin{align*}
    \|(1-\hbar^2\Delta)^{k+1}f\|_{L^2}&\leq
                                        \|(V-\hbar^2\Delta)^{k+1}f\|_{L^2}+C_k\|(V-\hbar^2\Delta)^{k}f\|_{L^2}\\
                                      &\leq C_k\|(V-\hbar^2\Delta)^{k+1}f\|_{L^2},
  \end{align*}
  where we used the fact that $V-\hbar^2\Delta\geq 1$ as
  operators and we upgraded the constant $C_k$ from line to line.
\end{preuve}

\begin{rem} \label{rk:ellpiticity2}
The argument of Lemma~\ref{prop:elliptic_reg} also shows that if $f\in
  \mathcal{S}'(\R^n\times\R^n)$, then 
  \begin{align*}
\| f\|_{C^{2k}(\R^{2n})} =& \sum_{\alpha , \beta \in\N_0^n : |\alpha|+ |\beta| \le 2k} \sup_{x,y\in\R^n}  \big| \partial_x^\alpha  \partial_y^\beta f (x,y) \big|
\\ \leq &C_k\hbar^{-2k-n}
\sum_{i,j \in \N_0 : i+j = k}
\|(1-\hbar^2\Delta_x)^{i+\left\lfloor
        \frac{n}{4}\right\rfloor+1} (1-\hbar^2\Delta_y)^{j+\left\lfloor
        \frac{n}{4}\right\rfloor+1}f\|_{L^2(\R^n)}. 
  \end{align*}
  Then, by adjusting the proof of Proposition~\ref{prop:elliptic_reg_V}, it is straightforward to show that under the same assumptions, for every $f\in  \mathcal{S}'(\R^n\times \R^n)$ with compact support in $\Omega$,
  \[
\| f\|_{C^{2k}(\R^{2n})}
\le C_k(V)\hbar^{-2k-n}\|(V-\hbar^2\Delta)^{k+\lfloor \frac n4\rfloor+1}_x (V-\hbar^2\Delta)^{k+\lfloor \frac n4\rfloor+1}_yf\|_{L^2}.
\]
The only difference is that for a test function $f\in
C^\infty(\R^n\times \R^n)$, for any $i,j\in\N_0$ with $i+j\le k$,
there holds
    \[
      (V-\hbar^2\Delta)_x^{i}  (V-
     \hbar^2\Delta)_y^{j} f=(1-\hbar^2\Delta)_x^i (1-\hbar^2\Delta)_y^jf+L_{i,j}f,
  \]
  where $L_{i,j}$ is an $\hbar$-differential operator, with bounded
  coefficients (by polynomials in
  $\|V\|_{C^{2k}(\overline{\Omega})}$ and $\hbar$), of order  $2i,2j$ with respect to 
 $x$ and $y$ respectively, and of total order $2(i+j-1)$.
Hence, since $x,y$ are independent variables, we can bound 
  \[
    \|L_{i,j}\|_{L^2}\leq C_k(V)
    \left(\|(1-\hbar^2\Delta)_x^{i}(1-\hbar^2\Delta)_y^{j-1}\|_{L^2}+\|(1-\hbar^2\Delta)_x^{i-1}(1-\hbar^2\Delta)_y^{j}\|_{L^2}\right). 
  \]
The proof now follows by the same induction as in Proposition~\ref{prop:elliptic_reg_V}.
\end{rem}

\subsection{Determinantal point processes.}
\label{sec:dpp}

This class of point processes was introduced in \cite{Macchi75} under the name \emph{Fermion processes}. 
Macchi's fundamental contributions were physically motivated by electron-interference experiments and the goal was to provide a mathematical model for point processes which obey the Pauli exclusion principle; hence the name  \emph{Fermion processes}. 
In contrast to Poisson processes which describes independent particles, determinantal processes are well-known to exhibit repulsion and are typically \emph{hyperuniform} \cite{torquato_point_2008}. 
The mathematical theory of determinantal processes has been developed largely by Soshnikov \cite{Soshnikov_01} and Shirai-Takahashi  \cite{ST03a,ST03b}. We also refer to \cite{HKPV06} for a survey with more probabilistic perspectives. 
There are many applications beyond the context of quantum mechanics discussed in the introduction which include random matrices and Coulomb gases, asymptotic representation theory, certain random tilings and 2-dimensional growth processes, zeros of Gaussian analytic functions, etc. We refer to \cite{Soshnikov_01,Johansson06, Borodin11} for comprehensive reviews of these examples.
In this section, we explain some basic concepts of the theory of determinantal point processes.
The  results are stated for the Hilbert space $L^2(\R^n)$ but they hold true for $L^2(\mathscr{E})$ if $\mathscr{E}$ is a locally compact, completely separable, Hausdorff space. 
We only focus on the Euclidean case for concreteness.

\subsubsection{Point processes on $\R^n$.}
Let us recall that a (simple) point process is a random measure of the form $\Xi = \sum_{\lambda \in \Lambda} \boldsymbol\delta_{\lambda} $ 
where $\Lambda\subset \R^n$ is a countable set with no accumulation points.
We refer to  \cite[Section 2]{Johansson06}  for the construction of such random measures. 
The law of a point process is usually characterized by its Laplace functional:
\[
\psi_\Xi(f) = \E\big[ e^{-\Xi(f)} \big]
 \qquad\text{for all function } f \in C^\infty_c\big(\R^n , \R_+\big).
\]
We define the correlation functions $\big( R_k \big)_{k=1}^{+\infty}$ of the point process $\Xi$ through its Laplace functional by
\begin{equation*} 
\psi_\Xi(f) =  1 + \sum_{k=1}^{+\infty} \frac{1}{k!} \int_{\R^{n\times k}} \prod_{i=1}^k \big( e^{-f(z_i)} -1 \big) R_k(z_1,\cdots, z_k)  \d z_1\cdots\d z_k ,
\end{equation*}
when this expansion makes sense.
For instance, if  $\Xi = \sum_{j=1}^N \boldsymbol\delta_{x_j} $  and $\{x_1,\dots, x_N\}$ has a (symmetric) joint probability density function $\P_N$, then we verify that the correlation functions of $\Xi$ are given by the marginals for $k\in\N$,
\begin{equation*} 
R_k( z_1,\cdots, z_k) = \1_{k\le N} \frac{N!}{(N-k)!} \int_{\R^{n\times(N-k)}} \P_N[z_1,\dots, z_k ,  \d z_{k+1}, \dots , \d z_N] .
\end{equation*}

\subsubsection{Determinantal processes.} In this article, we rely  on the following definition.

\begin{defn} \label{def:dp}
A point process $\Xi$ is called determinantal if its Laplace functional is of the form 
\begin{equation} \label{psiXi}
\psi_\Xi(f) = \det\big[\I - (1-e^{-f})  K\big]
\end{equation}
where $K$ is a locally trace-class operator on $L^2(\R^n)$ and the RHS of \eqref{psiXi} is a Fredholm determinant -- see \cref{prop:top} for the relevant definitions. 
Then, we say that the determinantal process $\Xi$ is associated with the operator $K$ and the correlation functions of $\Xi$ are given by 
\[
R_k(z_1, \dots , z_k) = {\textstyle \det_{k\times k}\big[ K(z_i,z_j)\big] }  , \qquad k\in\N .
\]
\end{defn}

\cref{def:dp}  is equivalent to the usual definition of determinantal processes in terms of its correlation kernel $K$, see e.g. \cite[Section 2]{Johansson06} or \cite[Theorem 2]{Soshnikov_01}. 
Moreover, while a determinantal process has several correlation kernels, the associated operator is uniquely defined.
In general, formula \eqref{psiXi} makes sense for any (measurable) function
$f:\R^n \to \R_+$. 
Moreover, if the operator $K$ is trace-class, then by \cref{prop:top} $iii)$, $\psi_\Xi$ is continuous with respect to the $L^\infty$-norm. 
 
\cref{def:dp}  has the following consequences for the distribution of a \emph{linear statistic} $\Xi(f)$.

\begin{lem} \label{lem:cov}
Let  $\Xi$ be a determinantal process associated with the operator $K$. Then for any
functions $f , g: \R^n\to\R_+$, 
\[
\E\, \Xi(f)  =\tr(fK) 
\]
and, if defined,
\begin{equation} \label{covariance}
 \operatorname{cov}\big(\Xi(g) ,\Xi(f) \big) = \tr \big( g(\I-K)f K\big)   = \tfrac12 \tr\big([g,K][K,f]\big)  +  \tr\big(gf K (\I-K)\big).
\end{equation}
\end{lem}

\begin{preuve}
By definition of the correlation functions, see \cref{def:dp}, 
\[
\E[\Xi(f)] =  \int_{\R^n} f(z) R_1(z) \d z =  \int_{\R^n} f(z)K(z,z)\d z   =  \tr[fK]  
\]
and for any $f, g \in L^\infty_c(\R^n)$, 
\[\begin{aligned}
 \E[\Xi(g) \Xi(f)] & =  \iint_{\R^{2n}}  g(z_1) f(z_2) R_2(z_1,z_2) \d z_1 \d z_2 +  \int_{\R^n} g(z)f(z) R_1(z) \d z \\
 & =  \tr[gK]\tr[gK] - \tr[gKfK]  + \tr[gfK] .  
\end{aligned}\]
This shows that the covariance between the random variables $\Xi(f)$ and $\Xi(g)$ is given by 
\begin{equation*}
 \operatorname{cov}\big(\Xi(g) ,\Xi(f) \big)   =  \E[\Xi(g) \Xi(f)] -  \E[\Xi(g)]\E[\Xi(f)]  = \tr[g(\I-K)f K] . 
\end{equation*}
This formula is actually symmetric and can be written using commutators as follows 
\[
 \tr[g(\I-K)f K]  = \tfrac12  \tr\big([g,\I-K][f,K]\big)  + \tr(g fK(\I-K))
\]
where we used cyclicity of $\tr[\cdot]$. 
To complete the proof, it remains to observe that $[g,\I-K] =[K,g]$. 
\end{preuve} 

Note that formula \eqref{covariance} implies that for any $g\in L^\infty_c(\R^n)$, the variance of the linear statistics $\Xi(g)$ is well-defined and given in terms of Hilbert-Schmidt norm by 
\begin{equation} \label{variance}
 \var \Xi(g)   = \big\| \sqrt{K} g \sqrt{\I-K}\big\|_{\J^2}^2 = \tfrac12 \big\| [g,K] \big\|_{\J^2}^2 +  \big\|  g \sqrt{K(\I-K)}\big\|_{\J^2}^2 .
\end{equation}

In the context of this article, it is relevant to recall the following  \cite[Theorem 3]{Soshnikov_01}.

\begin{prop} \label{thm:Sos}
If $K$ is a self-adjoint locally trace-class operator on $L^2(\R^n)$, then it $($uniquely$)$ defines a determinantal processes if and only if $0 \le K \le \I$. 
\end{prop}

For instance, consider $H$ an (unbounded) self-adjoint with a domain
$\mathscr{D}(H)$ which is dense in $L^2(\R^n)$ and let $K_S =
\1_{S}(H)$ where $S \subset \R$ is a Borel set. Then, by
\cref{thm:Sos}, if $K_S$ is locally trace-class on $\R^n$, there
exists a determinantal process associated with $K_S$. In particular,
if  $H$ is bounded from below and $ \1_{(-\infty, \mu)}(H)$ is a
finite-rank projection (as in our setting, thanks to Proposition
\ref{prop:rough_upper_Weyl}), there is a determinantal process associated with this operator. Moreover, this process has almost surely $N=\rank\big(\1_{(-\infty, \mu)}(H)\big) $ particles -- see  \cite[Theorem 4]{Soshnikov_01}.

\medskip

It is also of interest to record the following basic expansion for the Laplace functional \eqref{psiXi} for small $f$. The proof of \cref{prop:Laplace} is inspired from that of \cite[Theorem 1]{Soshnikov_02}, except that we control directly the Laplace functional instead of the cumulants of the linear statistic~$\Xi(f)$. 
This simplifies the argument and makes the result stronger. 
In particular, these asymptotics could certainly be of independent interest. 

\begin{prop} \label{prop:Laplace}
Under the assumptions of \cref{thm:Sos}, it holds uniformly for all  $f\in L^\infty_c(\R^n)$ with $f \le .69$, 
\begin{equation*}
\psi_\Xi(-f) = \exp\Big(\E[\Xi(f)] + \var\big(\Xi(e^{f}-1)\big)\big(\tfrac12+ \O(\|f_+\|_{L^\infty})\big) \Big) . 
\end{equation*}
\end{prop}

\begin{preuve}
By functional calculus,  it holds for \old{and}  $g\in L^\infty_c(\R^n)$ with $\|g\|_{L^\infty} <1$, 
\begin{equation} \label{laplace0}
\log \det\left(1- g K\right) = \tr\big[ \log(1- g  K)\big]
= - \sum_{\ell\in\N}  \tfrac{1}{\ell} \tr\big[(g K)^\ell \big]  . 
\end{equation}
 Now, define $A_{k} = g (\I-K) g^kK$ for $k\in \N_0$ and observe that for any $\ell \ge 2$, 
\begin{align}
\tr[(gK)^\ell] &\notag =   \tr[(gK)^{\ell-2} g^2K]  - \tr[(gK)^{\ell-2} A_1]  \\
&\notag =   \tr[(gK)^{\ell-3} g^3K]  - \tr[(gK)^{\ell-3}  A_2]   - \tr[(gK)^{\ell-2} A_1]   ,\qquad\text{if }\ell\ge3,  \\
& \label{induction} =  \tr[g^\ell K] - \sum_{j=2}^{\ell} \tr[(gK)^{\ell-j}  A_{j-1}]  . 
\end{align} 
Moreover, using that $0\le K \le 1$ and  cyclicity of $\tr[\cdot]$, we have for any $j, k\in\N$, 
\[
 \tr[(gK)^{k}  A_{j}]  =   \tr[ \sqrt{K}(gK)^{k-1}g\sqrt{K}   \sqrt{K} g (\I-K) g^j \sqrt{K}]  .
\]
This implies that for $j, k\in\N$, 
 \begin{align} \notag
\big| \tr[(gK)^{k}  A_{j}] \big| &\le \|(gK)^{k-1}g\| \big\| \sqrt{K} g (\I-K) g^j \sqrt{K} \big\|_{\J^1} \\
&\notag \le \|g\|_{L^\infty}^k \sqrt{\big\| \sqrt{K} g \sqrt{\I-K}\big\|_{\J^2} \big\| \sqrt{\I-K} g^j \sqrt{K} \big\|_{\J^2}} \\
& \label{covbd1} = \|g\|_{L^\infty}^k \sqrt{ \var \Xi(g)  \var \Xi(g^j) }
\end{align}
according to \eqref{variance}. 
Moreover, 
$ \var \Xi(g^j)  =  \tfrac12  \big\| [g^j,K] \big\|_{\J^2}^2  + \tr[g^{2j}K(\I-K)] $ for $j\in\N$  and we can bound
\[\tr\big[ g^{2j-2} g K(\I-K) g\big] \le \|g\|_{L^\infty}^{2(j-1)}\| g K(\I-K) g \|_{\J^1}
= \|g\|_{L^\infty}^{2(j-1)} \tr[gK(\I-K)g]   .
\]
Since $[K,g^j] = \sum_{k=1}^{j} g^{k-1} [K,g] g^{j-k}$, we also have  $\big\| [K,g^j]  \big\|_{\J^2}\le j\|g\|_{L^\infty}^{j-1} \big\| [K,g] \big\|_{\J^2}$ so that 
\begin{equation} \label{covbd2}
\var \Xi(g^j)   \le j^2 \|g\|_{L^\infty}^{2(j-1)} \big( \tfrac12 \big\| [K,g] \big\|_{\J^2}+ \tr[ g^2 (\I-K) K]  \big) 
=  j^2 \|g\|_{L^\infty}^{2(j-1)} \var \Xi(g) . 
\end{equation}
By \eqref{covbd1},  this shows that  for any $j, k\in\N$, 
\[
\big| \tr[(gK)^{k}  A_{j}] \big|  \le  j \|g\|_{L^\infty}^{k+j-1} \var \Xi(g) .
\]
Hence, by \eqref{induction}, we have shown that for any $\ell \ge 3$, 
\begin{equation} \label{laplace1}
\big| \tr[(gK)^\ell] -  \tr[g^\ell K] \big| \le   \|g\|_{L^\infty}^{\ell-2} \sum_{j=2}^{\ell} ( j-1) \var \Xi(g)
=    \|g\|_{L^\infty}^{\ell-2}  \tfrac{\ell(\ell-1)}{2} \var \Xi(g) . 
\end{equation}
Observe that  if $g=1-e^{-f}$, then by linearity of $\tr[\cdot]$,
\begin{equation} \label{mean1}
 \sum_{\ell\in\N}  \tfrac{1}{\ell} \tr[ g^\ell K] = \tr\bigg[ \bigg( \sum_{k\in\ell} \tfrac{1}{\ell} g^\ell \bigg) K \bigg] =- \tr\big[ \log(1- g )K\big] = \tr[f K] . 
\end{equation}

By combining \eqref{laplace0}, \eqref{mean1} with the bound \eqref{laplace1} for  $g=e^{f}-1$ and $\ell\ge 3$, 
this shows that if $\|g\|_{L^\infty} <1$, 
\[
\big| \log \det\left(1- g K\right) - \tr[f K] - \tfrac12 \var \Xi(g)] \big| \le   \tfrac12 \var \Xi(g)\sum_{\ell\ge 3}  (\ell-1)  \|g\|_{L^\infty}^{\ell-2} =  \var \Xi(g)\frac{ \|g\|_{L^\infty} }{(1- \|g\|_{L^\infty})^2} ,
\]
where we used that for $\ell=2$,
$\tr[(gK)^2]- \tr[g^2 K] = - \tr[A_1] = \var \Xi(g)$, according to formulas \eqref{variance}--\eqref{induction}. 
By \eqref{psiXi}, we conclude that if $f \le  .69 <\log 2$, then 
\[
\psi_\Xi(-f) = \exp\Big( \tr[fK] + \var \Xi(g) \big(\tfrac12+ \O(\|f_+\|_{L^\infty}) \big)\Big)
\]
where we used that under our assumptions:  $ \frac{\|g\|_{L^\infty}}{(1-\|g\|_{L^\infty})^2} \le C \|f_+\|_{L^\infty}$ for a numerical constant $C>0$. 
\end{preuve}

Like in \cite{Soshnikov_02}, one can immediately deduce from the asymptotics of \cref{prop:Laplace} a CLT for linear statistics of general determinantal processes.
We give the proof of this result since, compared to  \cite[Theorem 1]{Soshnikov_02}, we can remove the (technical) conditions on the mean and variance of the linear statistic~$\Xi_N(f_N)$.

 \begin{corr} \label{clt_dpp}
 Let $(K_N)_{N\in\N}$ be a sequence of $($self-adjoint$)$  trace-class operators on $L^2(\R^n)$ with $0\le K_N \le 1$ and let  $(\Xi_N)_{N\in\N}$ be the associated determinantal processes on $\R^n$. Let $f_N \in L^\infty(\R^n)$ with $\|f_N\|_{L^\infty} \le C$ and assume that  $\sigma_N^2 = \operatorname{var}[\Xi_N(f_N)]\to \infty$ as $N\to+\infty$. Then it holds uniformly for all $t\in \R$ with $|t| \le .69 \sigma_N/C$, 
 \[
 \psi_{\Xi_N}(t \sigma_N^{-1}f_N) \exp\Big(  t\sigma_N^{-1}\E\big[\Xi_N(f_N)\big] \Big) = 
   \exp\Big(  t^2  \big(\tfrac12+ \O\big(t \sigma_N^{-1}\big) \Big) .
    \]
In particular, this implies that
\[
\frac{\Xi_N(f_N) - \E[\Xi_N(f_N)]}{\sigma_N} \Rightarrow  \mathcal{N}_{0,1} , 
\]
where $\mathcal{N}_{0,1}$ denotes a standard Gaussian random variables. 
 \end{corr}
 
 \begin{preuve}
By the Cauchy-Schwarz's inequality, one has a crude bound
\[
\big| \var[\Xi(1-e^{-f})]  - \var[\Xi(f)]   \le  \sum_{\substack{ k,j \in \N \\ k+j >2}}  \tfrac{1}{k! j!} \sqrt{\var[\Xi(f^k)] \var[\Xi(f^j)]} . 
\]
Using the bound \eqref{covbd2}, this implies that 
\[ \begin{aligned}
\big| \var[\Xi(1-e^{-f})]  - \var[\Xi(f)]    \big| \le \frac{ \var[\Xi(f)] }{2}  \sum_{\substack{ k,j \in \N_0 \\ k+j >0}} \tfrac{1}{k! j!} \|f\|_{L^\infty}^ {k+j}  
& =  \frac{ \var[\Xi(f)] }{2} \big( e^{2\|f\|_{L^\infty}}-1\big) \\
& \le \|f\|_{L^\infty} \var[\Xi(f)] e^{2\|f\|_{L^\infty}} . 
\end{aligned}\]
Hence, by applying \cref{prop:Laplace}, it shows that if $\|f\|_{L^\infty} \le .69$, 
\[
\psi_\Xi(f) e^{\E[\Xi(f)] }= \exp\Big( \var[\Xi(f)]  \big(\tfrac12+ \O(\|f\|_{L^\infty})\big) \Big).
\]
Making the change of variable $f\leftarrow t \sigma_N^{-1}f_N$ with $t\in\R$, we conclude that as $N\to+\infty$,
\[
\psi_{\Xi_N}(t\sigma_N^{-1}f_N) e^{t \sigma_N^{-1}\E[\Xi_N(f_N)]} = \exp\Big(  t^2  \big(\tfrac12+ \O\big(t \sigma_N^{-1}\big) \Big) . 
\]
Since the convergence of the Laplace transform of the random variable $\frac{\Xi_N(f_N) - \E[\Xi_N(f_N)]}{\sigma_N}$ implies its convergence in distribution and 
$\E[e^{t   \mathcal{N}(0,1)}] = e^{t^2/2}$ for all $t\in\R$, this proves the claims. 
 \end{preuve}

\subsubsection{Weak convergence.} In the context of Theorem~\ref{prop:micro_limit_process}, we also recall the concept of convergence in distribution for point processes. 

\begin{defn} \label{def:wcvg}
We say that a sequence $\Xi_N$ of point processes on $\R^n$ converges in distribution to a point process $\Xi$ if the Laplace functionals $\psi_{\Xi_N}(f) \to \psi_{\Xi}(f)$  as $N\to+\infty$ for all $f\in C^\infty_c\big(\R^n,\R_+\big)$.
\end{defn}

In the context of \cref{thm:Sos},  this has the following consequence.

\begin{prop}[Weak convergence for determinantal processes] \label{prop:wcvg}
Let $(K_N)_{N\in\N}$ be a sequence of self-adjoint operator on $L^2(\R^n)$ with $0 \le K_N \le \I$. 
Suppose that as $N\to\infty$,  $K_N \to K$ locally uniformly where $K$ is locally  trace-class, then there exits a determinantal process $\Xi$ associated with the operator $K$ and the determinantal processes $\Xi_N$ associated with $K_N$ converge weakly to $\Xi$ as $N\to+\infty$.  
\end{prop}

\begin{preuve}
By assumptions, $K_N \to K$ in the weak operator topology as $n\to\infty$ (cf.~Proposition~\ref{prop:op}). Indeed, by density, it suffices to verify that 
$\langle \phi , K_N \phi \rangle  \to \langle \phi , K \phi \rangle$ for any $\phi \in C_c(\R^n)$.
This also implies that  $0 \le K \le \I$ and by \cref{thm:Sos}, there exists a determinantal process associated with the operator $K$. 
Moroever, using that $K_N\ge 0$ and \cref{prop:top}~$i)$, it holds as $N\to\infty$, $\| \chi K_N \|_{\J^1} =\tr[\chi K_N] \to \tr[\chi K] = \|\chi K \|_{\J^1}<\infty$ for any $\chi \in L^\infty_c(\R^n,\R_+)$. 
Since both $K_N, K$ are positive operators, by \cite[Theorem 2.19]{simon_trace_2005}, this implies that $\| \chi(K_N - K) \|_{\J^1}\to 0$   for any such $\chi$. 
Now,  by formula \eqref{def:dp} and \cref{prop:top} $iii)$,  the Laplace functionals of these point processes satisfy for every $f \in C^\infty_c\big(\R^n ,\R_+\big)$. 
\[
\big| \psi_{\Xi_N}(f)  - \psi_\Xi(f) \big| 
\le \big\|  \chi (K_N - K) \big\|_{\J^1}  e^{1+ \|\chi K_N\|_{\J^1}  +  \| \chi K\|_{\J^1}  } 
\]
 with $\chi = 1-e^{-f}$.
 Since $\chi\ge 0$ with compact support, 
\[
\limsup_{N\to \infty} \big| \psi_{\Xi_N}(f)  - \psi_\Xi(f) \big|  \le e^{1+2 \tr[\chi K]}  \limsup_{N\to \infty} \big(\|\chi(K_N - K) \|_{\J^1} e^{\|\chi(K_N - K) \|_{\J^1} } \big)=0. 
\]
According to \cref{def:wcvg}, this completes the proof. 
\end{preuve}

\subsection{Oscillatory integrals}
\label{sec:osc-int}

The technical core of this article relies on standard techniques of
semiclassical analysis, in particular, the stationary phase method to obtain the asymptotics of oscillatory integrals. 

\begin{prop}[Stationary phase lemma, Proposition 5.2 in \cite{dimassi_spectral_1999}]\label{prop:spl}
  Let $\Omega \subset \R^d$ be an open set, $A$ be a compact  and  
  $x\mapsto\Phi(x,y)\in C^{\infty}(\Omega,\R)$ for every $y\in A$. Suppose that for any $y\in A$, there is exactly one point $x_y\in \Omega$ such that $\partial_x \Phi(x_y,y)=0$, and that
 the Hessian matrix, ${\rm Hess}\, \Phi(x_y;y)$,  is non-degenerate. 
  
  Then, there exists a sequence of differential operators $(L_k)_{k\geq
    0}$ on $\R^d$, depending on $y\in A$ such that $L_k$ is of degree $2k$ and
  for every $\ell\geq 0$ and  $\mathcal{K}\Subset \Omega$, 
  there is a constant $C_{\ell,\Phi,\mathcal{K}}$ such that, for
any $f\in C_c^{\infty}(\mathcal{K})$,
  \[
\sup_{y\in A}    \left|\int e^{\i\tfrac{\Phi(x,y)}\hbar}f(x)\dd x- e^{\i\tfrac{\Phi(x_y,y)}\hbar}
    (2\pi \hbar)^{\frac d2}\sum_{k=0}^{\ell}\hbar^k(L_kf)(x_y)\right|\leq
  C_{\ell,\Phi,\mathcal{K}}\, \hbar^{\frac d2+\ell+1}\|f\|_{C^{2\ell+d+1}}.
\]
 Moreover, one can also allow $f$ to depend smoothly on the parameter $y\in A$, in which case the error can be controlled with respect to $C^k(A)$ for every $k\in\N_0$.
  \end{prop}
  An explicit form for the operators $L_k$, depending on $\Phi$, can be found in Theorem
  7.7.5 of \cite{hormander_analysis_2003}. In particular, it holds
 \[
 L_0 = \frac{\i^{n_+-n_-}}{\sqrt{\det(|{\rm Hess}(\Phi)(x_y,y)|)}},
\]
where $n_+$, $n_-$ respectively denote the number of positive and
negative eigenvalues of ${\rm Hess}(\Phi)(x_y,y)$.

\medskip

In the special case where $d=2n$ and $\Phi:\R^{2n}\ni (x,\xi)\mapsto
x\cdot \xi \in \R$, one has
\begin{equation}
  \label{eq:statphas:xxi}
  L_k= \frac{\i^k}{k!}\sum_{|\alpha|=k}\partial^{\alpha}_x\partial^{\alpha}_{\xi}.
\end{equation}

In general, we apply Proposition~\ref{prop:spl} in the following standard way.
If  $a\in S^0$ is a classical symbol (cf.~Definition~\ref{def:class_symb}), then there exists another classical symbol $b\in S^0$ so that 
\[
\int_{\Omega}e^{\i\tfrac{\Phi(x,y)}\hbar}a(x,y;\hbar)\dd x = b(y,\hbar)  + \O(\hbar^\infty) . 
\]

The error being controlled by $\|f\|_{k}$ for some $k\in\N$ allows to apply Proposition~\ref{prop:spl} to
  $\hbar$-dependent functions satisfying, uniformly as $\hbar \to 0$, a control of the form
  \[
    \|f\|_{C^k}\leq C_k\hbar^{-k\delta} \qquad \text{for some $\delta<\frac 12$}. 
  \]
In this case, the terms in the expansion have decreasing magnitudes and the remainder is still $\O(\hbar^\infty)$ by choosing $\ell$ arbitrary large. 

\medskip
  
The twin proposition of Proposition \ref{prop:spl} is the \emph{non-stationary}
phase lemma.
\begin{prop}[Non-stationary phase lemma, Theorem 7.7.1 in \cite{hormander_analysis_2003}]\label{prop:nspl}
  Let $\Omega \subset \R^d$ be an open set, $A$ be a compact  and  
  $x\mapsto\Phi(x,y)\in C^{\infty}(\Omega,\R)$ for every $y\in A$. 
 Suppose that $ (x,y)  \mapsto \partial_x\Phi(x,y)$ is bounded
  away from $0$ on $\Omega \times A$.
  Then, for every $f\in C^{\infty}_c(\Omega)$, as $\hbar\to0$,
  \[
    \int_{\Omega}e^{\i\Phi(x)/\hbar}f(x)\dd x=\O(\hbar^{\infty}).
    \]
  \end{prop}
  An explicit estimate (involving derivatives of $f$ and a lower bound
  on $|\partial_x \Phi|$) is equation (7.7.1)' in
  \cite{hormander_analysis_2003}.

\medskip

In the remainder of this section, we give more precise estimates in the case $\phi:(x,y)\mapsto
x\cdot y$.

\begin{prop}\label{prop:bound-osc-lowreg-easy}
Let $f:\R^n\times \R^n\to \R$ be a measurable function and denote by $\hat{f}$ its Fourier transform with respect to the second variable. 
Define for $s\ge 0$
\[
\|f\|^2_{L^{\infty}H^s}:=\int \sup_{x\in \R^n}|\hat{f}(x,\xi)|^2(1+|\xi|)^{2s}\dd \xi 
\]
and let $L^{\infty}H^s$  be the corresponding function space. 
If $f\in L^{\infty}H^s$ with $s>\frac n2$, then for any $\hbar\in(0,1]$, 
\[
\bigg| \frac{1}{\hbar^n}\int e^{\i \tfrac{x\cdot y}\hbar}f(x,y)\dd x \dd y \bigg| \leq
 C_s\|f\|_{L^{\infty}H^s}.
\]
\end{prop}

\begin{preuve}
It holds 
\[
\frac{1}{\hbar^n}\int e^{-\i \tfrac{x\cdot y}\hbar}f(x,y)\dd x \dd y
= \frac{(2\pi)^{\frac n2}}{\hbar^n} \int \hat{f}(x,x/\hbar) \d x
=  (2\pi)^{\frac n2} \int \hat{f}(\hbar \xi, \xi)\d \xi
\]
Since $C_s = (2\pi)^{\frac n2} \displaystyle\int (1+|\xi|)^{-2s}\dd \xi  <\infty$ if $s>n/2$, by Cauchy-Schwarz's inequality, we obtain the bound
\[
\bigg| \frac{1}{\hbar^n}\int e^{\i \tfrac{x\cdot y}\hbar}f(x,y)\dd x \dd y \bigg| 
\le C_s \int \sup_{x\in \R^n}|\hat{f}(x,\xi)|^2(1+|\xi|)^{2s}\dd \xi  . \qedhere
\]
\end{preuve}
  
By assuming more regularity on $f$, we can bootstrap the previous proposition to obtain higher order expansion of such oscillatory integral with a similar control of the remainder.

   \begin{prop}\label{prop:spl_xxi}
     Let
     $\ell,n\in \N$ and let $s>\ell+\frac n2$. 
     Let $(L_k)_{k\in \N}$ be the sequence of differential operators
     given by \eqref{eq:statphas:xxi}. 
     There exists constants $C_{\ell,s}$ such that for any function $f:\R^{n}_x\times
     \R^n_{\xi}\to \R$ with  $\partial_x^{\alpha}f\in L^{\infty}_xH^{s}_{\xi}$ for every $|\alpha|\leq \ell$,  
     it holds for any $\hbar\in (0,1]$,
\begin{equation} \label{spl_xxi}
       \left|\frac{1}{(2\pi\hbar)^{n}}\int e^{\i\frac{x\cdot \xi}{\hbar}}f(x, \xi)\dd
         x \dd \xi - \sum_{k=0}^{\ell-1}\hbar^kL_kf(0,0)\right|\leq
       C_{\ell,s}\hbar^\ell\sum_{|\alpha|\leq
         \ell}\|\partial_x^{\alpha}f\|_{L^{\infty}_xH^s_{\xi}}.
       \end{equation}
     \end{prop}
     This statement, more precise than Proposition \ref{prop:spl}, is
     tailored to the case where $f$ is smooth but oscillates rapidly with respect to the first variables: given $g\in C^{\infty}_c(\R^{2n},\R)$,
     we can apply Proposition \ref{prop:spl_xxi} to
     \[
       f:(x,\xi)\mapsto g(\hbar^{-\delta}x,\xi)
     \]
     for any $\delta <1$ (since in this case,
     $\|\partial_x^{\alpha}f\|_{L^{\infty}_xH^s_{\xi}}=\O(\hbar^{-\delta\alpha\ell})$
     for any $|\alpha| \le \ell$),  whereas \new{one needs
       $\delta<\frac 12$ to use the result of }Proposition \ref{prop:spl}\old{ is limited
     to $\delta<\frac 12$}.
     
\medskip     
     
 \begin{preuve}
   We proceed by induction over $\ell\in\N_0$ utilizing that the case $\ell=0$ follows from Proposition \ref{prop:bound-osc-lowreg-easy}.
Hence, we assume that the bound \eqref{spl_xxi} holds for a given $\ell \in \N_0$.  
Define 
\[
F_1(x, \xi) = \frac{f(x,\xi) - f(x,\xi)|_{x_1=0}}{x_1} , \qquad x , \xi \in \R^{n} . 
\]
Integrating by parts,
\[
 \frac{1}{\hbar^n}\int e^{\i\frac{x\cdot
     \xi}{\hbar}}x_1F_1(x,\xi)\dd x\dd \xi = \frac{\i\hbar}{\hbar^n}\int e^{\i\frac{x\cdot
     \xi}{\hbar}}\partial_{\xi_1}F_1(x,\xi)\dd x\dd \xi.
\]
Note also that by \eqref{eq:statphas:xxi} and Taylor's theorem, for every $k\in \N_0$, 
\[
\i L_k(\partial_{\xi_1}F_1)(0,0)= \frac{\i^{k+1}}{(k+1)!}\sum_{|\alpha|=k}\partial_x^{\alpha}\partial_\xi^{\alpha}\partial_{x_1}\partial_{\xi_1}f(0,0) =   \frac{\i^{k+1}}{(k+1)!}\sum_{\substack{ |\alpha|=k+1 \\ \alpha_1>0}} \partial_x^{\alpha}\partial_\xi^{\alpha} f(0,0) 
\]
that is, all terms in $L_{k+1}f(0,0)$ such that $\alpha_1>0$.

Continuing in this fashion, one can write for $j\in\{1,\dots, n\}$, 
\[
F_j(x, \xi) = \frac{f(x,\xi)|_{x_1=\cdots=x_{j-1}=0} - f(x,\xi)|_{x_1=\cdots=x_{j}=0}}{x_j} , \qquad x , \xi \in \R^{n} 
\]
and verify that for every $k\in \N_0$, 
\begin{equation} \label{induction}
\i L_k(\partial_{\xi_j}F_j)(0,0) = \frac{\i^{k+1}}{(k+1)!}\sum_{\substack{|\alpha|=k+1,\, \alpha_j>0 \\ 
{\alpha_1=\cdots=\alpha_{j}=0}}} \partial_x^{\alpha}\partial_\xi^{\alpha} f(0,0)  .
\end{equation}

Hence, in the end,
\[
\frac{1}{(2\pi\hbar)^{n}}\int e^{\i\frac{x\cdot \xi}{\hbar}}f(x,\xi)\dd x = 
 \frac{1}{(2\pi\hbar)^n}\int e^{\i\frac{x\cdot
     \xi}{\hbar}}f(0,\xi)\dd \xi\dd x +  \frac{\i \hbar}{(2\pi\hbar)^n} \sum_{j=1}^n  
\int e^{\i\frac{x\cdot
     \xi}{\hbar}}\partial_{\xi_j}F_j(x,\xi)\dd x\dd \xi.
\]

For the first term, $\xi \mapsto f(0,\xi)$ belongs to $H^s$ for
$s>\frac n2 + 1$,  which is a subset of $C^0\cap L^2$ (by \new{Proposition \ref{prop:elliptic_reg}}). 
Hence, by a double Fourier transform,
\[
 \frac{1}{(2\pi\hbar)^n}\int e^{\i\frac{x\cdot
     \xi}{\hbar}}f(0,\xi)\dd
 \xi\dd x= f(0,0) . 
\]

Note also that by assumptions, the functions $\partial_{\xi_j}F_j$ are such that, for every
$|\alpha|\leq \ell$, one has $\partial_x^{\alpha}\partial_{\xi_j}F_j\in
L^{\infty}_xH^{s-1}_{\xi}$ with for instance,
\begin{equation}\label{inductionbound}
 \|\partial_x^{\alpha}\partial_{\xi_1}F_1\|_{L^{\infty}_xH^{s-1}_{\xi}}\leq \|\partial_x^{\alpha+(1,0,\ldots,0)}f\|_{L^{\infty}_xH^{s}_{\xi}}.
\end{equation}

Hence, using the induction hypothesis, this implies that if $s-1>\ell + \frac n2$, 
\[
\bigg| \frac{1}{(2\pi\hbar)^{n}}\int e^{\i\frac{x\cdot \xi}{\hbar}}f(x,\xi)\dd x -f(0,0) 
- \i \hbar \sum_{k=0}^{\ell-1} \sum_{j=1}^n \hbar^k L_k(\partial_{\xi_j}F_j)(0,0)  \bigg| 
\le  C_{\ell,s-1}\hbar^{\ell+1} \sum_{j=1}^n \sum_{|\alpha| =
         \ell}\|\partial_x^{\alpha} \partial_{\xi_j}F_j \|_{L^{\infty}_xH^{s-1}_{\xi}}.
\]

Using \eqref{induction} and the bound \eqref{inductionbound}, we conclude that $s>\ell+1 + \frac n2$, 
\[
\bigg| \frac{1}{(2\pi\hbar)^{n}}\int e^{\i\frac{x\cdot \xi}{\hbar}}f(x,\xi)\dd x 
- \sum_{k=0}^{\ell}  \hbar^k L_k(\partial_{\xi_j}F_j)(0,0)  \bigg| 
\le   C_{\ell,s-1}\, \hbar^{\ell+1} \sum_{|\alpha| =
         \ell+1}\|\partial_x^{\alpha} f\|_{L^{\infty}_xH^{s}_{\xi}} , 
\]
which completes the proof. 
\end{preuve}

To conclude this section, we provide another useful estimate on oscillatory
integrals. 

\begin{prop}\label{prop:sineint}
Let $\mathcal{K} \Subset \R$ and $f : \mathcal{K}\times \R\to \R$ be  such that
\[
 \|f\|_{L^{\infty}C^{2}}:=\sup_{(\lambda,t)\in \mathcal{K}\times
 \R}(|f(\lambda,t)|+|\partial_tf(\lambda,t)|+|\partial_{t}^2f(\lambda,t)|)<+\infty
                        \]
Then for every $\gamma\in (0,1]$, there exists $C_{K,\gamma}$ such that, as $\delta\to 0$,
\[
 \left|\int_{\new{\R^2}} \frac{\sin(\frac{\lambda t}{\delta})}{t}  f(\lambda,t)
   \d t \d \lambda-\new{2}\pi \int f(\lambda,0)  \d \lambda\right|\leq
 C_{K,\gamma}\, \delta^{1-\gamma}\left(\|f\|_{L^{\infty}C^{2}}+\|(t,\lambda)\mapsto
   t^{\gamma-1}\partial_tf(\lambda,t)\|_{L^{\infty}L^1}\right).
 \]
\end{prop}
\begin{preuve}
  Let us observe that by making an integration by parts,  it holds for
  any $f\in C^{1,1}_c(\R)$ and every $\lambda \neq 0$,  
\begin{align*}
\int_{\new{\R}} \frac{\sin(\lambda t)}{t}  f(t)\d t
&=  - \frac{2}{\lambda} \int  \sin(\lambda t/2)^2 \partial_t\bigg(
                                                      \frac{f(t)}{t}
                                                      \bigg) \d t \\
  &
= \frac{2}{\lambda} f(0) \int_{\old{0}\new{-\infty}}^{\new{+}\infty}  \frac{\sin(\lambda t/2)^2}{t^2} \d t
+  \int_{\new{-\infty}}^{\new{+\infty}} \frac{1-\cos(\lambda t)}{\lambda} \partial_t\bigg( \frac{f(0)-f(t)}{t} \bigg) \d t
\end{align*}
where $\R \ni t\mapsto \partial_t\big( \tfrac{f(0)-f(t)}{t} \big)$ is
bounded. Indeed, by the Taylor integral theorem,
\[
  \left|\partial_t\big( \tfrac{f(0)-f(t)}{t}
  \big)\right|=\left|\frac{f(t)-f(0)-tf'(t)}{t^2}\right|=\left|\frac{1}{2t^2}\int_0^tsf''(s)\dd
  s\right|\leq \frac 14 \|f''\|_{L^{\infty}}. 
\]
Moreover, $\displaystyle \int_{\R}\left(\frac{\sin(t)}{t}\right)^2=\pi$ as shown by
a Fourier transform.

Let now $f$ be as in the claim and let $g:(\lambda,t)\mapsto
\partial_t\left(\tfrac{f(\lambda,0)-f(\lambda,t)}{t}\right)$. Then
\[
  \int_{\new{\R^2}} \frac{\sin(\lambda t/\delta)}{t}f(\lambda,t)\dd t \dd
  \lambda=\new{2}\pi \int f(0,\lambda)\dd
  \lambda+\int_{\new{\R^2}}\frac{1-\cos(\lambda\delta^{-1}t)}{\lambda\delta^{-1}}g(\lambda,t)\dd
  t\dd \lambda.
\]
Using the Hölder bound $|1-\cos(u)|\leq |u|^{\gamma}$, valid for all
$u\in \R,\gamma\in (0,1)$, one has
\begin{align*}
  \left|\int_{\new{\R^2}}\frac{1-\cos(\lambda\delta^{-1}t)}{\lambda\delta^{-1}}g(\lambda,t)\dd
  t\dd \lambda\right|&\leq \delta^{1-\gamma}\int_{\mathcal{K}\times \R}
                       |\lambda|^{\gamma-1}|t|^{\gamma}|g(\lambda,t)|\dd t\dd \lambda\\
  &\leq
    \delta^{1-\gamma}\left(\int_{\mathcal{K}}|\lambda|^{\gamma-1}\dd
    \lambda\right)\left(\int_{\R}|t|^{\gamma}\sup_{\lambda\in
    \mathcal{K}}|g(\lambda,t)|\dd t\right).
\end{align*}
It remains to decompose
\[
  \int_{\R}|t|^{\gamma}\sup_{\lambda\in
    \mathcal{K}}|g(\lambda,t)|\dd t=\int_{[-1,1]}|t|^{\gamma}\sup_{\lambda\in
    \mathcal{K}}|g(\lambda,t)|\dd t+\int_{\R\setminus [-1,1]}|t|^{\gamma} \sup_{\lambda\in
    \mathcal{K}}|g(\lambda,t)|\dd t.
\]
The integral on $[-1,1]$ is bounded by
$\frac{1}{4}\|f\|_{L^{\infty}C^{2}}$. On $\R\setminus [-1,1]$, we
can write
\[
  g(\lambda,t)=-\frac{f(\lambda,0)-f(\lambda,t)}{t^2}-\frac{\partial_tf(t)}{t}.
\]
and then
\[
  \int_{\R\setminus [-1,1]}|t|^{\gamma}\sup_{\lambda\in
    \mathcal{K}}|g(\lambda,t)|\dd t \leq
  C(\gamma)\|f\|_{L^{\infty}C^{2}}+\|(\lambda,t)\mapsto
  t^{\gamma-1}\partial_tf\|_{L^{\infty}L^1}.
\]
This concludes the proof.
\end{preuve}

  \subsection{The Airy function}
\label{sec:airy-type-integrals}

The (classical) Airy function can be defined as the oscillatory integral; for $x\in\R$, 
\begin{equation} \label{def:Airy}
\Ai(x)= \frac1\pi \int_0^\infty \cos(t^3/3+xt) \d t   = \frac{1}{2\pi} \int_{\R} e^{\i (t^3/3+xt)} \d t . 
\end{equation}
These improper integrals converge and the definition can be extended to $x\in\C$ by considering a contour in the complex plane instead. The Airy function gives raise to the decaying solutions of the Schrödinger equation:
\[
\big(-\Delta+x \big) \Ai(x-\lambda) = \lambda\, \Ai(x-\lambda) , \qquad  x\in \R ,\, \lambda\in\C.
\]
 
For this reason, the Airy function arises in semiclassical approximations at (generic) boundary points as in Theorem~\ref{thr:micro_edge_kernel}. 
The differential operator $-\Delta+x$ on $\R$ is essentially self-adjoint, absolutely continuous and limit point at both $\pm\infty$. Moreover, its \emph{projection-valued measure} can be computed explicitly. Namely, for any $\lambda\in\R$, its kernel is given by
\begin{equation}\label{eq:measure_kernel_Airy}
 \frac1\pi\lim_{\epsilon\to0}\Im\big(-\Delta+x-(\lambda+\i \epsilon) \big)^{-1} : (x,y) \in \R^2  \mapsto   \Ai(x-\lambda) \Ai(y-\lambda).
\end{equation}

\new{Before studying further the applications of the Airy functions to
  Schrödinger operators with linear potentials, let us prove a useful
  fact.}

\begin{lem} \label{lem:Airy}
Let $\chi \in C^\infty_c(\R,[0,1])$ be any cutoff with $\1_{[-1,1]}\leq \chi\leq \1_{[-2,2]}$. For any $k\in\N$ and $x\in\R$, 
\[
\bigg| \frac{1}{2\pi} \int_{\R} e^{\i (t^3/3+xt)} \chi(\epsilon t) \d t  -\Ai(x) \bigg| \le \frac{C_k}{(\epsilon^{-2}+x)_+^{k}} .
\] 
\end{lem}

\begin{preuve}
By integrations by parts, we can rewrite for $x \in\R$,
\[ \begin{aligned}
\Ai(x)  & = \frac{\i}{2\pi} \int_{\R} e^{\i (t^3/3+t)} \partial_t \bigg(\frac{e^{\i (x-1) t}}{t^2+1} \bigg) \d t \\
&=  \frac{1}{2\pi}\int_{\R} e^{\i (t^3/3+xt)} \chi(\epsilon t) \d t  +  \frac{\i}{2\pi} \int_{\R} e^{\i (t^3/3+t)} \partial_t \bigg(\frac{e^{\i (x-1) t}}{t^2+1} \big( 1-\chi(\epsilon t) \big)   \bigg) \d t
\end{aligned}\]
where these integrals converge in the usual sense. 

Let $\mathcal{D} = \partial_t \big(\frac{\cdot}{t^2+x} \big) $
By making repeated integration by parts, it also holds for $x\ge -\epsilon^{-1}$ and for any $k\in\N$, 
\[ \begin{aligned}
\int_{\R} e^{\i (t^3/3+t)} \partial_t \bigg(\frac{e^{\i (x-1) t}}{t^2+1} \big( 1-\chi(\epsilon t) \big)   \bigg) \d t
& =\int_{\R} e^{\i (t^3/3+t)}  \mathcal{D}^k \bigg( e^{\i (1-x) t} \partial_t \bigg(\frac{e^{\i (x-1) t}}{t^2+1} \big( 1-\chi(\epsilon t) \big)   \bigg) \bigg) \d t
\end{aligned}\]
where we verify by induction that 
\[
 \mathcal{D}^k \bigg( e^{\i (1-x) t} \partial_t \bigg(\frac{e^{\i (x-1) t}}{t^2+1} \big( 1-\chi(\epsilon t) \big)   \bigg) \bigg)
 = \begin{cases}
 0 &\text{if }  |t| \le \epsilon^{-1} \\
 \O_k\big( (t^2+1)^{-1}(\epsilon^{-2}+x)^{-k} \big) &\text{if } |t| \ge \epsilon^{-1}
 \end{cases} .
\]
This proves the claim. 
\end{preuve}
 
Note that taking $\epsilon=1$, Lemma~\ref{lem:Airy} and Proposition~\ref{prop:nspl} (applied with $\Phi(t) =t$ and $\hbar = 1/x$) imply that as $x\to+\infty$, 
\begin{equation} \label{Ai_tail}
\Ai(x) = \O(x^{-\infty}).
\end{equation}
In fact, applying the steepest descent method to the integral \eqref{def:Airy}, one obtains the well-known asymptotics $\Ai(x) \sim \frac{e^{-\frac23 x^{3/2}}}{2\sqrt{\pi}x^{1/4}}$ as $x\to+\infty$.

As a direct consequence of \eqref{Ai_tail}, the integral over
$\lambda \in (-\infty, \mu]$ of the \emph{resolvent kernel}
\eqref{eq:measure_kernel_Airy} is convergent for any $\mu\in\R$. 
This implies in particular that the operator $\1_{-\Delta+x \le \mu}$  is locally
trace-class and its (integral) kernel is given by
\[
  K_{\rm Airy,\mu}=\1_{-\Delta+x\leq \mu}:(x,y)\mapsto \int_{-\infty}^\lambda
  \Ai(x+\mu-\lambda)\Ai(y+\mu-\lambda)\dd \lambda.
  \]
see e.g.~\cite[Section~A.2]{bornemann_scaling_2016}. 

\medskip

From this Airy kernel and the bulk kernel \eqref{kernelPi}, one can
formally
construct an integral kernel for the edge operator $\1_{-\Delta+x_1\leq
  0}$ on $\R^n$ in arbitrary dimension $n\ge 2$ as in formula \eqref{eq:Kedge}. Indeed, one can decompose
  $-\Delta+x_1=H_1+H_2$, where
  $H_1=-\partial^2_{x_1}+x_1$ is the one-dimensional Airy operator and $H_2 = -\Delta_{x^\perp}$
  is the Laplacian on the orthogonal hyperplane (we use the coordinates $x=(x_1,x^\perp)$ with $x^\perp=(x_2,\cdots,x_n)$ for $x\in\R^n$ with $n\ge2$). These two operators
  commute, hence the \emph{projection-valued measure} for $-\Delta+x_1$ is a convolution whose kernel can be written as for $\lambda\in\R$, 
  \[
  (x,y) \in \R^{2n}  \mapsto   \int_\R  \frac{\dd}{\dd \mu} K_{\rm
      Airy,\mu}(x_1,y_1) \Big|_{\mu}\frac{\dd}{\dd \mu}K_{{\rm
        b}, \sqrt{\mu}}^{(n-1)}(x^{\perp},y^{\perp})\Big|_{\lambda-\mu}\dd \mu . 
  \]
By \eqref{kernelPi}, this integral can be restricted to $\{\mu\le \lambda\}$ since  $K_{{\rm b},\mu} =0$ for $\mu\le 0$. This shows that this integral converges since 
$ \big| \frac{\dd}{\dd \mu} K_{{\rm
        b},
      \sqrt{\mu}}^{(n-1)}(x^{\perp},y^{\perp})\big|_{\lambda-\mu}
    \big| \le C_{n} \lambda^{\frac{n-2}{2}}$  uniformly for
    $x,y\in\R^n$ and since 
  $\Ai(\mu) = \O(\mu^{-\infty})$ as $\mu\to+\infty$ (cf.~\eqref{Ai_tail}). 
 Then, by Fubini's Theorem, we also obtain
 \[ \begin{aligned}
     K_{\rm edge}^{(n)}=\1_{(-\infty,0]}(-\Delta+x_1):(x,y)\mapsto
    &  \int_{\R_-} \int_{-\infty}^\lambda   \frac{\dd}{\dd \mu} K_{{\rm
        b},\sqrt\mu}^{(n-1)}(x^{\perp},y^{\perp})\Big|_{\lambda-\mu} \Ai(x-\mu) \Ai(y-\mu) \d \mu \d\lambda \\
        & = \int_{\R_+} K_{{\rm
        b},\sqrt\mu}^{(n-1)}(x^{\perp},y^{\perp}) \Ai(x+\mu) \Ai(y+\mu) \d \mu
  \end{aligned} \]
This formula defines a locally trace-class operator, and allows
  us to conjugate explicitly $-\Delta+x_1$ into a multiplication
  operator with a conjugation kernel that is well-defined and
  $L^2$-unitary for functions in Schwartz space. In particular,
  $-\Delta+x_1$ is essentially self-adjoint, and using the explicit formula
\eqref{Picomputation}, this concludes the proof of \eqref{eq:Kedge}.
  
\bibliographystyle{abbrv}


\end{document}